\documentclass[fleqn,usenatbib,trackchanges,linenumbers]{mnras}
\usepackage[T1]{fontenc}
\usepackage{ae,aecompl}
\usepackage{graphicx}	
\usepackage{amsmath}	
\usepackage{amssymb}	
\usepackage{comment}
\usepackage{float}
\usepackage{xcolor,soul}
\usepackage{tablefootnote}
\usepackage[switch]{lineno}
\title[The VVV Open Cluster Project II]{The VVV Open Cluster Project II. Near-infrared sequences of 37 open clusters on eight-dimensional parameter space}

\author[Pe\~na Ram\'irez et al.]{
K. Pe\~na Ram\'irez,$^{1}$\thanks{E-mail: karla.pena@uantof.cl}  
L.C. Smith,$^{2}$  
S. Ram\'irez Alegr\'ia,$^{1}$  
A.-N. Chen\'e,$^{3}$ 
C. Gonz\'alez-Fern\'andez,$^{2}$  \and
P. W. Lucas,$^{4}$ 
and D. Minniti$^{5,6}$
\\
$^{1}$ Centro de Astronom\'ia (CITEVA), Universidad de Antofagasta, Av. Angamos 601, Antofagasta, Chile.\\
$^{2}$ Institute of Astronomy, University of Cambridge, Madingley Road, Cambridge CB3 0HA, UK.\\
$^{3}$ Gemini Observatory/NSF’s NOIRLab, 670 N. A`ohoku Place, Hilo, Hawai`i, 96720, USA.\\
$^{4}$ Centre for Astrophysics Research, University of Hertfordshire, Hatfield  AL10 9AB, UK.\\
$^{5}$ Departamento de Ciencias F\'isicas, Universidad Andr\'es Bello, Rep\'ublica 220, 8320000, Santiago, Chile.\\ 
$^{6}$ Vatican Observatory, V00120 Vatican City State, Italy.\\
}

\date{Accepted 2022 May 4. Received 2022 May 4; in original form 2022 March 23}
\pubyear{2022}

\begin{document}
\label{firstpage}
\pagerange{\pageref{firstpage}--\pageref{lastpage}}
\maketitle

\begin{abstract}
Open clusters are key coeval structures that help us understand star formation, stellar evolution and trace the physical properties of our Galaxy. In the past years, the isolation of open clusters from the field has been heavily alleviated by the access to accurate large-scale stellar parallaxes and proper motions along a determined line of sight. Still, there are limitations regarding their completeness since large-scale studies rely on optical wavelengths. Here we extend the open clusters sequences towards fainter magnitudes complementing the Gaia photometric and astrometric information with near-infrared data from the VVV survey. We performed a homogeneous analysis on 37 open clusters implementing two coarse-to-fine characterization methods: extreme deconvolution Gaussian mixture models coupled with an unsupervised machine learning method on 8-dimensional parameter space. The process allowed us to separate the clusters from the field at near-infrared wavelengths. We report an increase of $\sim$47\% new member candidates on average in our sample (considering only sources with high membership probability p$\geqq$0.9). This study is the second in a series intended to reveal open cluster near-infrared sequences homogeneously. 

\end{abstract}



\section{Introduction}
With the advancement of large-scale photometric and state-of-the-art astrometric instruments, the field of open clusters has been revisited in recent years with significant progress mainly in two aspects: accounting for large cluster samples while ensuring a uniform spatial distribution of the clusters and homogeneity in terms of observations and data processing. Building upon the first large-scale seminal works on open clusters \citep{dias02, kharchenko13} and the Gaia satellite releases (Gaia DR1/DR2/EDR3, \citealt{gaia16,gaia18, gaiaedr3}), several authors have revisited the characterization of open clusters following a plethora of techniques leading to the most complete and accurate cluster membership determination scenario to date \citep[among others]{cantatgaudin18, castroginard18, cantatgaudin19perseus, castroginard19, castroginard20, cantatgaudin20, newcantatgaudin20, olivares19, miretroig19, galli20, tarricq22, dias22}. What is more, recently, \citet{jackson20, jackson21} has built upon the kinematic space, including spectroscopic information from the GES internal data releases 5 and 6, to assign membership probabilities on more than 60 open clusters.

There is, however, a front in which the field is still lacking a proper analysis. Since most of the large-scale studies have been done at optical wavelengths, the near-infrared picture of the open clusters have been analyzed either on specific single cluster studies  \citep[e.g.][]{bonatto06,rangwal19,bisht20_4337,bisht20} or under the scrutiny of specific stellar populations \citep[among others]{mauerhan11,borissova16,borissova18,borissova20,delafuente21}. In the case of studies with a moderate number of clusters \citep{bica05,santossilva12,kharchenko16}, the available near-infrared data  was the Two Micron All-Sky Survey (2MASS, \citealt{skrutskie06}).  In dense fields, such studies lead to source confusion issues. 

The work presented here follows on from \citet[][hereafter PR21]{penaramirez21} where refined open cluster census via Gaia DR2, 2MASS data, and high-quality multi-epoch near-infrared data from the VISTA Variables in the V\'ia L\'actea Survey (VVV, \citealt{vvv}) was combined to provide membership probabilities for six open clusters based on their three-dimensional kinematics and their near-infrared photometry. Here, we refine the methodology and extend the study to a new sample of 37 open clusters on the VVV footprint. 

We present the dataset used for the membership analysis in Section \ref{sec:data}. The methodology implemented to select the cluster members and the derivation of the membership probabilities is covered in Section \ref{sec:method}. Section \ref{sec:discusion} presents our main findings unveiling the open cluster near-infrared sequences of 37 open stellar clusters down to $K_s\,\sim\,14.0$\,mag, and a summary and conclusions are outlined in Section \ref{sec:conclusions}. Appendix A shows a set of plots per each of the clusters studied. 


\section{Observational data}\label{sec:data}
\subsection{Open cluster sample}
We selected a set of 37 clusters out of the list of clusters recovered from the literature and revisited by \citet{cantatgaudin18} and \citet{cantatgaudin20} using an unsupervised membership assignment ensemble (see Section\,\ref{sec:method}). These clusters fall within the VVV footprint, with a known considerable distance and age span, and were located at first instance through the coordinates and distances reported by \citet{cantatgaudin20}. The surveys involved are all-sky, and therefore our spatial coverage within the VVV footprint is homogeneous, and the photometric ones are only constrained by magnitude completeness. Table \ref{table:clusters} presents the cluster names and their fundamental parameters recovered from the literature.

\begin{table*}
\centering
\caption{Literature-based parameters \citep{newcantatgaudin20, kounkel20} for our cluster sample. Columns 2-11 contain the spatial location, proper motion, parallax, age, extinction, and the number of members with a  membership probability above or equal the 90\%.}
\resizebox{\textwidth}{!}{
\begin{tabular}{lcccccccccc}
  \hline
 Name & $\alpha$ & $\delta$ & $\mu_{\alpha}*$ & $\mu_{\delta}$ & $\varpi$ & log(Age/yr)\textsuperscript{a} & log(Age/yr)\textsuperscript{b} & $A_\text{v}$\textsuperscript{a} & $A_\text{v}$\textsuperscript{b} &  Number \\ 
 & [deg] & [deg] & [mas\,yr$^{-1}$] & [mas\,yr$^{-1}$] & [mas] & & & [mag]  & [mag] & p$\geq$0.9  \\ 
  \hline
  \hline
 Alessi\,Teutsch\,8 & 180.649 & -60.935 & -6.625\,$\pm$\,0.136 & 1.630\,$\pm$\,0.136 & 0.967\,$\pm$\,0.049 & 8.07 & 7.95\,$\pm$\,0.12 & 0.49 & 0.67\,$\pm$\,0.10 & 229\\
 ASCC\,88 & 256.886 & -35.564 & 0.919\,$\pm$\,0.164 & -3.282\,$\pm$\,0.142 & 1.097\,$\pm$\,0.060 & 8.39 & 7.88\,$\pm$\,0.22 & 1.45 & 1.71\,$\pm$\,0.15 & 69\\
 BH\,202 & 253.779 & -40.947 & -1.869\,$\pm$\,0.125 & -3.802\,$\pm$\,0.110 & 0.551\,$\pm$\,0.052 & 8.58 & 8.74\,$\pm$\,0.19 & 1.84 & 2.10\,$\pm$\,0.19 & 19\\
 Basel\,18 & 201.967 & -62.306 & -5.008\,$\pm$\,0.080 & -2.050\,$\pm$\,0.096 & 0.508\,$\pm$\,0.057 & 7.44 & 7.72\,$\pm$\,0.18 & 0.65 & 1.25\,$\pm$\,0.19 & 14\\
 Hogg\,21 & 251.414 & -47.747 & -0.937\,$\pm$\,0.132 & -2.206\,$\pm$\,0.102 & 0.366\,$\pm$\,0.060 & 7.95 & 8.03\,$\pm$\,0.37 & 1.54 & 2.16\,$\pm$\,0.29 & 7\\
 Lynga\,6 & 241.218 & -51.960 & -1.866\,$\pm$\,0.144 & -2.757\,$\pm$\,0.098 & 0.383\,$\pm$\,0.050 & 6.49 & 7.71\,$\pm$\,0.34 & 3.51 & 4.08\,$\pm$\,0.26 & 3\\
 Lynga\,9 & 245.170 & -48.523 & -2.559\,$\pm$\,0.177 & -2.415\,$\pm$\,0.124 & 0.359\,$\pm$\,0.073 & 8.80 & 8.72\,$\pm$\,0.26 & 2.50 & 3.36\,$\pm$\,0.33 & 216\\
 NGC\,4103 & 181.628 & -61.245 & -6.184\,$\pm$\,0.117 & 0.075\,$\pm$\,0.107 & 0.473\,$\pm$\,0.053 & 7.32 & 7.68\,$\pm$\,0.15 & 0.85 & 0.95\,$\pm$\,0.12 & 190\\
 NGC\,4349 & 186.048 & -61.866 & -7.827\,$\pm$\,0.159 & -0.296\,$\pm$\,0.119 & 0.490\,$\pm$\,0.036 & 8.50 & 8.49\,$\pm$\,0.09 & 1.01 & 1.18\,$\pm$\,0.12 & 11\\
 NGC\,4463 & 187.466 & -64.800 & -5.326\,$\pm$\,0.090 & -0.435\,$\pm$\,0.094 & 0.523\,$\pm$\,0.041 & 7.46 & 7.76\,$\pm$\,0.15 & 1.53 & 1.43\,$\pm$\,0.12 & 28\\
 NGC\,4609 & 190.582 & -62.995 & -4.870\,$\pm$\,0.127 & -1.045\,$\pm$\,0.146 & 0.660\,$\pm$\,0.053 & 7.91 & 7.72\,$\pm$\,0.23 & 0.88 & 1.20\,$\pm$\,0.13 & 163\\
 NGC\,5269 & 206.147 & -62.907 & -4.490\,$\pm$\,0.078 & -1.918\,$\pm$\,0.075 & 0.460\,$\pm$\,0.036 & 8.12 & 8.41\,$\pm$\,0.15 & 0.99 & 1.50\,$\pm$\,0.12 & 8\\
 NGC\,5316 & 208.516 & -61.883 & -6.297\,$\pm$\,0.099 & -1.515\,$\pm$\,0.119 & 0.665\,$\pm$\,0.055 & 8.22 & 8.28\,$\pm$\,0.19 & 0.92 & 1.08\,$\pm$\,0.22 & 246\\
 NGC\,5381 & 210.205 & -59.578 & -6.034\,$\pm$\,0.116 & -2.932\,$\pm$\,0.122 & 0.379\,$\pm$\,0.052 & 8.55 & 8.59\,$\pm$\,0.25 & 1.62 & 1.90\,$\pm$\,0.24 & 213\\
 NGC\,5606 & 216.946 & -59.632 & -4.880\,$\pm$\,0.082 & -2.897\,$\pm$\,0.103 & 0.374\,$\pm$\,0.040 & 7.26 & 7.47\,$\pm$\,0.14 & 1.69 & 1.82\,$\pm$\,0.09 & 16\\
 NGC\,5715 & 220.859 & -57.578 & -3.487\,$\pm$\,0.123 & -2.324\,$\pm$\,0.114 & 0.435\,$\pm$\,0.045 & 8.73 & 8.66\,$\pm$\,0.19 & 1.68 & 2.15\,$\pm$\,0.17 & 95\\
 NGC\,5925 & 231.847 & -54.515 & -4.319\,$\pm$\,0.163 & -5.142\,$\pm$\,0.150 & 0.679\,$\pm$\,0.050 & 8.69 & 8.47\,$\pm$\,0.15 & 1.34 & 1.64\,$\pm$\,0.13 & 239\\
 NGC\,5999 & 238.046 & -56.482 & -3.389\,$\pm$\,0.094 & -4.216\,$\pm$\,0.097 & 0.326\,$\pm$\,0.039 & 8.30 & 8.54\,$\pm$\,0.12 & 1.40 & 1.50\,$\pm$\,0.15 & 78\\
 NGC\,6134 & 246.953 & -49.161 & 2.184\,$\pm$\,0.211 & -4.483\,$\pm$\,0.184 & 0.846\,$\pm$\,0.062 & 8.99 & 8.91\,$\pm$\,0.15 & 0.87 & 1.45\,$\pm$\,0.18 & 664\\
 NGC\,6192 & 250.077 & -43.355 & 1.653\,$\pm$\,0.190 & -0.188\,$\pm$\,0.138 & 0.571\,$\pm$\,0.058 & 8.38 & 8.34\,$\pm$\,0.15 & 1.57 & 2.06\,$\pm$\,0.14 & 390\\
 NGC\,6204 & 251.538 & -47.027 & -0.690\,$\pm$\,0.189 & -0.596\,$\pm$\,0.182 & 0.805\,$\pm$\,0.064 & 7.97 & 7.95\,$\pm$\,0.19 & 1.13 & 1.51\,$\pm$\,0.15 & 113\\
 NGC\,6268 & 255.524 & -39.721 & 0.908\,$\pm$\,0.125 & -0.781\,$\pm$\,0.119 & 0.634\,$\pm$\,0.073 & 8.30 & 8.47\,$\pm$\,0.10 & 1.19 & 1.56\,$\pm$\,0.11 & 34\\
 NGC\,6568 & 273.192 & -21.612 & 0.593\,$\pm$\,0.144 & -1.377\,$\pm$\,0.131 & 0.936\,$\pm$\,0.043 & 8.94 & 8.72\,$\pm$\,0.10 & 0.67 & 1.08\,$\pm$\,0.13 & 60\\
 NGC\,6583 & 273.962 & -22.143 & 1.303\,$\pm$\,0.103 & 0.110\,$\pm$\,0.097 & 0.413\,$\pm$\,0.045 & 9.08 & 8.95\,$\pm$\,0.14 & 1.52 & 2.37\,$\pm$\,0.32 & 101\\
 Pismis\,19 & 217.666 & -60.889 & -5.460\,$\pm$\,0.142 & -3.247\,$\pm$\,0.210 & 0.255\,$\pm$\,0.086 & 8.92 & 8.94\,$\pm$\,0.21 & 3.66 & 3.84\,$\pm$\,0.26 & 296\\
 Ruprecht\,121 & 250.436 & -46.159 & -1.073\,$\pm$\,0.137 & -2.569\,$\pm$\,0.116 & 0.462\,$\pm$\,0.041 & 8.25 & 8.51\,$\pm$\,0.30 & 2.54 & 2.39\,$\pm$\,0.26 & 138\\
 Ruprecht\,128 & 266.063 & -34.879 & 1.889\,$\pm$\,0.169 & -1.262\,$\pm$\,0.137 & 0.498\,$\pm$\,0.070 & 8.98 & 9.01\,$\pm$\,0.16 & 1.92 & 3.82\,$\pm$\,0.41 & 83\\
 Ruprecht\,130 & 266.900 & -30.098 & 0.477\,$\pm$\,0.116 & -1.795\,$\pm$\,0.102 & 0.389\,$\pm$\,0.041 & 8.60 & 8.77\,$\pm$\,0.28 & 2.76 & 2.83\,$\pm$\,0.26 & 20\\
 Ruprecht\,134 & 268.184 & -29.537 & -1.653\,$\pm$\,0.085 & -2.432\,$\pm$\,0.100 & 0.391\,$\pm$\,0.057 & 9.22 & 8.28\,$\pm$\,0.18 & 1.15 & 2.79\,$\pm$\,0.40 & 10\\
 Teutsch\,84 & 256.090 & -42.070 & -1.738\,$\pm$\,0.246 & -1.109\,$\pm$\,0.244 & 0.406\,$\pm$\,0.160 & 9.02 & 8.83\,$\pm$\,0.21 & 3.48 & 3.88\,$\pm$\,0.21 & 108\\
 Trumpler\,25 & 261.125 & -39.006 & 0.358\,$\pm$\,0.150 & -2.120\,$\pm$\,0.132 & 0.402\,$\pm$\,0.059 & 8.36 & 8.65\,$\pm$\,0.23 & 2.47 & 2.75\,$\pm$\,0.30 & 284\\
 Trumpler\,26 & 262.126 & -29.487 & -0.873\,$\pm$\,0.129 & -3.082\,$\pm$\,0.095 & 0.623\,$\pm$\,0.064 & 8.40 & 8.33\,$\pm$\,0.15 & 1.78 & 2.24\,$\pm$\,0.20 & 16\\
 Trumpler\,29 & 265.347 & -40.158 & 0.489\,$\pm$\,0.122 & -2.308\,$\pm$\,0.101 & 0.673\,$\pm$\,0.063 & 7.71 & 7.93\,$\pm$\,0.16 & 0.72 & 0.92\,$\pm$\,0.15 & 51\\
 Trumpler\,30 & 269.182 & -35.298 & 1.243\,$\pm$\,0.131 & -2.189\,$\pm$\,0.117 & 0.707\,$\pm$\,0.071 & 8.64 & 8.37\,$\pm$\,0.10 & 0.81 & 1.25\,$\pm$\,0.13 & 66\\
 UFMG\,1 & 236.593 & -56.792 & -1.630\,$\pm$\,0.089 & -3.196\,$\pm$\,0.110 & 0.360\,$\pm$\,0.054 & 8.43 & $\cdots$  & 2.38 & $\cdots$ &  2\\
 UFMG\,2 & 237.585 & -55.961 & -4.425\,$\pm$\,0.143 & -3.048\,$\pm$\,0.125 & 0.357\,$\pm$\,0.082 & 9.11 & $\cdots$  & 2.86 & $\cdots$ & 132\\
 UFMG\,3 & 238.115 & -55.419 & -1.037\,$\pm$\,0.118 & -2.225\,$\pm$\,0.120 & 0.461\,$\pm$\,0.061 & 8.29 & $\cdots$ & 2.25 & $\cdots$ & 23\\
   \hline
\multicolumn{11}{l}{\footnotesize{\textsuperscript{a} Values from \citet{cantatgaudin20}. \textsuperscript{b} Values from \citet{kounkel20}.}}
\end{tabular}
}
\label{table:clusters}
\end{table*}


\subsection{Multi-dimensional data sets}
We kept our original purpose as outlined in PR21: to identify each open cluster sequence with the most extensive dynamical range possible in the near-infrared. We have access to the brightest sources using 2MASS photometry in the $J$, $H$ and $K_s$ bandpasses. For fainter targets we exploited the dynamic range of the VVV survey in the $J$, $H$ and $K_s$ bandpasses. 

The VVV survey, and its temporal and spatial extension the VVVX, are ESO Public Surveys targeting the inner disc and the bulge of the Milky Way. The VVV and VVVX utilise the VISTA telescope and its main imager, the VISTA InfraRed CAMera (VIRCAM, \citealt{vista_2}), to obtain multi-epoch, infrared photometry. Observations taken with VISTA are reduced at the Cambridge Astronomy Survey Unit (CASU; \url{http://casu.ast.cam.ac.uk/}) as part of the VISTA Data Flow System (VDFS; \citealt{irwin04}). The latest data release uses v1.5 of their pipeline. At the bright end the VVV survey manages $\sim$[11.0, 11.0, 11.5, 12.0, 11.0]\,mag in [$Z$, $Y$, $J$, $H$, $K_s$], at which point detector non-linearity brings photometric errors over 0.1\,mag. At the faint end it reaches $\sim$[18.5, 18.0, 17.5, 16.5, 16.0]\,mag in those same bandpasses, at which point sky noise is the limiting factor for a typical observing sequence.
More details about the performance and photometric properties of the data can be obtained in \citet{gonzalez18}.

Regarding astrometric measurements, we also combined data from two surveys. For the brightest sources, the near-infrared photometric data sets were combined with astrometric information using a pre-computed nearest-neighbor, proper motion aware, cross-match between 2MASS and the latest release of Gaia\footnote{Using the Q3C software: \citet{Q3C}.} to retrieve  $J$, $H$, and $K_s$ magnitudes from 2MASS and the 5-parameter astrometric solution from Gaia EDR3 (right ascension, declination, proper motions in right ascension and declination, and parallaxes). For the faintest sources, we relied on preliminary data from the VVV Infrared Astrometric Catalogue version 2 (VIRAC2; Smith et al. in prep; see \citealt{smith18} for details of v1). VIRAC2 used PSF fitting to obtain astrometry and photometry of sources in VVV observations and VVVX observations of the original VVV area. The preliminary VIRAC2 data used in this work was calibrated astrometrically against Gaia DR2 (taking into account Gaia proper motions and parallaxes), and photometrically against the catalogues of \citet{alonsogarcia18}. Individual detections were mapped to unique stars using a bespoke pipeline, which involved an iterative process of 5-parameter astrometric solution fitting and re-mapping until convergence.

As in PR21, our final working product is an astronomical dataset in eight dimensions (five astrometric parameters and three photometric ones, $J$, $H$, and $K_s$) for each cluster covering an area of five times their published \textit{r50} value\footnote{The $r50$ value from \citet{cantatgaudin20} is the radius (in degrees) from the cluster center that encompasses 50\% of the members identified by the authors.} In the 2MASS/VVV magnitude overlap, sources present in both catalogs are combined photometrically with an optimal inverse variance weighting. In the VVV/Gaia overlap the astrometric parameters of duplicated stars were also combined using an inverse variance weighted average. Since the magnitude range is broad, there is a significant difference in precision between the sources, from the brightest to the faintest. On the bright end, the nominal uncertainty reaches the Gaia precisions of 0.02--0.03\,mas (yr$^{-1}$) in parallax and proper motions \citep{gaiaedr3}, while at $K_s\sim15.0$\,mag, the uncertainties are under $\sim$1.0\,mas (yr$^{-1}$) in parallax and proper motions.


\section{Methodology}\label{sec:method}
The analysis used to determine membership probability here is similar to that described in PR21; however, the cluster tagging and their sequence determination have changed in some key aspects to tackle a broader mix of cluster ages, distances, and shapes. In our previous work, we applied the Gaussian Mixture Model \citep[GMM; e.g.][]{everitt11} technique to isolate the clusters from the field as a first instance. This technique is based on the assumption that the star's distribution within an overdensity can be described by a superposition of multivariate Gaussian distributions \citep{desouza17}. Nevertheless, the technique does not account for the uncertainties on the parameter space at work. As noted by \citet{jaehnig21} it is ideal for a GMM to fit an intrinsic, underlying distribution rather than noisy discrete data. In that line, we implemented the method developed by \citet{bovy11}
in which the observed measurements are deconvolved from their uncertainties and fitting the intrinsic distribution with a gaussian mixture model. The Extreme Deconvolution Gaussian Mixture Model (XDGMM) has been implemented successfully for the identification and characterization of specific open clusters \citep{olivares19, pricewhelan19, jaehnig21}, globular clusters \citep{vasiliev21} and even supernova and host galaxy populations \citep{holoien17}. 

The clusters were initially identified purely from their published coordinates, proper motions, and parallaxes as on \citet{cantatgaudin20}. All the subsequent analyses covered the entire area of study (5$\times$ $r50$) per cluster. Once spatially identified each cluster ($\alpha$, $\delta$ parameter space), the XDGMM was implemented in the 3-dimensional parameter space of proper motions and parallaxes ($\mu_{\alpha}*$, $\mu_{\delta}$, $\varpi$). In the pre-processing stage, we removed clear outlier sources in the proper motion space by discarding sources with $\mu_{\alpha}*$ or $\mu_{\delta}$ measurements more than 10$\sigma$ from the cluster proper motion. The central parallax values and their uncertainties were recorded for each cluster. At this stage, we also discard the sources with photometric uncertainties above the $0.5$ magnitudes in $J$ and $K_s$ to remove obvious spurious sources. We used the \textsc{astroML} \citep{astroML} and \citet{bovy11} implementations developed by \citep{holoien17}\footnote{\url{https://github.com/tholoien/XDGMM}}. XDGMM takes into account the full covariance matrix representation on $\mu_{\alpha}*$, $\mu_{\delta}$, and $\varpi$ where we considered ten Gaussian components (and their respective covariance matrices) to describe the above-mentioned parameter space. This approach gives us results insensitive to the exact number of Gaussians, as long as it is large enough to disentangle the background from the members. 

For selecting the gaussian component that traces the open cluster, we followed the prescription given in \citet{jaehnig21} to obtain the differential entropy information metric \citep{ahmed89}. It is related to the components covariance as in PR21 but extended to account for the dimension of the parameter space, therefore giving us a measure of how compact a distribution is within a volume. The component with the lowest differential entropy, i.e., the most compact Gaussian component, was selected as the cluster component. XDGMM resampling calculates the individual membership probabilities via bootstrap. We recompute component assignments for each star for a total of 200 iterations. Only sources with membership probability (p$\geq0.9$) to the selected cluster component were considered to our refinement stage. 

At this point, the coherent motion of the conglomerates is assured since the input data was already clustered in the 3-dimensional proper motion-parallax parameter space ($\mu_{\alpha}*$, $\mu_{\delta}$, $\varpi$). Therefore, our refinement stage was focused on group stars according to their spatial and photometric distribution using the $J$, $H$, $K_s$ photometry, and the spatial positions. To do so, we employed the Unsupervised Photometric Membership Assignment in Stellar Clusters \textsc{UPMASK} \citep{upmask14} on each cluster component. This approach has been successfully applied to identify and characterize open clusters \citep{cantatgaudin18, cantatgaudin19perseus, penaramirez21} and also accounts for the uncertainties in the explored parameter space. We used the \textsc{k-means} clustering \citep{kmeans} partition algorithm with a large k to the dataset size guaranteeing at least 25 objects per group. For testing whether the distribution of stars within each group is more concentrated than what is expected for a random fluctuation in a uniform distribution, we use the total length of a minimum spanning tree \cite[e.g.][]{mst} and we iterate 100 times per cluster field. At each iteration, the photometry data were randomly sampled from the probability distribution function of each star's positional parameters while taking into account uncertainties for each variable. The clustering score (i.e., membership probabilities) for a given source is derived directly from the number of iterations during which it was a member of a concentrated group and can also be interpreted as a membership probability. 

Our final membership probabilities are then based on the 8-dimensional parameter space ($\alpha$, $\delta$, $\mu_{\alpha}*$, $\mu_{\delta}$, $\varpi$, $J$, $H$, $K_s$) information of each star and their associated nominal uncertainties. The figures in Appendix\,\ref{sec:anex} show the near-infrared sequence for all the explored clusters, together with their spatial distributions and proper motions. The full membership list per cluster is available in electronic form (see Section \ref{sec:catalogs}).



\section{Discusion}\label{sec:discusion}
\subsection{Cluster parameters}
As pointed out in PR21, data-driven and unbiased membership probabilities are required to derive a reliable list of cluster members and pursue further studies. High precision characterization of star cluster's fundamental parameters, such as age, distance, reddening, and total mass, depends on the quality of the cluster membership determination. Therefore we need datasets that are as diverse as they are accurate. The decontamination of background interlopers must ensure that no additional biases are imposed and that whatever cluster members remain are robust and representative of the cluster itself.

Using only sources with a membership probability larger than or equal to 90\%, we have re-derived the median spatial position, proper motion, and parallax for each cluster. We used the mean absolute deviation (from the median), or MAD values \citep{feigelson12}, to estimate the dispersion of those parameters. The offset between our recalculated central positions and those from literature are minimal and range from -0.04 to 0.04\,arcmin. We obtained average proper motion values that agree within the uncertainties with the values published in the literature. The same applies to the parallax values. 

\begin{figure}
\includegraphics[scale=0.1]{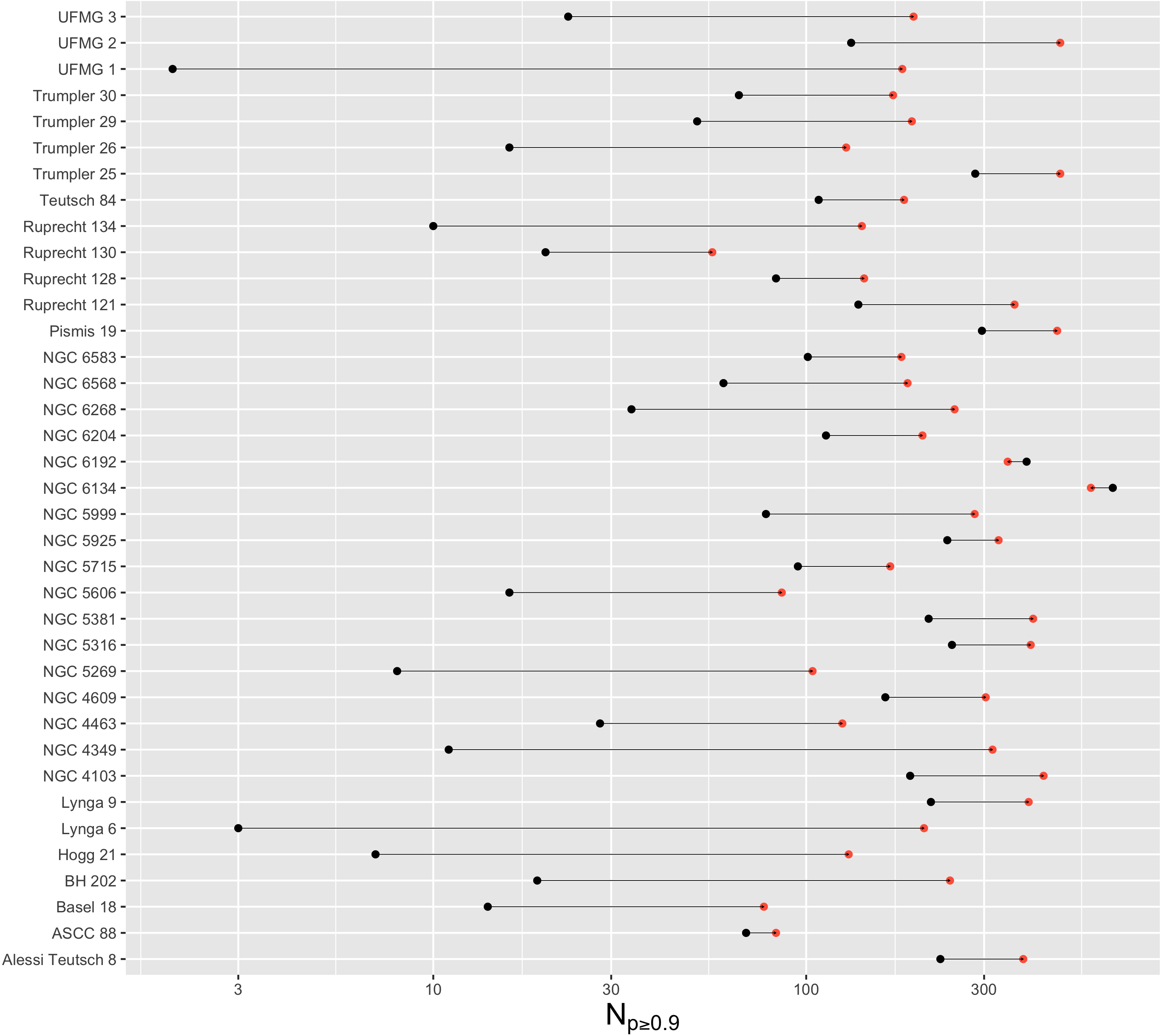}
\caption{Number of high probability members (membership probabilities $p\geq$90\%) per studied cluster in reverse alphabetical order. The black points represent the number density of members from the literature \citep{cantatgaudin20}. The red points represent the number densities we are reporting here. Horizontal axis in logarithmic scale.
}
\label{fig:gain}
\end{figure}

\begin{figure*}
\includegraphics[scale=0.08]{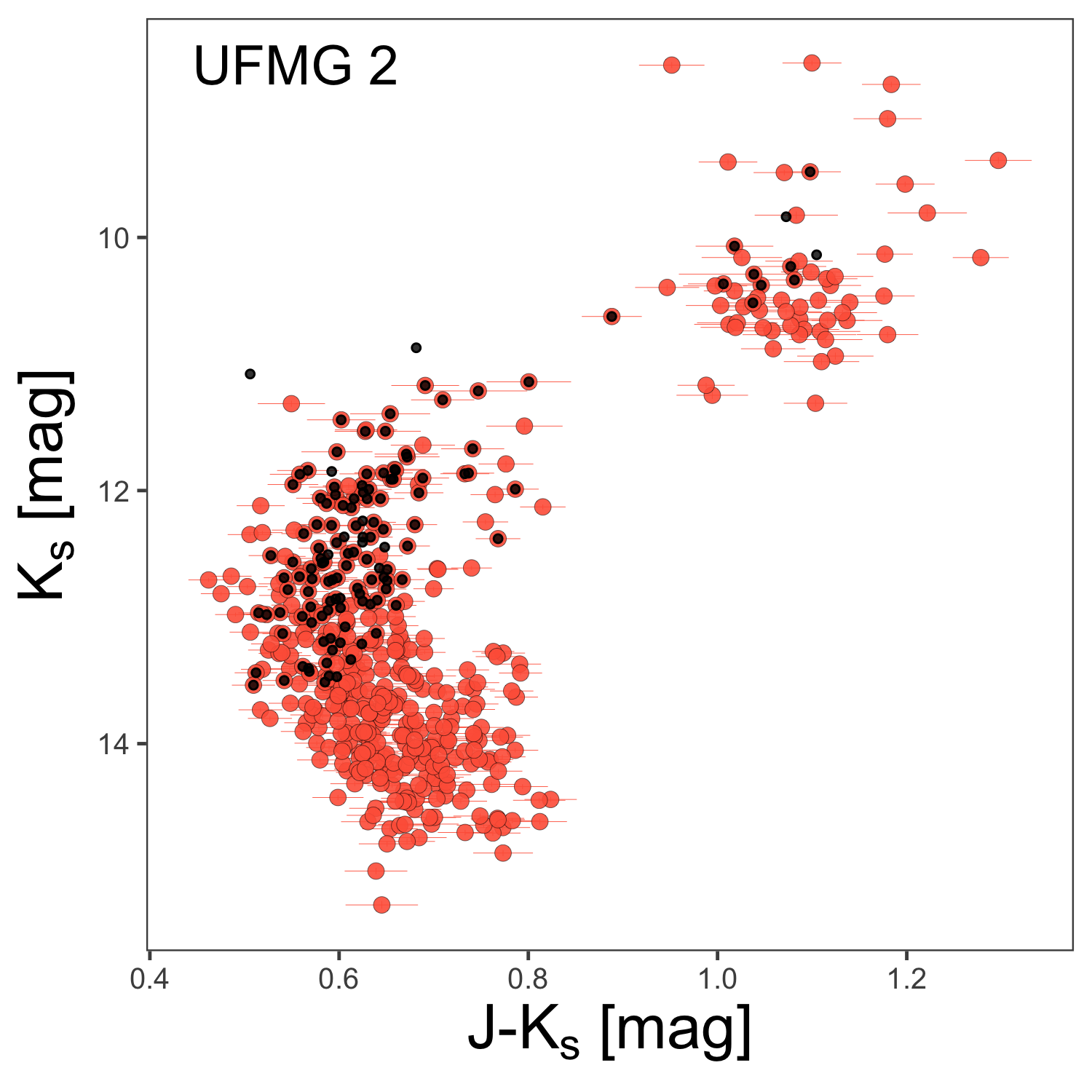}
\includegraphics[scale=0.08]{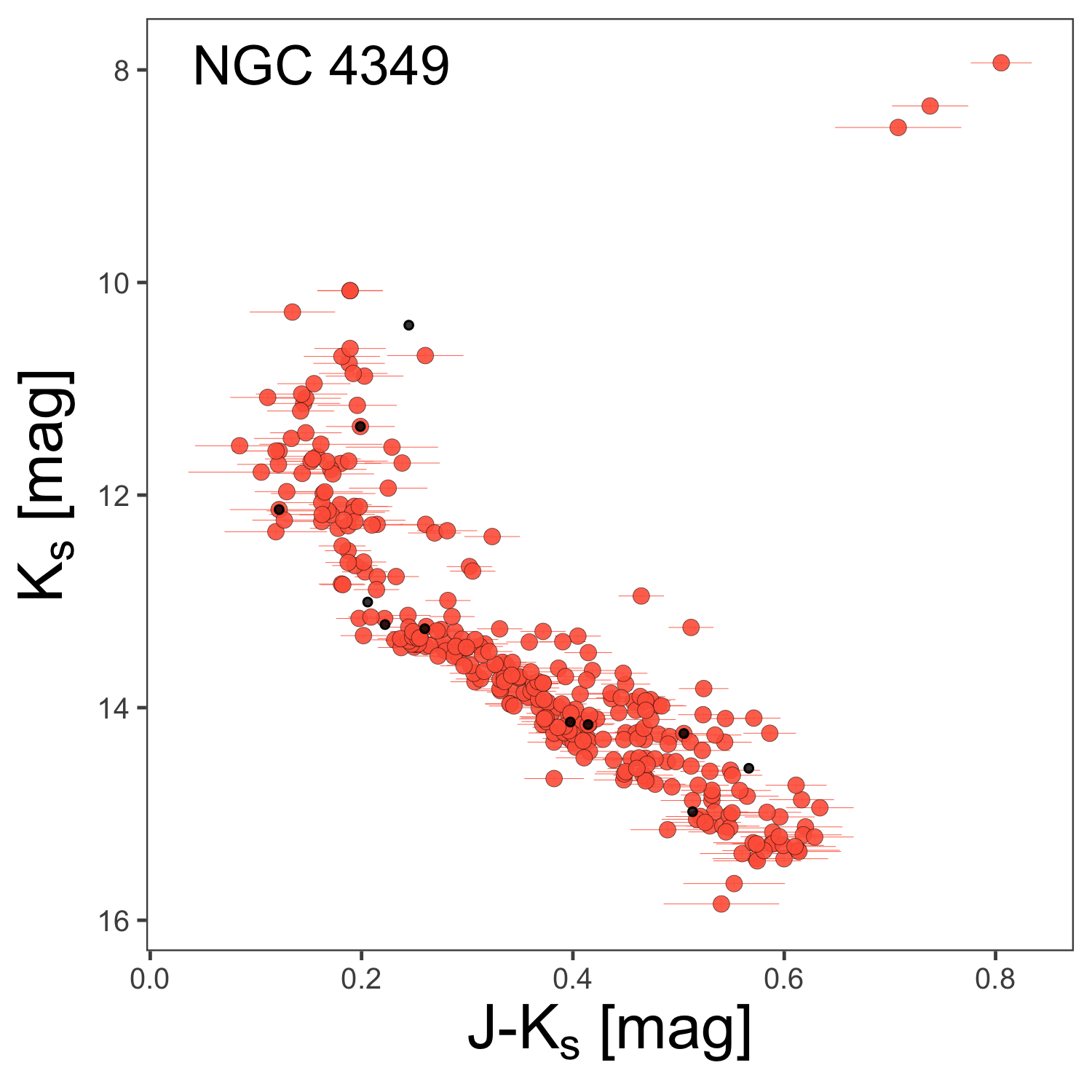}
\includegraphics[scale=0.08]{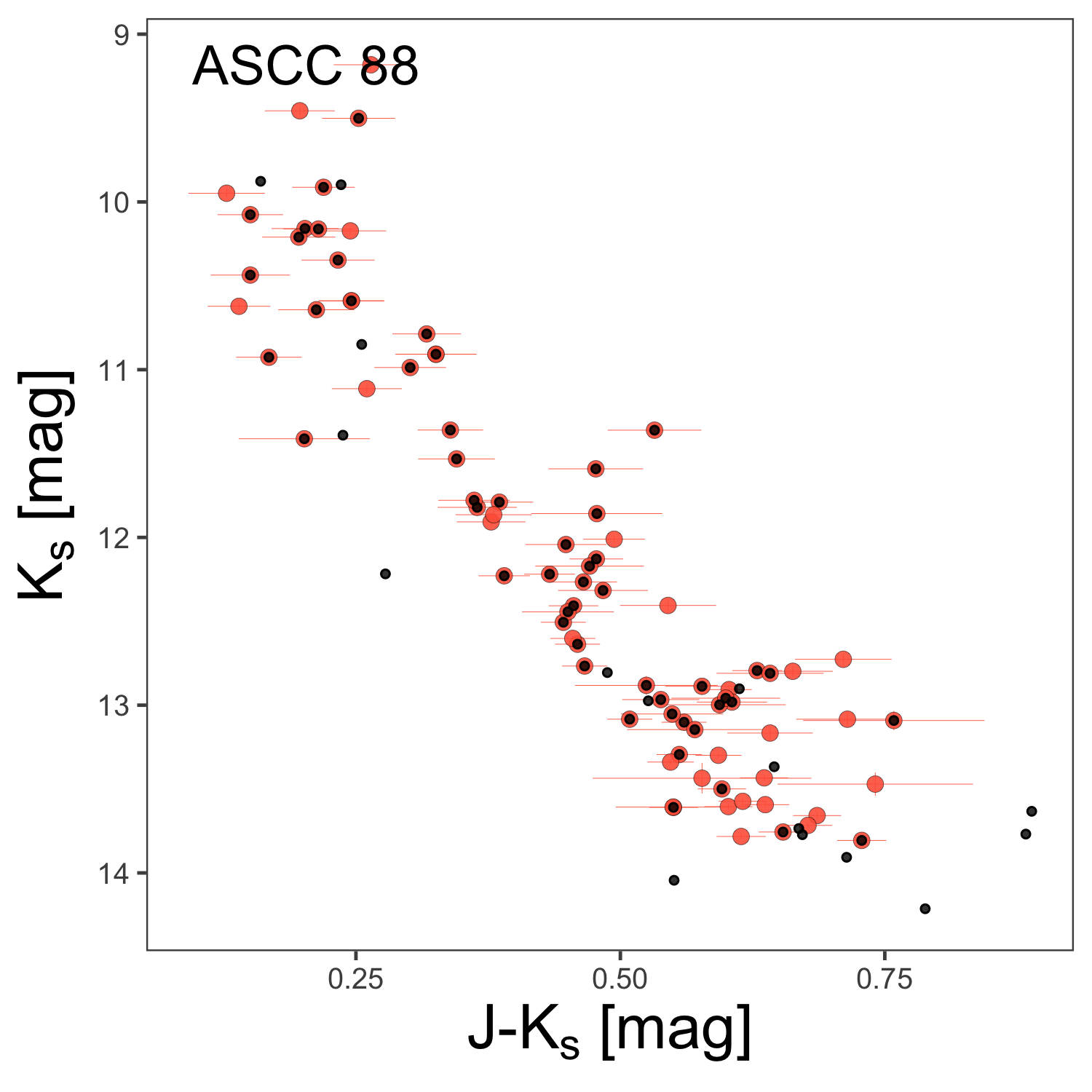}
\includegraphics[scale=0.08]{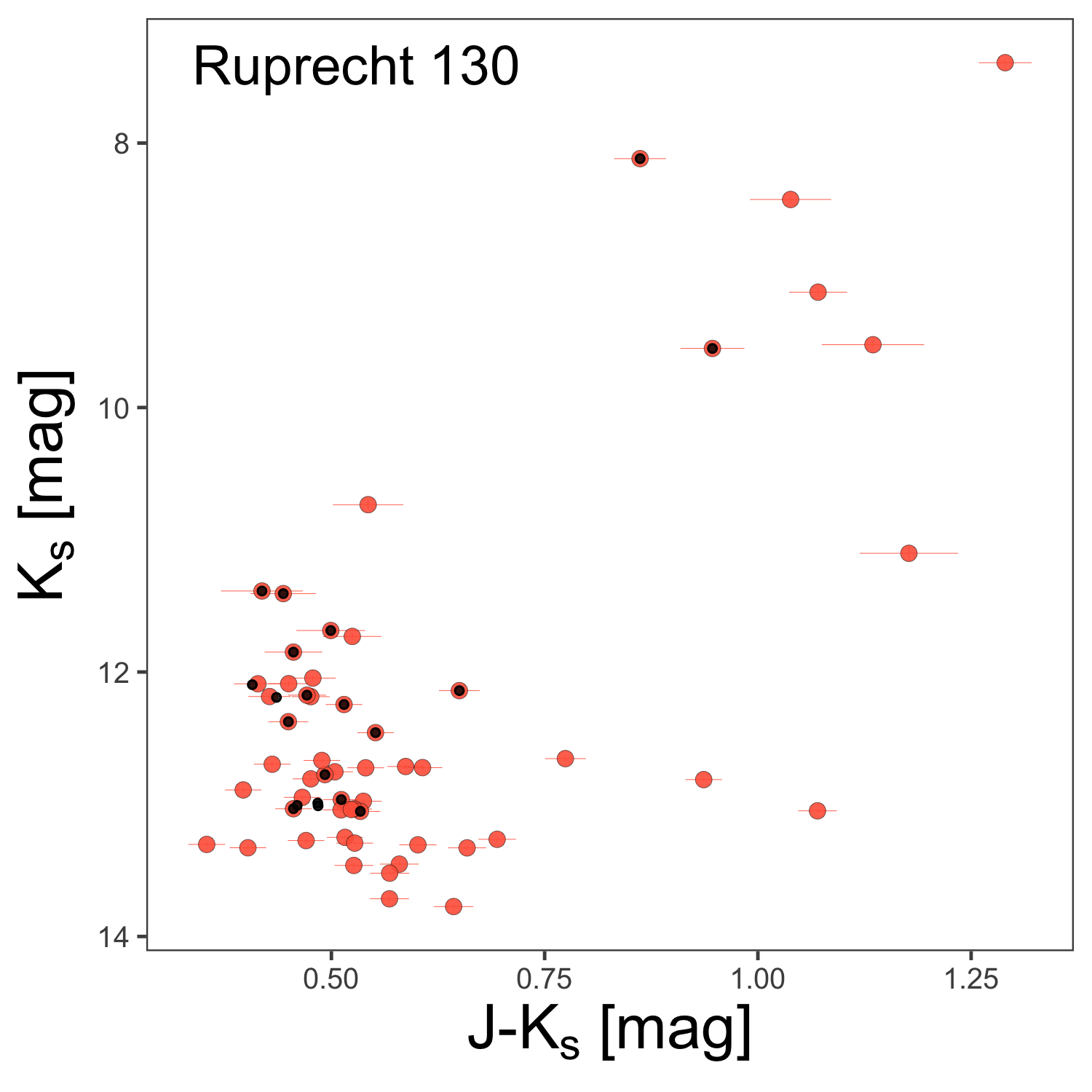}
\caption{$K_s$ vs. $J-K_s$ color-magnitude diagrams of the selected open stellar clusters. UFGM\,2 and NGC\,4349 are the clusters most highly impacted by this study (largest number of new highly probable members). The opposite for ASCC\,88 and Ruprecht\,130.  Black dots correspond to the \citet{cantatgaudin20} members with reported probability $p\geq$90\%. whereas the high probability members presented here are represented with red circles. Text on the plots shows the cluster name.
}
\label{fig:gainexamples}
\end{figure*}

Due to the levity of interstellar extinction in the near-infrared, we can in this study map the cluster sequences better than the previous work based on optical data. From the 37 clusters analyzed, the literature reports 4435 high probability members. Here we surpass that number, reaching a total of 9357 high probability members. Therefore, we report an increase of $\sim$47\% new candidate members on average in our sample. This number is in total agreement with the increase of new member candidates reported in PR21. The improvement in quality of the near-infrared sequences of the clusters is evident compared with the available near-infrared pass-bands of the 2MASS survey. Figure\,\ref{fig:gain} shows the difference in the number of high probability members on the studied clusters. As can be seen, most of the clusters show an increase in their populations except for two clusters (NGC\,6134 and NGC\,6192, where we identify less sources at intermediate magnitudes but increasing the number of highly probable members towards fainter magnitudes). It is important to note that the 90\% threshold compared with literature members relies on fairly similar methodological procedures. We present in Figure\,\ref{fig:gainexamples} the color-magnitude diagrams of the clusters with the highest and lowest change in number density of high probability members. For the most impacted clusters, we report 348 new high probability members for UFGM\,2, whereas for NGC\,4349 at near-infrared wavelengths, we change from 11 highly probable members to 316. On the contrary, for ASCC\,88, we report an addition of only 11 high probability members, and for Ruprecht\,130, the addition of 36 new high probability members on a very diluted cluster sequence. All the cluster sequences are available on Appendix\,\ref{sec:anex}.

For each cluster, we compare the absolute sequences in the VISTA photometric system to the models of \textsc{PARSEC-COLIBRI} \citep{marigo17}. We selected the model corresponding to the age derived by \citet{newcantatgaudin20}. We applied the distance modulus and the reddening vector based on the distance and the median extinction $A_V$ from the same study. The reddening vector was transformed to $A_{K_s}$ using $A_{K_s}/A_V=0.11802$\footnote{\url{http://stev.oapd.inaf.it/cgi-bin/cmd}}), assuming the total-to-selective extinction ratios $R_J$ and $R_{K_s}$ of \citet{gonzalez18} for VISTA data. Similarly, we have placed the model tied to the cluster parameters reported by \citep{kounkel20}. 

Our reference studies, \citet{newcantatgaudin20} and \citet{kounkel20} used artificial neural networks to derive ages, distances and extinctions per cluster. This technique maps the input observables to the
target output quantities (age, extinction, and distance modulus/log d) through a series of nodes. The nodes are arranged in layers that communicate among themselves through non-linear functions. The mentioned studies differ in the artificial neural network architecture, i.e., the way the nodes and the layers are designed refers to the neural network architecture. In the case of \citet{newcantatgaudin20} they used a rectified linear unit (ReLU, see their Figure 2), whereas the work of \citep[see their Appendix A]{kounkel20} uses a convolutional artificial neural network based on the MNIST model \citep{hoffer18}. Their trained model (Auriga\footnote{\url{https://github.com/mkounkel/Auriga}}) is publicly available and considers inputs that were scattered by the errors as statistically comparable. Each realization from the neural network on the same dataset is independent \citep{olney20}, making it possible to measure scatter between them; therefore, it reports uncertainties on the output parameters. 

\begin{figure}
\includegraphics[scale=0.09]{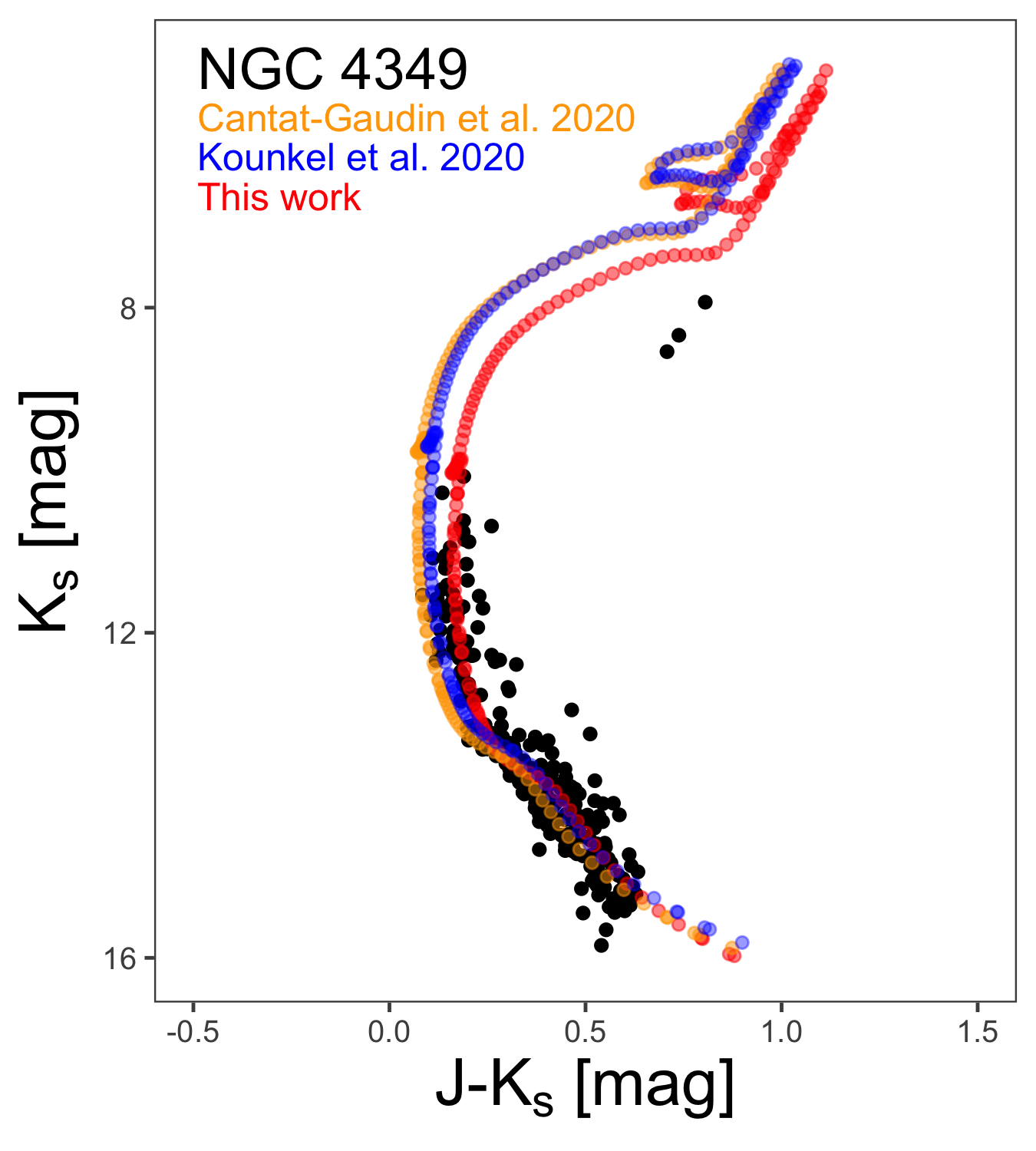}
\includegraphics[scale=0.09]{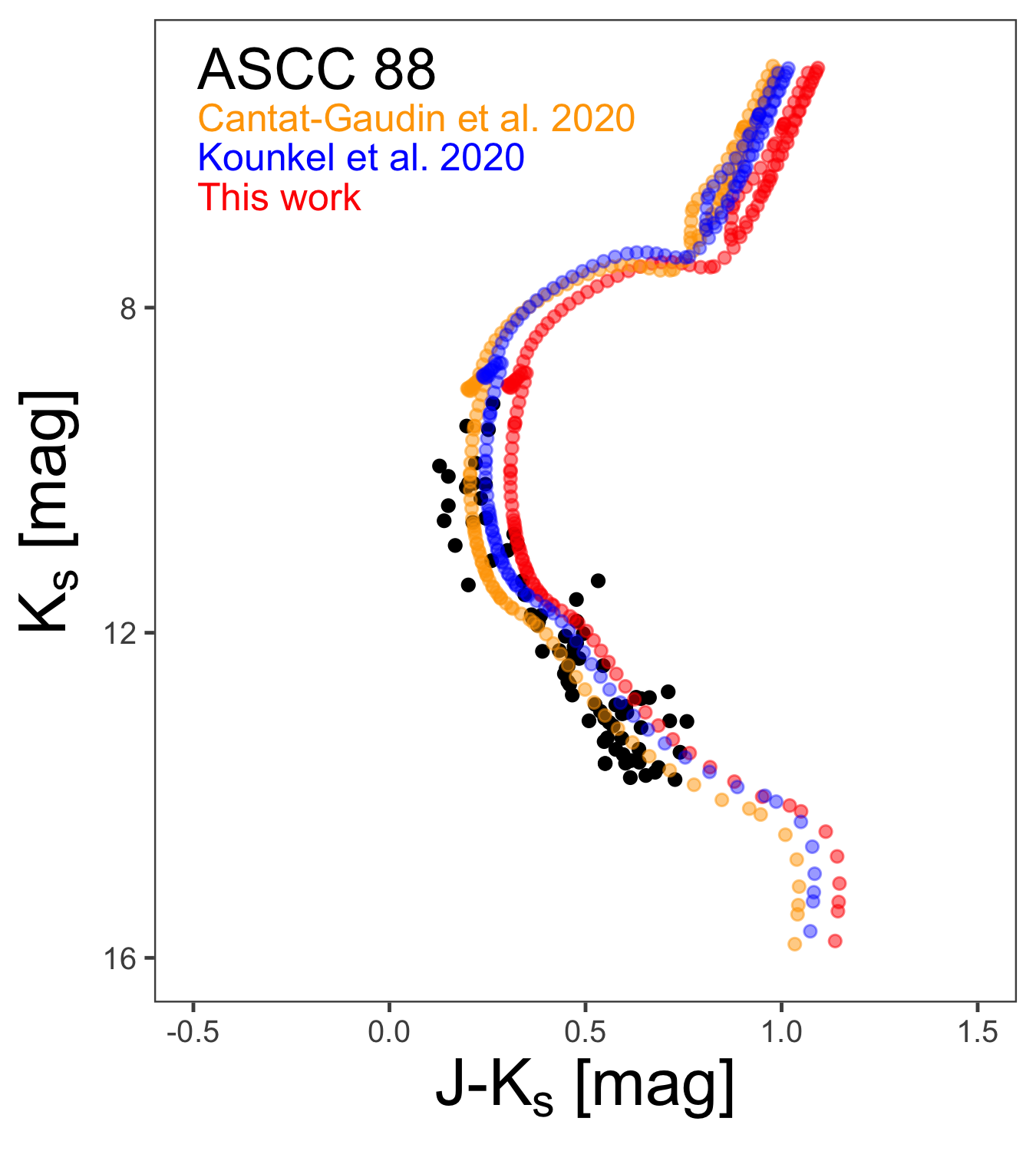}
\caption{$K_s$ vs. $J$-$K_s$ color-magnitude diagrams for NGC\,4349 and ASCC\,88. The theoretical isochrones are located based on the cluster parameters from \citet{newcantatgaudin20, kounkel20} and this work over our catalog members as black points.
}
\label{fig:iso}
\end{figure}

The resulting isochrones (plotted in orange and blue in the figures of Section\,\ref{sec:anex}) are not always well aligned with our observed sequences. Therefore, we run the Auriga artificial neural network on our highly probable members per cluster. One hundred iterations of each cluster were passed through the Auriga model to generate the parameter errors. We also present the isochrones using the updated Auriga distance and age determinations and adjusting the reddening. The result is presented as red isochrones in Figure\,\ref{fig:iso} for NGC\,4349 and ASCC\,88 as an example, the complete figure set can be found in Section\,\ref{sec:anex}. The derived age, distance, reddening and extinction values are presented in Table\,\ref{table:values}.

 
\begin{table*}
\centering
\caption{Derived parameters of our cluster sample. Columns 2-10 contain the spatial location, distance, extinction, age, reddening, mass and the number of members with a  membership probability above or equal the 90\% derived in this study.}
\resizebox{\textwidth}{!}{%
\begin{tabular}{lcccccccccc}
  \hline
Name & $\alpha$ & $\delta$ & d & $A_\text{v}$ & log(Age/yr) & E($J-K_s$) & $A_{K_s}$ & M &  Number \\ 
     & [deg]    & [deg]    & [pc]& [mag]      &             & [mag]      & [mag]     & [M$_\odot$] & p$\geq$0.9\\ 
\hline
\hline
Alessi\,Teutsch\,8 & 180.650 & -60.93 & 999\,$\pm$\,19 & 0.67\,$\pm$\,0.04 & 8.51\,$\pm$\,.08 & 0.10 & 0.08 & 453\,$\pm$\,14 & 399 \\ 
  ASCC\,88 & 256.900 & -35.58 & 950\,$\pm$\,27 & 2.13\,$\pm$\,0.07 & 8.60\,$\pm$\,0.12 & 0.32 & 0.26 & 123\,$\pm$\,3 & 84 \\ 
  BH\,202 & 253.792 & -40.94 & 1877\,$\pm$\,53 & 2.29\,$\pm$\,0.07 & 8.31\,$\pm$\,0.10 & 0.35 & 0.28 & 489\,$\pm$\,96 & 336 \\ 
  Basel\,18 & 201.979 & -62.33 & 2160\,$\pm$\,96 & 1.31\,$\pm$\,0.06 & 7.16\,$\pm$\,0.12 & 0.20 & 0.16 & 170\,$\pm$\,17 & 77 \\ 
  Hogg\,21 & 251.380 & -47.73 & 2837\,$\pm$\,104 & 2.13\,$\pm$\,0.05 & 7.40\,$\pm$\,0.11 & 0.32 & 0.26 & 367\,$\pm$\,231 & 168 \\ 
  Lynga\,6 & 241.220 & -51.96 & 2569\,$\pm$\,127 & 4.42\,$\pm$\,0.11 & 7.39\,$\pm$\,0.10 & 0.67 & 0.53 & 587\,$\pm$\,118 & 238 \\ 
  Lynga\,9 & 245.173 & -48.52 & 2394\,$\pm$\,87 & 3.26\,$\pm$\,0.08 & 8.51\,$\pm$\,0.10 & 0.50 & 0.39 & 799\,$\pm$\,25 & 429 \\ 
  NGC\,4103 & 181.634 & -61.24 & 2051\,$\pm$\,92 & 1.20\,$\pm$\,0.05 & 7.72\,$\pm$\,0.11 & 0.18 & 0.14 & 893\,$\pm$\,16 & 438 \\ 
  NGC\,4349 & 186.091 & -61.86 & 2234\,$\pm$\,45 & 1.59\,$\pm$\,0.08 & 8.11\,$\pm$\,0.11 & 0.24 & 0.19 & 636\,$\pm$\,56 & 395 \\ 
  NGC\,4463 & 187.473 & -64.80 & 2064\,$\pm$\,48 & 1.85\,$\pm$\,0.07 & 7.84\,$\pm$\,0.14 & 0.28 & 0.22 & 254\,$\pm$\,8 & 127 \\ 
  NGC\,4609 & 190.554 & -62.99 & 1512\,$\pm$\,33 & 1.32\,$\pm$\,.06 & 7.96\,$\pm$\,0.08 & 0.20 & 0.16 & 466\,$\pm$\,6 & 306 \\ 
  NGC\,5269 & 206.146 & -62.91 & 2106\,$\pm$\,66 & 1.72\,$\pm$\,0.07 & 7.74\,$\pm$\,0.14 & 0.26 & 0.21 & 211\,$\pm$\,7 & 105 \\ 
  NGC\,5316 & 208.508 & -61.88 & 1507\,$\pm$\,32 & 1.04\,$\pm$\,0.04 & 8.23\,$\pm$\,0.05 & 0.16 & 0.12 & 603\,$\pm$\,32 & 429 \\ 
  NGC\,5381 & 210.209 & -59.58 & 2777\,$\pm$\,82 & 2.03\,$\pm$\,0.09 & 8.20\,$\pm$\,0.13 & 0.31 & 0.24 & 876\,$\pm$\,58 & 453 \\ 
  NGC\,5606 & 216.932 & -59.64 & 2717\,$\pm$\,122 & 2.14\,$\pm$\,0.10 & 6.85\,$\pm$\,0.12 & 0.33 & 0.26 & 282\,$\pm$\,38 & 94 \\ 
  NGC\,5715 & 220.874 & -57.57 & 2279\,$\pm$\,54 & 2.58\,$\pm$\,0.07 & 8.17\,$\pm$\,0.12 & 0.39 & 0.31 & 349\,$\pm$\,11 & 173 \\ 
  NGC\,5925 & 231.806 & -54.52 & 1470\,$\pm$\,27 & 1.89\,$\pm$\,0.07 & 8.40\,$\pm$\,0.10 & 0.29 & 0.23 & 517\,$\pm$\,19 & 344 \\ 
  NGC\,5999 & 238.058 & -56.49 & 3106\,$\pm$\,101 & 2.12\,$\pm$\,0.06 & 7.69\,$\pm$\,0.10 & 0.32 & 0.25 & 856\,$\pm$\,33 & 300 \\ 
  NGC\,6134 & 246.941 & -49.16 & 1141\,$\pm$\,36 & 1.49\,$\pm$\,0.05 & 8.93\,$\pm$\,0.05 & 0.23 & 0.18 & 681\,$\pm$\,19 & 594 \\ 
  NGC\,6192 & 250.085 & -43.35 & 1844\,$\pm$\,57 & 2.32\,$\pm$\,0.07 & 8.17\,$\pm$\,0.09 & 0.35 & 0.28 & 633\,$\pm$\,46 & 400 \\ 
  NGC\,6204 & 251.543 & -47.02 & 1218\,$\pm$\,47 & 1.56\,$\pm$\,0.14 & 8.37\,$\pm$\,0.18 & 0.24 & 0.19 & 288\,$\pm$\,22 & 231 \\ 
  NGC\,6268 & 255.528 & -39.71 & 1699\,$\pm$\,47 & 1.56\,$\pm$\,0.07 & 8.21\,$\pm$\,0.09 & 0.24 & 0.19 & 422\,$\pm$\,13 & 257 \\ 
  NGC\,6568 & 273.198 & -21.62 & 1033\,$\pm$\,19 & 1.44\,$\pm$\,0.07 & 8.53\,$\pm$\,0.06 & 0.22 & 0.17 & 283\,$\pm$\,14 & 196 \\ 
  NGC\,6583 & 273.960 & -22.14 & 2607\,$\pm$\,80 & 2.87\,$\pm$\,0.07 & 8.61\,$\pm$\,0.08 & 0.44 & 0.35 & 420\,$\pm$\,17 & 204 \\ 
  Pismis\,19 & 217.670 & -60.89 & 3194\,$\pm$\,182 & 3.76\,$\pm$\,0.15 & 8.81\,$\pm$\,0.12 & 0.57 & 0.45 & 955\,$\pm$\,130 & 563 \\ 
  Ruprecht\,121 & 250.428 & -46.17 & 2524\,$\pm$\,176 & 2.78\,$\pm$\,0.14 & 7.37\,$\pm$\,0.32 & 0.42 & 0.33 & 1194\,$\pm$\,100 & 416 \\ 
  Ruprecht\,128 & 266.064 & -34.88 & 2062\,$\pm$\,111 & 3.04\,$\pm$\,0.11 & 7.86\,$\pm$\,0.19 & 0.46 & 0.37 & 346\,$\pm$\,20 & 150 \\ 
  Ruprecht\,130 & 266.900 & -30.09 & 3092\,$\pm$\,184 & 3.91\,$\pm$\,0.08 & 7.29\,$\pm$\,0.10 & 0.60 & 0.47 & 243\,$\pm$\,24 & 56 \\ 
  Ruprecht\,134 & 268.181 & -29.54 & 1939\,$\pm$\,165 & 2.66\,$\pm$\,0.12 & 8.31\,$\pm$\,0.27 & 0.40 & 0.32 & 331\,$\pm$\,32 & 151 \\ 
  Teutsch\,84 & 256.097 & -42.06 & 3083\,$\pm$\,160 & 5.36\,$\pm$\,0.14 & 7.53\,$\pm$\,0.15 & 0.82 & 0.64 & 562\,$\pm$\,7 & 184 \\ 
  Trumpler\,25 & 261.125 & -39.01 & 2238\,$\pm$\,108 & 3.01\,$\pm$\,0.10 & 8.26\,$\pm$\,0.17 & 0.46 & 0.36 & 1172\,$\pm$\,229 & 666 \\ 
  Trumpler\,26 & 262.122 & -29.48 & 1679\,$\pm$\,72 & 2.43\,$\pm$\,0.06 & 7.52\,$\pm$\,0.14 & 0.37 & 0.29 & 321\,$\pm$\,14 & 134 \\ 
  Trumpler\,29 & 265.343 & -40.14 & 1582\,$\pm$\,50 & 1.30\,$\pm$\,0.05 & 7.93\,$\pm$\,0.09 & 0.20 & 0.16 & 331\,$\pm$\,20 & 200 \\ 
  Trumpler\,30 & 269.176 & -35.30 & 1336\,$\pm$\,26 & 1.63\,$\pm$\,0.06 & 8.18\,$\pm$\,0.09 & 0.25 & 0.20 & 304\,$\pm$\,30 & 171 \\ 
  UFMG\,1 & 236.599 & -56.79 & 2841\,$\pm$\,134 & 2.92\,$\pm$\,0.11 & 8.03\,$\pm$\,0.16 & 0.44 & 0.35 & 475\,$\pm$\,11 & 189 \\ 
  UFMG\,2 & 237.590 & -55.96 & 2326\,$\pm$\,76 & 3.55\,$\pm$\,0.11 & 8.63\,$\pm$\,0.09 & 0.54 & 0.43 & 909\,$\pm$\,62 & 520 \\ 
  UFMG\,3 & 238.124 & -55.41 & 2044\,$\pm$\,58 & 2.87\,$\pm$\,0.08 & 8.00\,$\pm$\,0.22 & 0.44 & 0.35 & 389\,$\pm$\,47 & 208 \\ 
   \hline
\end{tabular}}
\label{table:values}
\end{table*}

Adding the new high probability members helps the isochrone models to trace most of the cluster sequences better. As noted in PR21, for the vast majority of the clusters, even the fainter end of the sequences match the theoretical isochrones. Since we are still in the stellar regime at those magnitudes and for the range of cluster ages explored, we do not expect deviations due to the uncertainties in the models. That indicates a low rate of interlopers across the whole magnitude range covered in this study. In various cases (BH\,202, Lynga\,9, NGC\,4349, NGC\,5381, NGC\,5715, NGC\,5999, NGC\,6568, UFMG\,1, UFGM\,2), it is reproduced the phenomenon found in PR21 for NGC\,6259,  where we identify a group of high probability members at $K\sim$11\,mag. Most sources are not identified as high probability members at optical wavelengths. On the other hand, the isochrone location for all Ruprecht clusters plus Teutsch\,84 and Trumpler\,26 is rather poor compared with the outlined cluster sequences. As discussed below, this is partly induced by the very dispersed sequences with large error bars on distance, extinction, and age. All the clusters mentioned earlier are also located in the IV Galactic quadrant (Sagittarius-Carina (Sag-Car) and Scutum arms). Recently, \citep{poggio21} made a comparison between their map of density variations (built with upper main sequence stars, open clusters, and classical Cepheids) and the Galactic models of \citet{taylor93} and \citet{reid19}. In quadrant IV, the authors find a possible inter-arm gap between Sag-Car and the inner Scutum arm. This inter-arm would be closer to the Sun than \citet{taylor93} and \citet{reid19} models suggest. Alternatively, the authors mentioned the possibility of an additional branch off the Sag-Car arm. Given this scenario, we avoid manually reallocating the cluster sequences on that specific region, and a dedicated study will address the phenomenon in future work.

Since our membership determination isolates a unique compact Gaussian distribution among the field, we rule out the inclusion of other known clusters and cluster candidates in the surveyed cluster area. Moreover, given that the cluster-selected Gaussian component was allowed full covariance, we did not restrict the study to a specific cluster shape. Recent studies have explored the open cluster shapes extensively in our Galaxy \citep{pang21,hu21}, leaving evident the enormous diversity of cluster forms.  

\begin{figure*}
\includegraphics[scale=0.15]{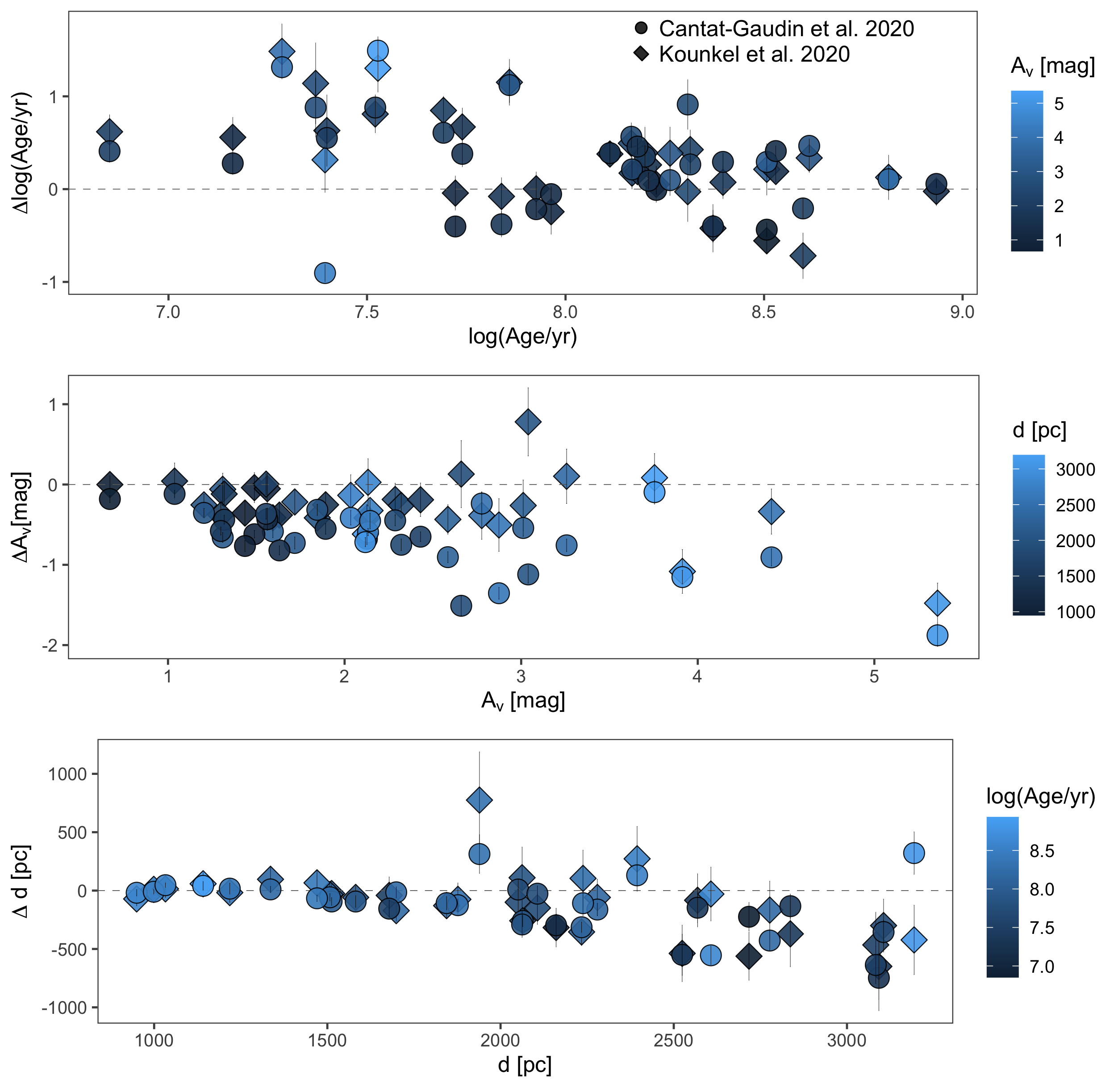}
\caption{Difference between cluster literature values (age, extinction, distance) and the calcuted parameters presented here. Filled circles relate to \citep{newcantatgaudin20} values, filled diamonds for those from \citep{kounkel20}. The symbols are color coded based on the explored parameters. UFGM\,1, UFGM\,2, and UFGM\,3, are not included.
}
\label{fig:comparison}
\end{figure*}

The following presents the comparison of our results with those in the literature. For the broad comparison of our results with the literature, we have focused on the studies of \citet{newcantatgaudin20} and \citet{kounkel20}. We must first point out that our membership probabilities are skewed towards high values since we adopt only the highest members probabilities from the XDGMM procedure as input to the spatial/photometric membership refinement. The overall parameter comparison can be seen in Figure\,\ref{fig:comparison}. The bulk of the Auriga ages ranges from 14\,Myr to 870\,Myr. Only NGC\,5606 reports a young age of about 7\,Myr where the Auriga model can overestimate the age. As can be seen, the most discrepant ages are linked to the most extincted clusters or to those where there is no agreement among published values (e.g., Ruprecht\,130, Ruprecht\,134, Lynga\,6, Teutsch\,84, ASCC\,88). For sources below $\sim$ 100\,Myr we report younger ages with the inclusion of the newly identified members. In terms of extinction values, the most distant clusters have the most distinct values, which is expected given the near-infrared nature of our dataset. The general trend is that we report larger extinction values than those available in the literature by about 0.5\,mag. We cover cluster distances ranging from about 0.9 to 3.2\,kpc. We found larger distance values for those clusters beyond $\sim$2.5\,kpc. The overall distance values differ for the clusters with the literature's most significant uncertainties (e.g., Ruprecht\,130, Ruprecht\,134, Pismis\,19). 


\subsection{Cluster total mass}
Following the methodology outlined on PR21, each cluster’s density profile was integrated to obtain the total number of stars contained in the cluster’s main sequence. We determined the completeness magnitude for each cluster as the faintest magnitude at which the number of sources per interval of magnitude does not deviate from an increasing distribution. Assuming the central values of Kroupa mass function \citep{kroupa01} and the PARSEC-COLIBRI isochrone on the rederived Auriga parameters, we extended the counting, adding all the masses of stars down to the stellar/substellar frontier (0.08\,M$_\odot$). The evolved sources out from the main sequence were also linked to the best-fitting position on the corresponding isochrone. The membership probability of each source weighted the mass values.

The studied clusters have total masses in the range $\sim$120 -- 2000\,M$_\odot$, and the values are presented in Table\,\ref{table:values}. The adopted uncertainties consider the density profiles with and without weighting the mass values by each source’s membership probability. 

\section{Conclusions}\label{sec:conclusions}
This study unveils the near-infrared sequences of 37 open clusters located between $\sim$0.9 and $\sim$3.2\,kpc. We employed a data-driven approach tailored toward disentangling the cluster population from field stars. The cluster members search comprises two main steps: a distribution-aware tagging of the central values and uncertainties in the astrometric space and a cluster membership assignment based on positional and photometric similarity.
Our methodology has allowed us to disentangle the cluster populations from their dense backgrounds. Our study increased cluster membership by $\sim$47\% on average, this increase in efficiency at recovering cluster members builds up our knowledge on the near-infrared cluster sequences. The measured physical parameters of the clusters generally agree with the literature values. However, we update their extinction values and total masses by unveiling a large portion of the low-to-intermediate mass population with a native near-infrared data-driven approach. In the near future a similar method can be applied to large samples of clusters to be optically discovered by the Vera Rubin telescope \citep{LSST}.

The so-called Gaia revolution has changed the picture of our Galaxy in just a few years. Now we count in the thousands the number of over-densities in our Galaxy that diverse machine learning methods have tagged as open clusters. Here we show that the synergies with other instruments/data-sets, such as photometry and astrometry from VISTA, allow us to unveil the near-infrared counterpart of each studied cluster. We have taken advantage of high quality parallaxes and deep near-infrared photometry, and machine learning methods to enable us to obtain clean near-infrared color-magnitude diagrams from which astrophysical parameters can be inferred.

\section*{Acknowledgements}
We thank Marina Kounkel for her comments on the implementation of the Auriga model and Laurent Chemin for his feedback on the spiral Galactic structure. We also thank the referee for the received comments. This work was supported by ANID FONDECYT Iniciaci\'on 11201161, and ANID FONDECYT Regular 1201490. D.M. gratefully acknowledges support by the ANID BASAL projects ACE210002 and FB210003, and by Fondecyt Project No. 1220724. This paper made use of the Whole Sky Database (wsdb) created by Sergey Koposov and maintained at the Institute of Astronomy, Cambridge by Sergey Koposov, Vasily Belokurov and Wyn Evans with financial support from the Science \&
Technology Facilities Council (STFC) and the European Research Council (ERC). This work has made use of data from the European Space Agency (ESA) mission
{\it Gaia} (\url{https://www.cosmos.esa.int/gaia}), processed by the {\it Gaia} Data Processing and Analysis Consortium (DPAC, \url{https://www.cosmos.esa.int/web/gaia/dpac/consortium}). Funding for the DPAC has been provided by national institutions, in particular the institutions participating in the {\it Gaia} Multilateral Agreement.

The preparation of this work has made extensive use of Topcat \citep{topcat}, of NASA’s Astrophysics Data System
Bibliographic Services, of the open-source Python packages Astropy \citep{astropy}, NumPy \citep{numpy}, and scikit-learn \citep{scikitlearn} as well as the R packages under the tidyverse suite \citep{tidyverse}. The figures in this paper were produced with ggplot2 \citep{ggplot2}.

\section{Data availability}\label{sec:catalogs}
The data underlying this article are available in the Open Software Foundation, at \url{https://dx.doi.org/10.17605/OSF.IO/8D2HJ}.

\newpage
\bibliographystyle{mnras}
\bibliography{references}

\appendix
\renewcommand\thefigure{\thesection.\arabic{figure}}
\section{Diagrams of studied clusters} \label{sec:anex}
Figures produced using the supplementary online material (see Section\,\ref{sec:catalogs}).
\setcounter{figure}{0}   

\begin{figure*}
\includegraphics[scale=0.075]{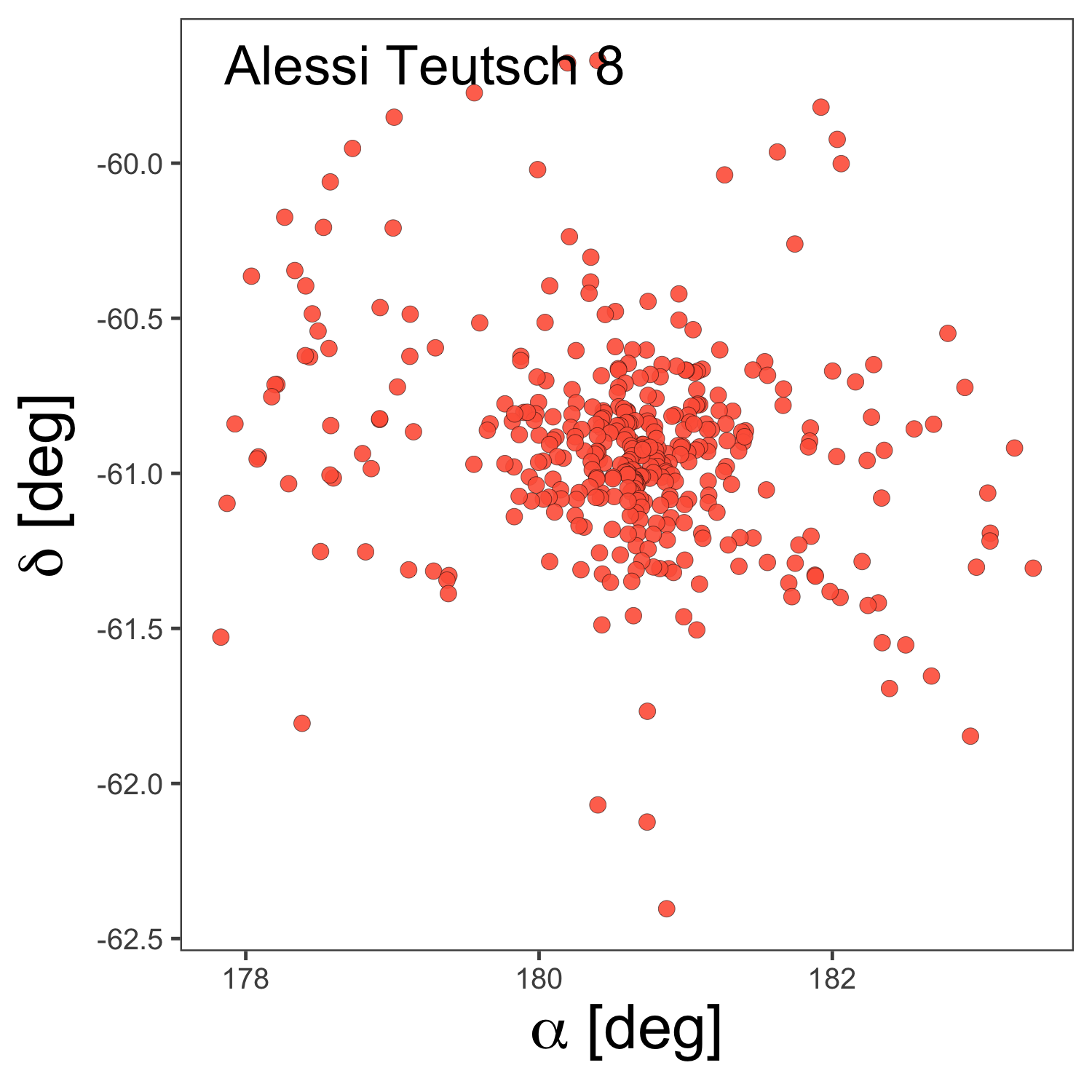}
\includegraphics[scale=0.075]{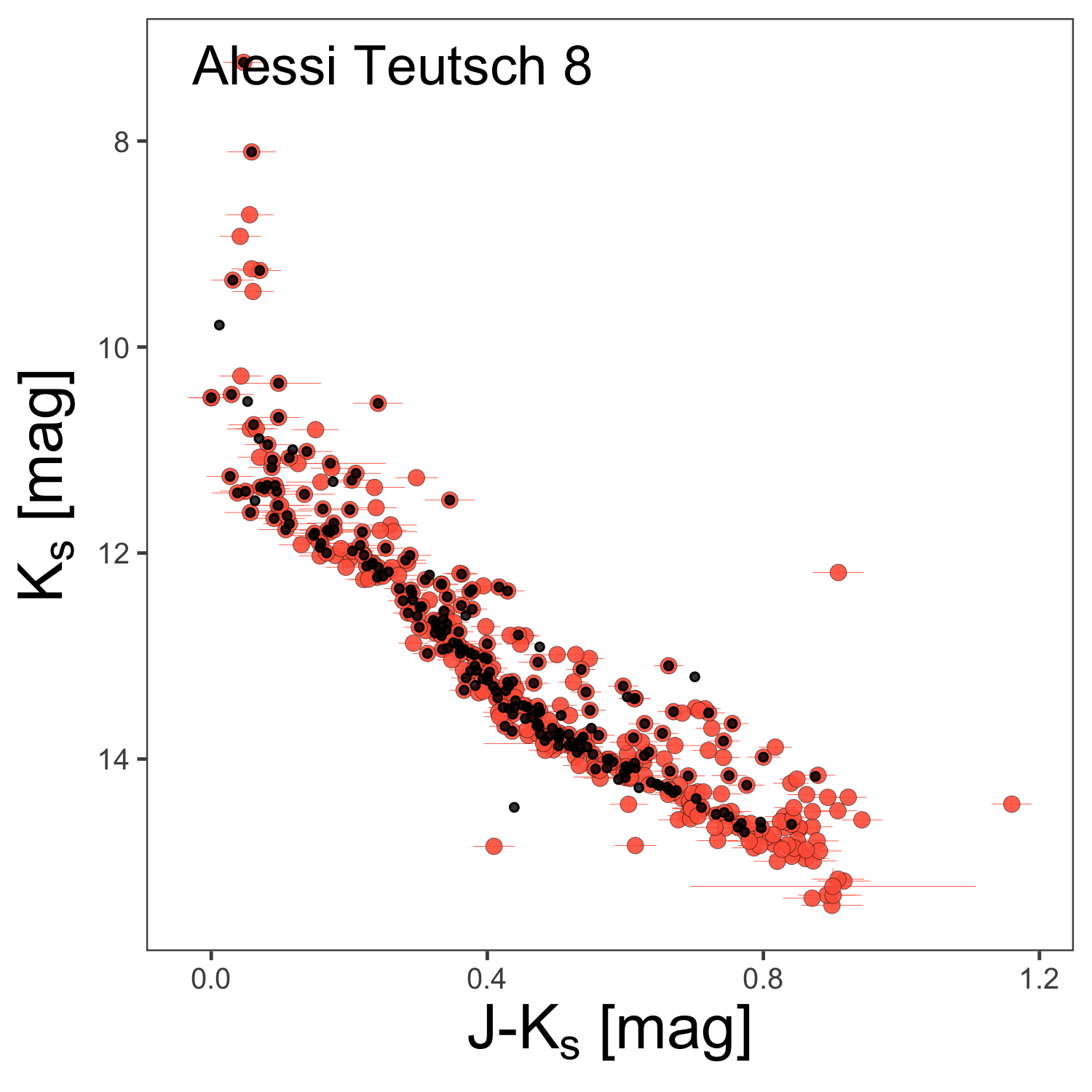}
\includegraphics[scale=0.075]{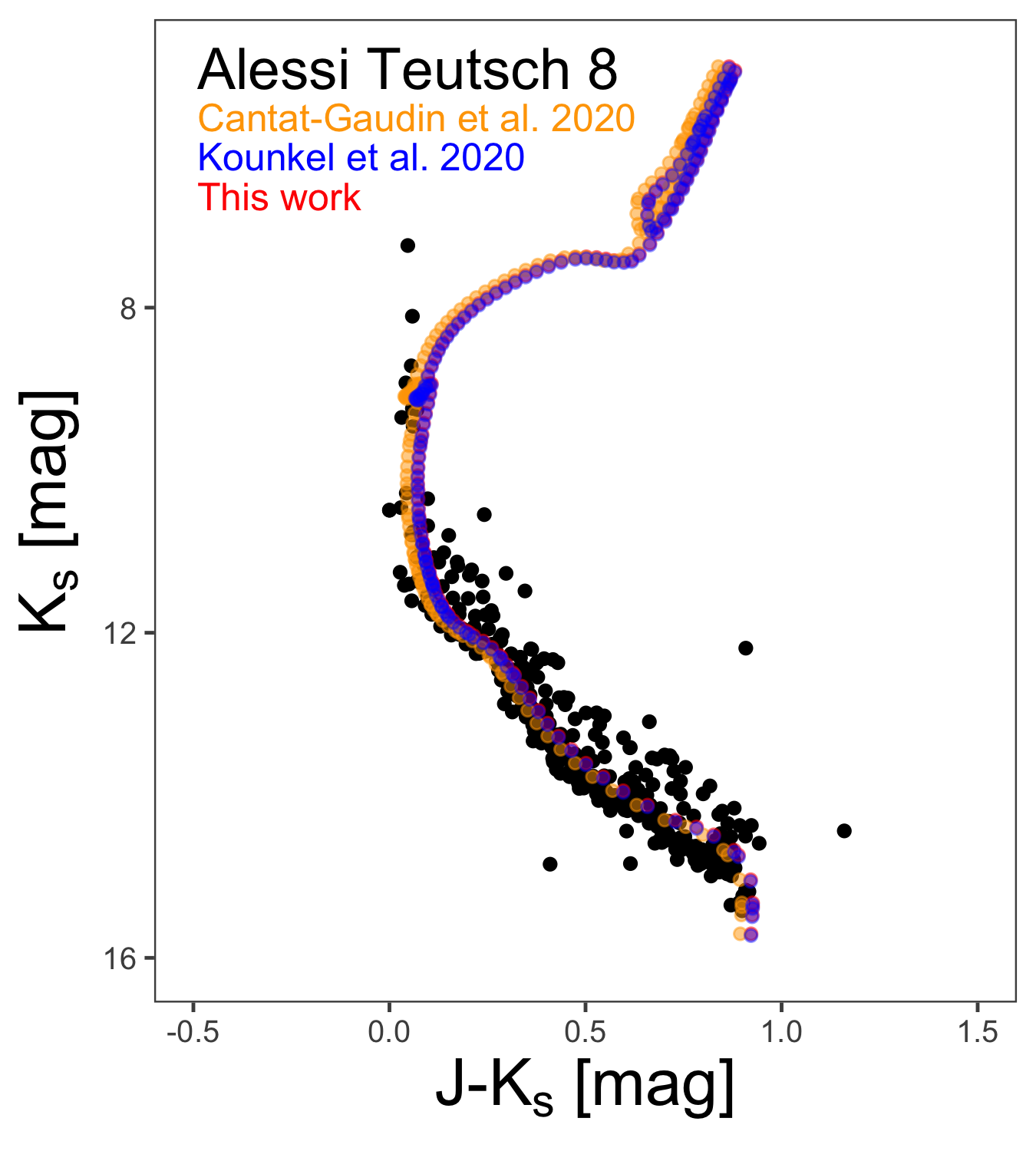}
\includegraphics[scale=0.075]{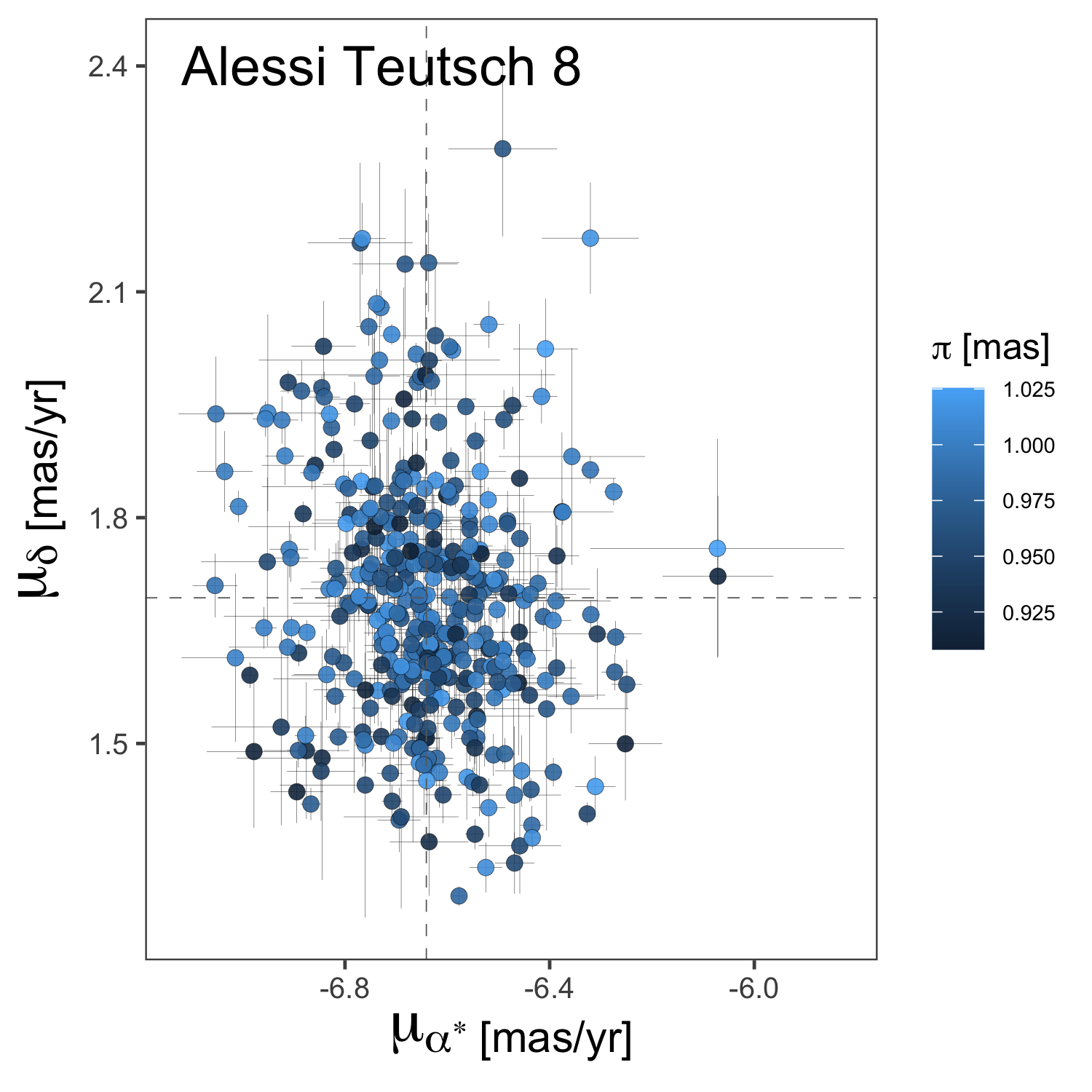}
\includegraphics[scale=0.075]{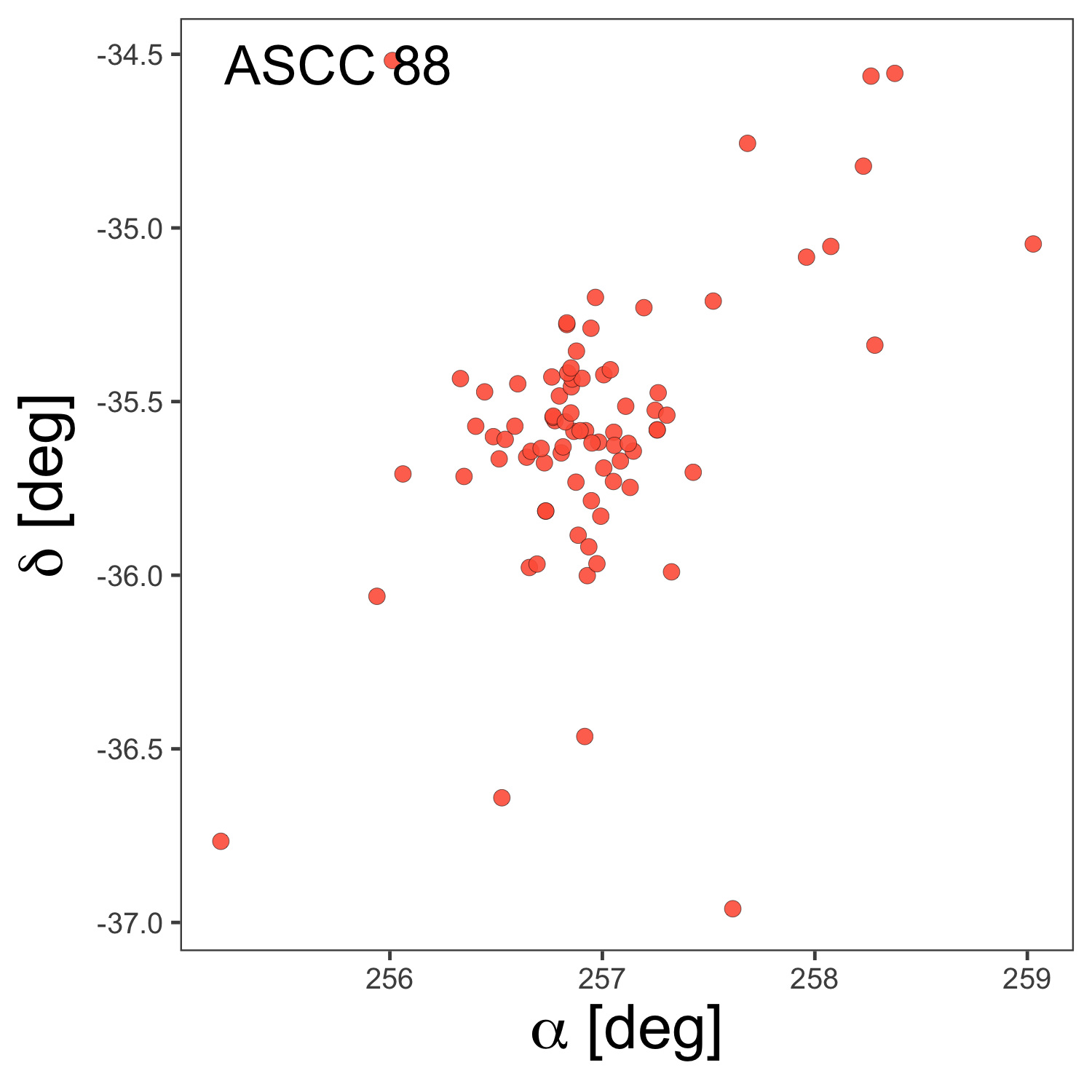}
\includegraphics[scale=0.075]{ASCC_88pub_cmd.png}
\includegraphics[scale=0.075]{ASCC_88_iso.png}
\includegraphics[scale=0.075]{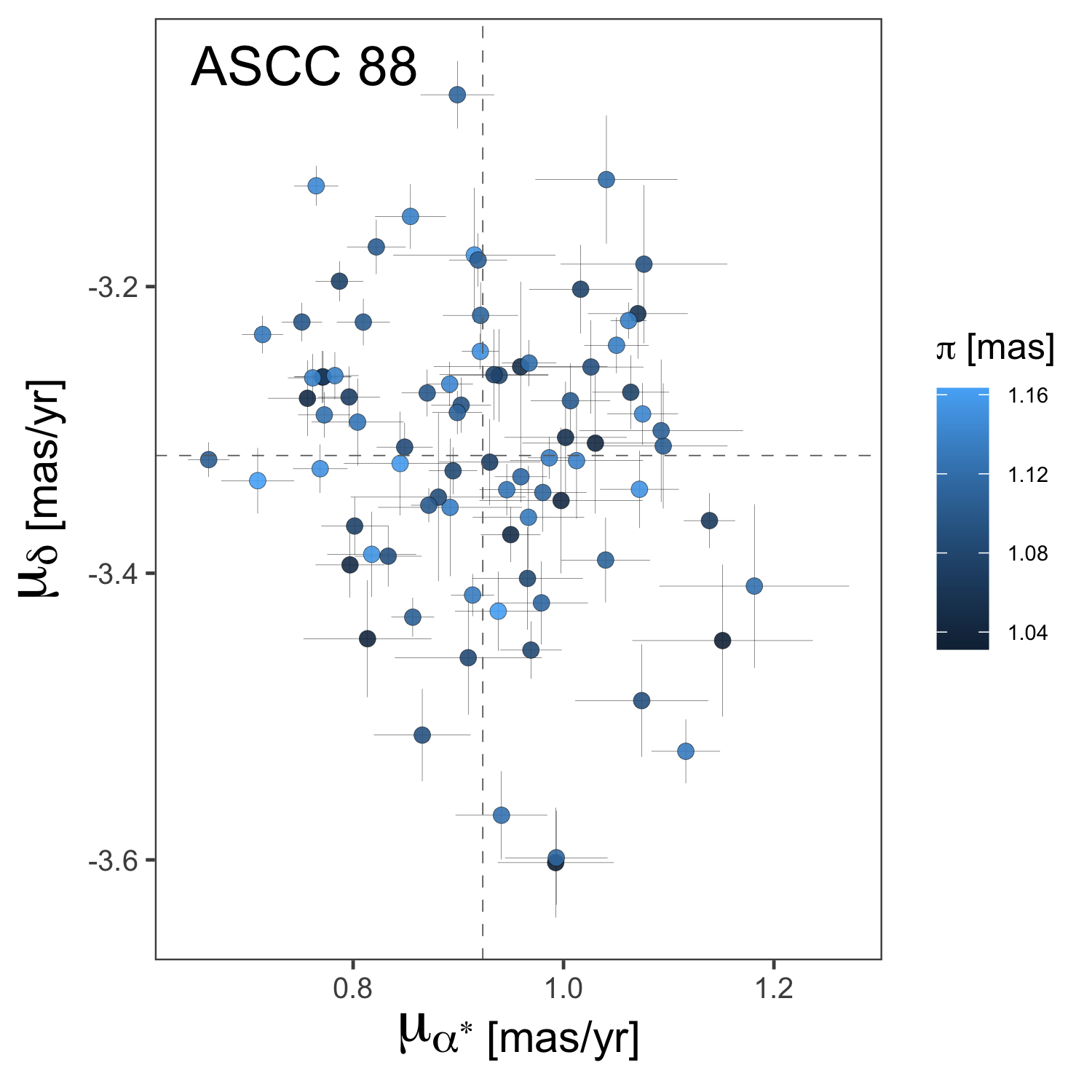}
\includegraphics[scale=0.075]{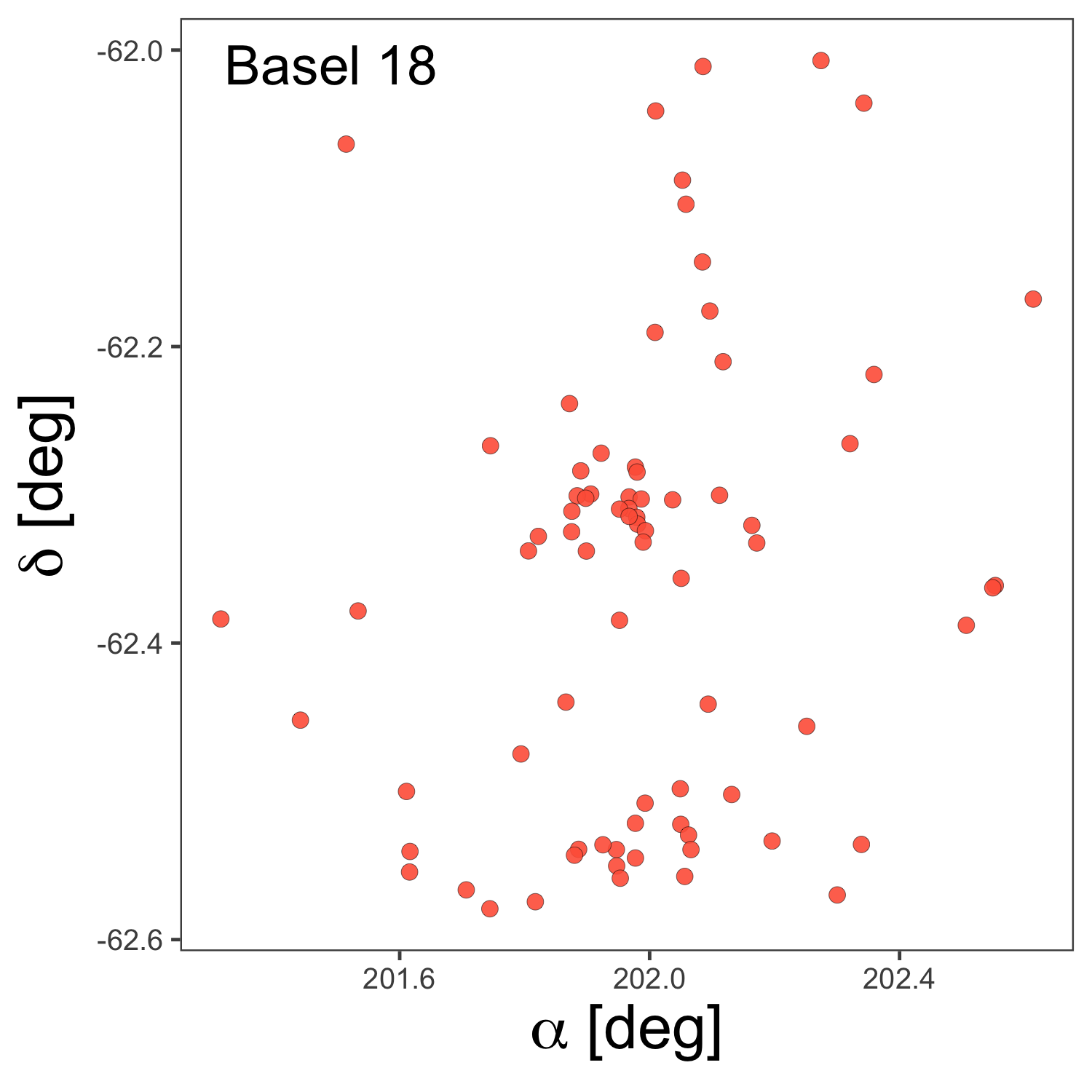}
\includegraphics[scale=0.075]{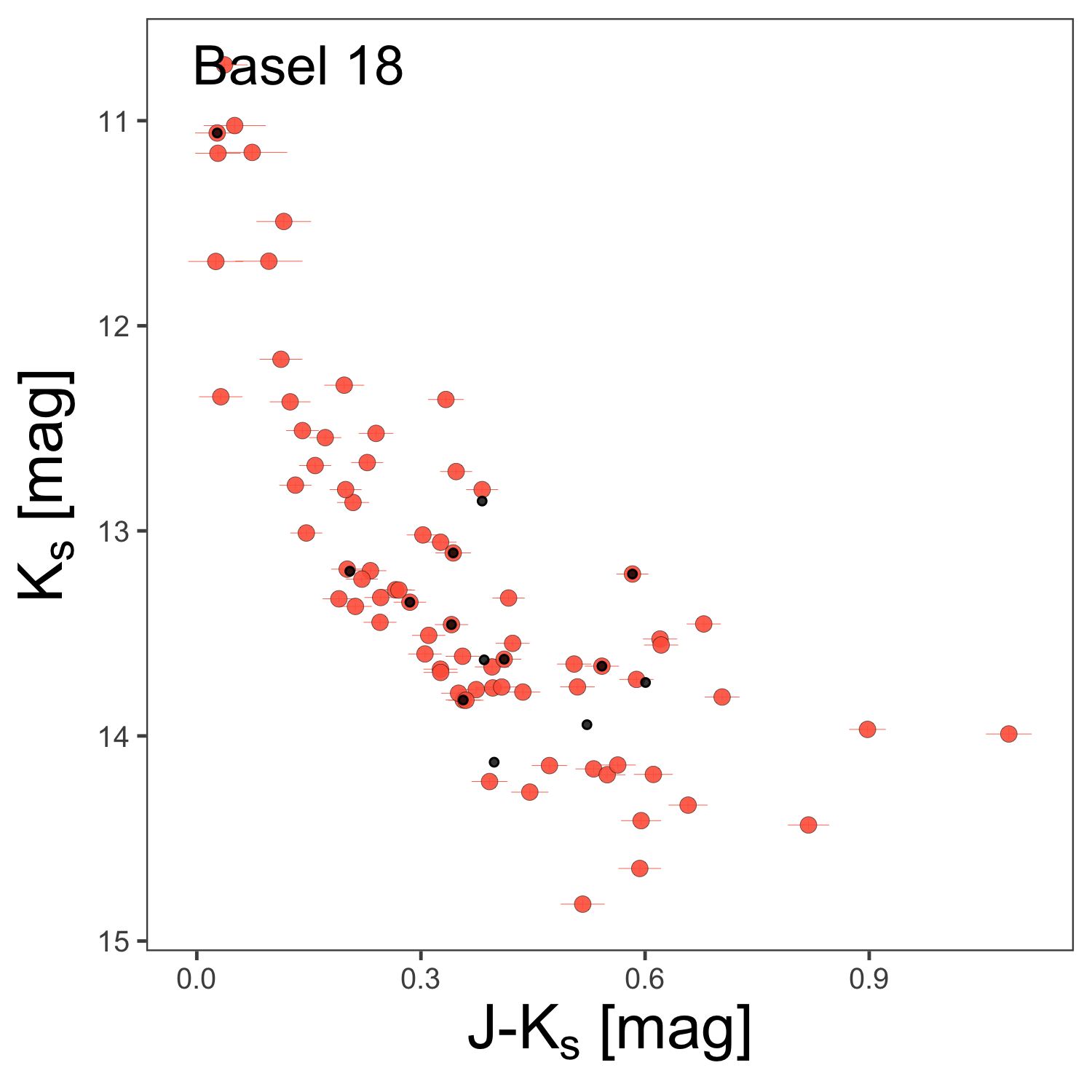}
\includegraphics[scale=0.075]{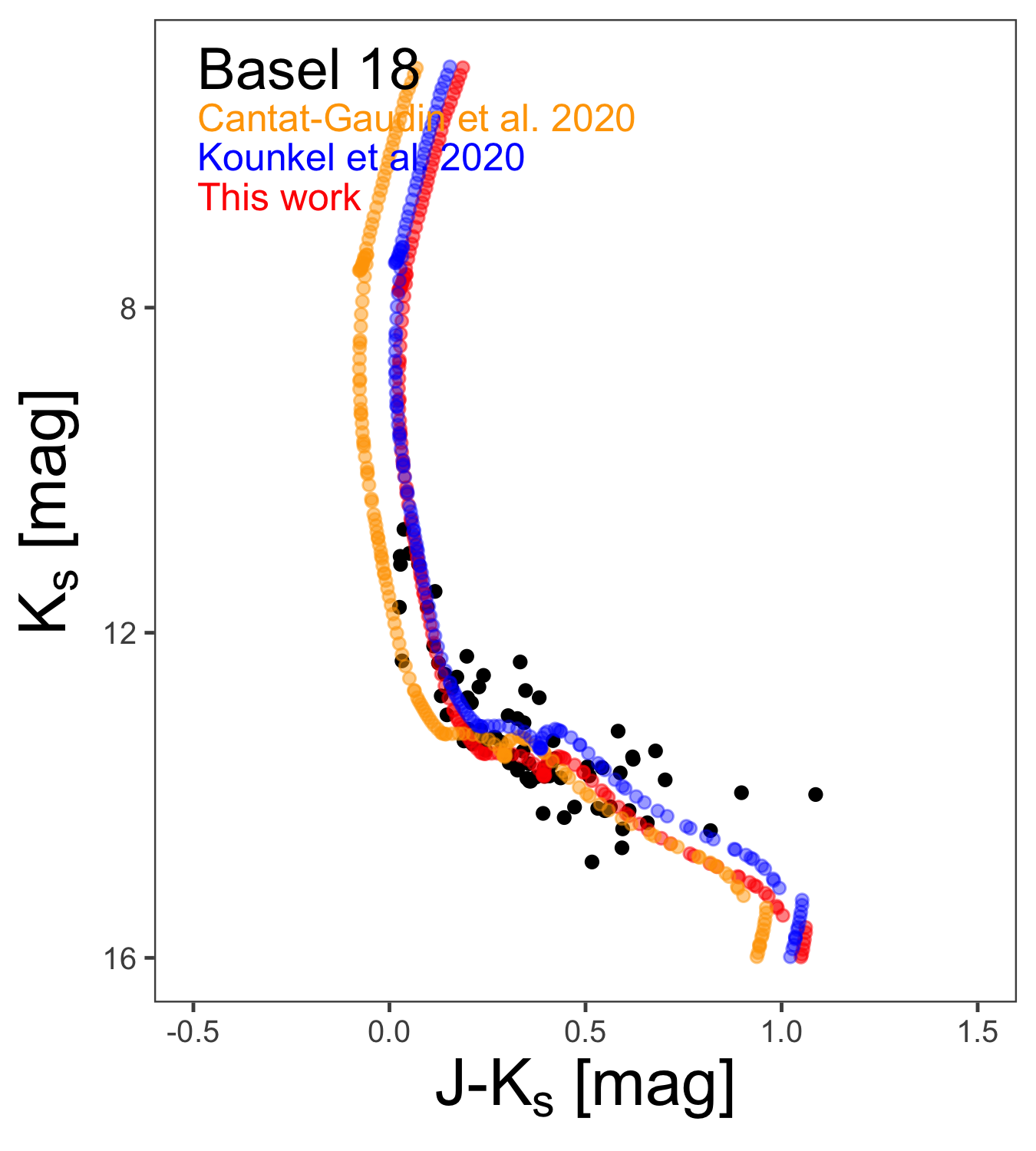}
\includegraphics[scale=0.075]{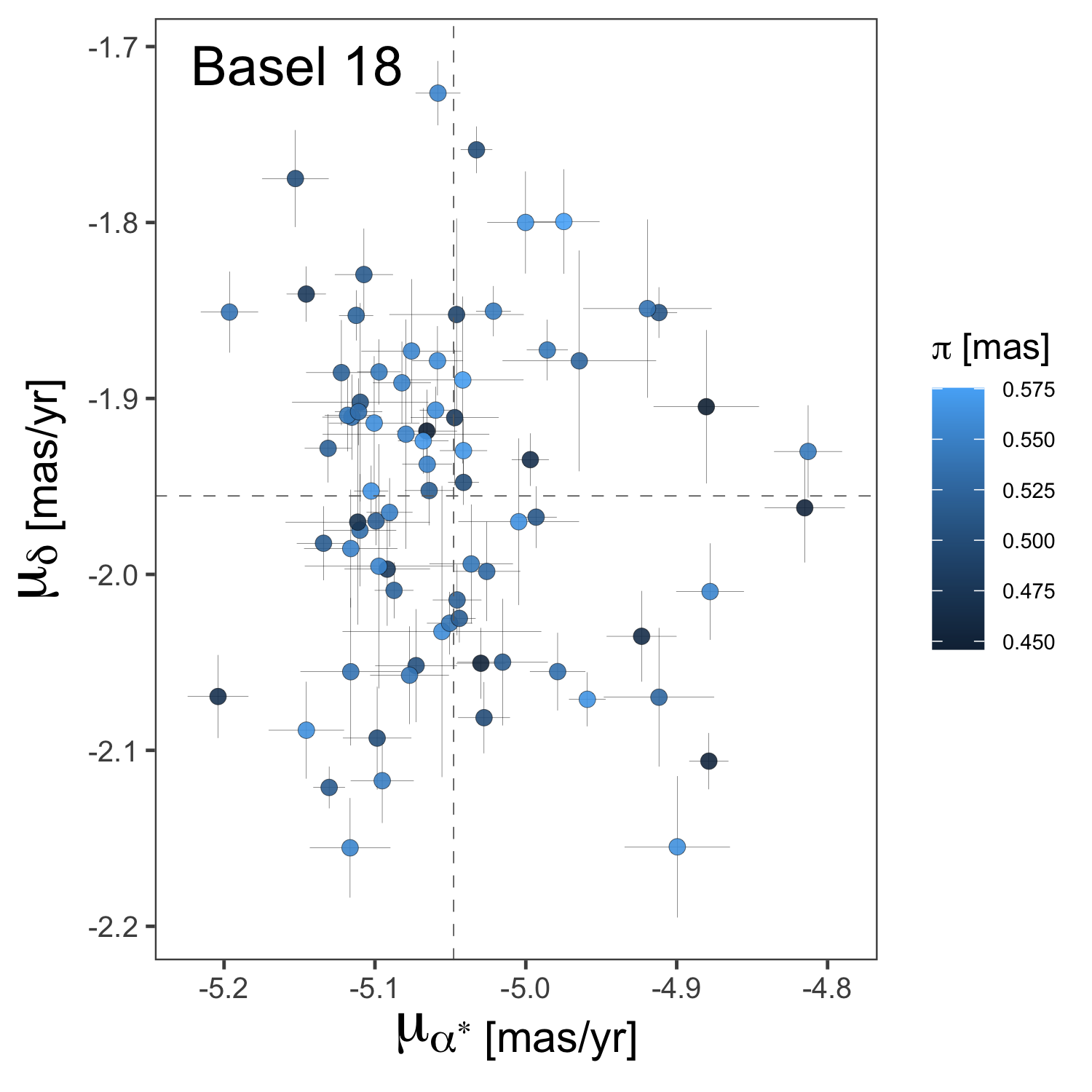}
\includegraphics[scale=0.075]{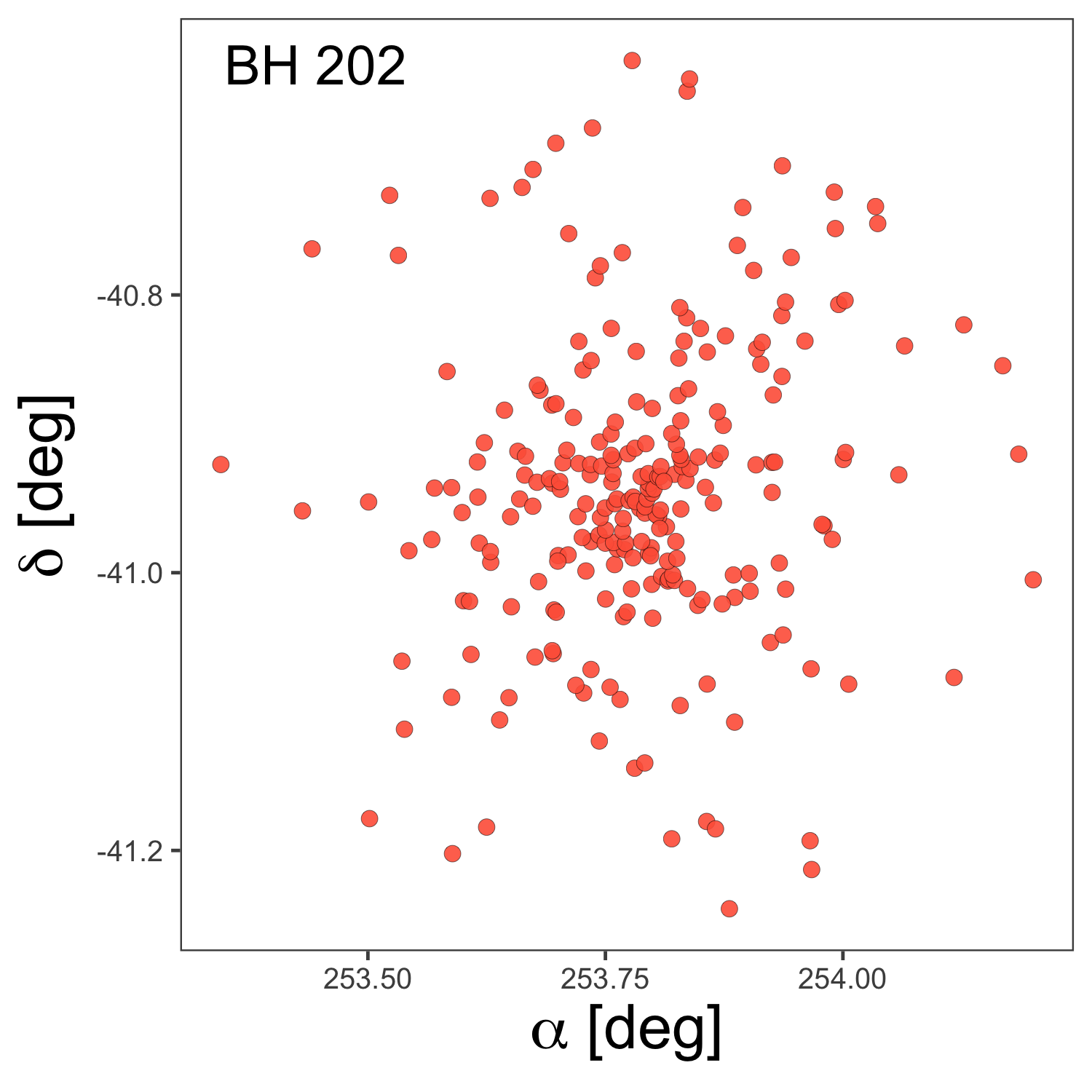}
\includegraphics[scale=0.075]{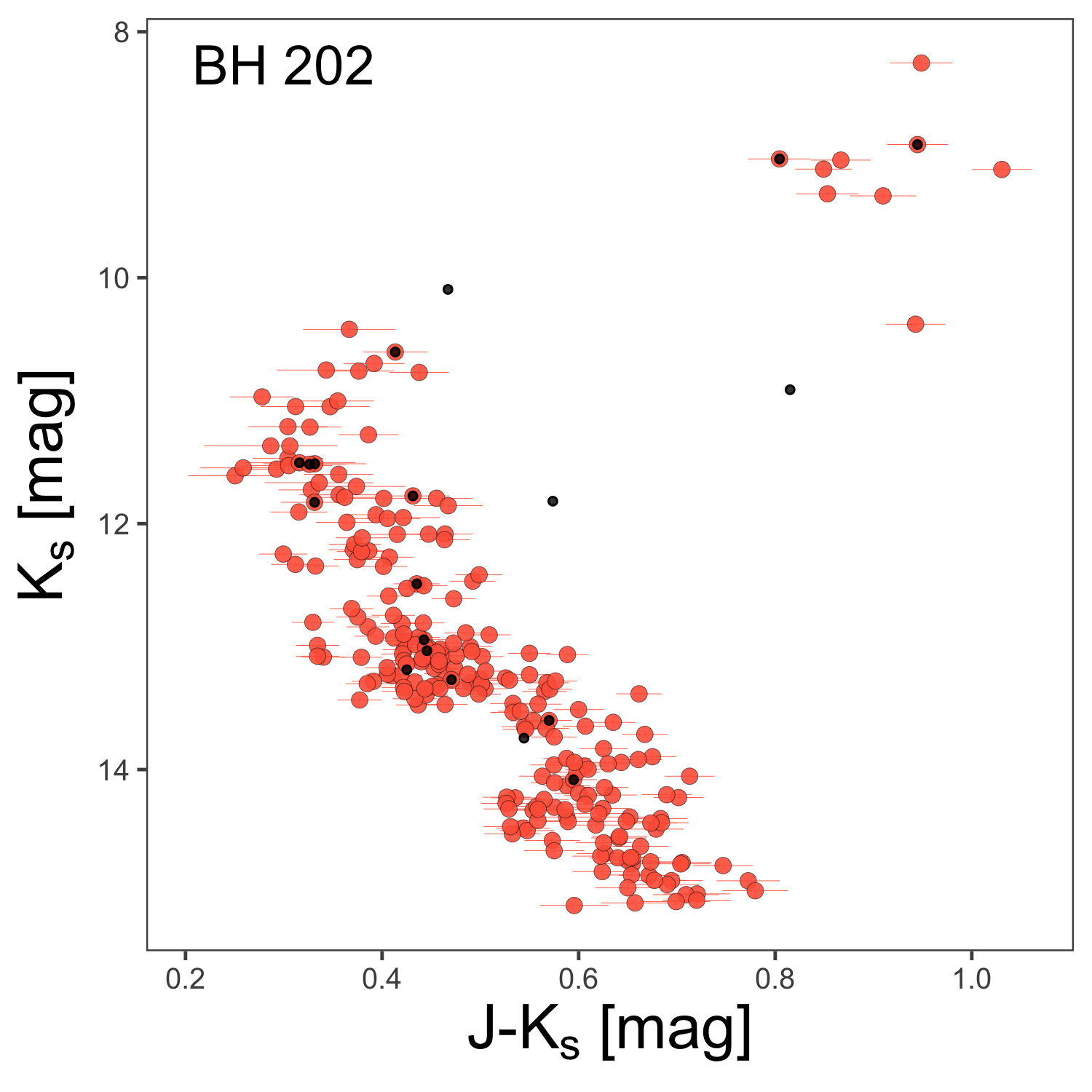}
\includegraphics[scale=0.075]{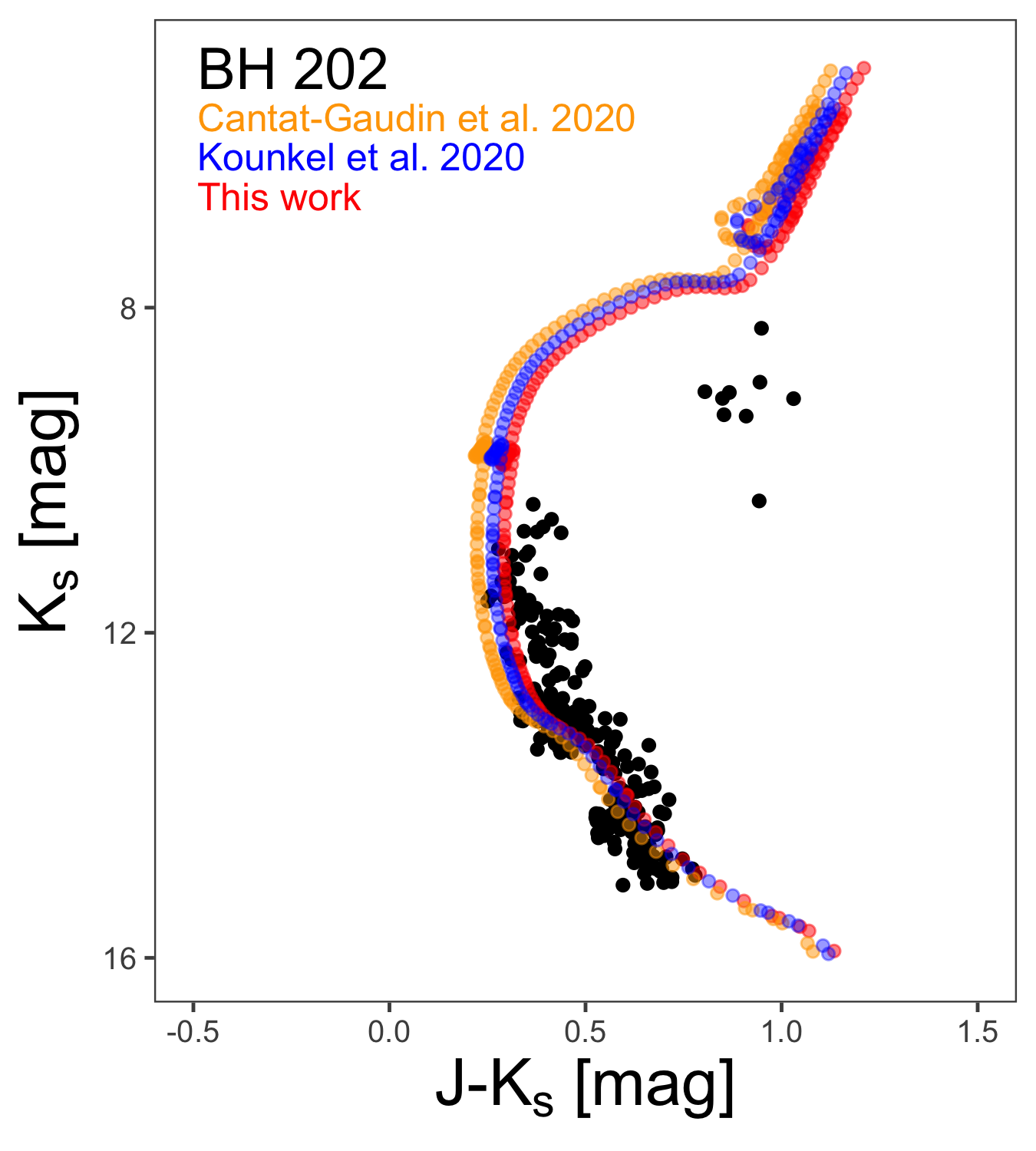}
\includegraphics[scale=0.075]{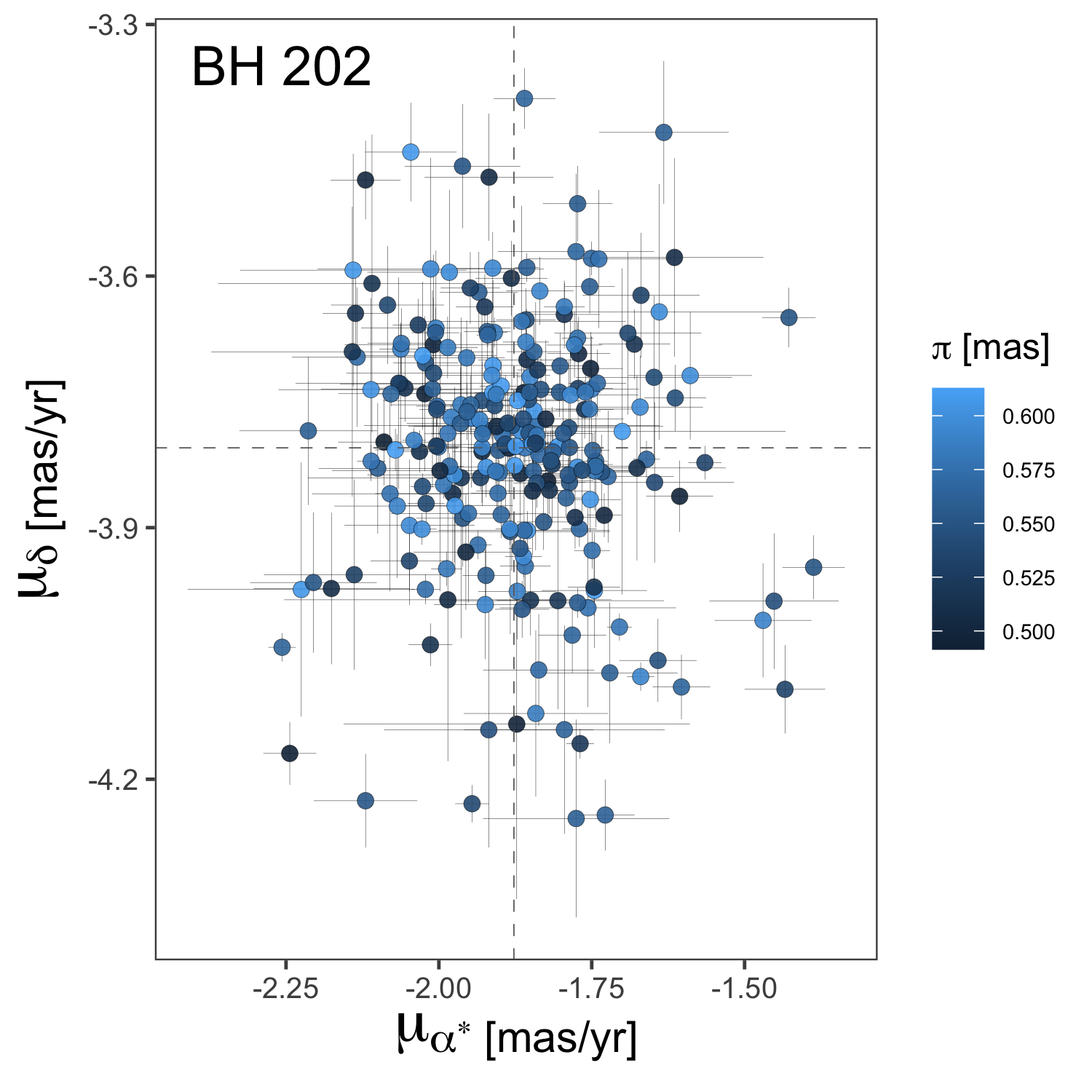}
\includegraphics[scale=0.075]{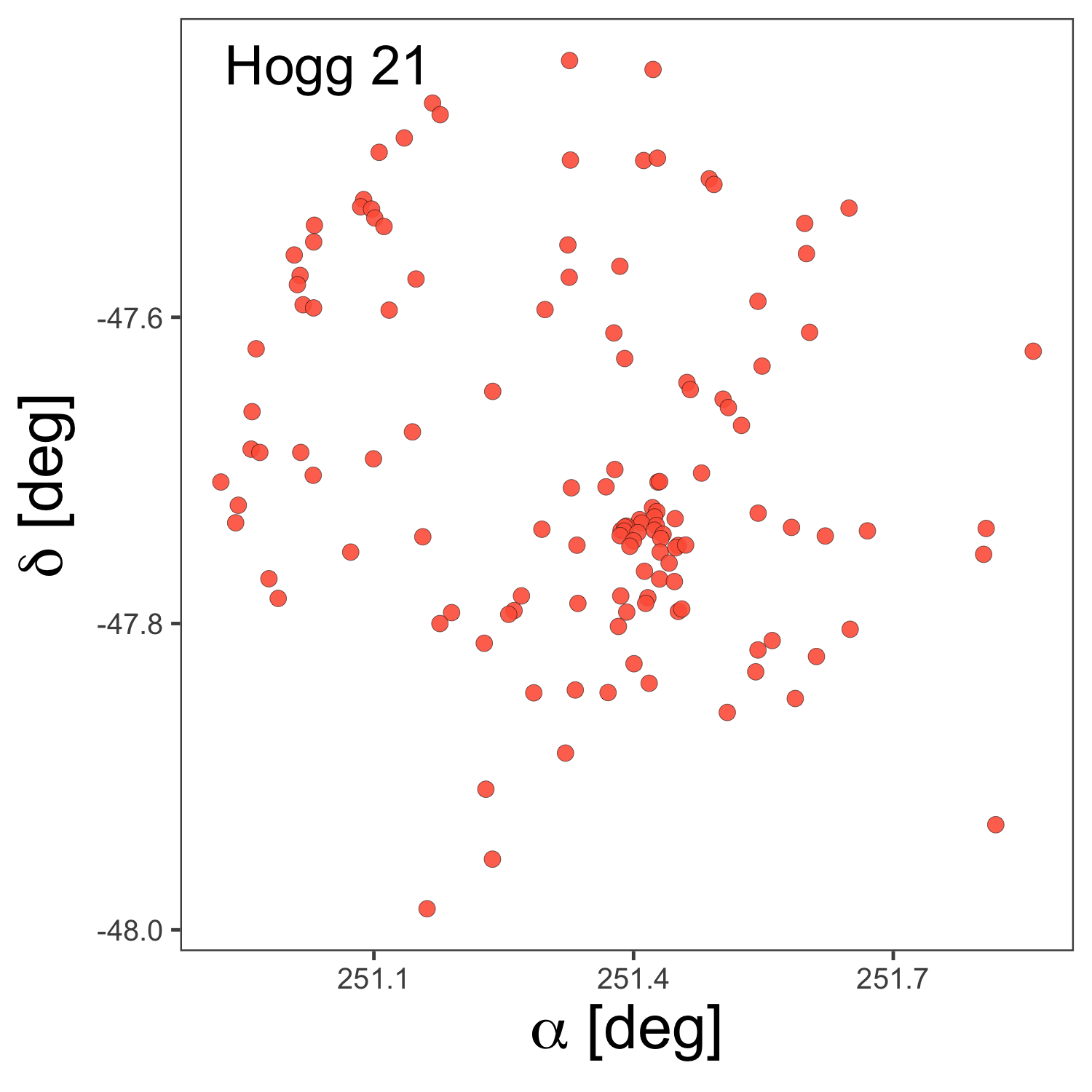}
\includegraphics[scale=0.075]{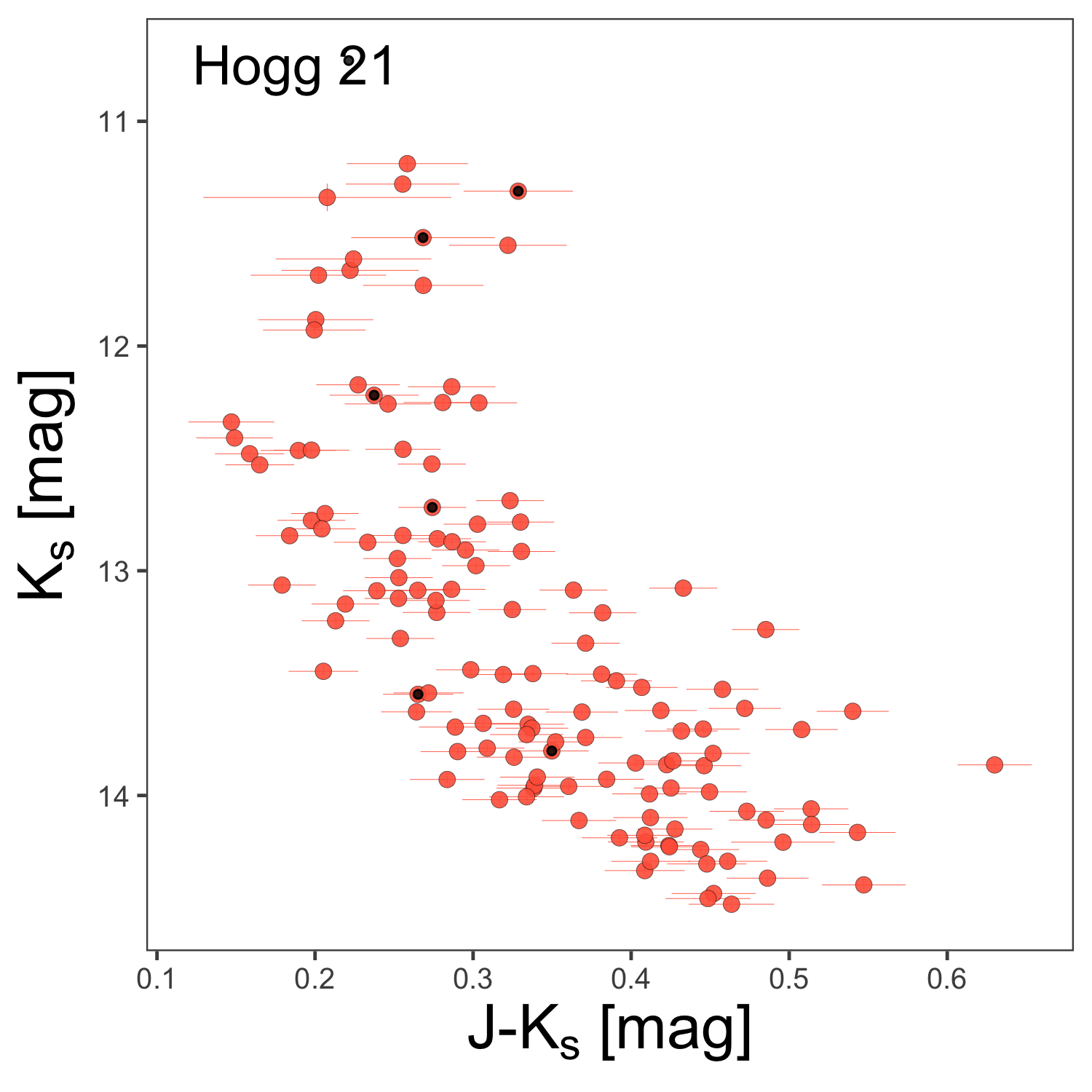}
\includegraphics[scale=0.075]{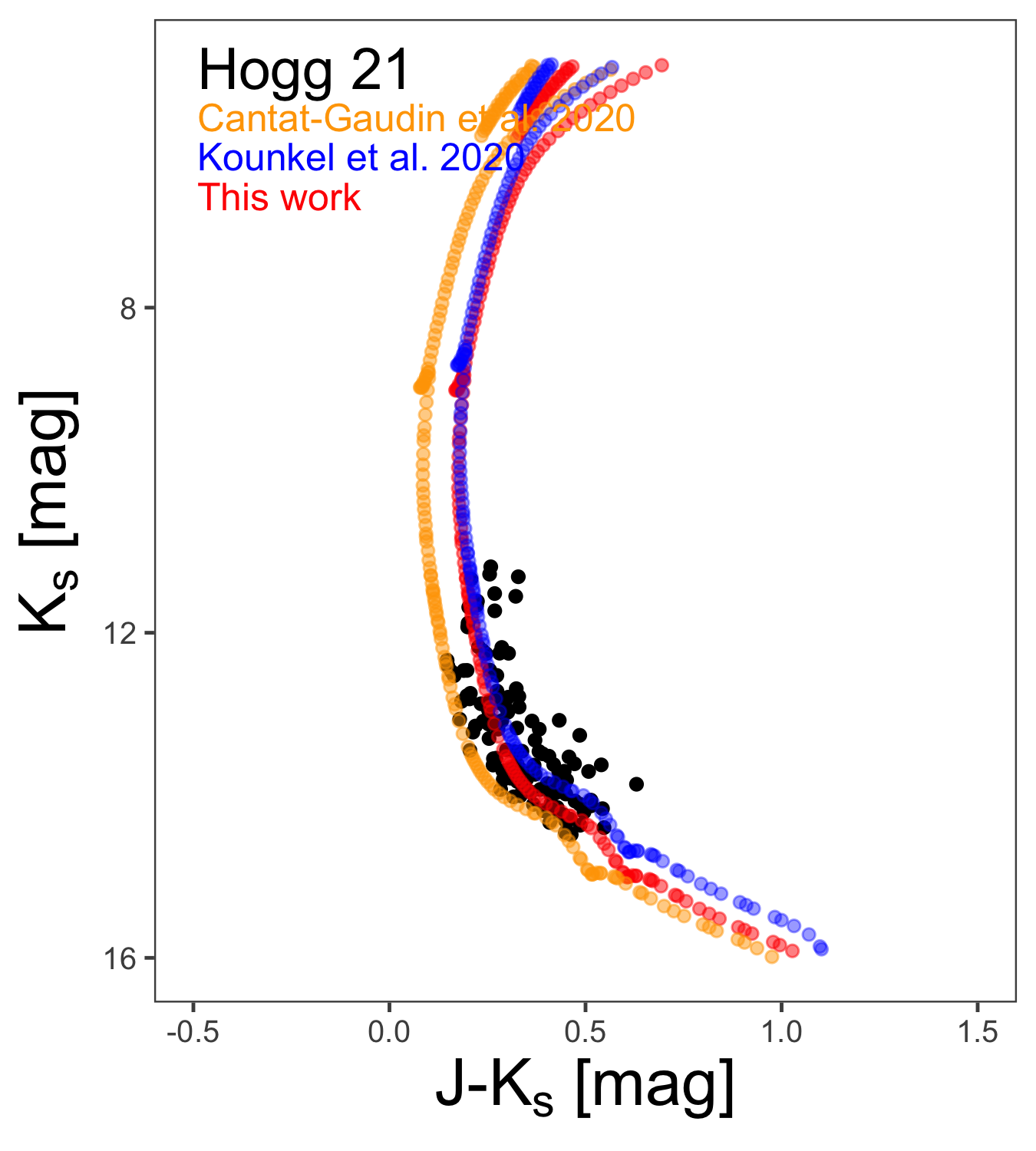}
\includegraphics[scale=0.075]{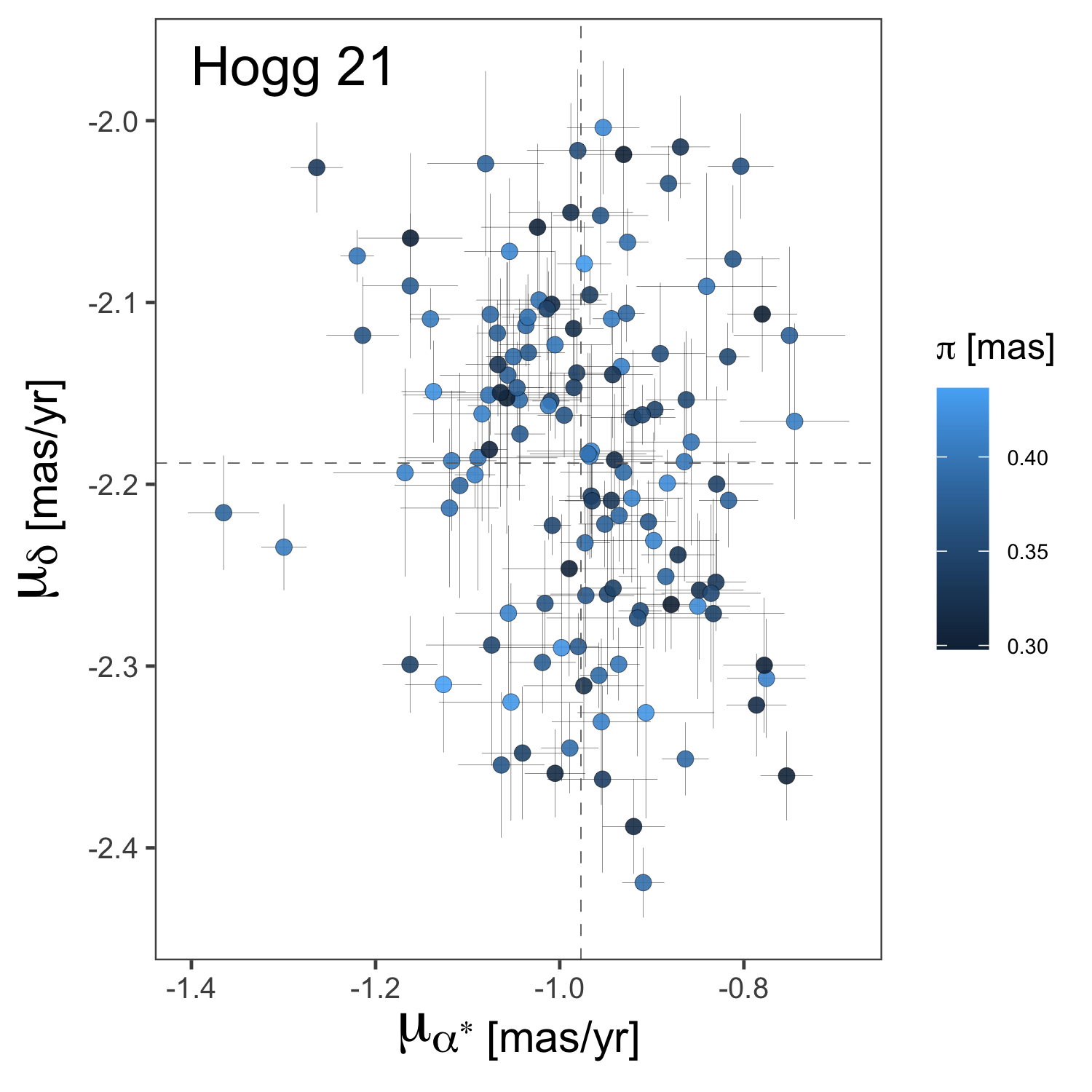}
\caption{\textit{Left:} Spatial distribution of the cluster members. \textit{Center-left:} $K_s$ vs. $J$-$K_s$ color-magnitude diagram of the studied open stellar clusters. Black dots correspond to the \citet{cantatgaudin20} members with reported probability $p\geq$90\%. whereas the sources presented here are represented with red circles. \textit{Center-right:} $K_s$ vs. $J$-$K_s$ color-magnitude diagrams along with the theoretical isochrones  located based on the cluster parameters from \citet{cantatgaudin20, kounkel20} and this work over our catalog members as black points. All the sequences preserved a common scale. \textit{Right:} Members proper motion distribution color-coded based on their parallax value. Mean proper motions in right ascension and declination are plotted with dotted lines. Only sources with membership probabilities $p\geq$90\% are shown in all the diagrams. Text on the plots shows the cluster name.
}
\label{fig:clusters}
\end{figure*}

\begin{figure*}
\includegraphics[scale=0.075]{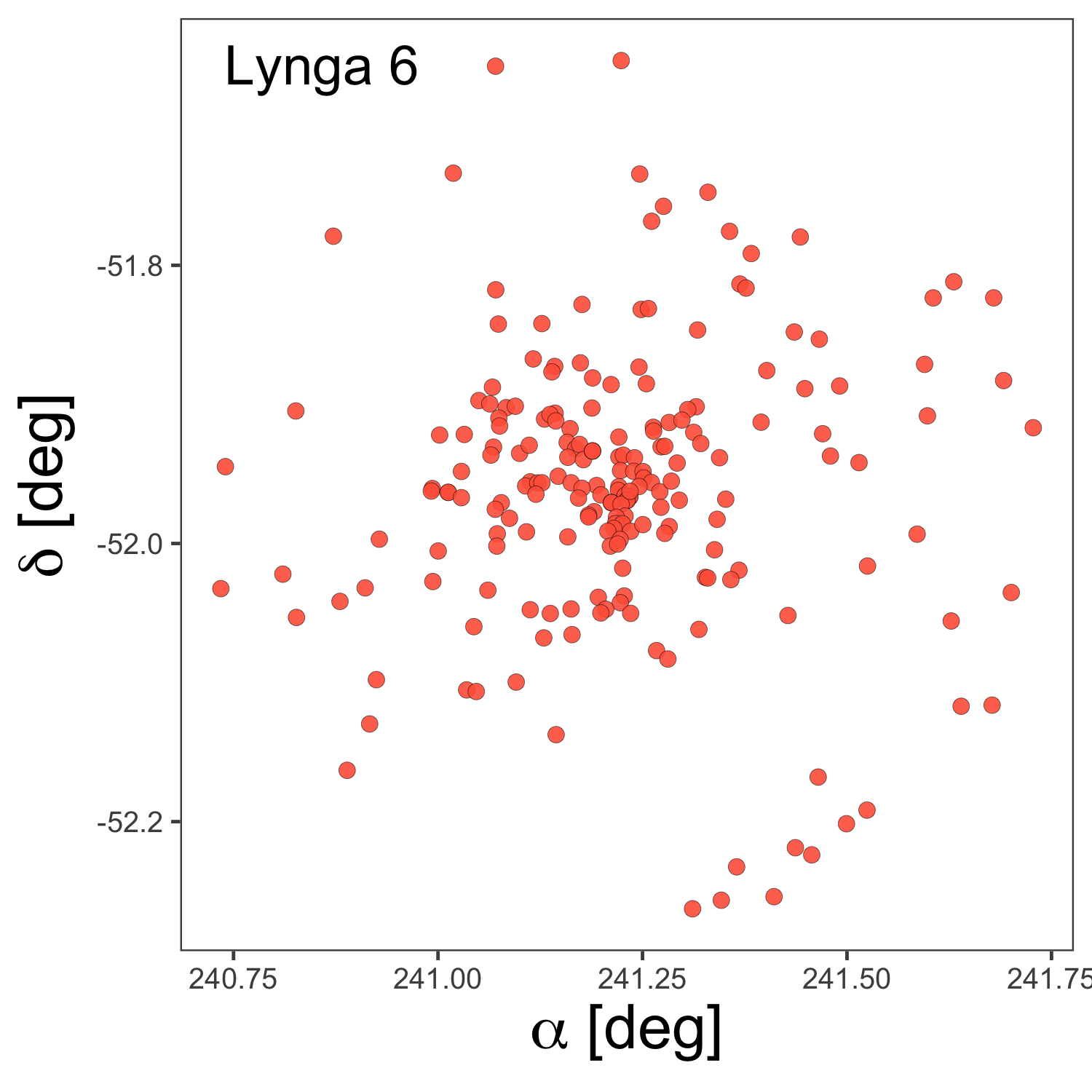}
\includegraphics[scale=0.075]{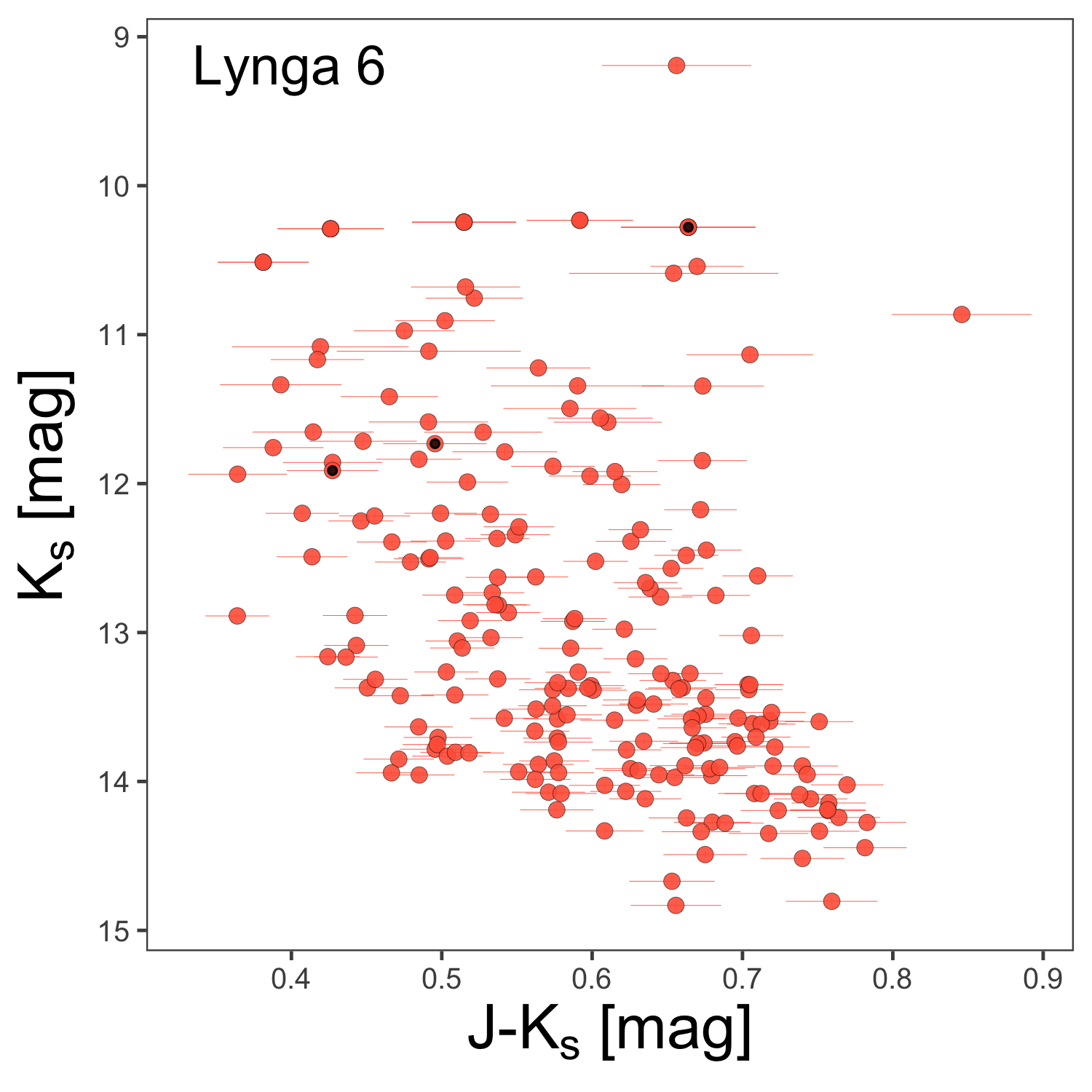}
\includegraphics[scale=0.075]{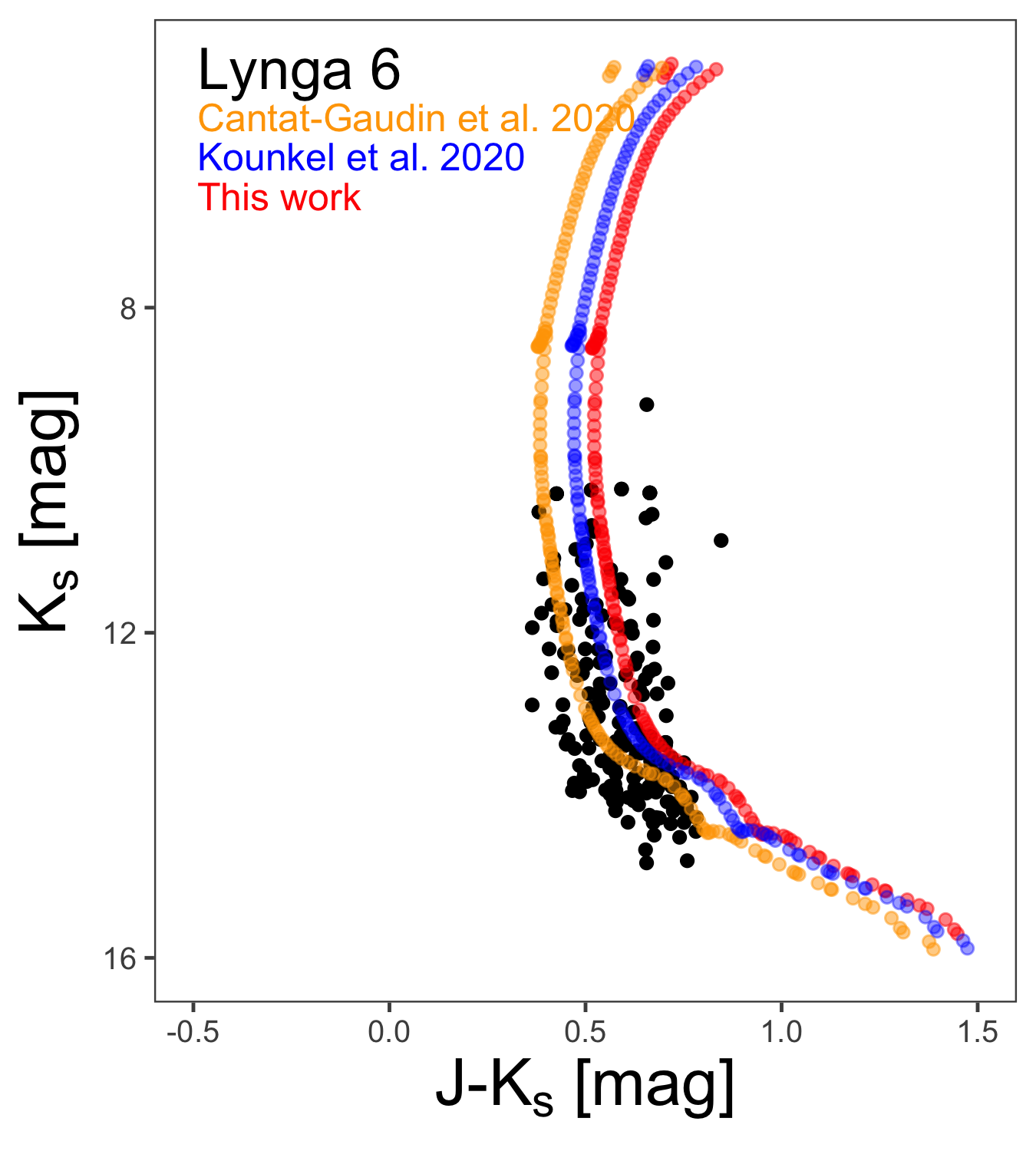}
\includegraphics[scale=0.075]{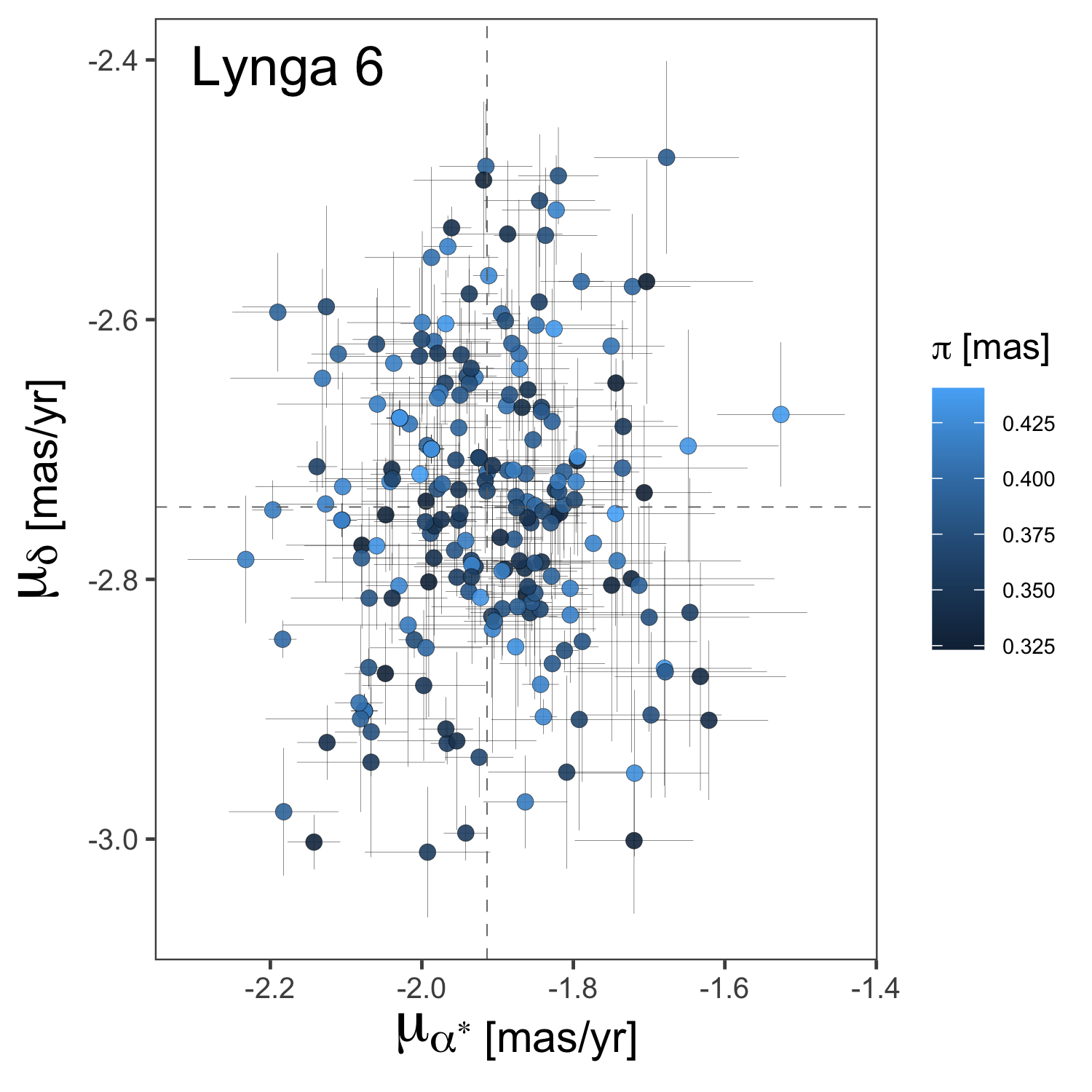}
\includegraphics[scale=0.075]{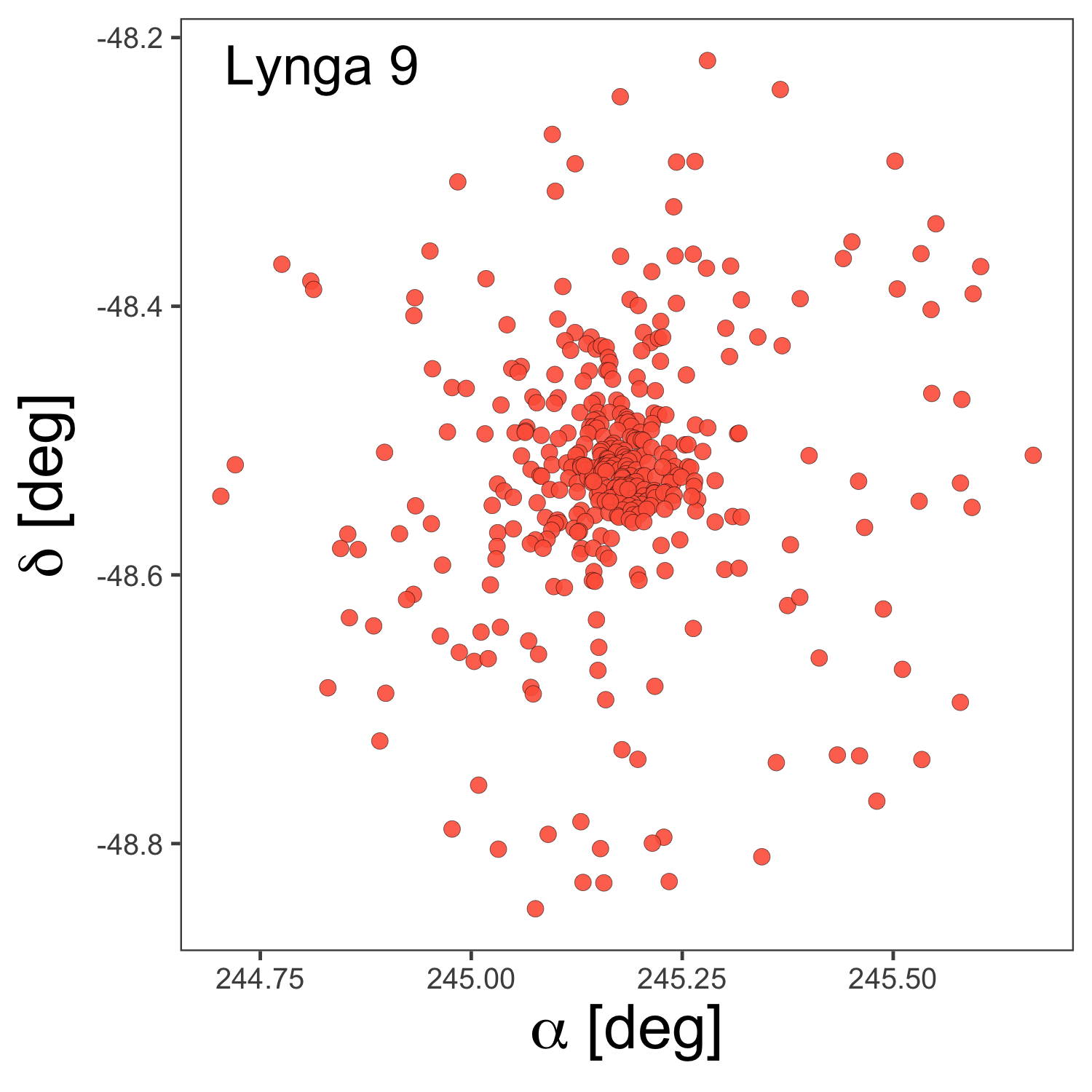}
\includegraphics[scale=0.075]{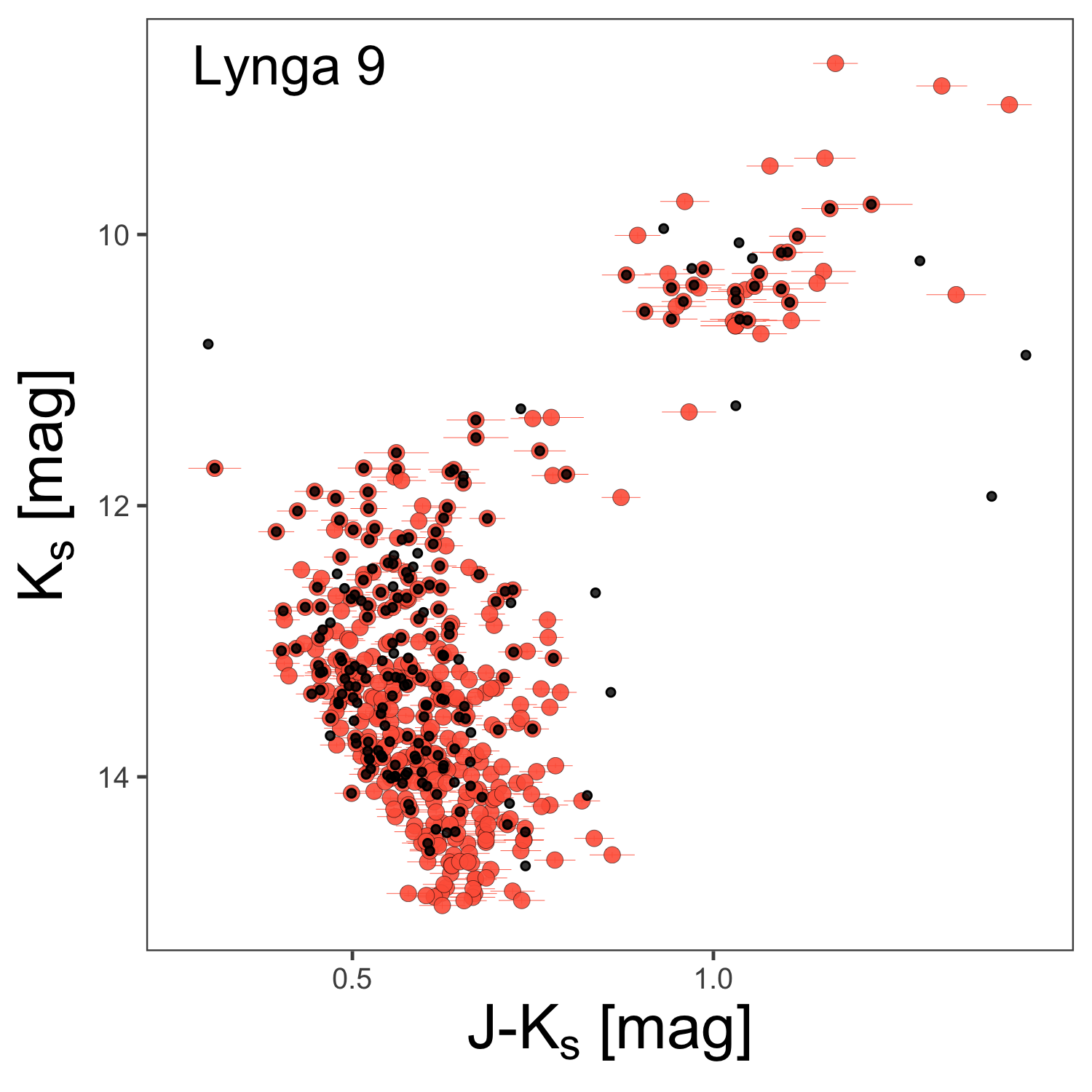}
\includegraphics[scale=0.075]{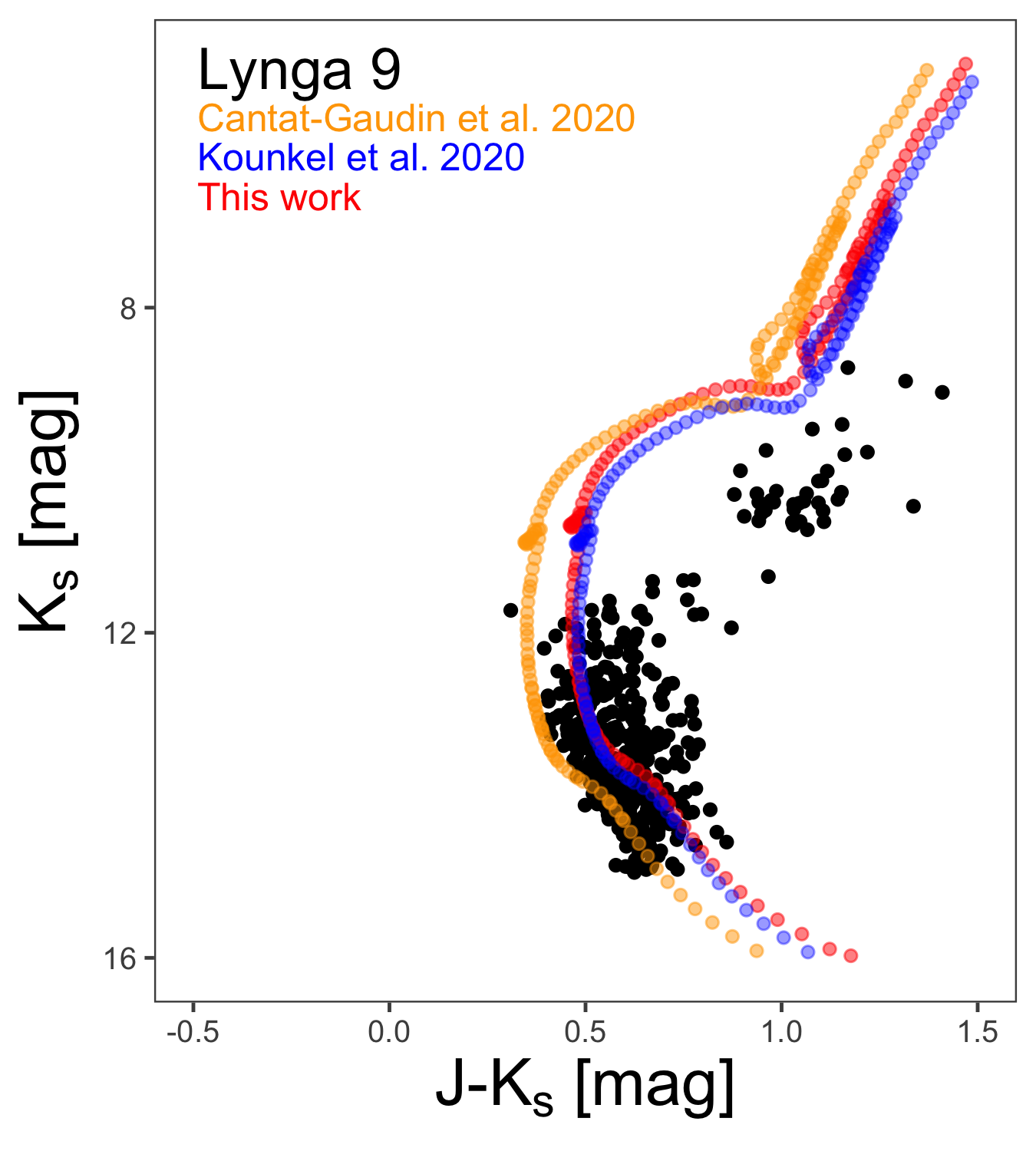}
\includegraphics[scale=0.075]{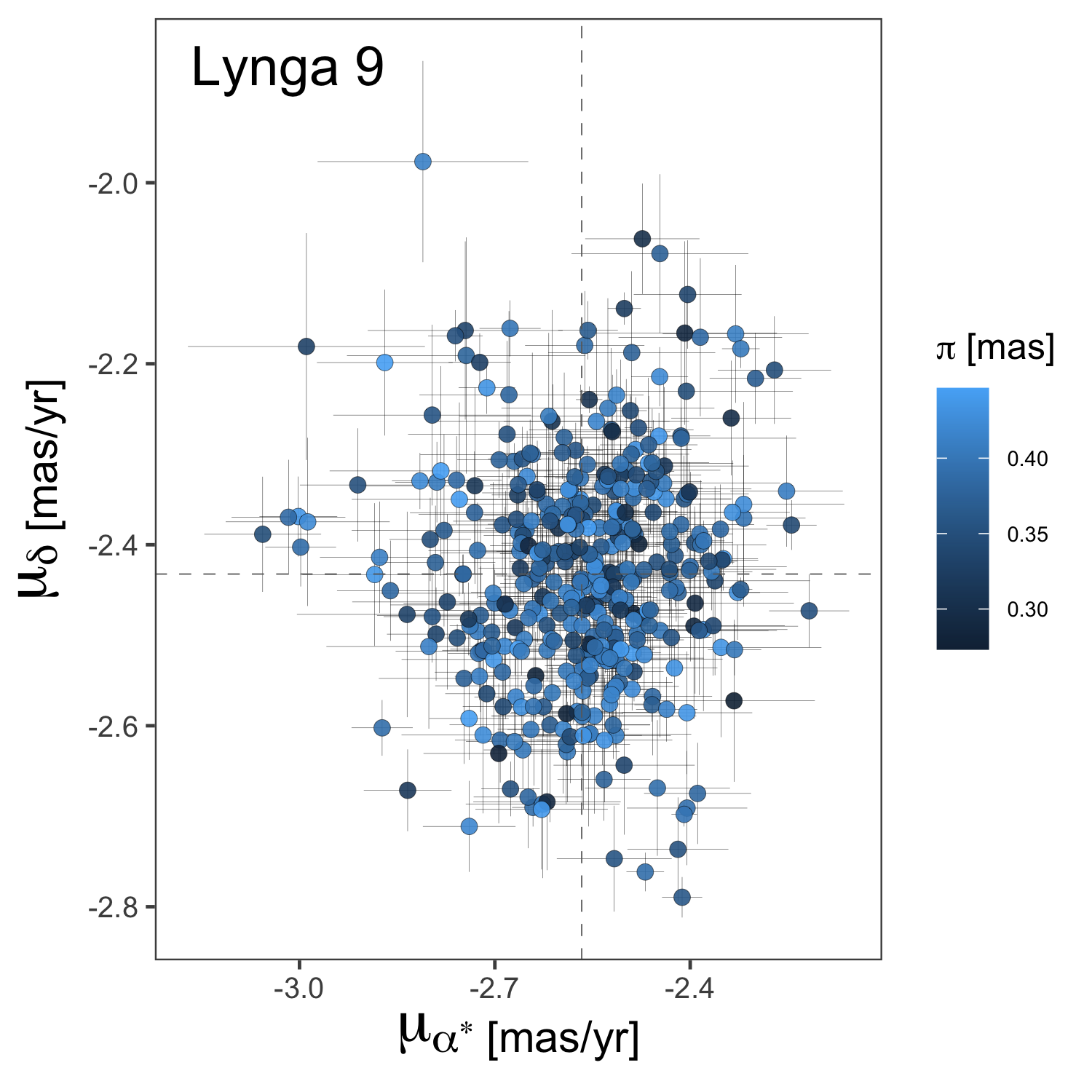}
\includegraphics[scale=0.075]{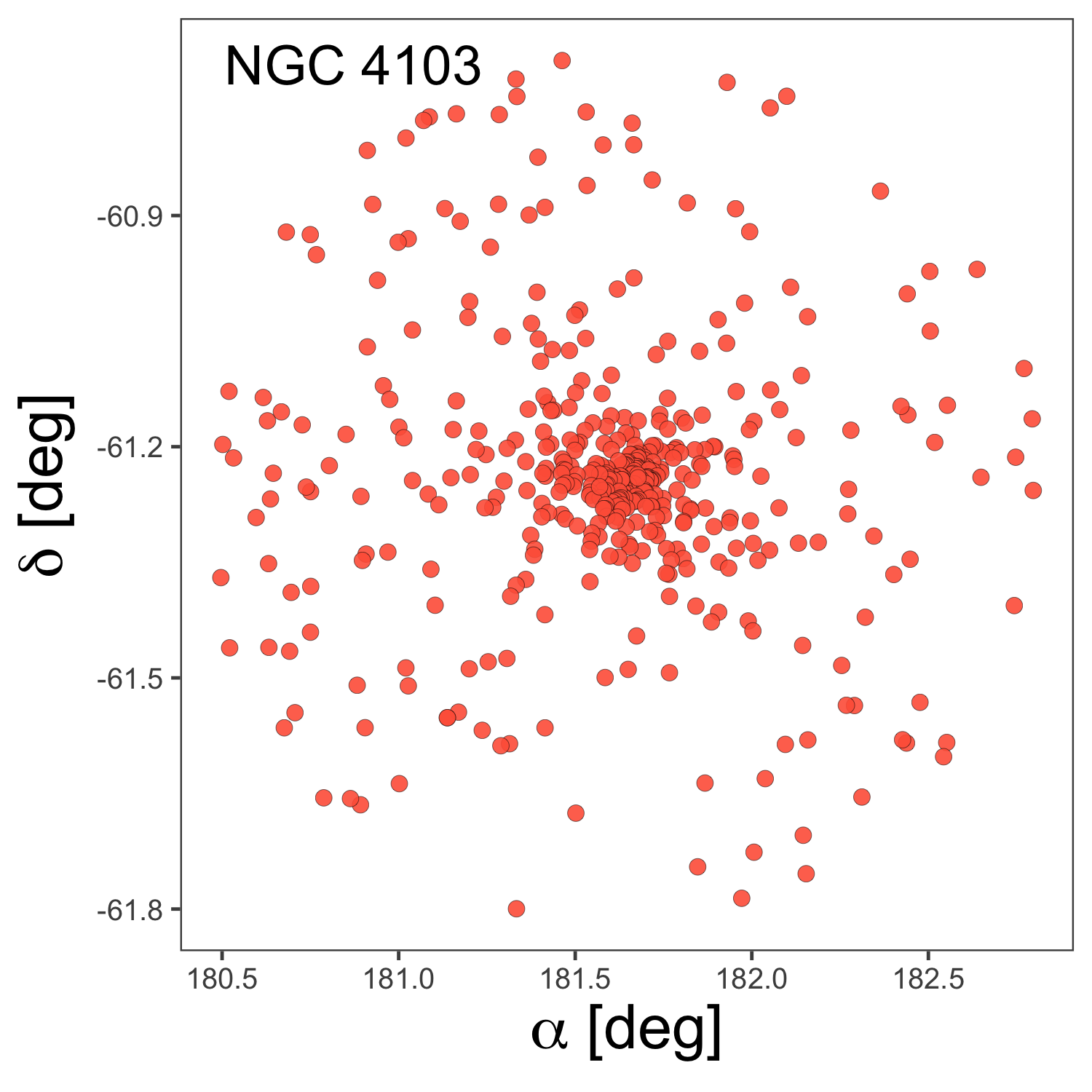}
\includegraphics[scale=0.075]{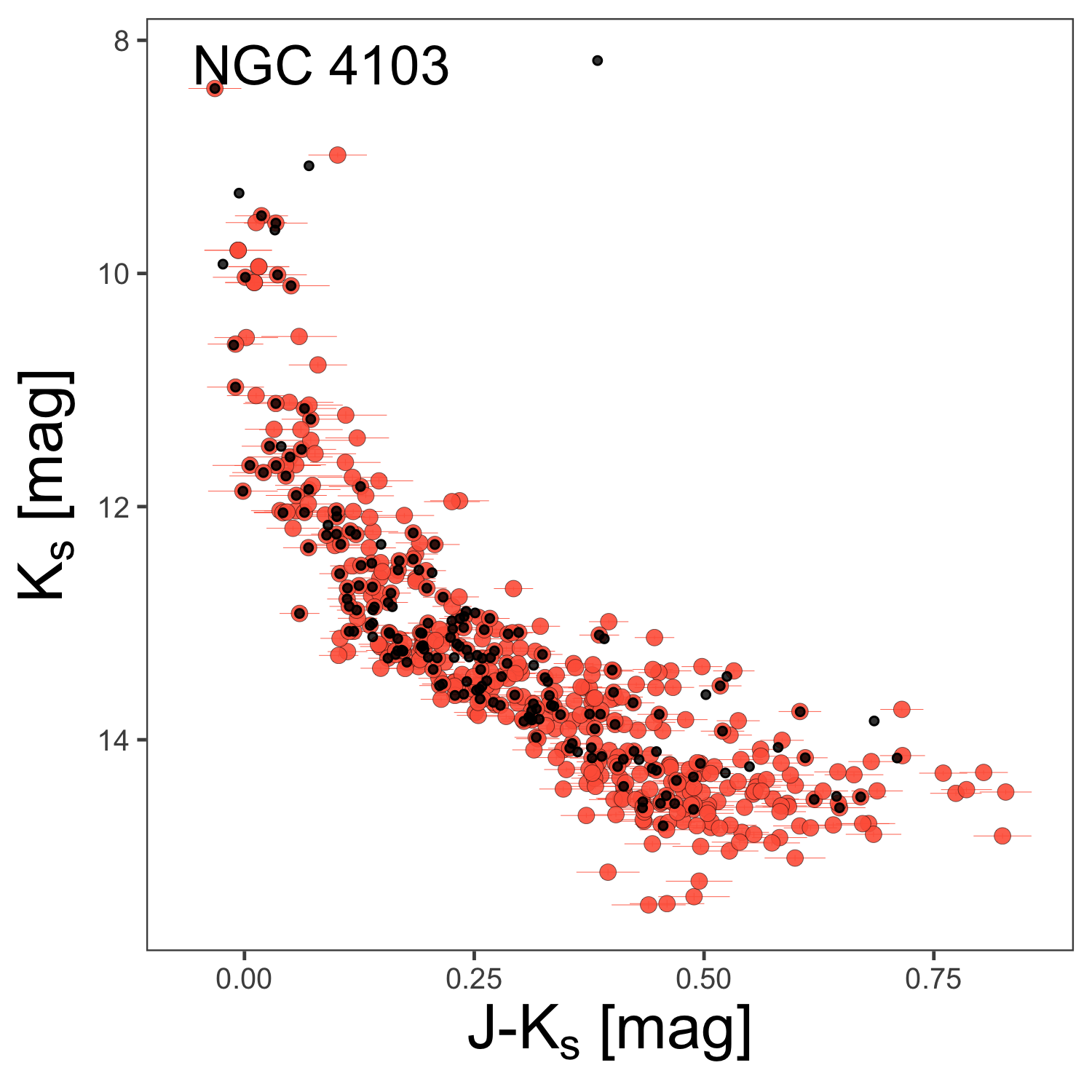}
\includegraphics[scale=0.075]{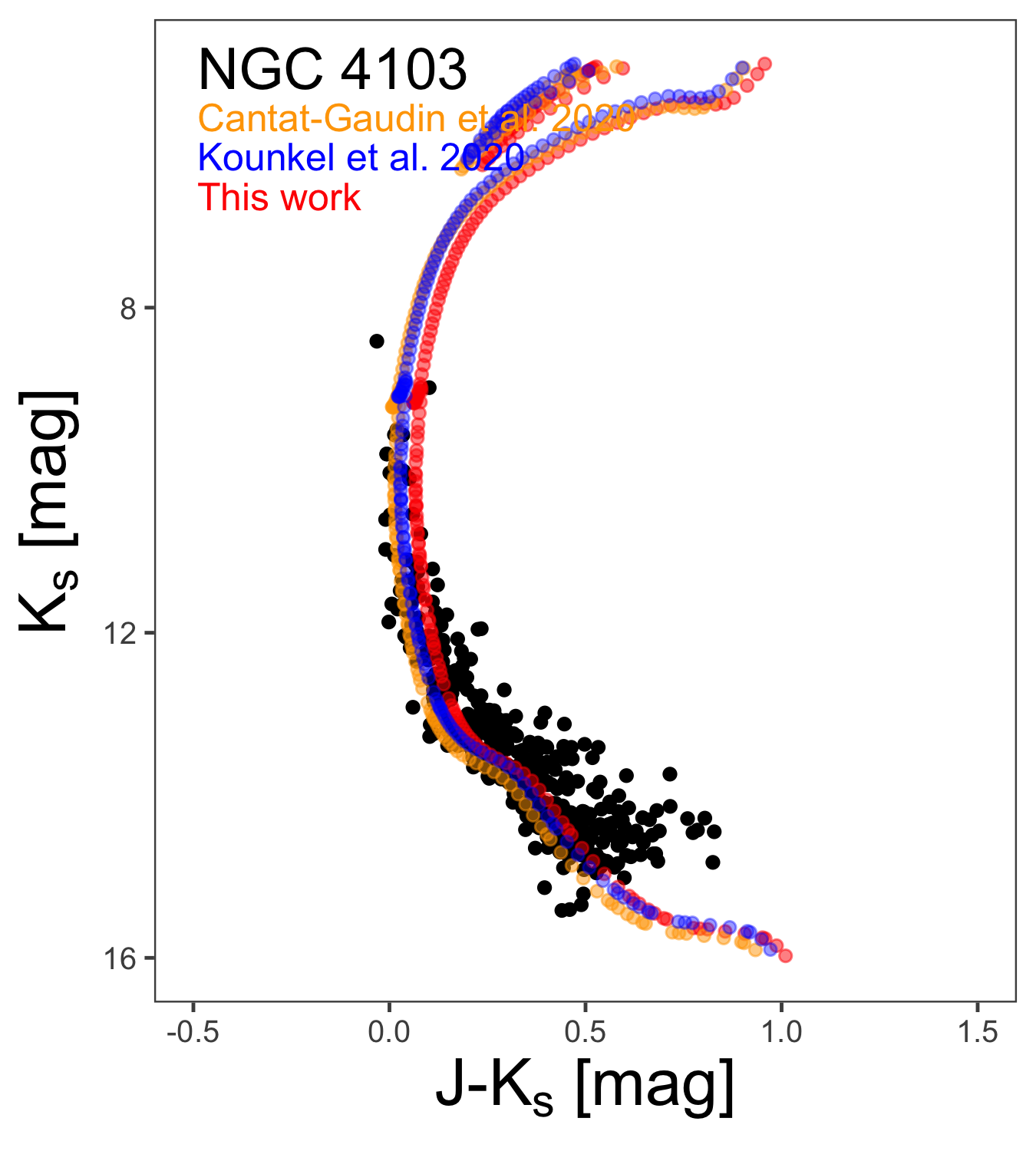}
\includegraphics[scale=0.075]{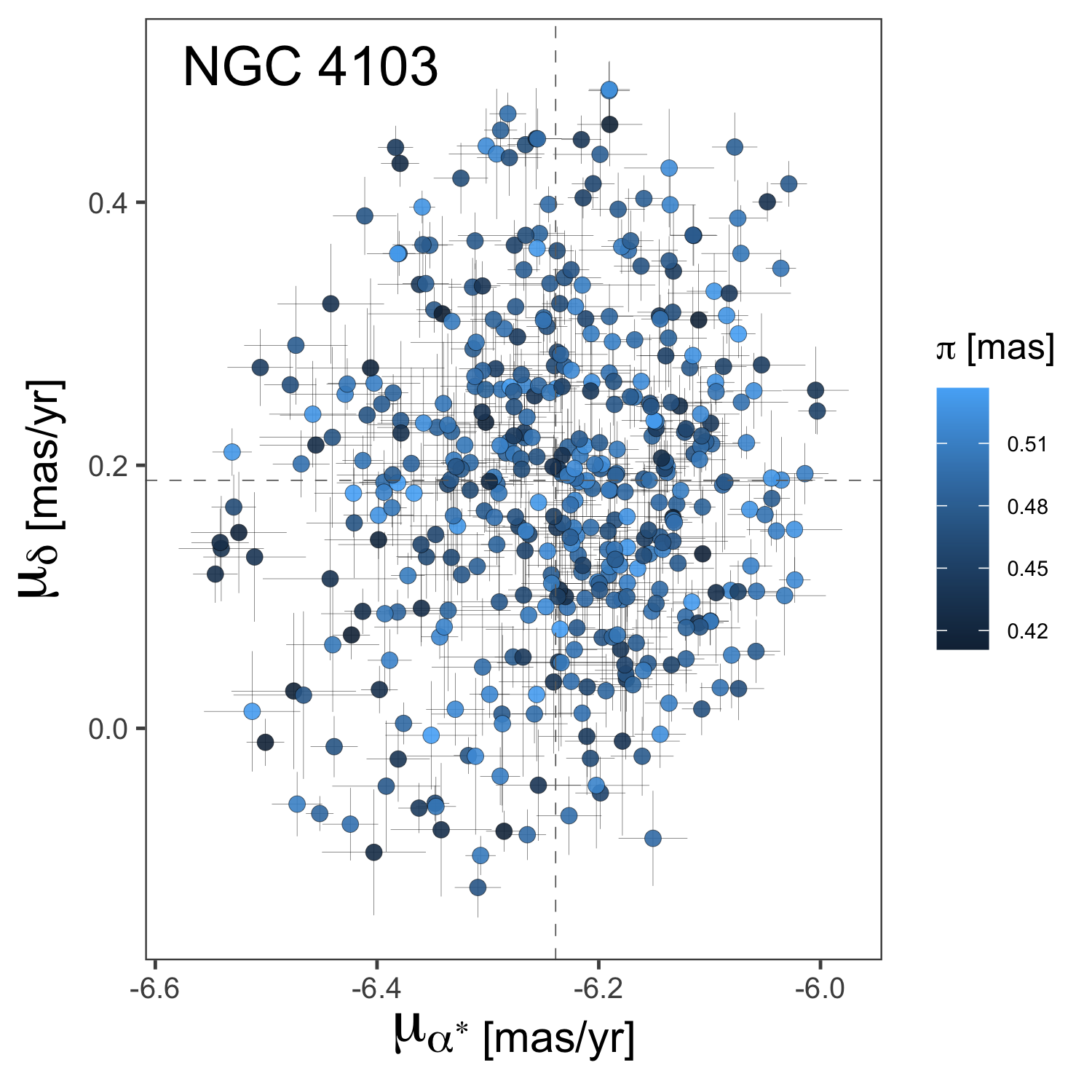}
\includegraphics[scale=0.075]{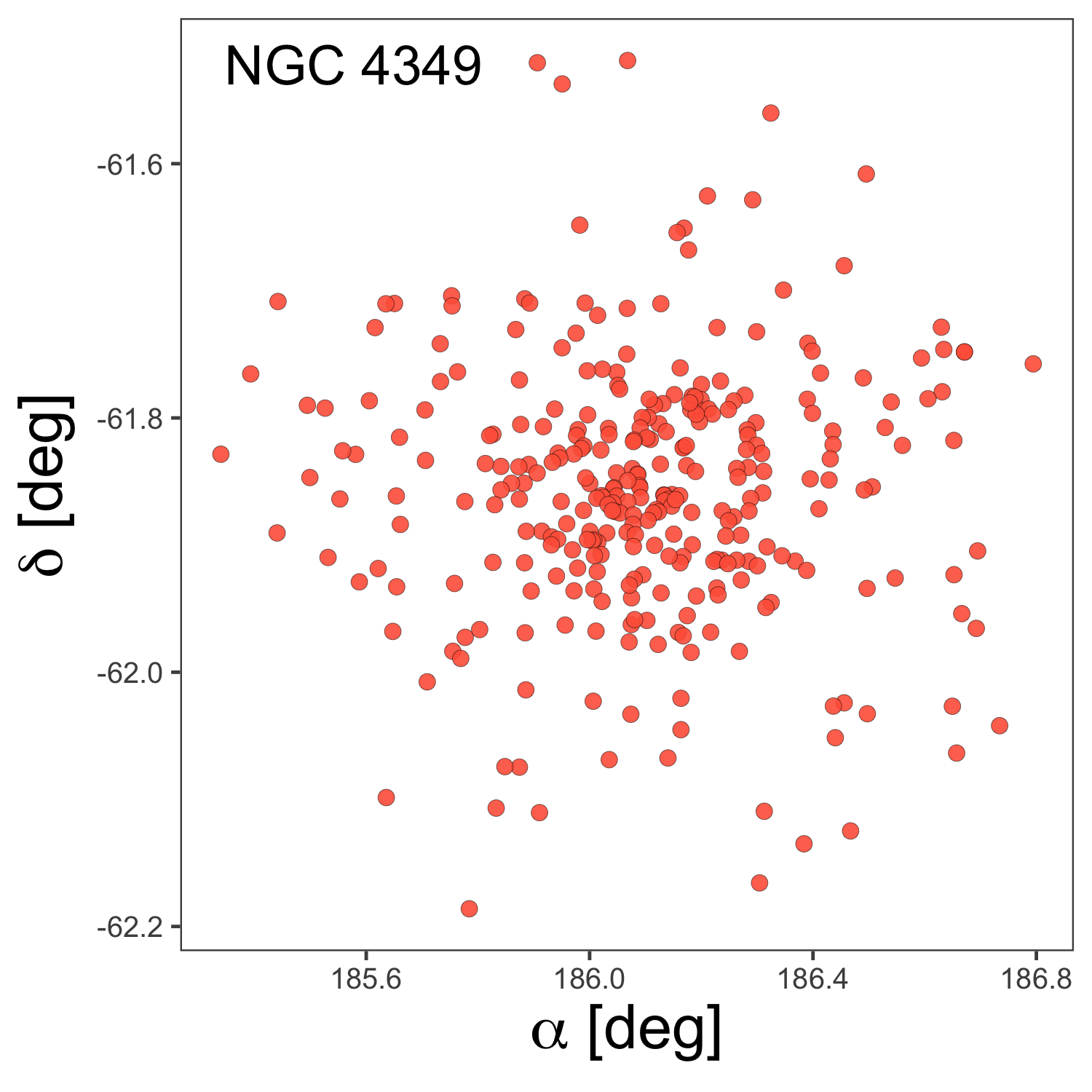}
\includegraphics[scale=0.075]{NGC_4349pub_cmd.png}
\includegraphics[scale=0.075]{NGC_4349_iso.png}
\includegraphics[scale=0.075]{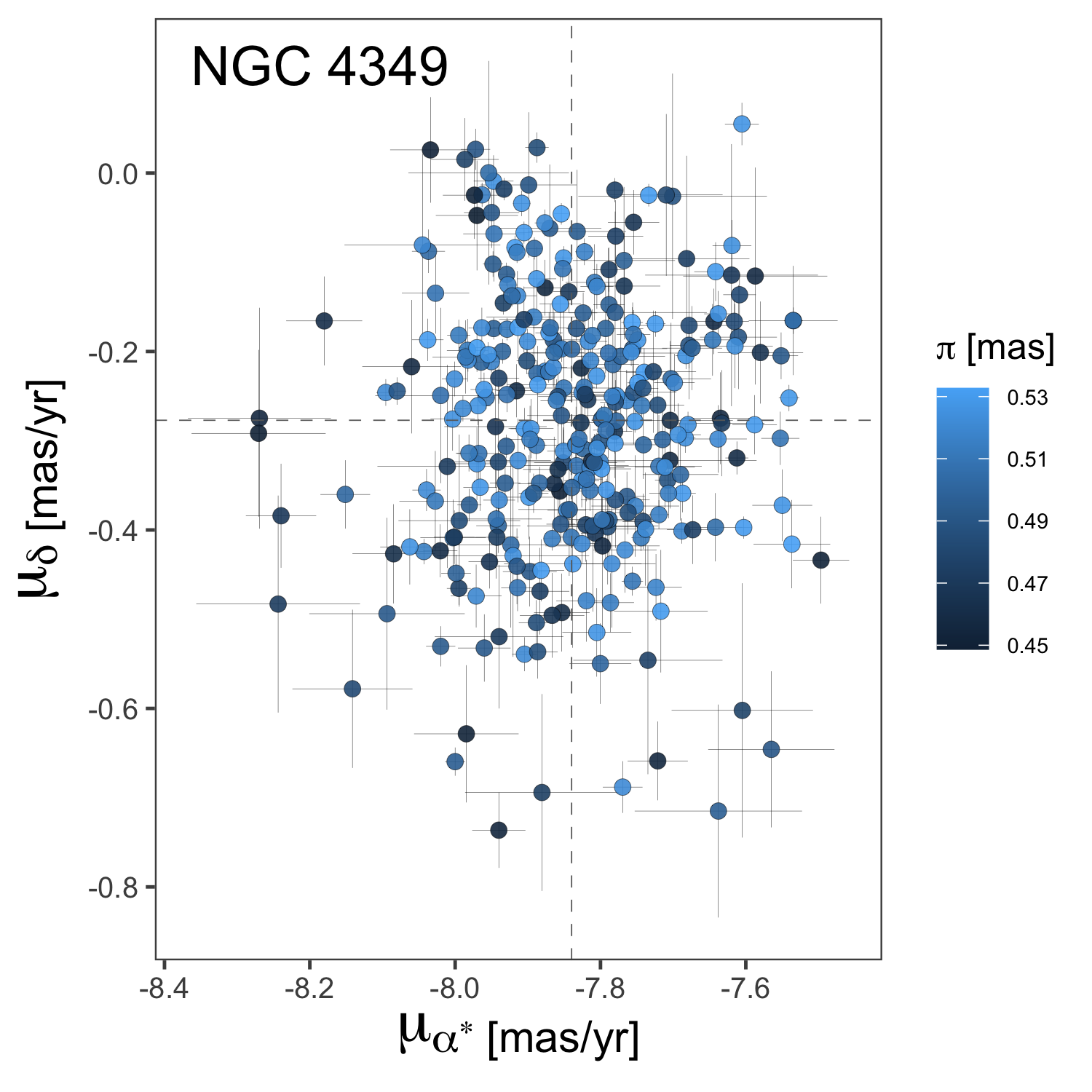}
\includegraphics[scale=0.075]{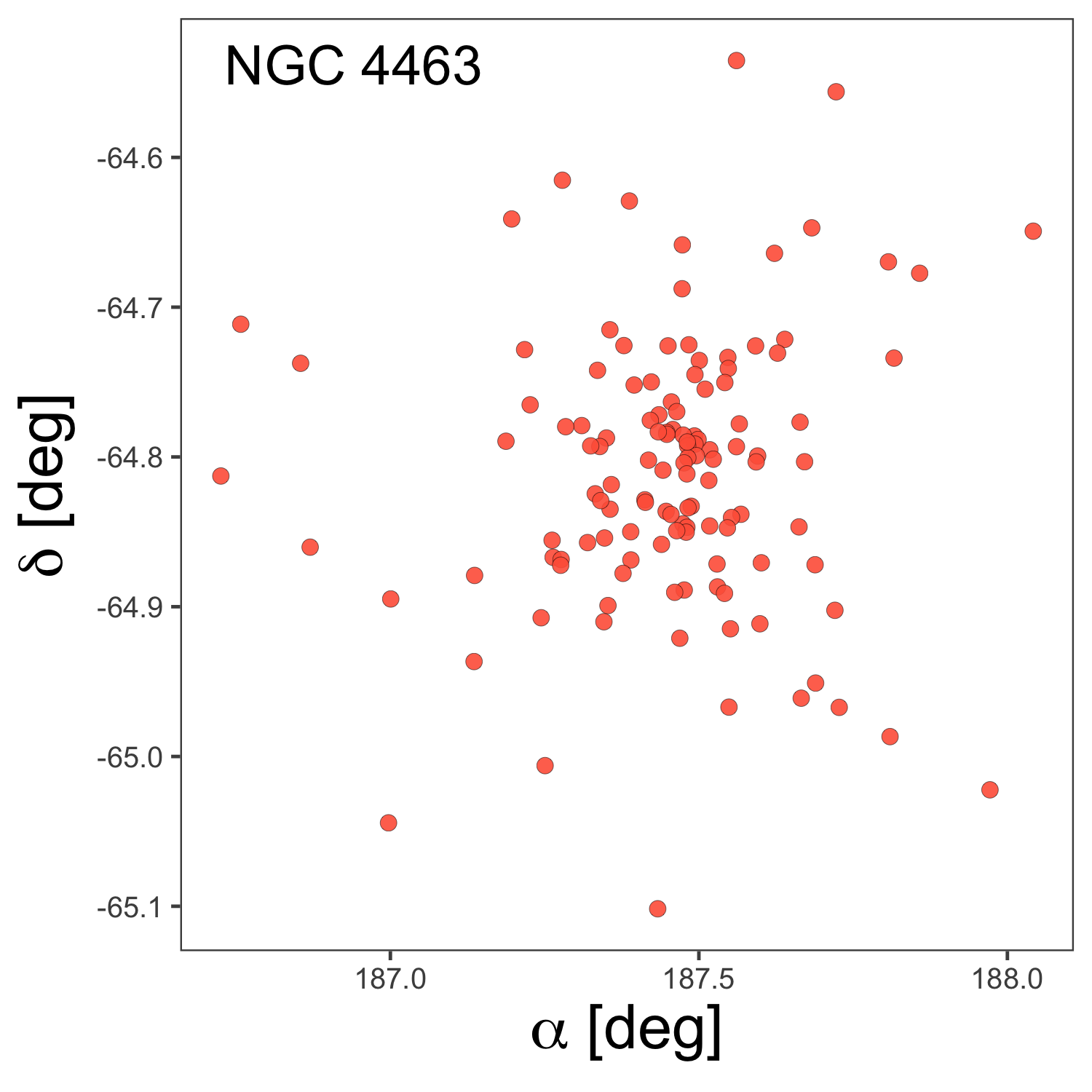}
\includegraphics[scale=0.075]{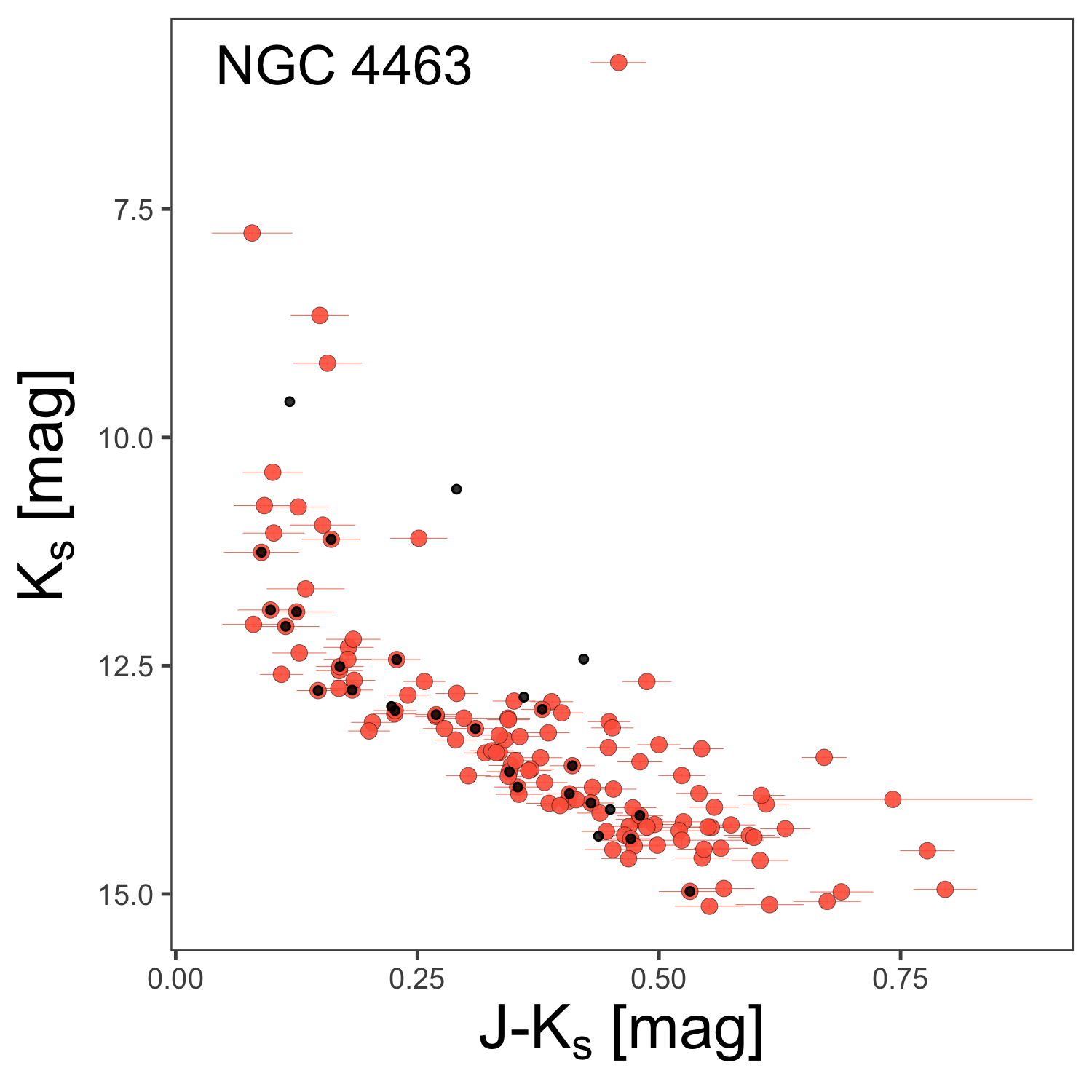}
\includegraphics[scale=0.075]{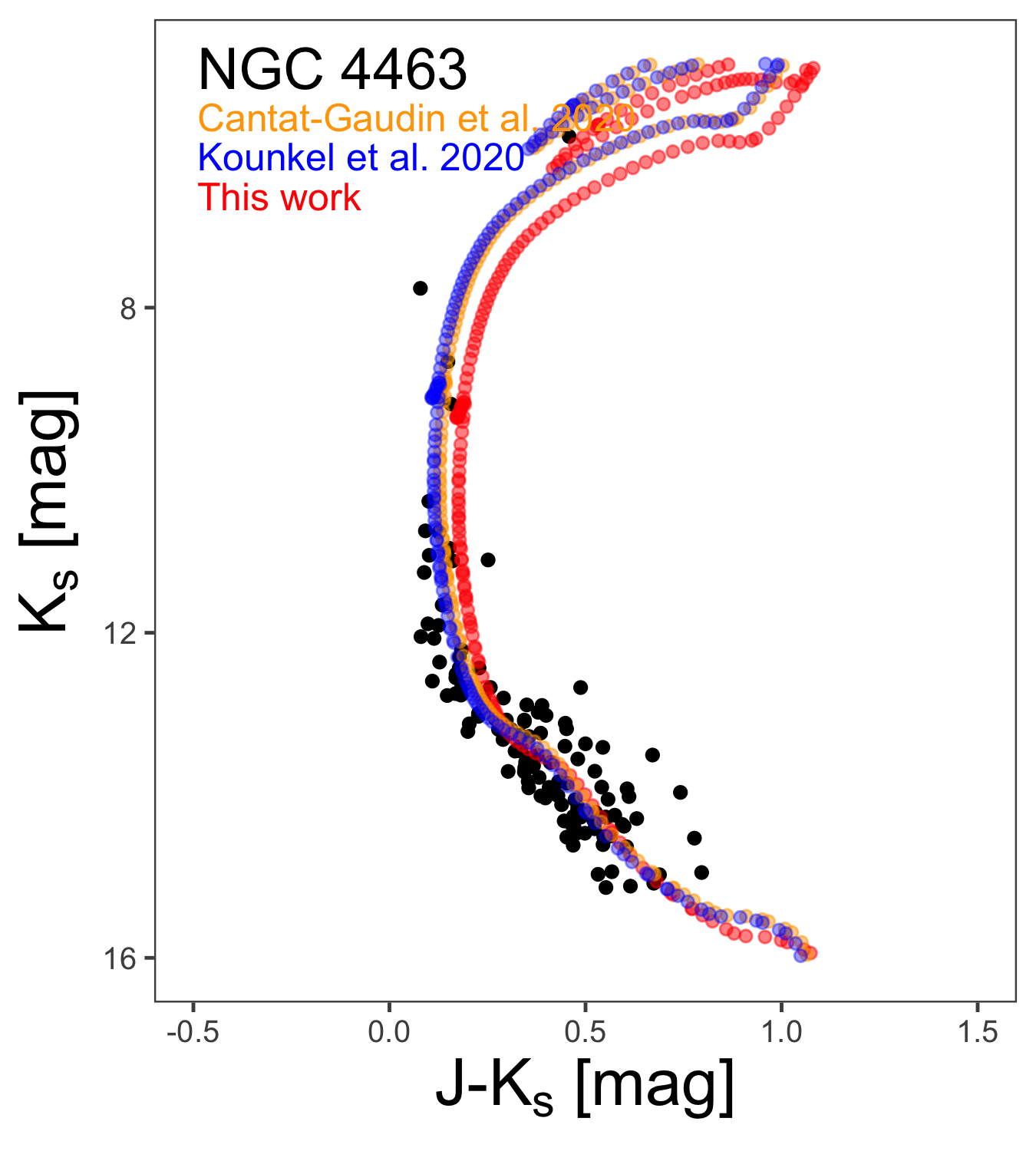}
\includegraphics[scale=0.075]{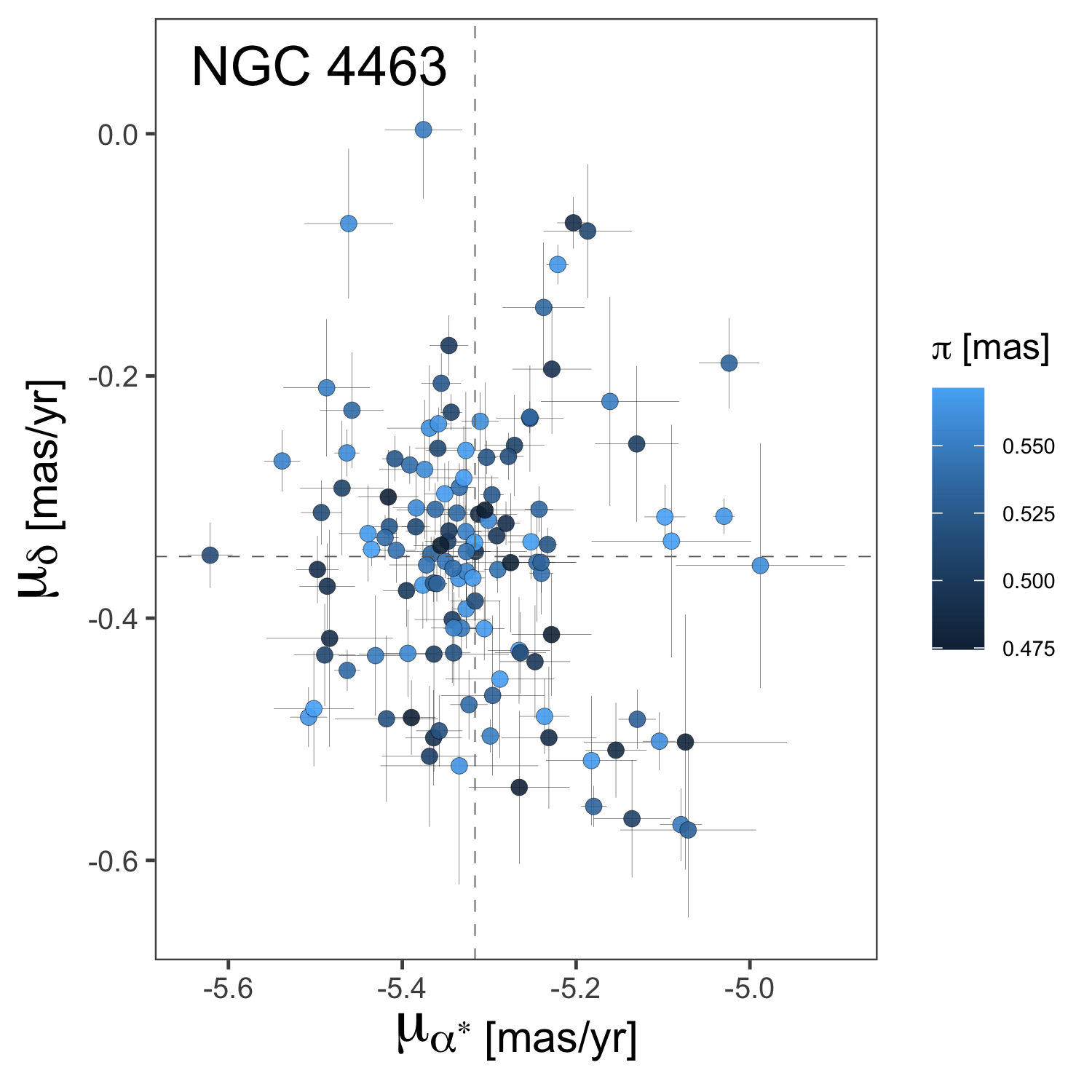}
\includegraphics[scale=0.075]{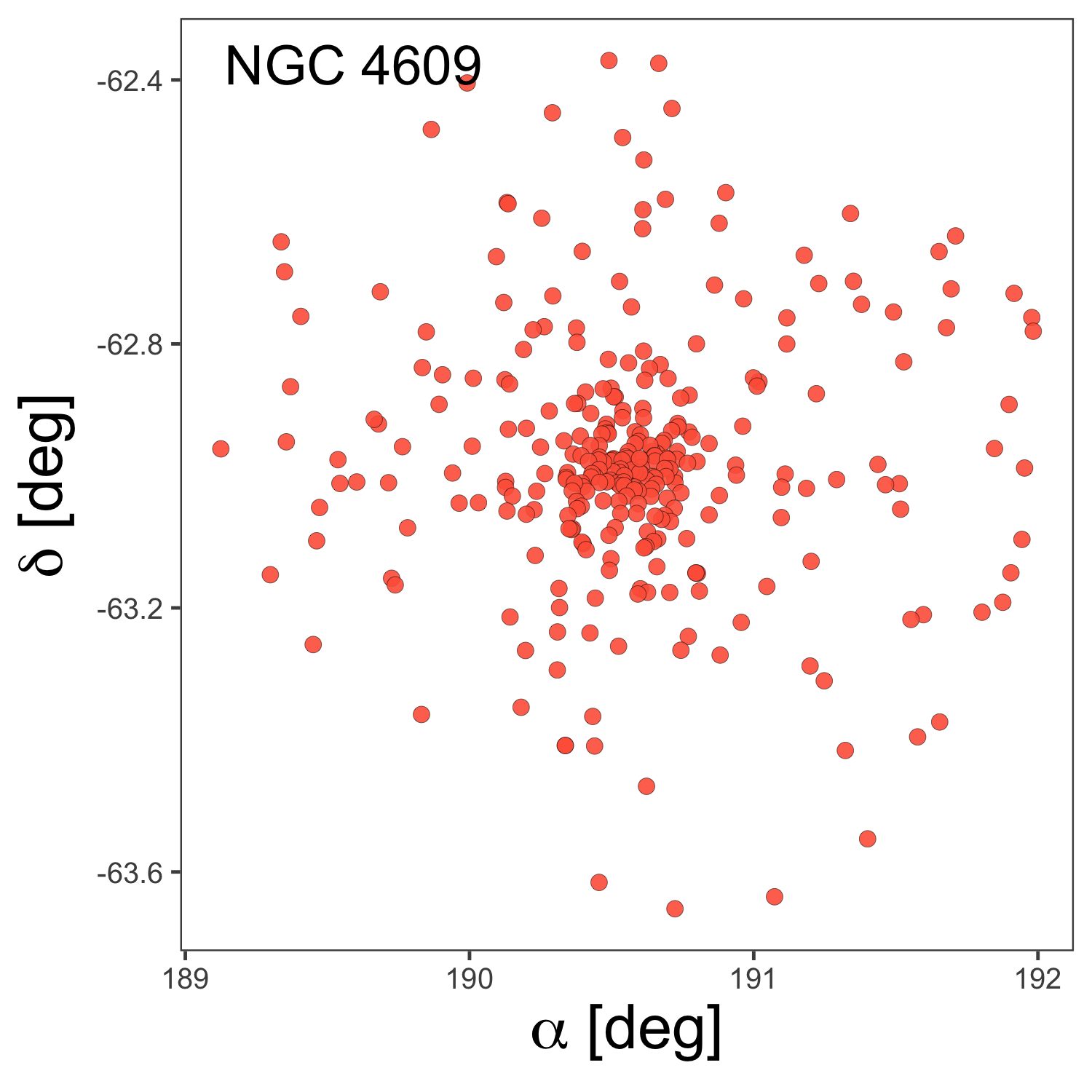}
\includegraphics[scale=0.075]{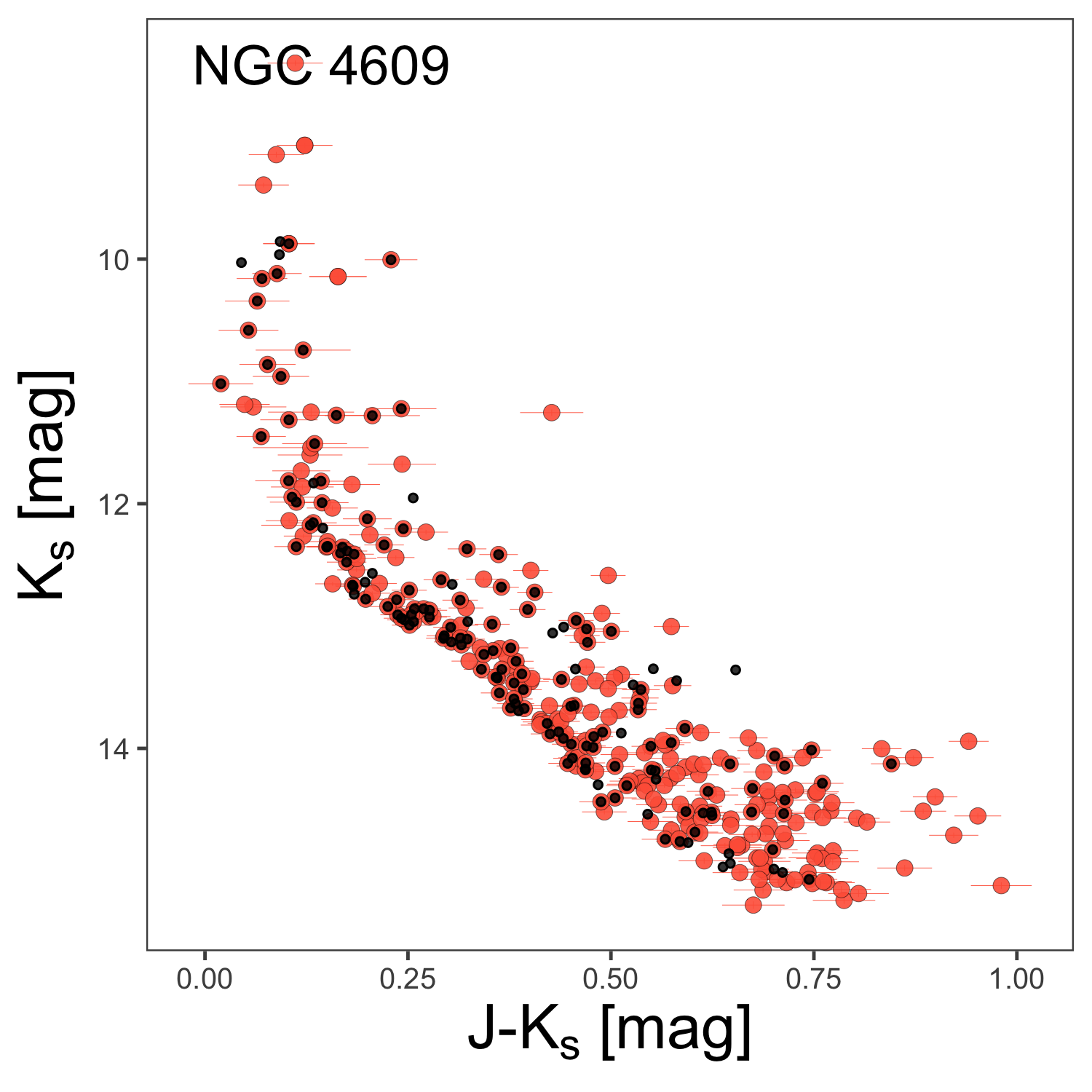}
\includegraphics[scale=0.075]{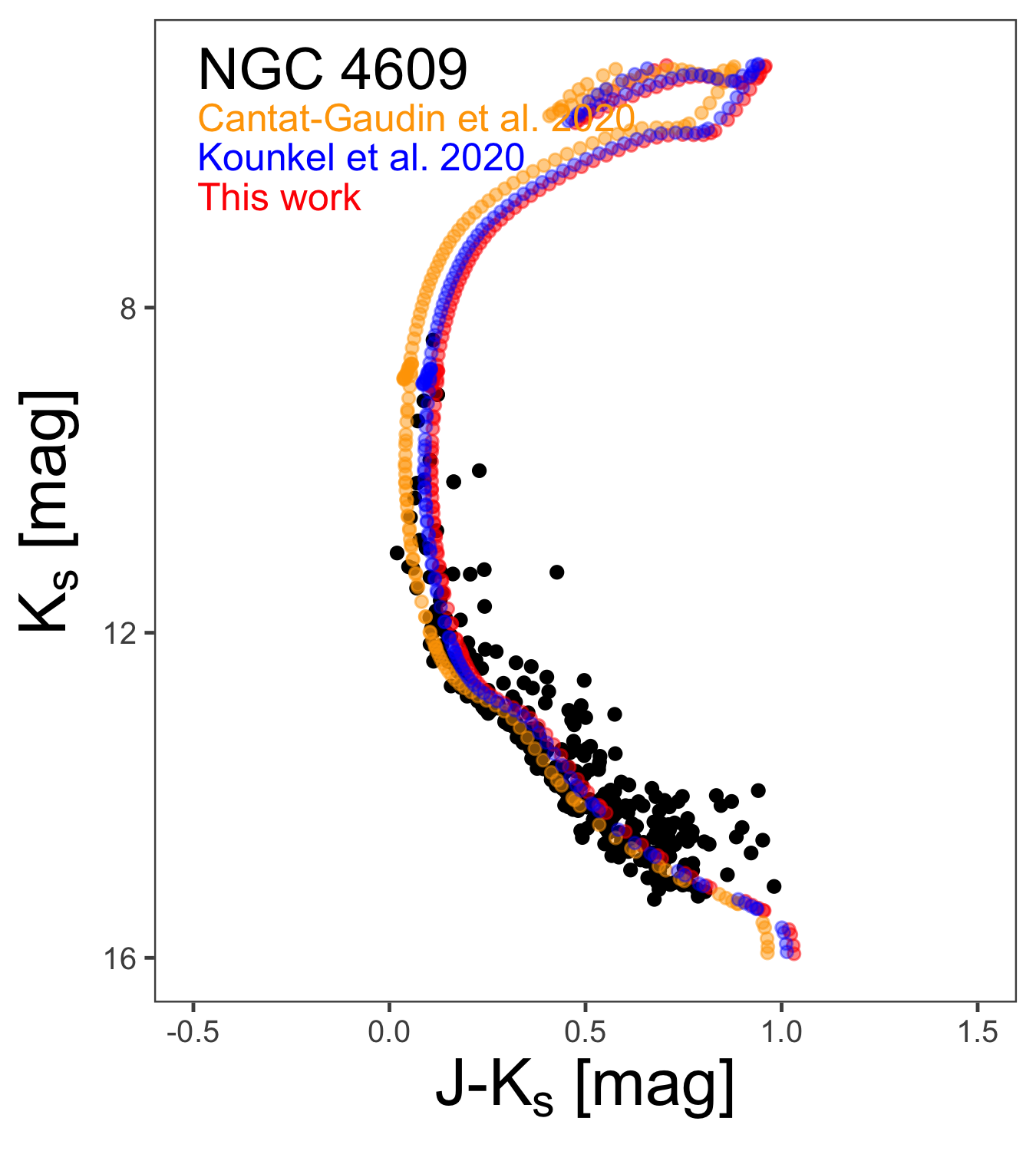}
\includegraphics[scale=0.075]{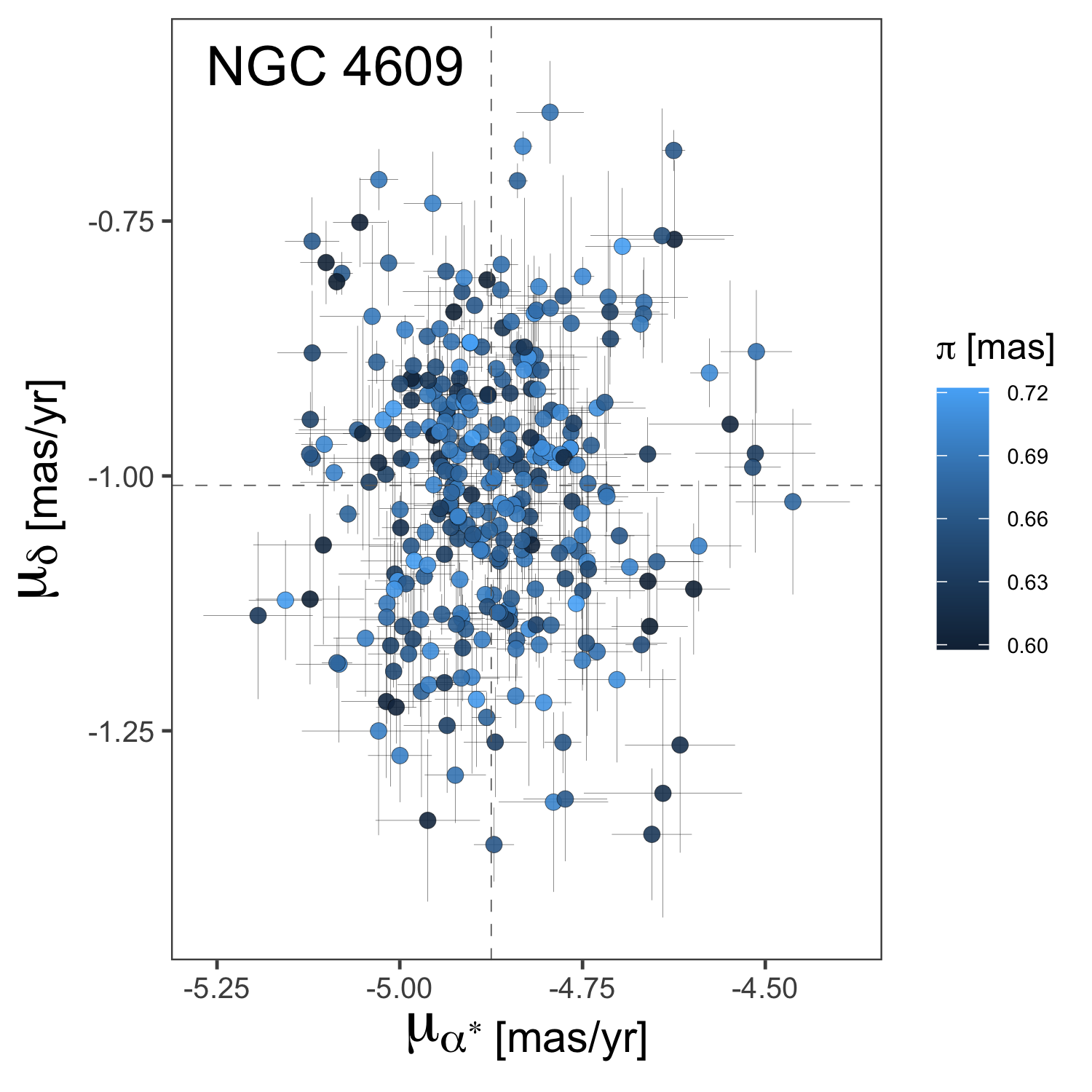}
\caption{Same as Figure\,\ref{fig:clusters}}
\end{figure*}

\begin{figure*}
\includegraphics[scale=0.075]{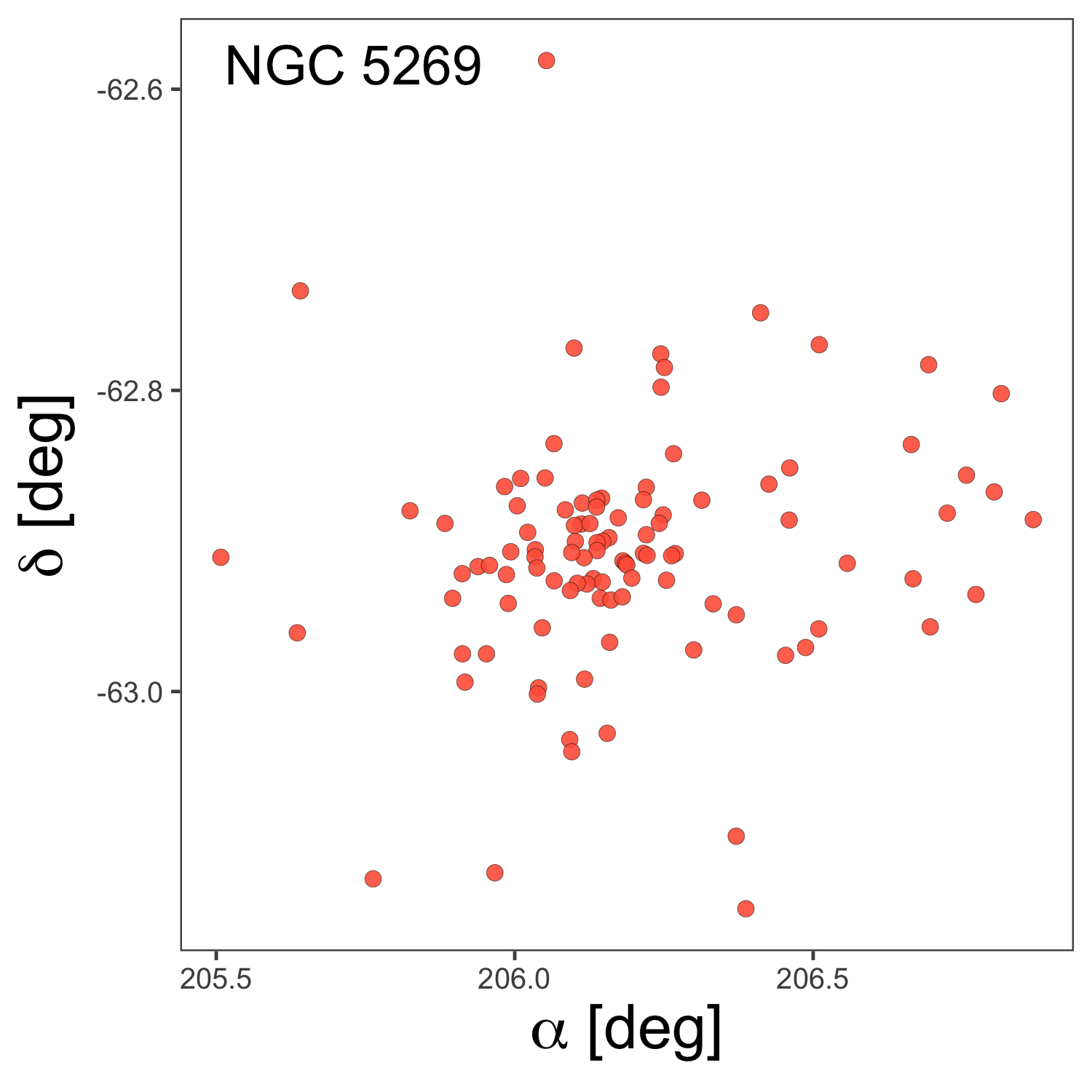}
\includegraphics[scale=0.075]{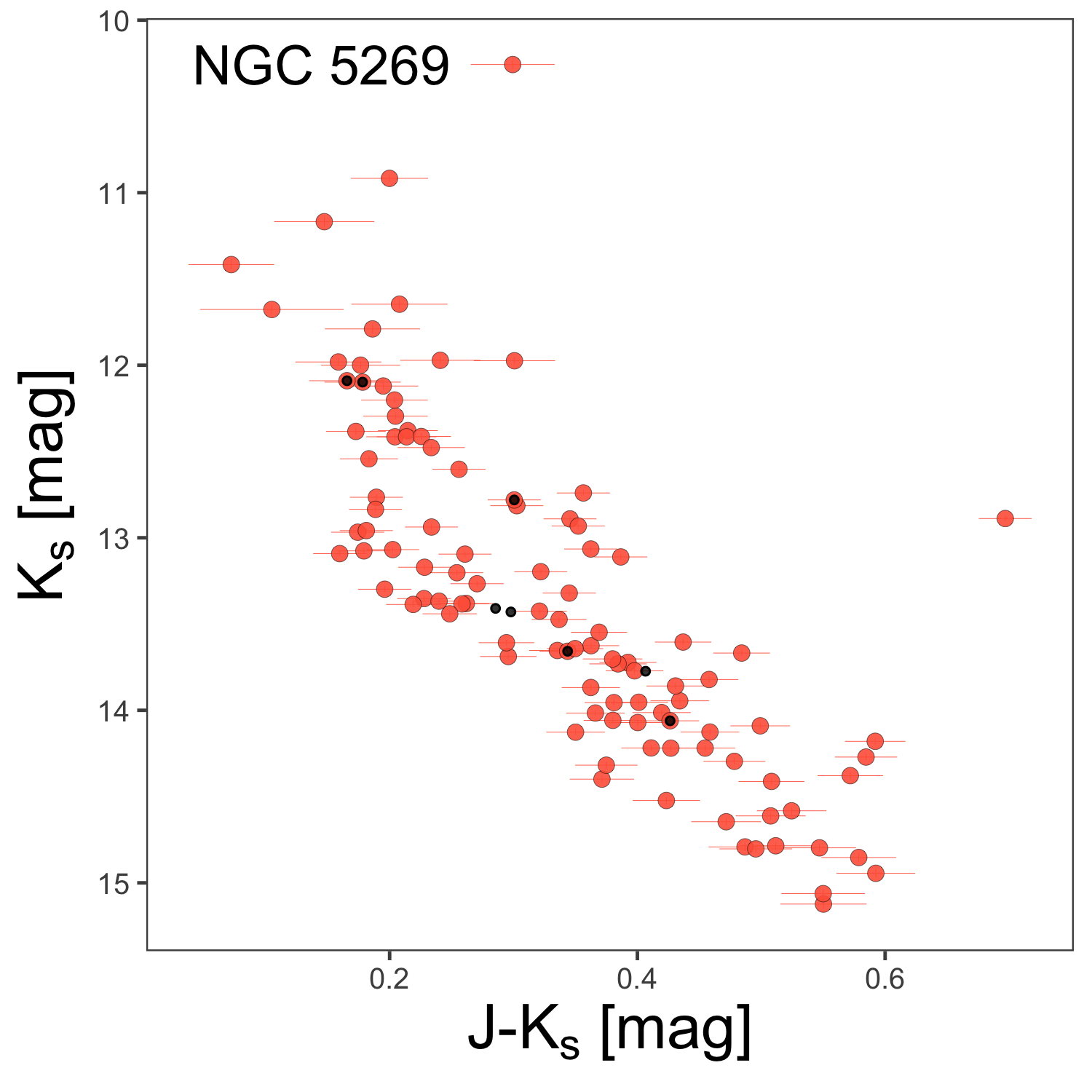}
\includegraphics[scale=0.075]{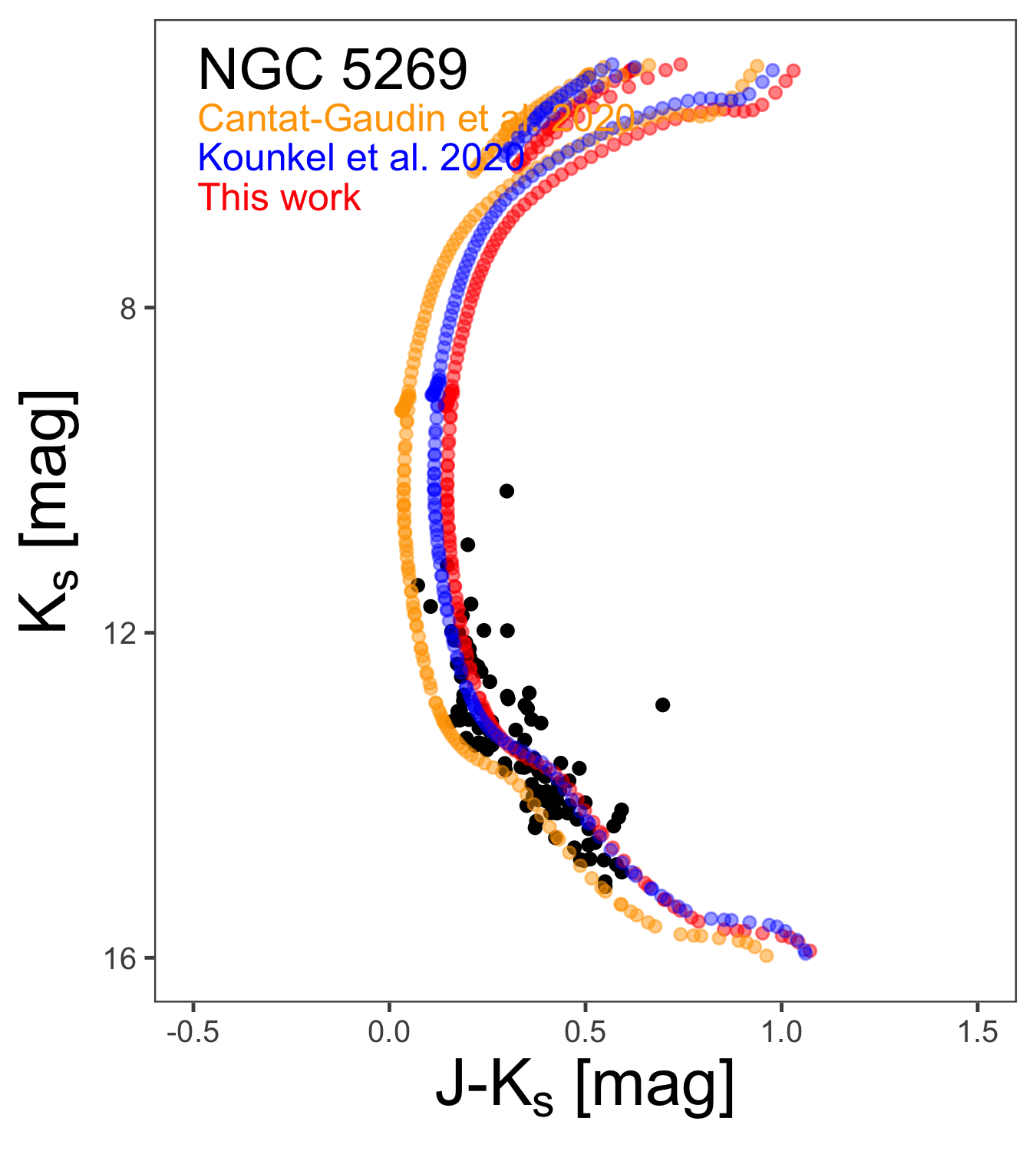}
\includegraphics[scale=0.075]{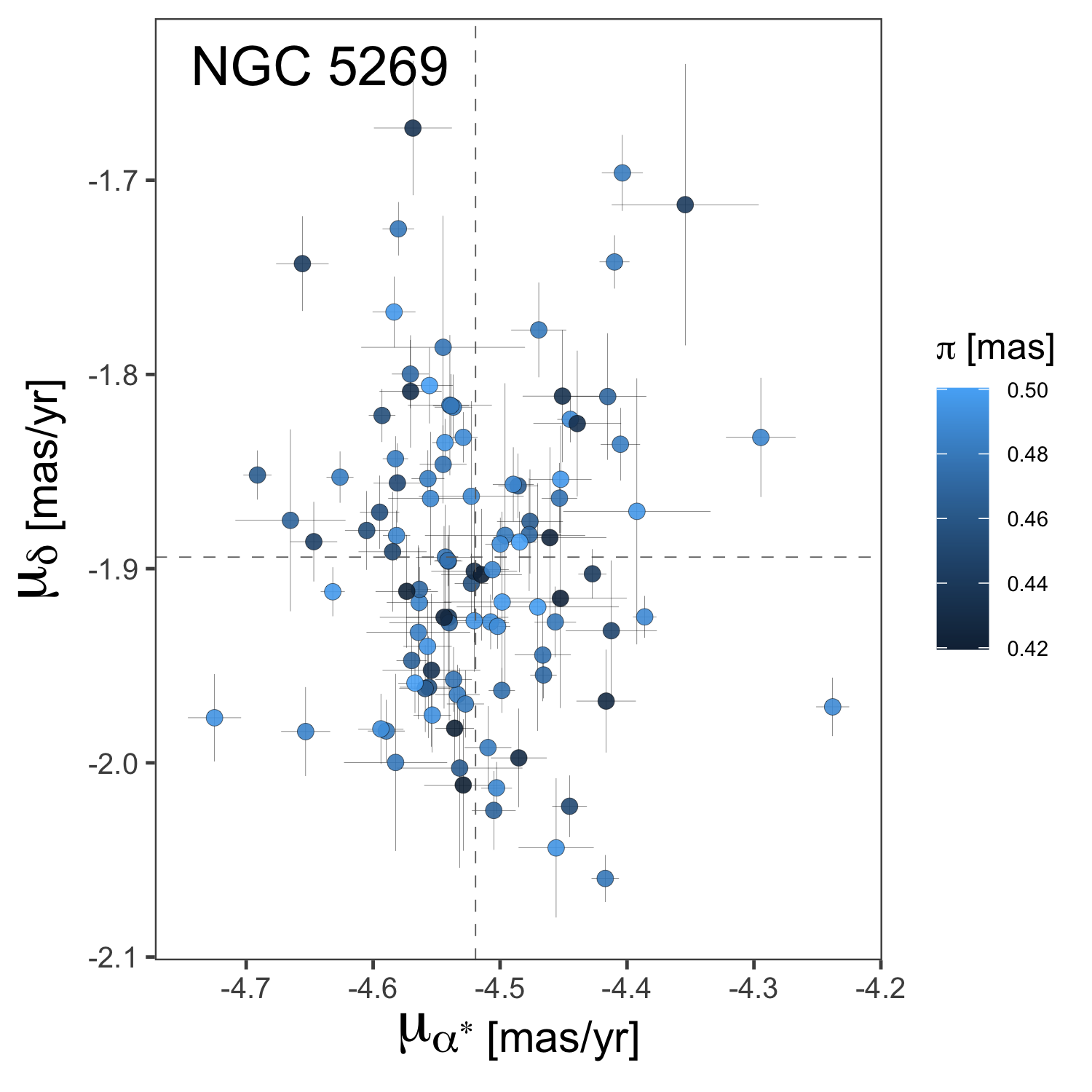}
\includegraphics[scale=0.075]{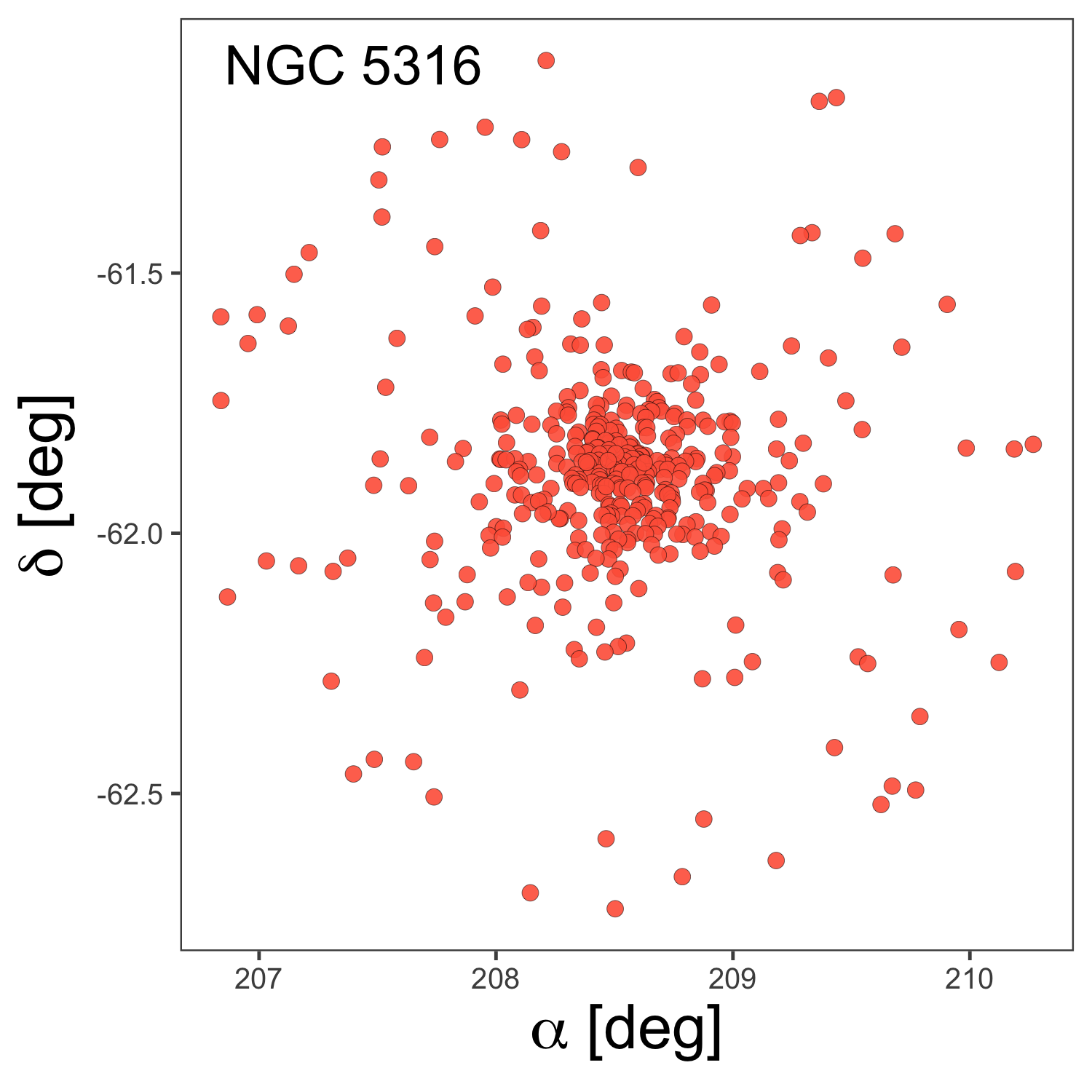}
\includegraphics[scale=0.075]{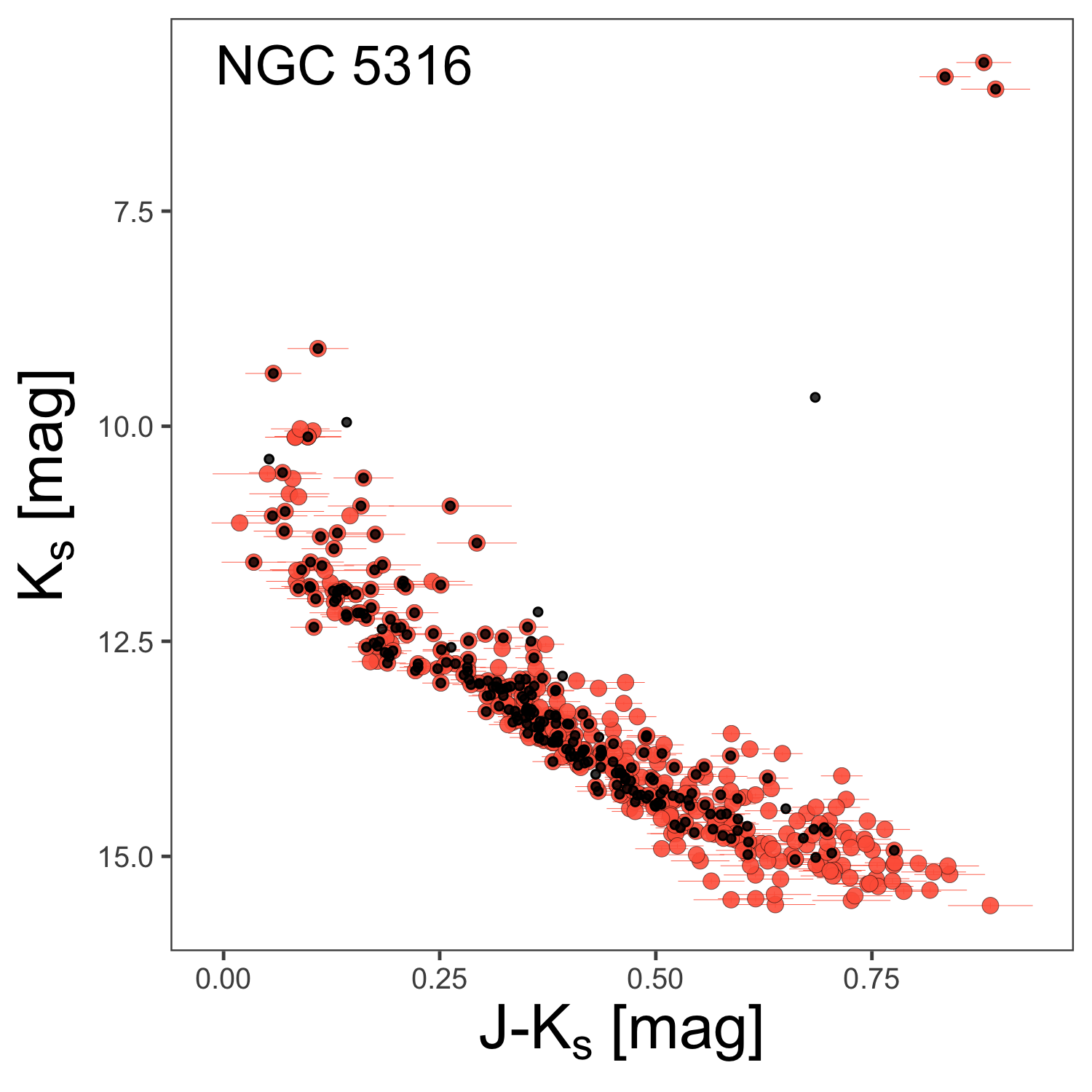}
\includegraphics[scale=0.075]{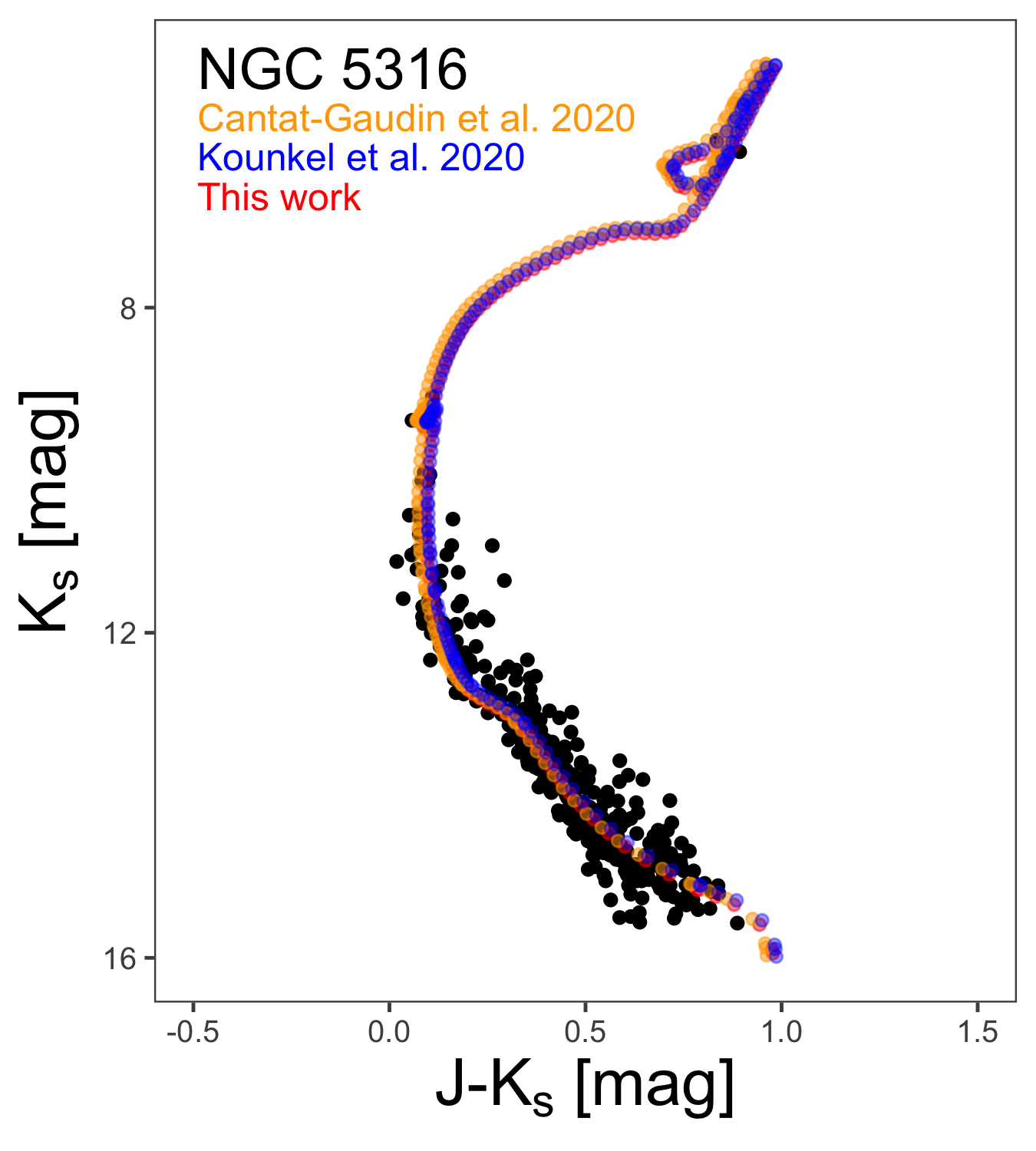}
\includegraphics[scale=0.075]{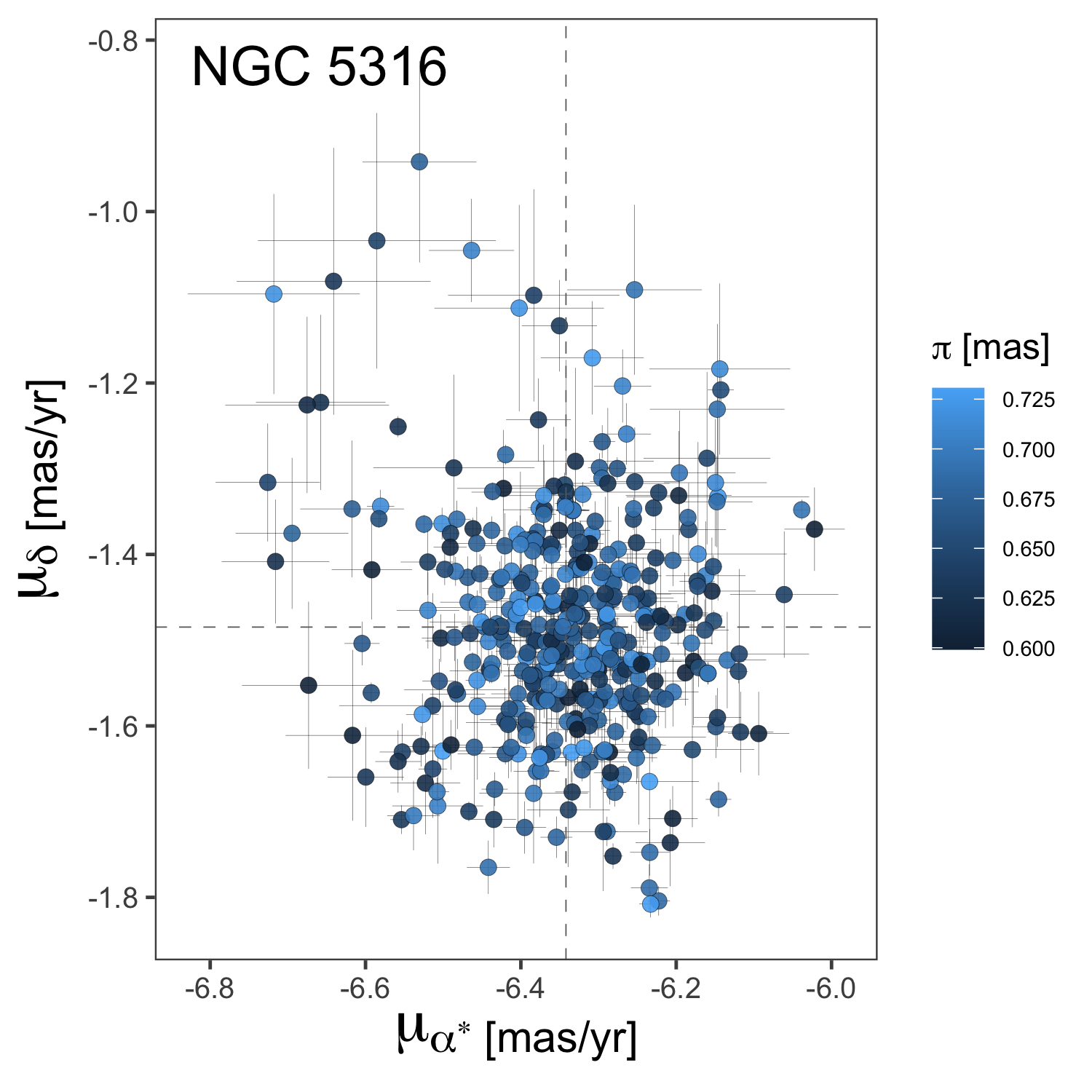}
\includegraphics[scale=0.075]{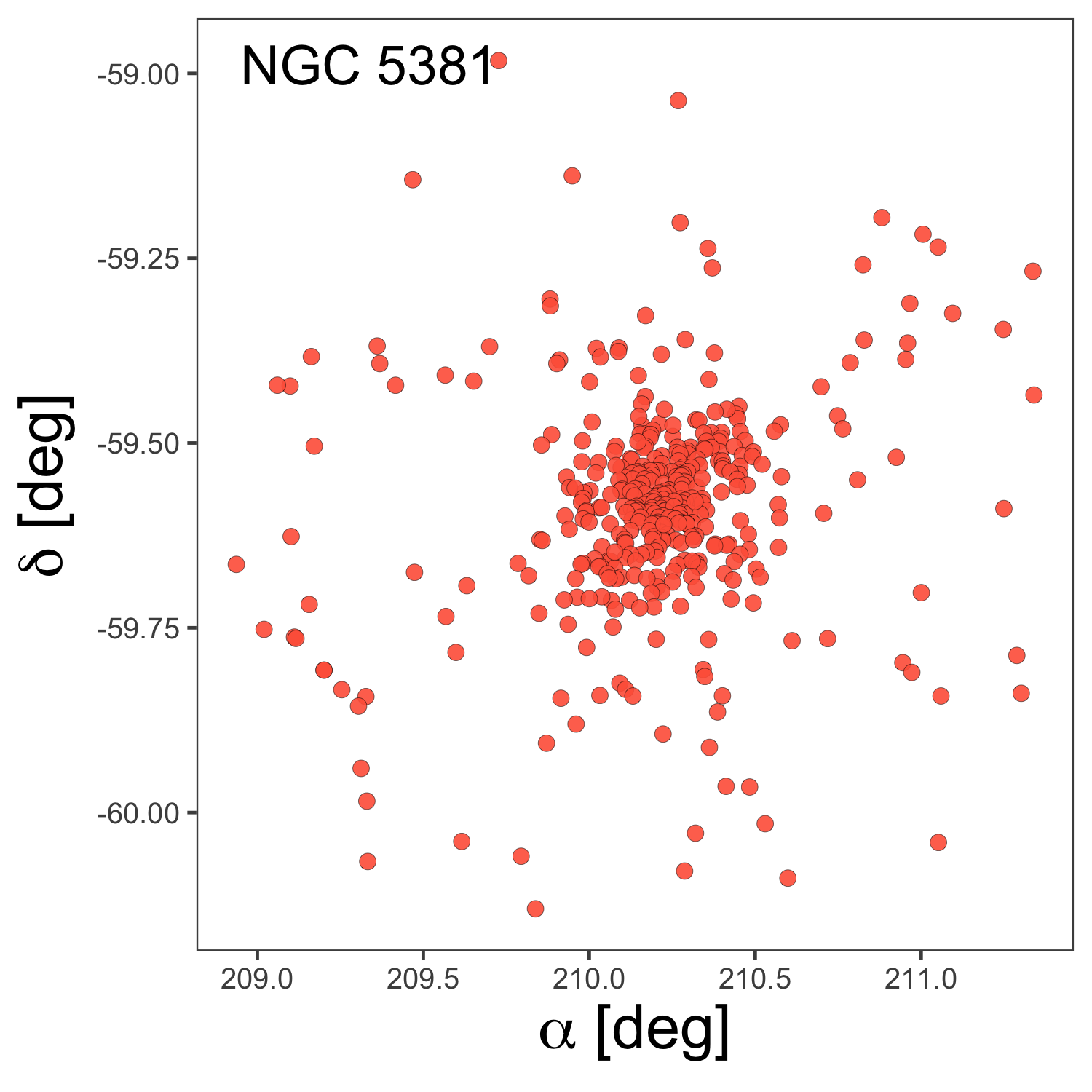}
\includegraphics[scale=0.075]{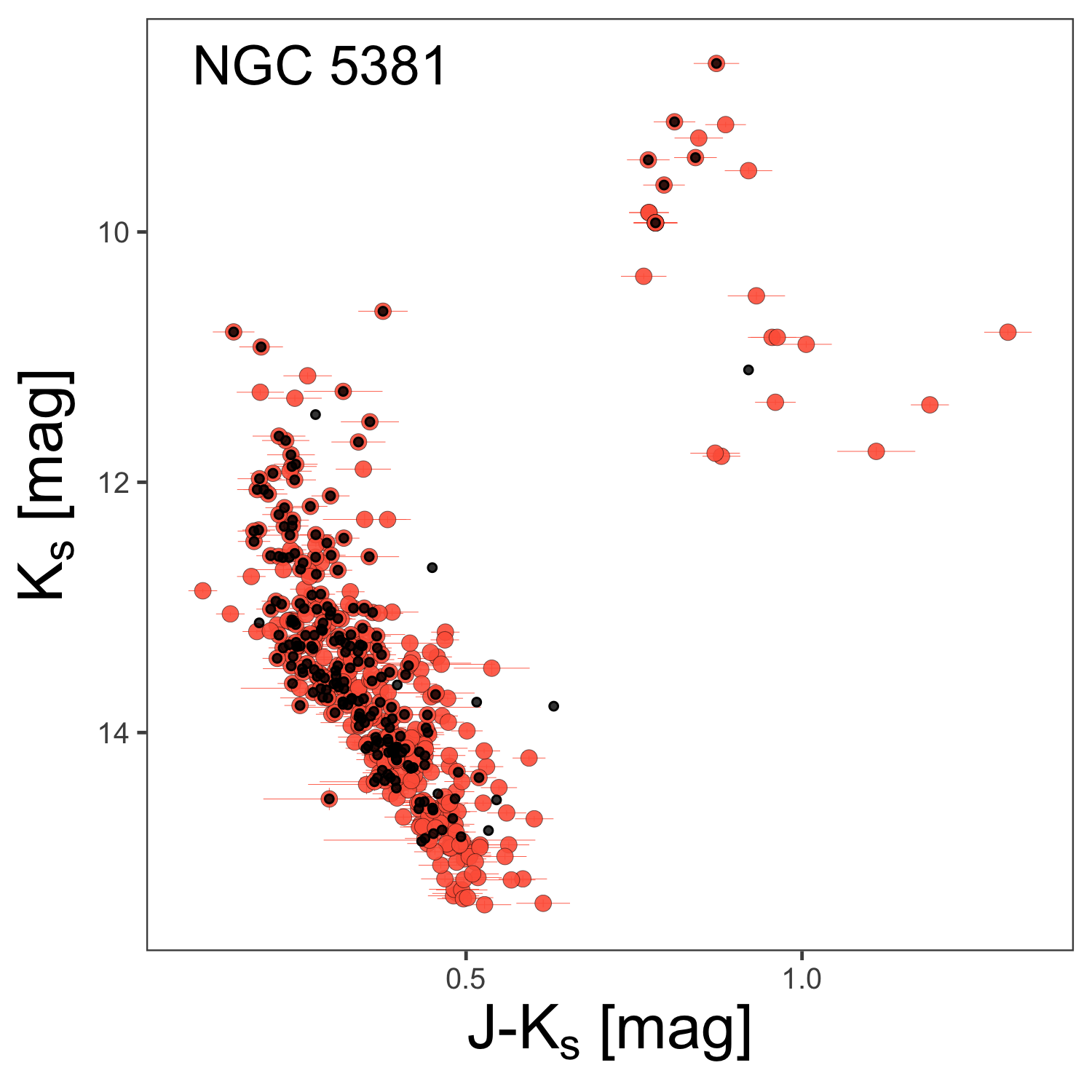}
\includegraphics[scale=0.075]{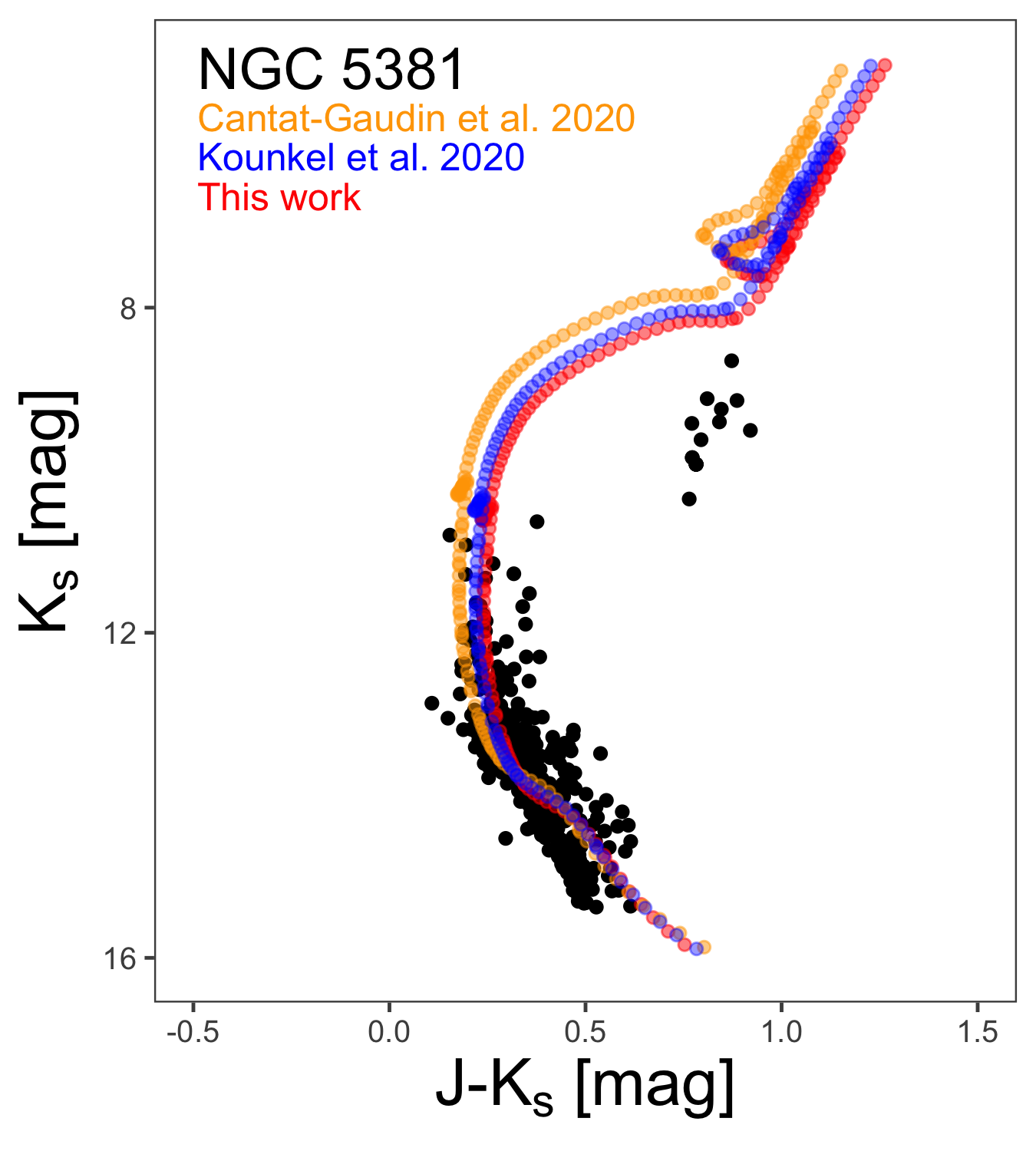}
\includegraphics[scale=0.075]{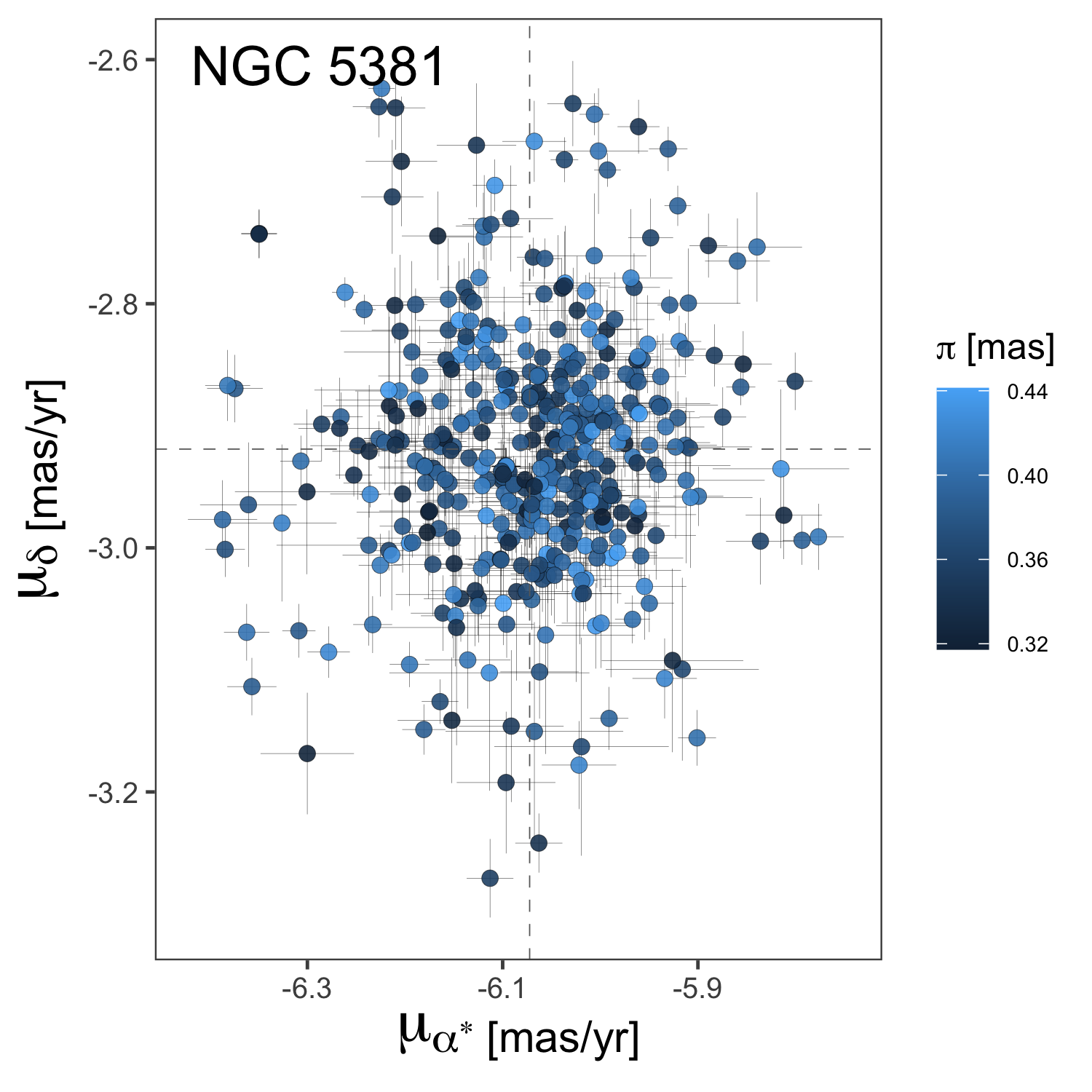}
\includegraphics[scale=0.075]{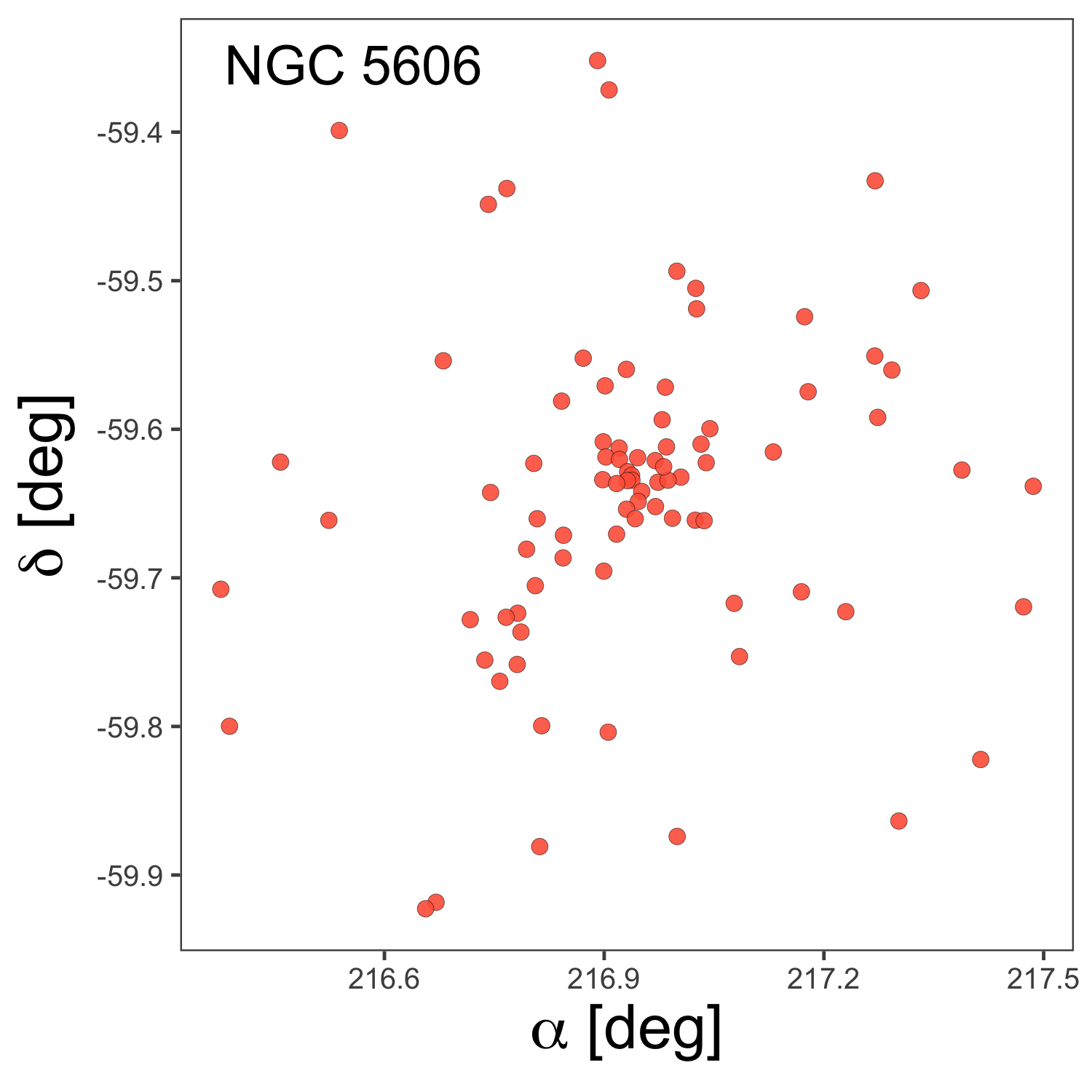}
\includegraphics[scale=0.075]{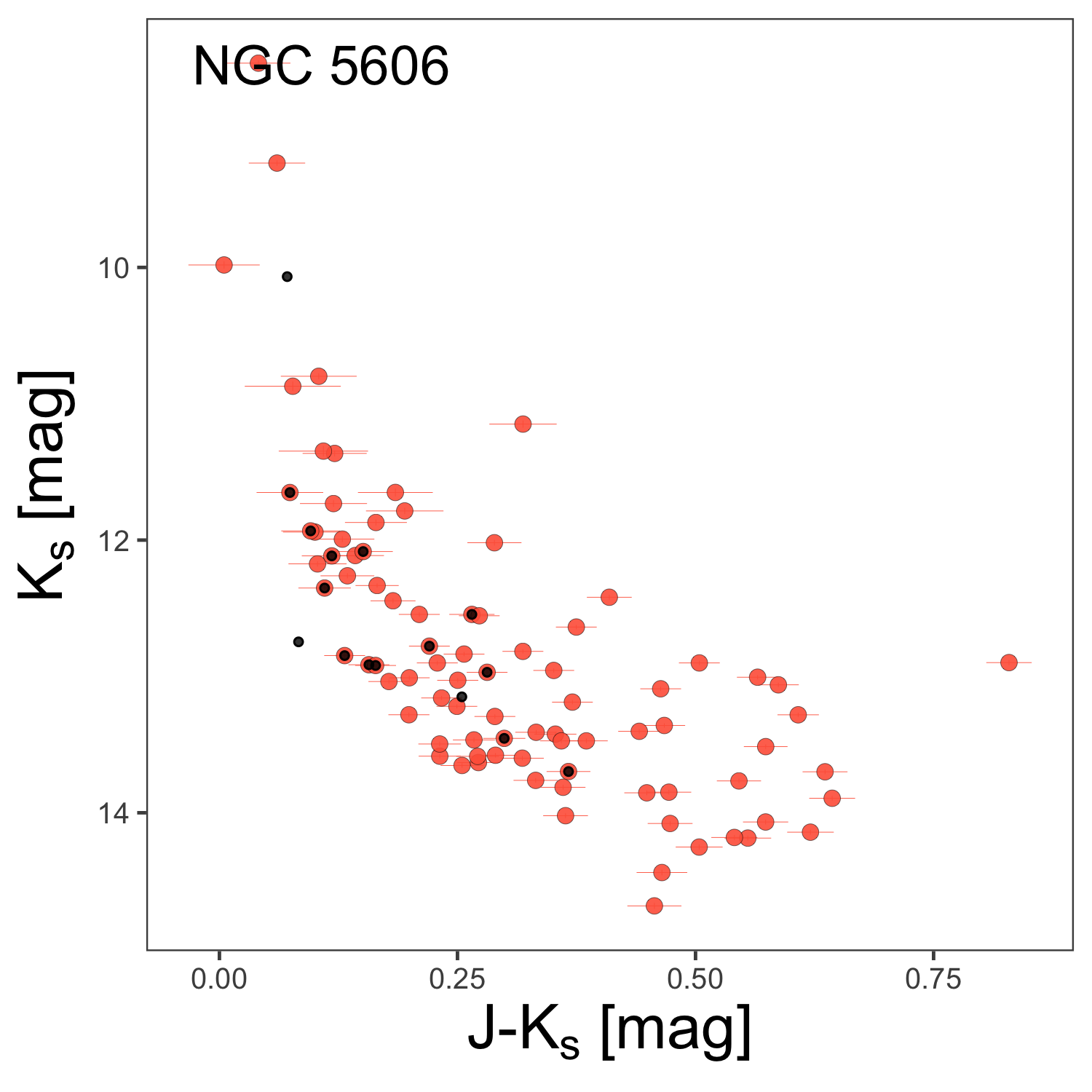}
\includegraphics[scale=0.075]{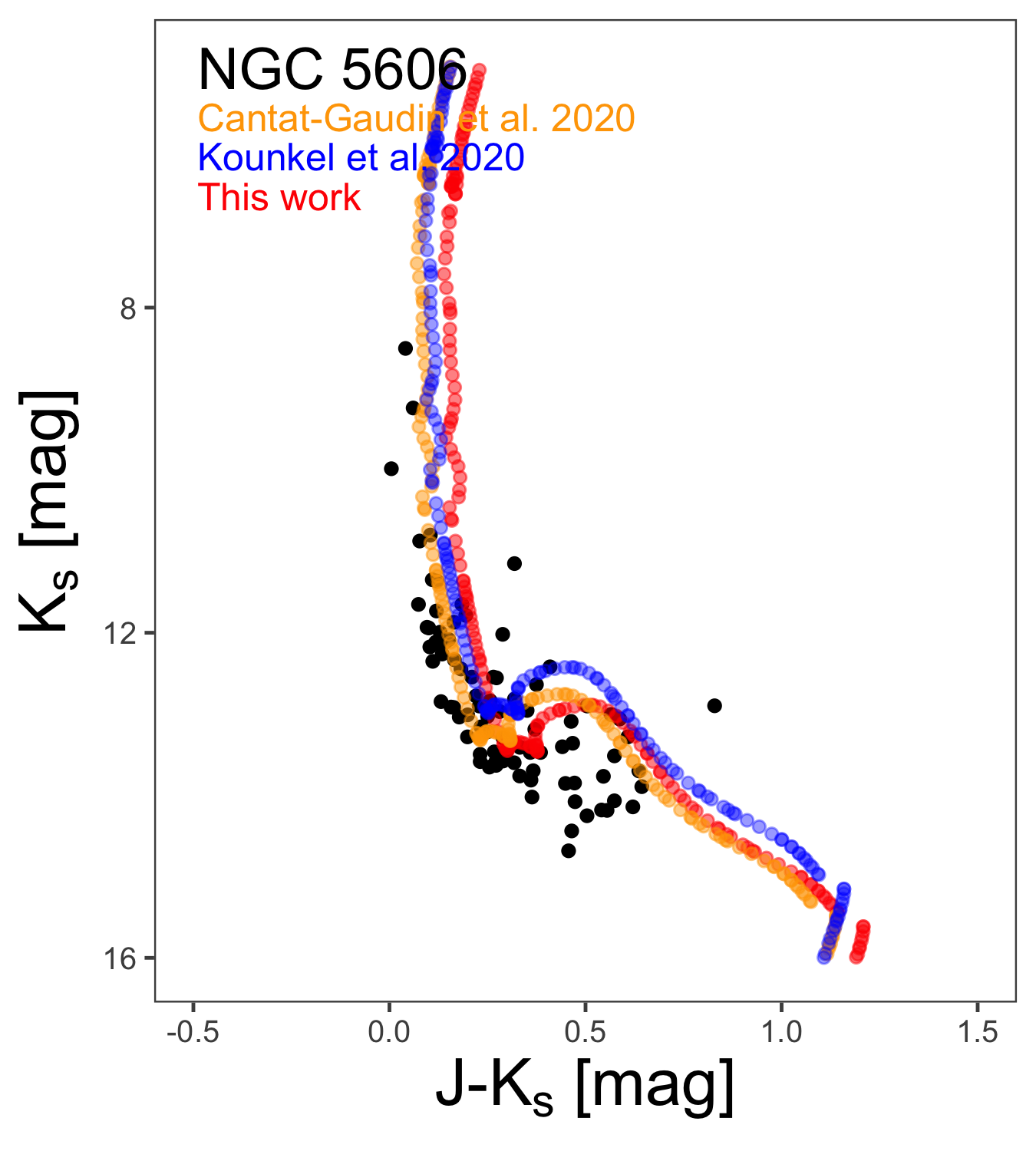}
\includegraphics[scale=0.075]{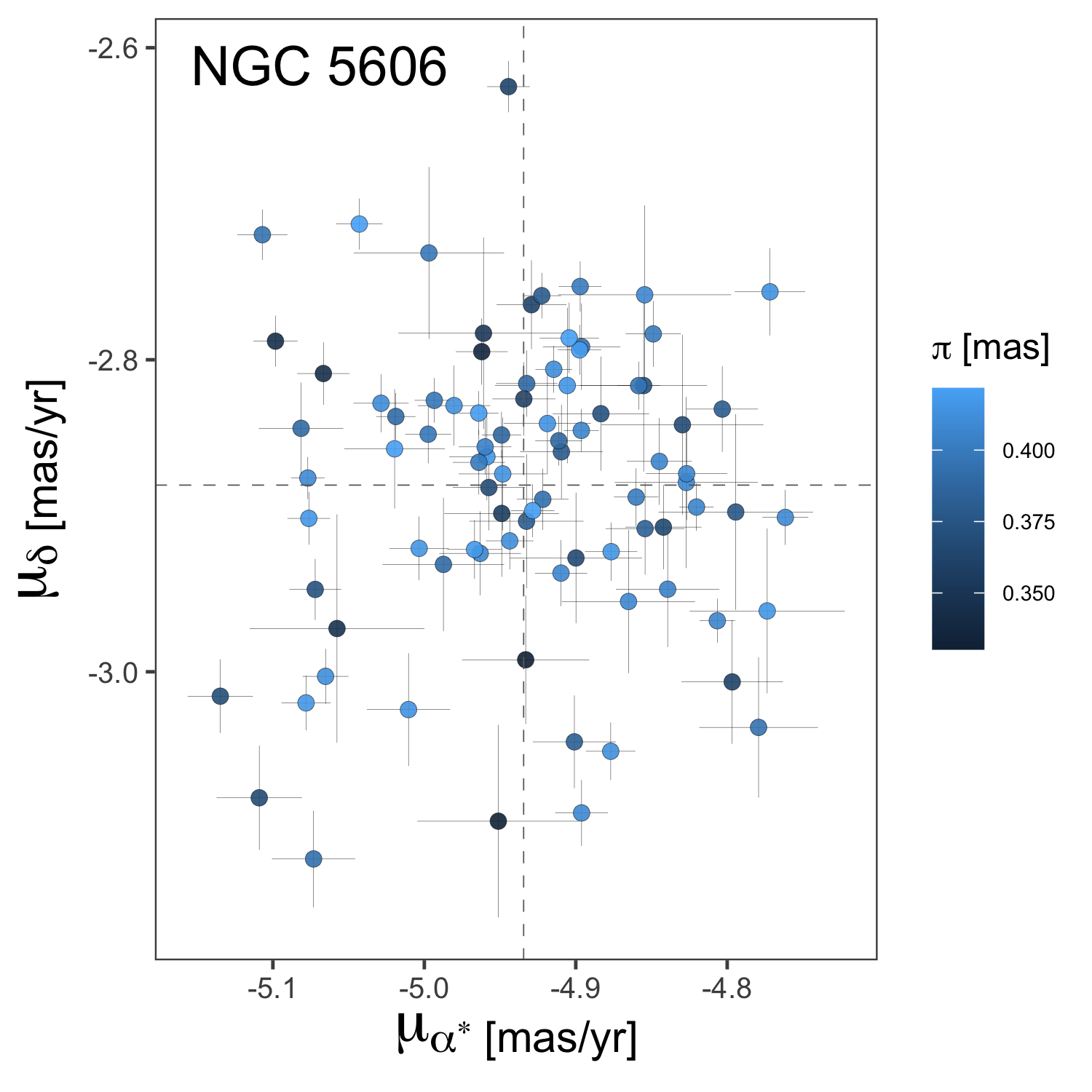}
\includegraphics[scale=0.075]{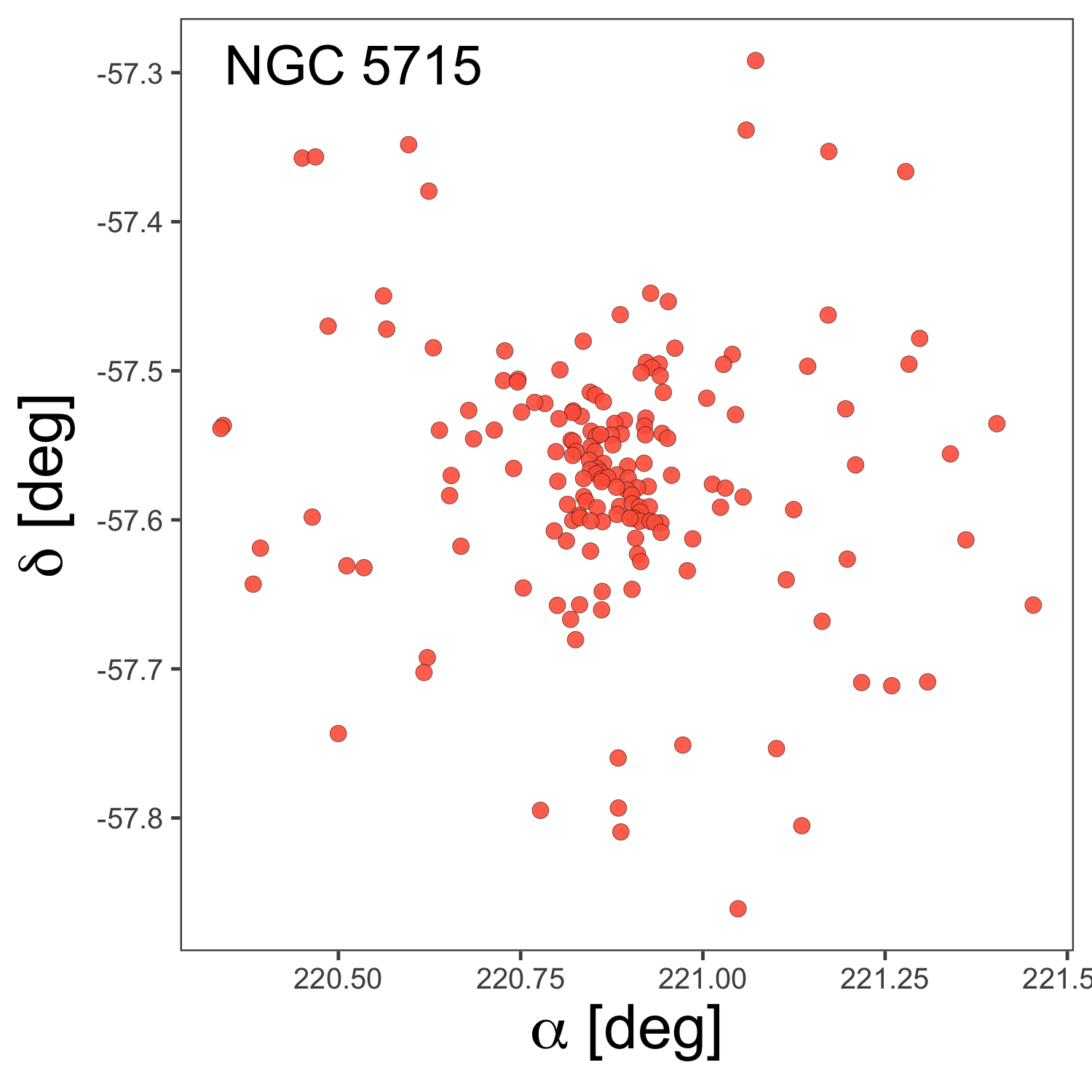}
\includegraphics[scale=0.075]{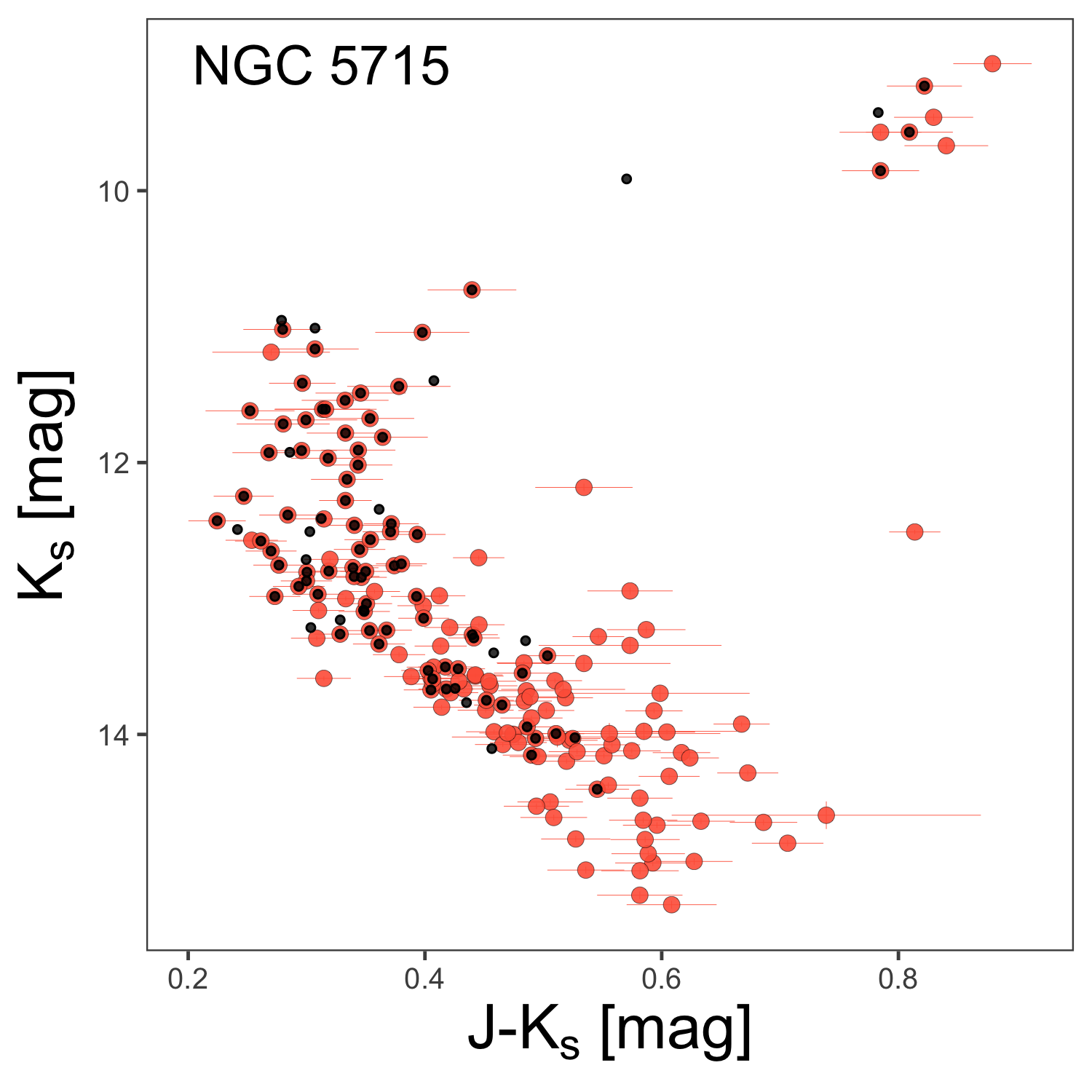}
\includegraphics[scale=0.075]{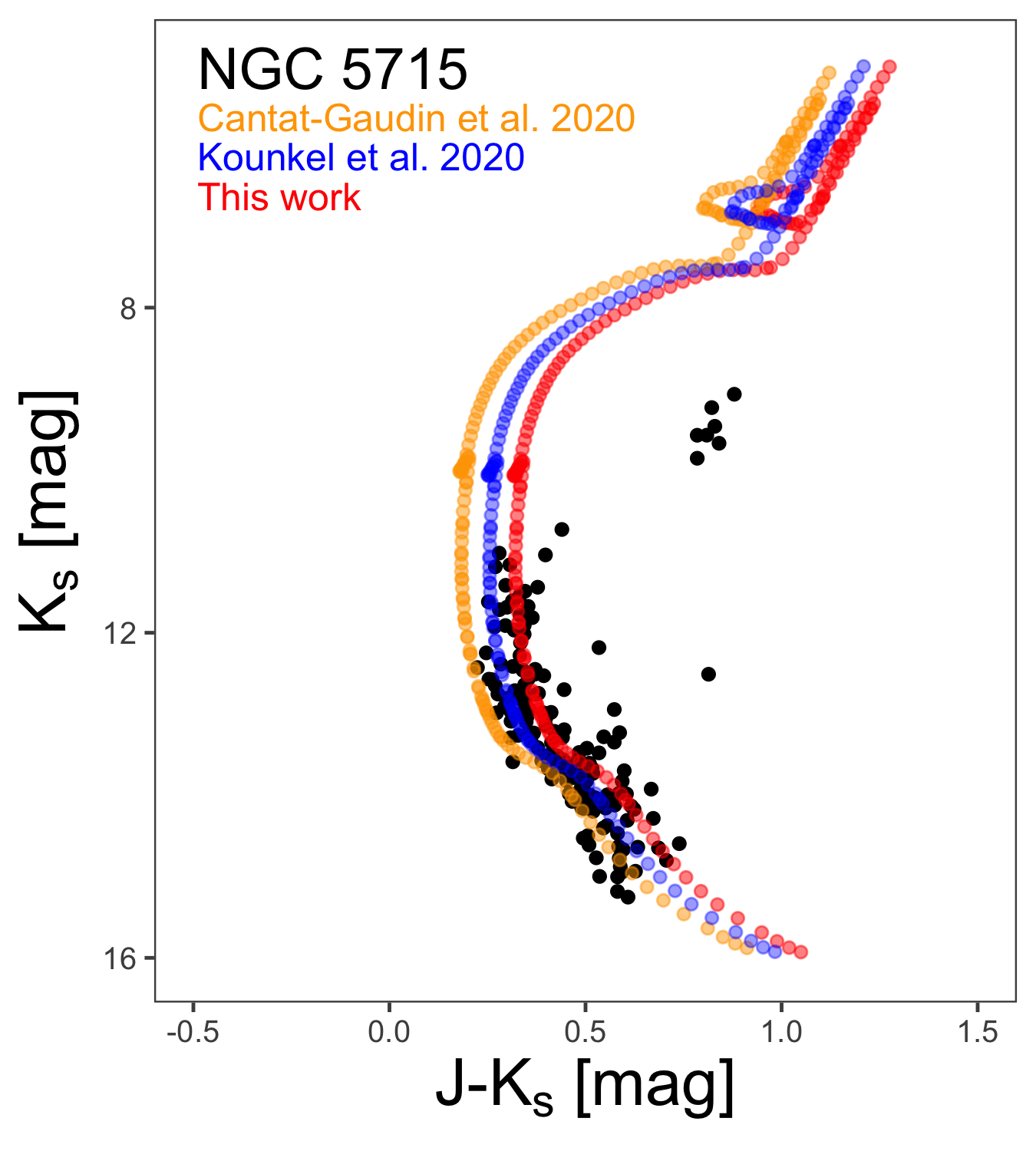}
\includegraphics[scale=0.075]{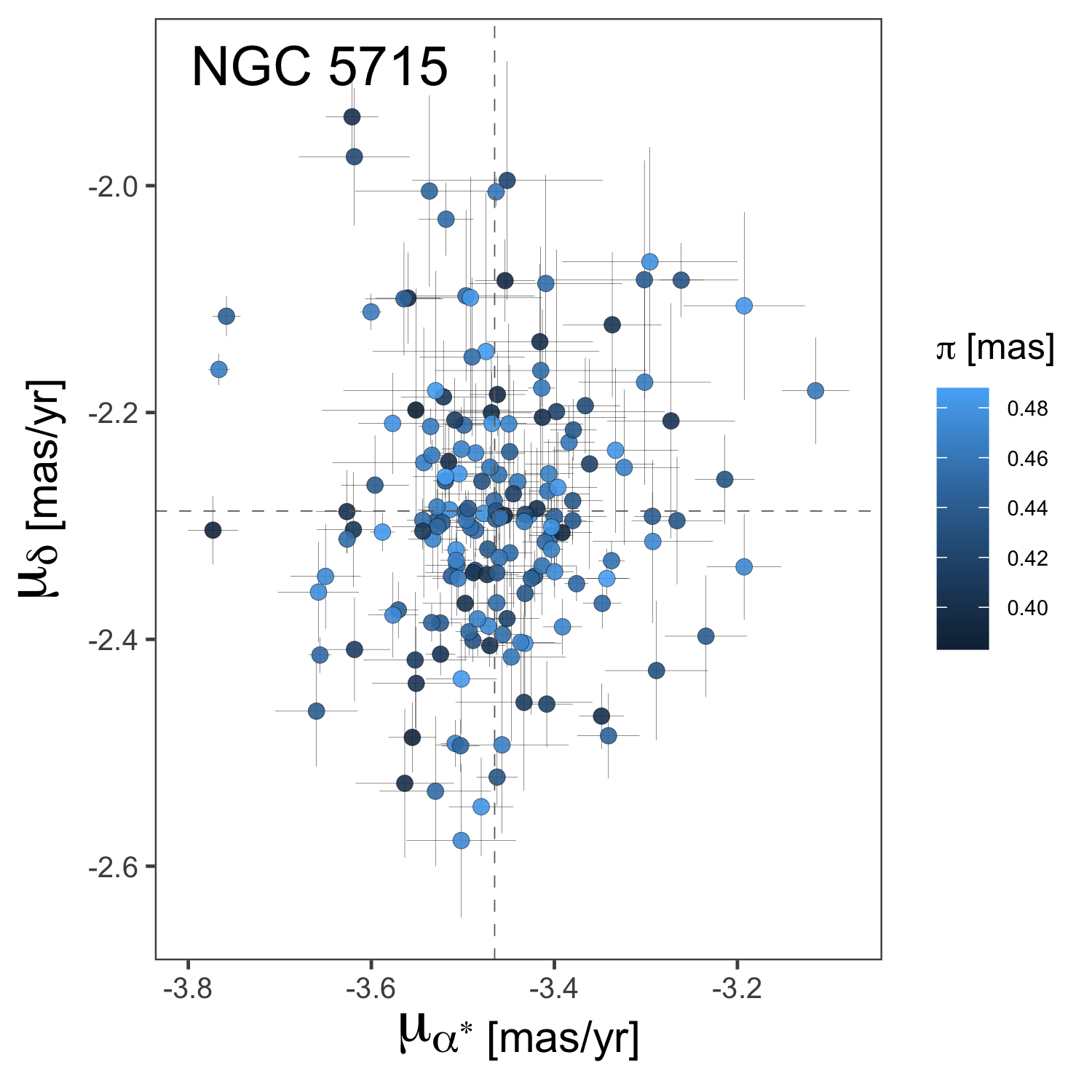}
\includegraphics[scale=0.075]{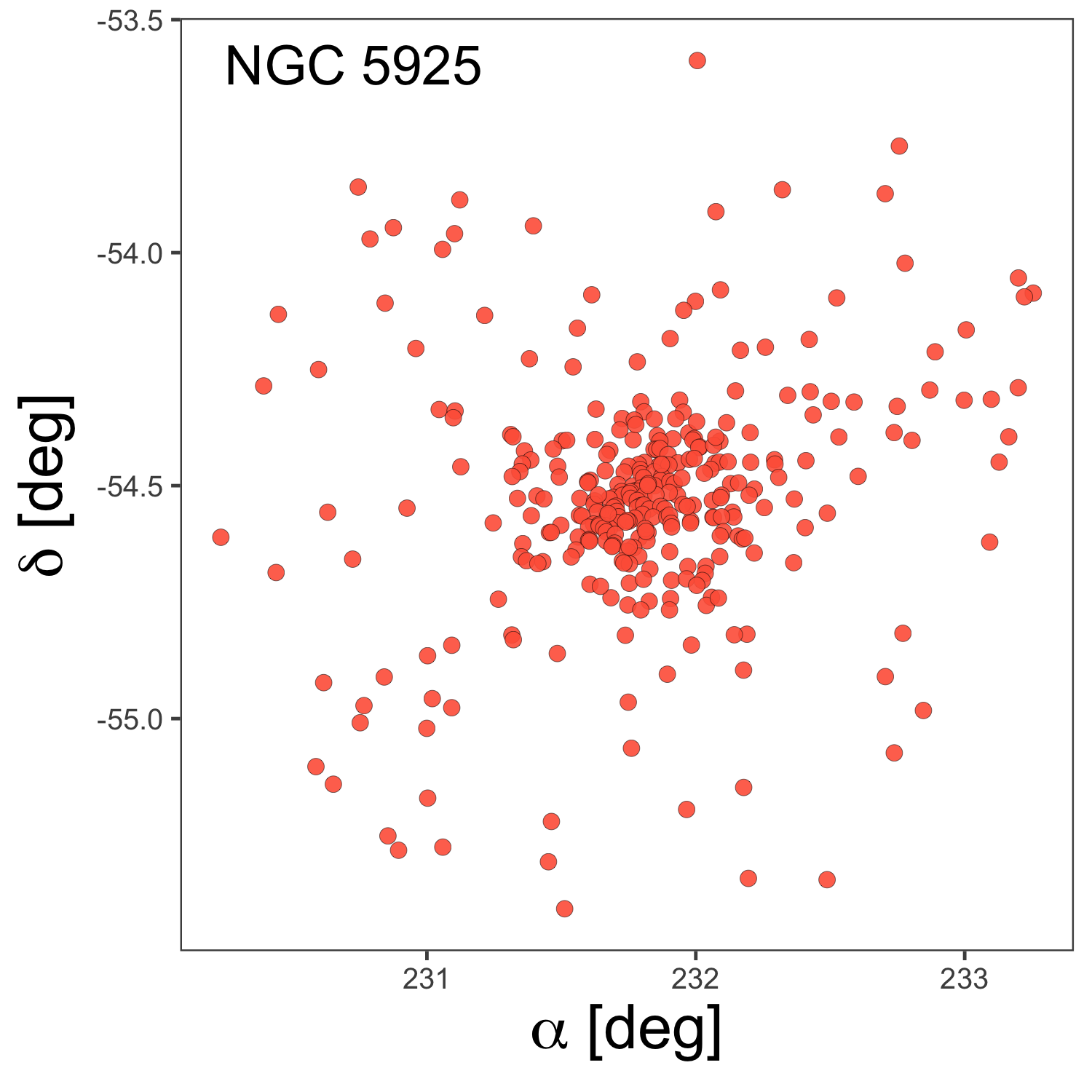}
\includegraphics[scale=0.075]{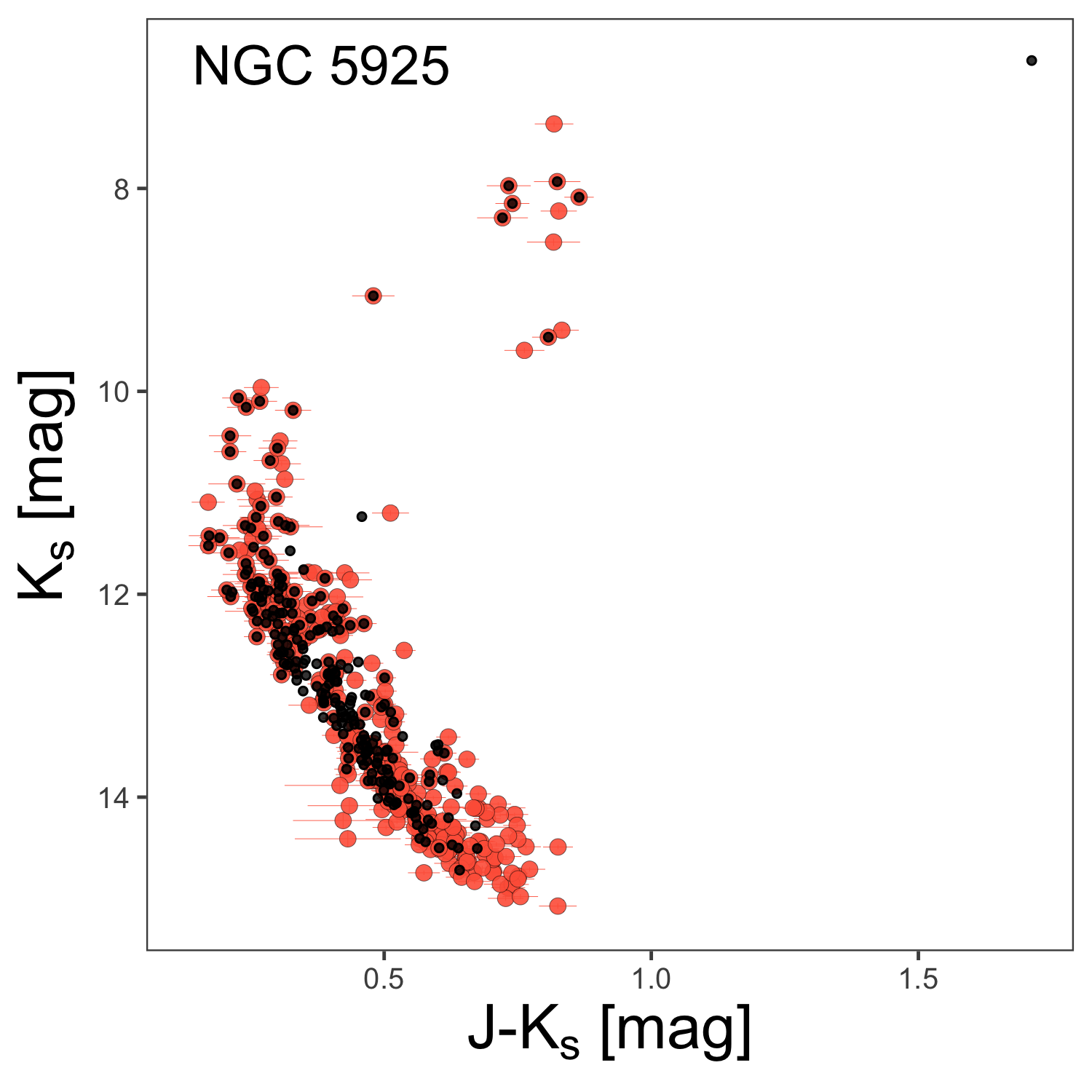}
\includegraphics[scale=0.075]{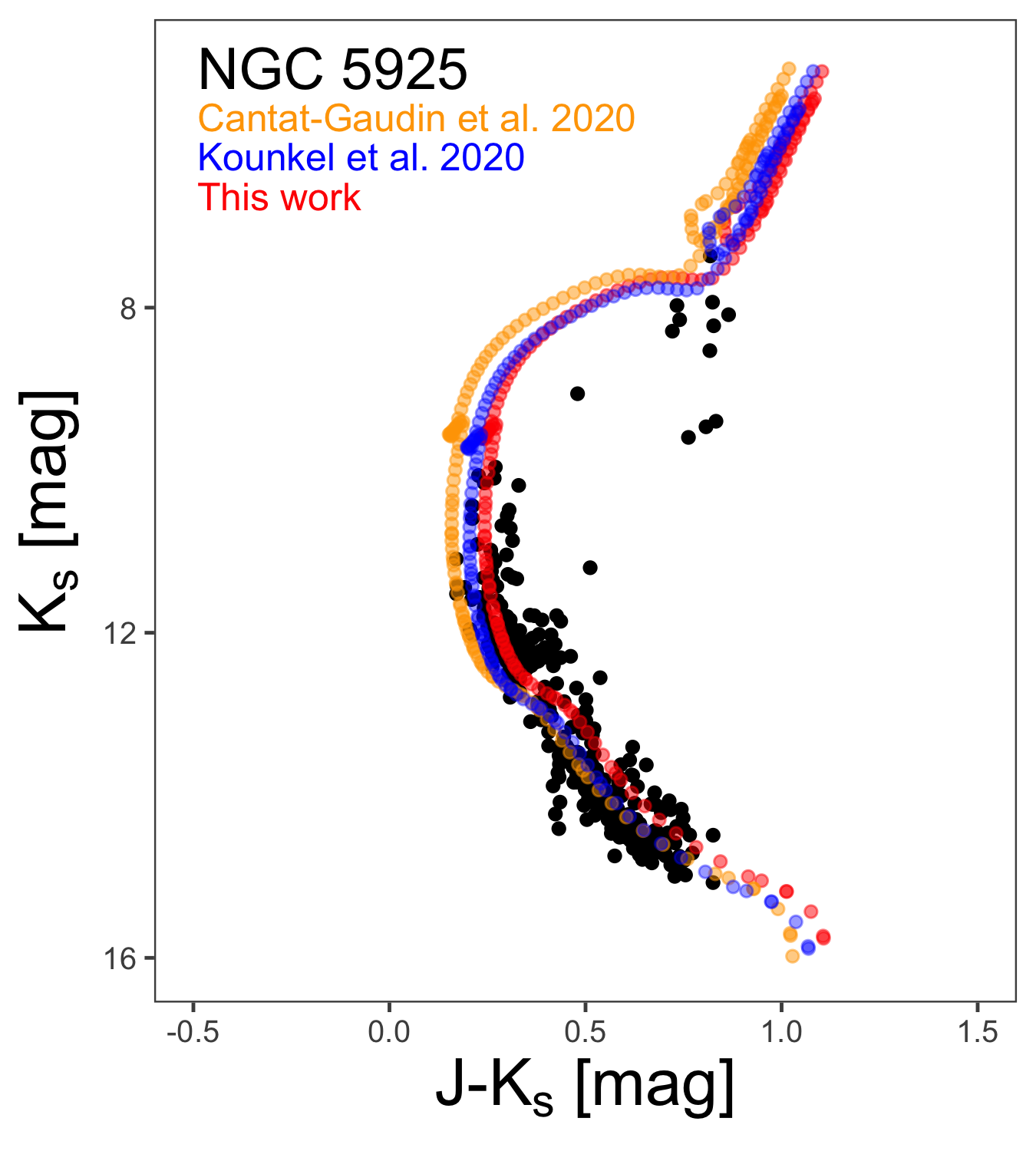}
\includegraphics[scale=0.075]{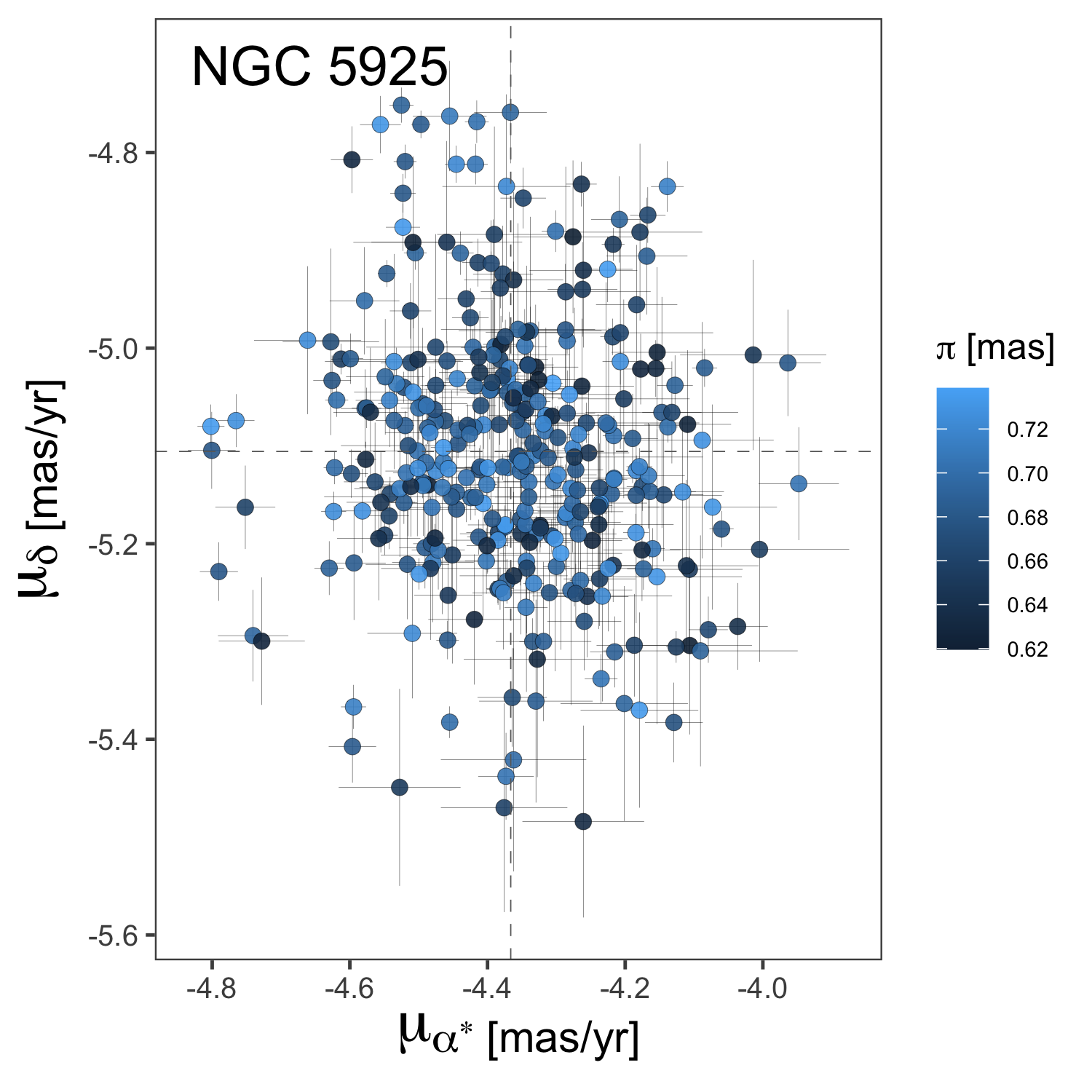}
\caption{Same as Figure\,\ref{fig:clusters}}
\end{figure*}

\begin{figure*}
\includegraphics[scale=0.075]{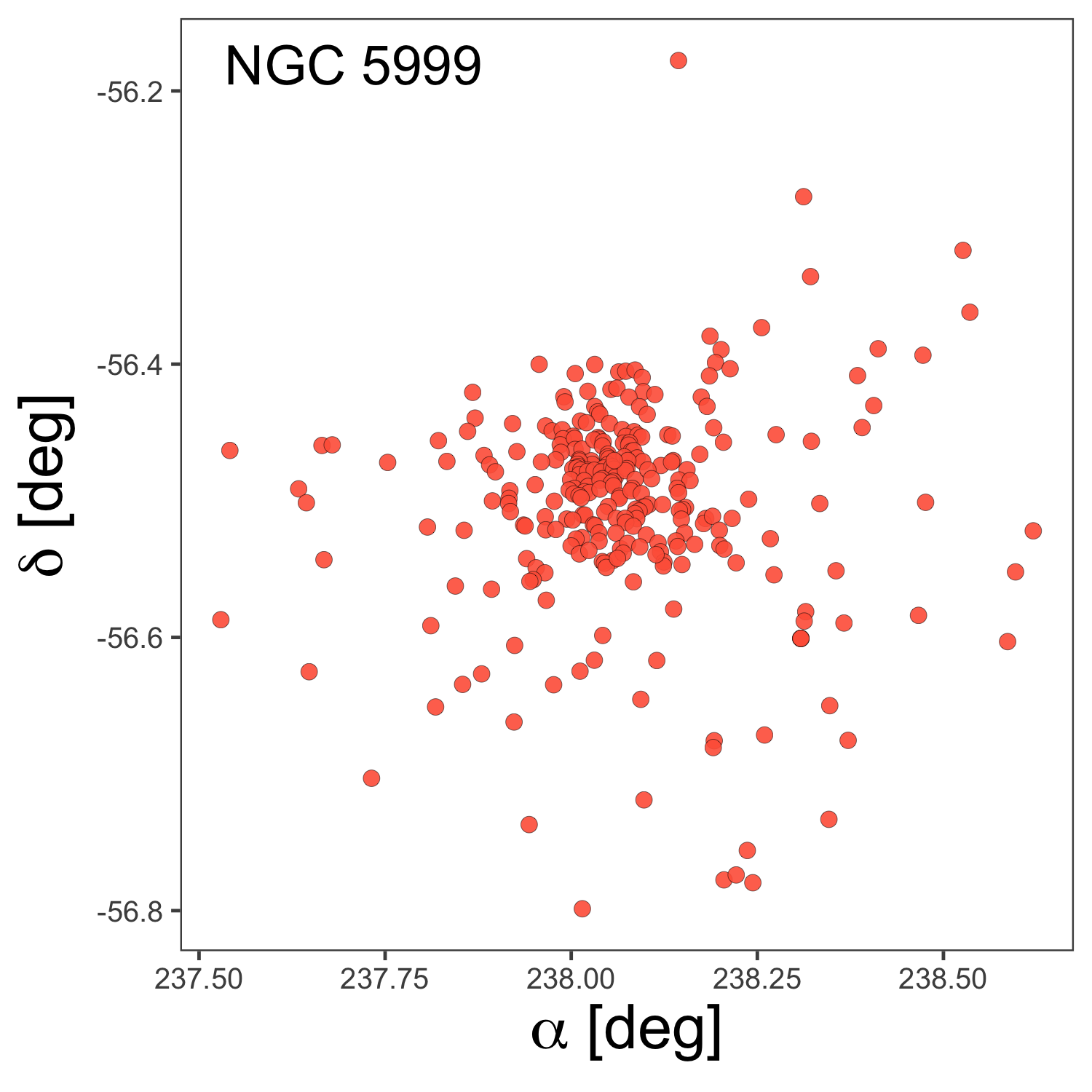}
\includegraphics[scale=0.075]{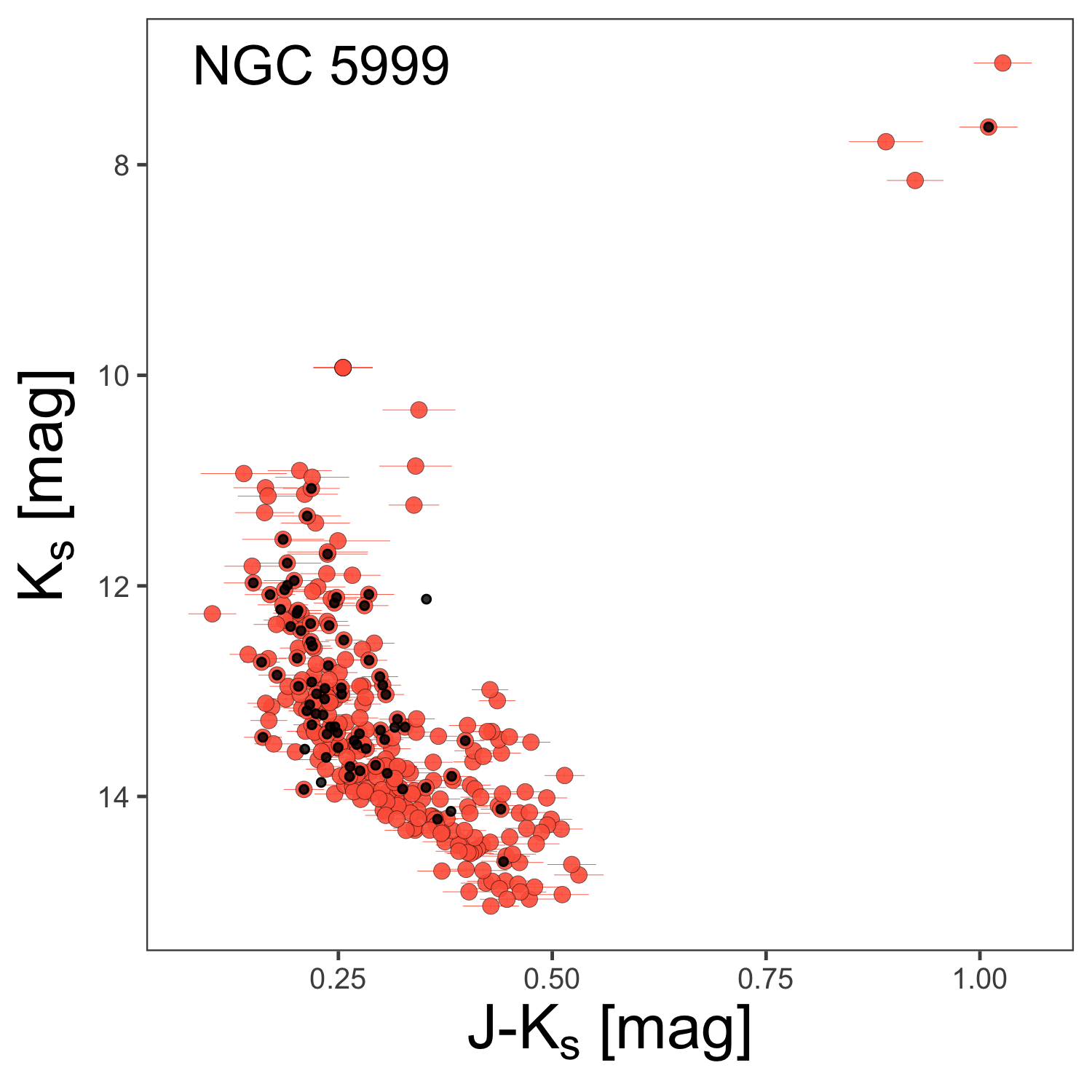}
\includegraphics[scale=0.075]{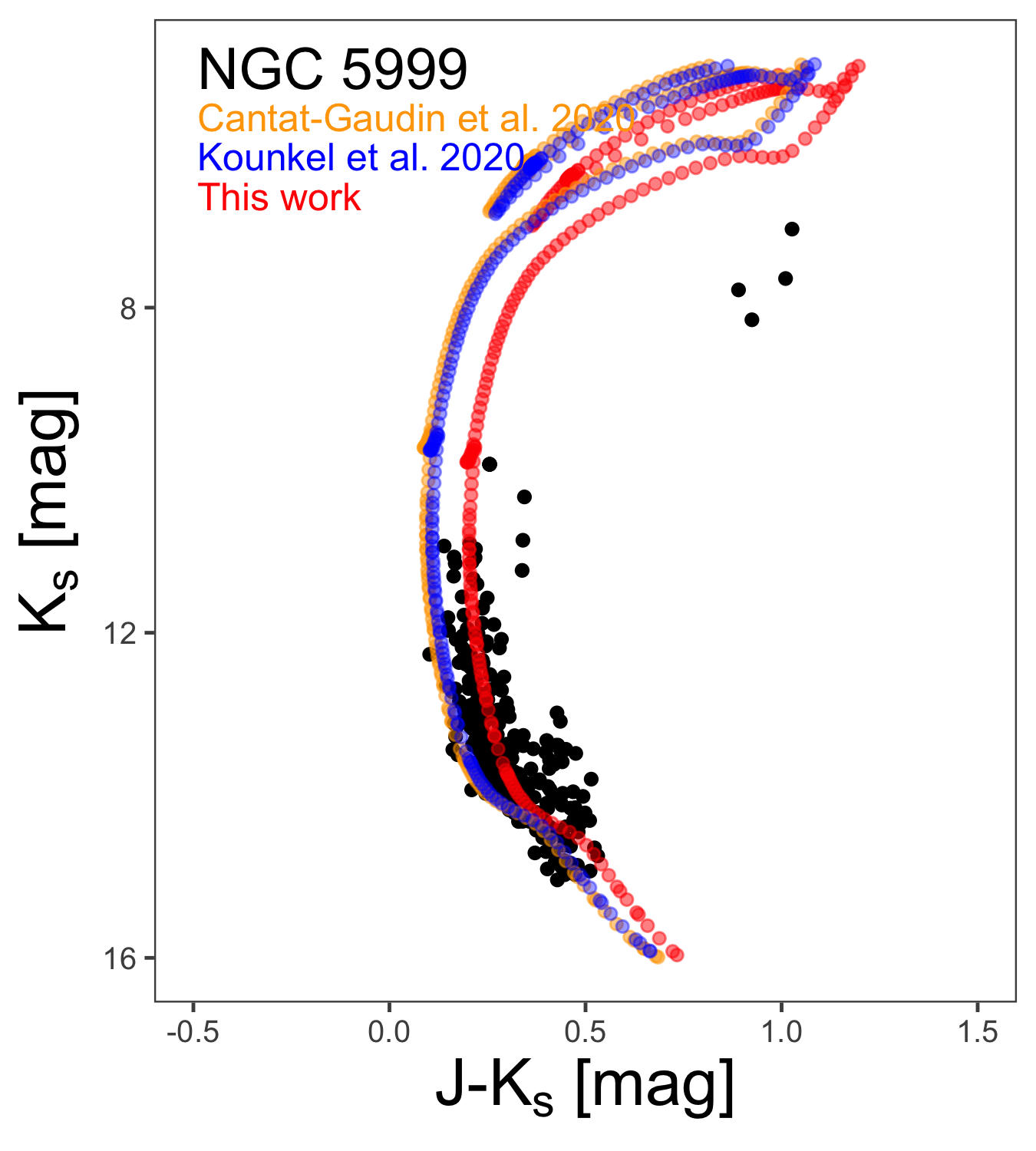}
\includegraphics[scale=0.075]{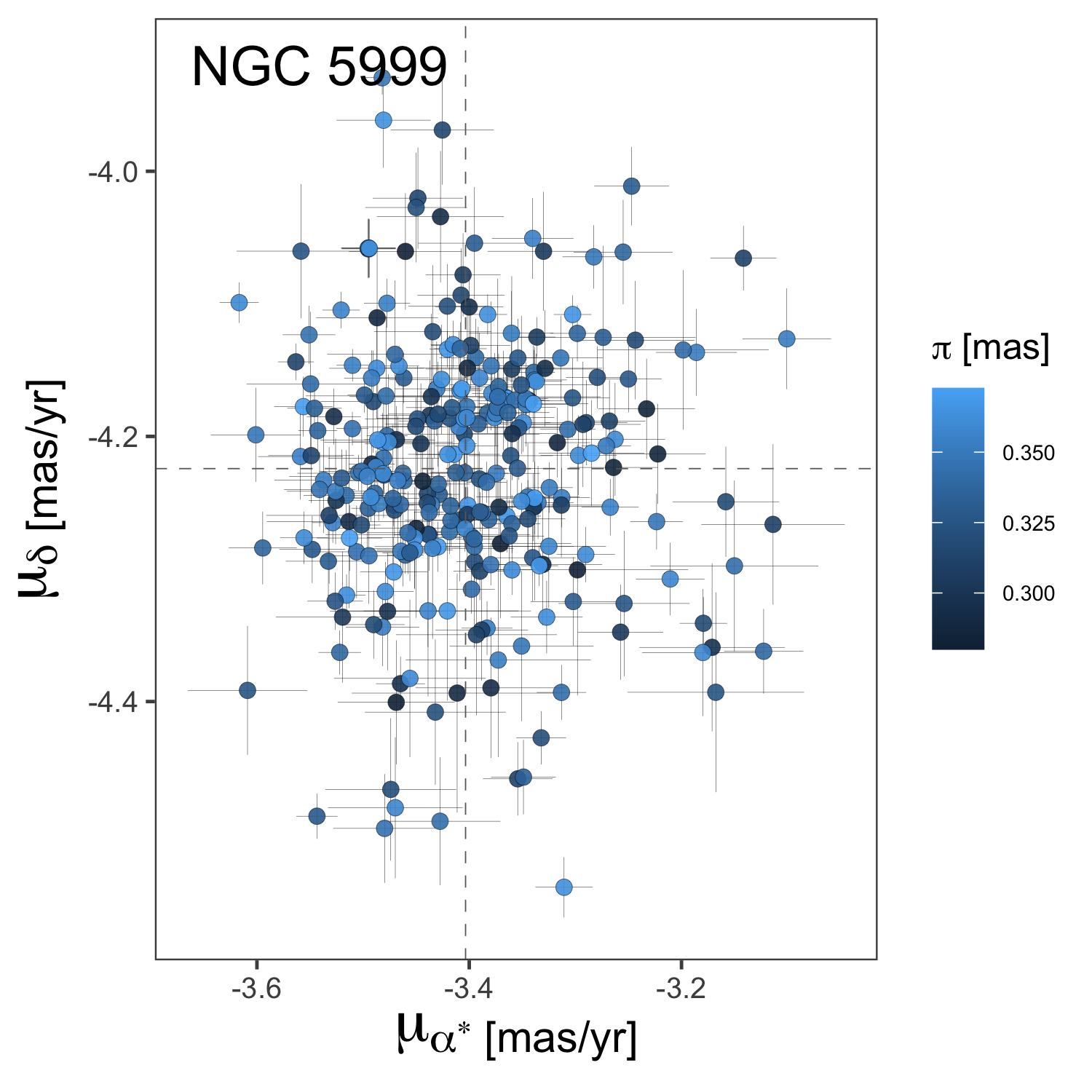}
\includegraphics[scale=0.075]{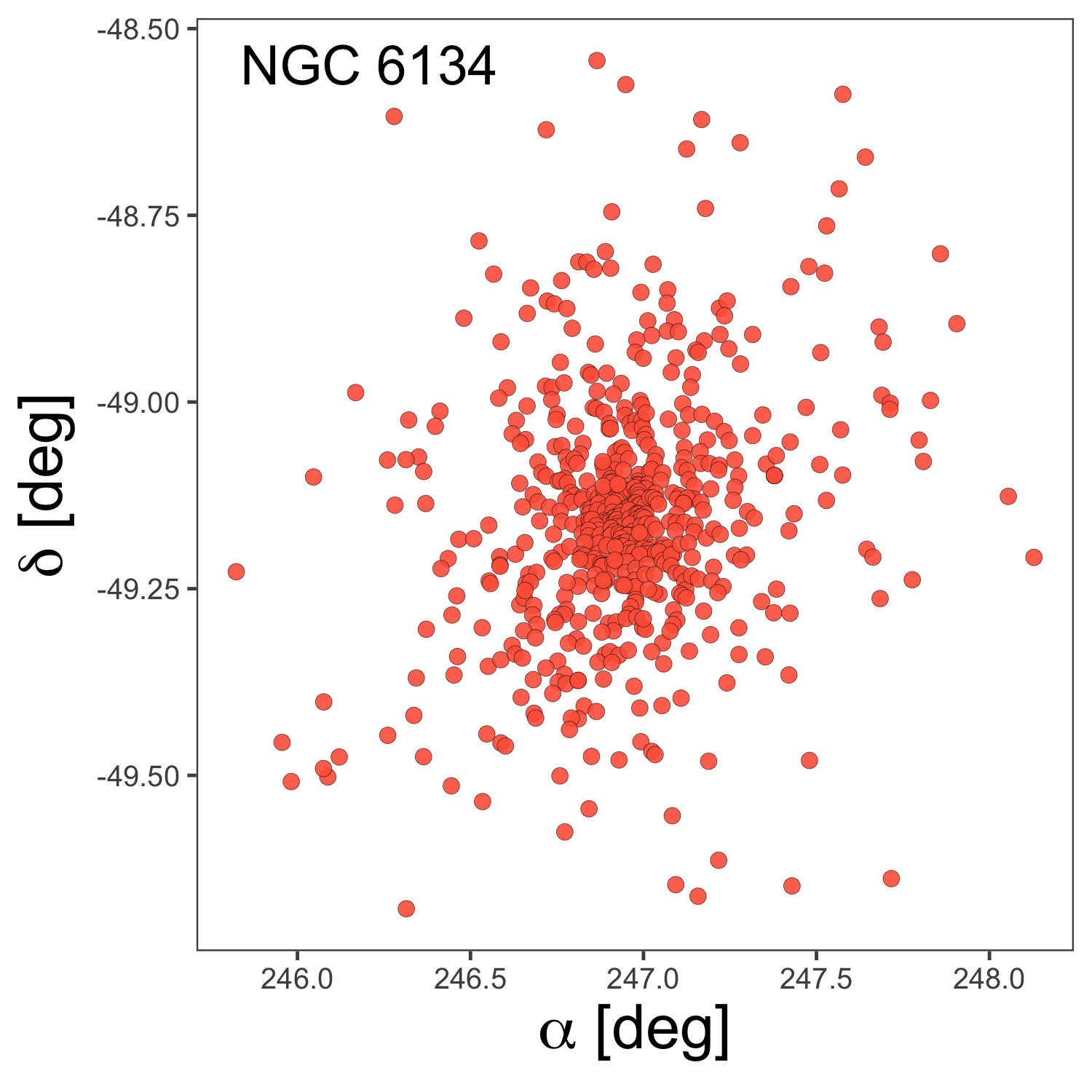}
\includegraphics[scale=0.075]{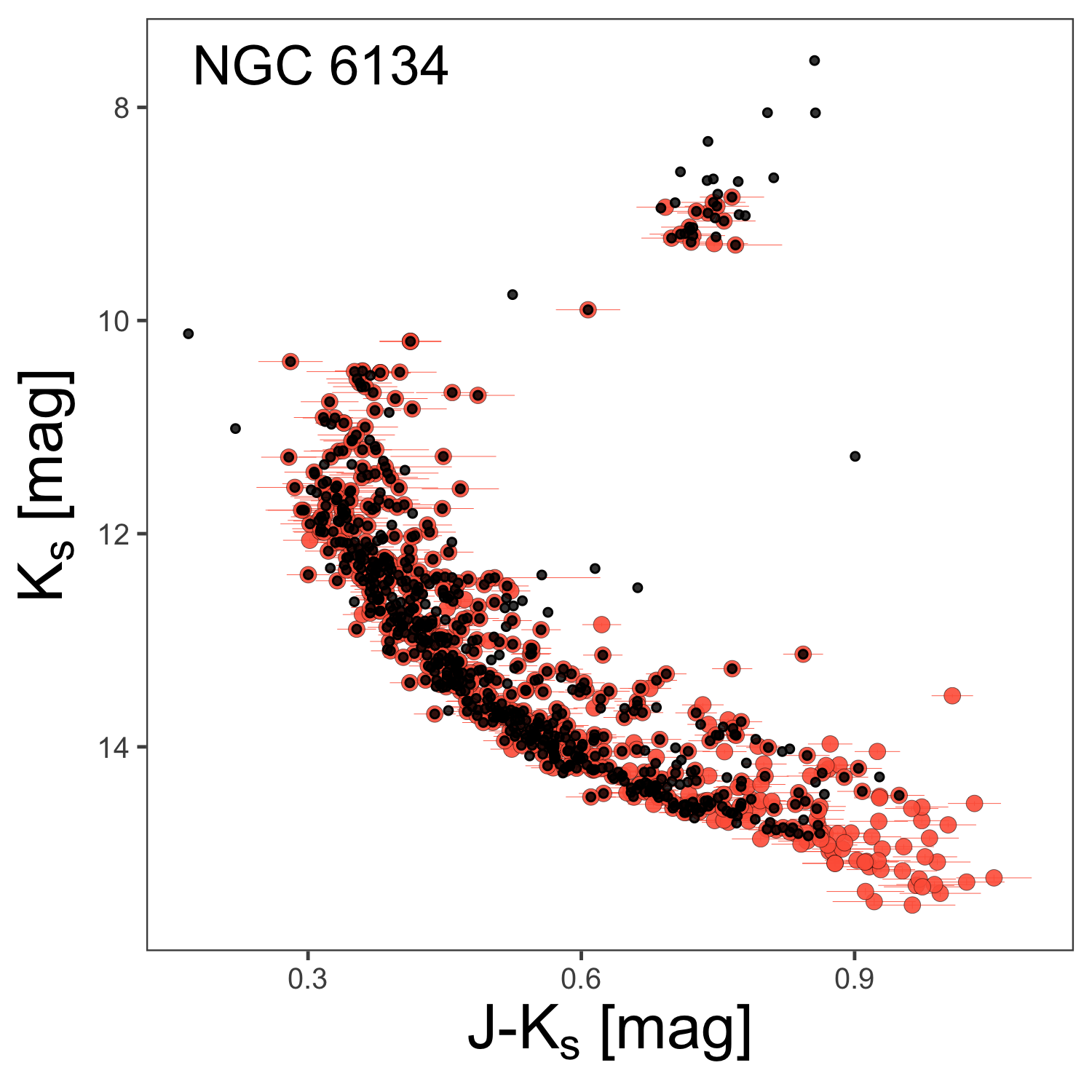}
\includegraphics[scale=0.075]{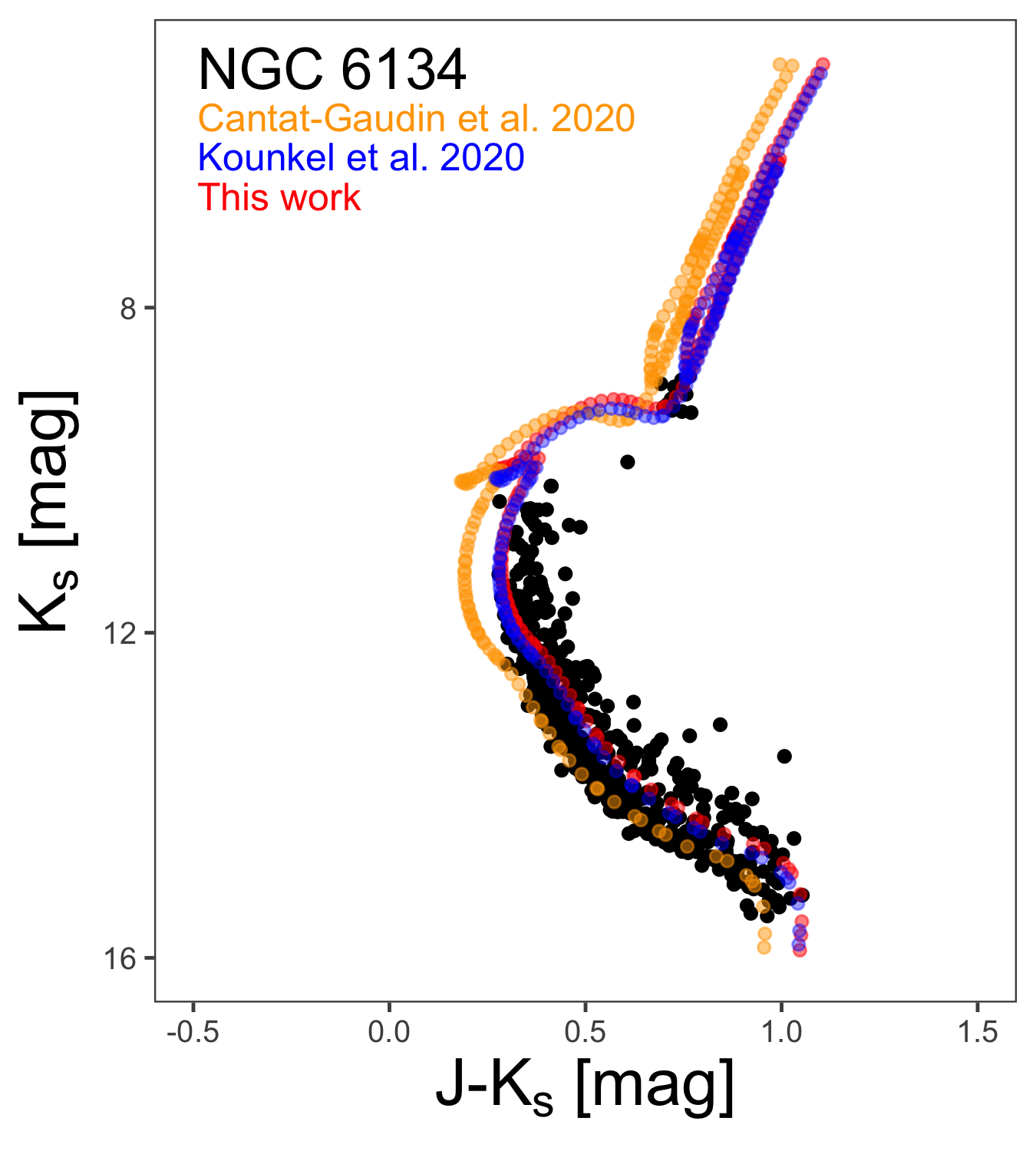}
\includegraphics[scale=0.075]{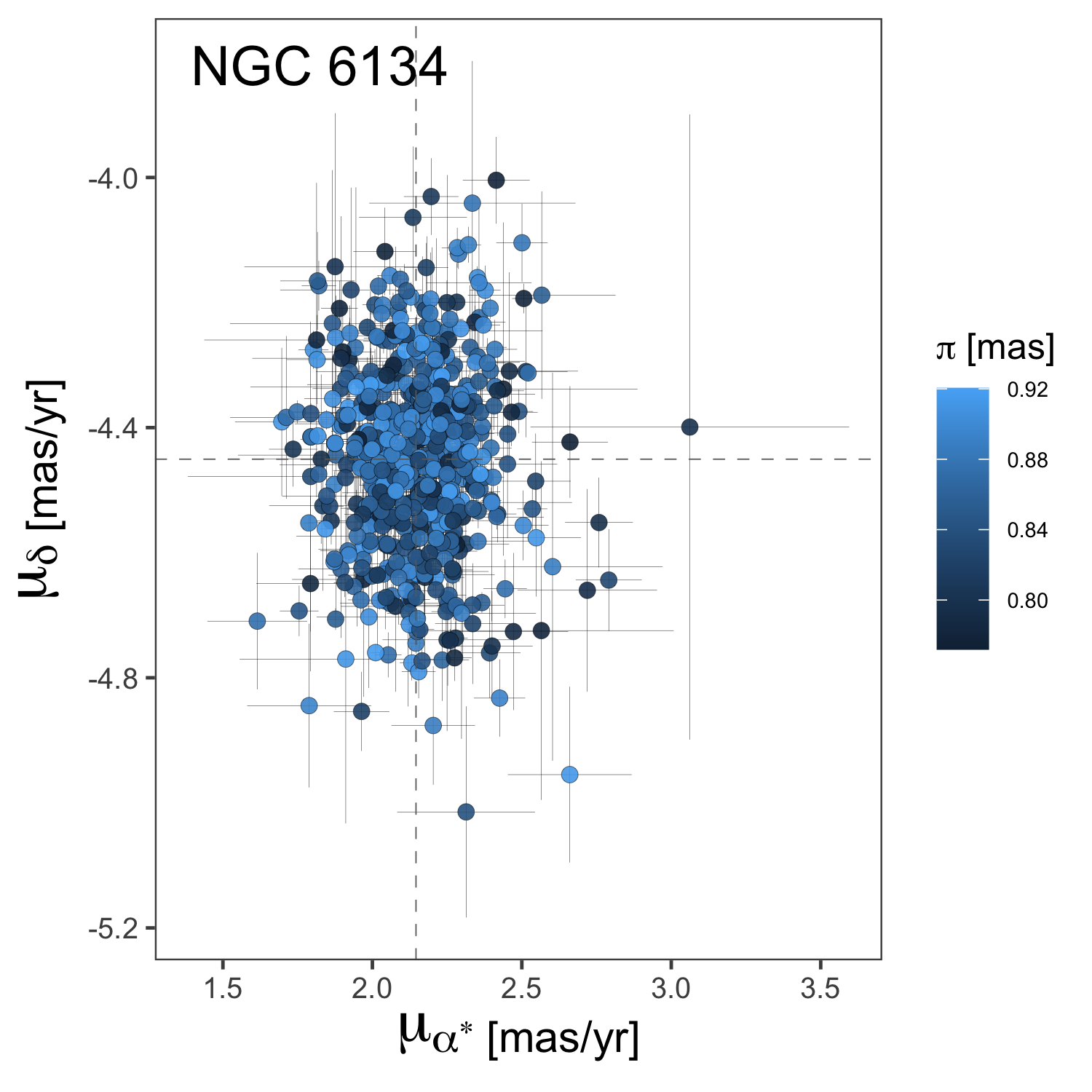}
\includegraphics[scale=0.075]{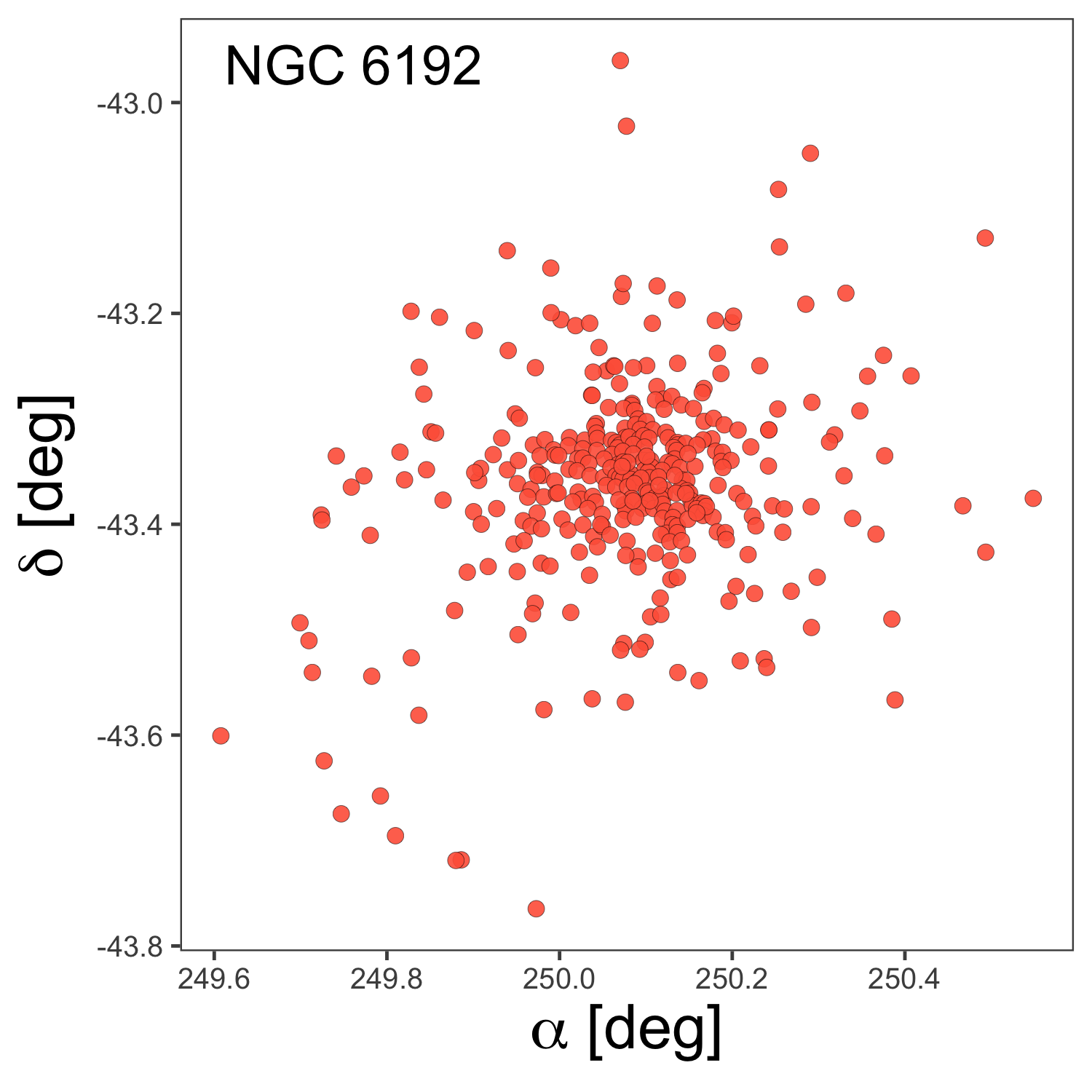}
\includegraphics[scale=0.075]{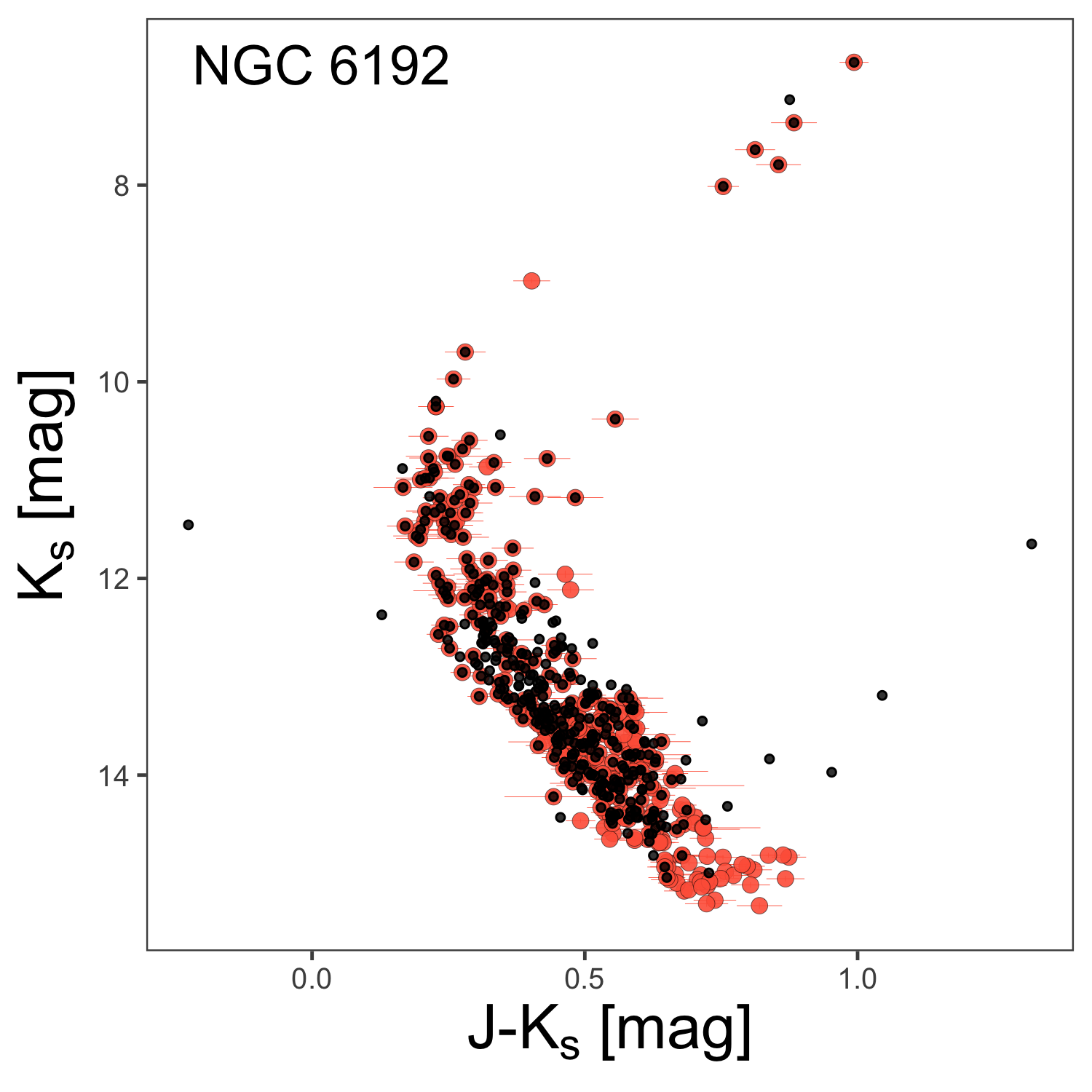}
\includegraphics[scale=0.075]{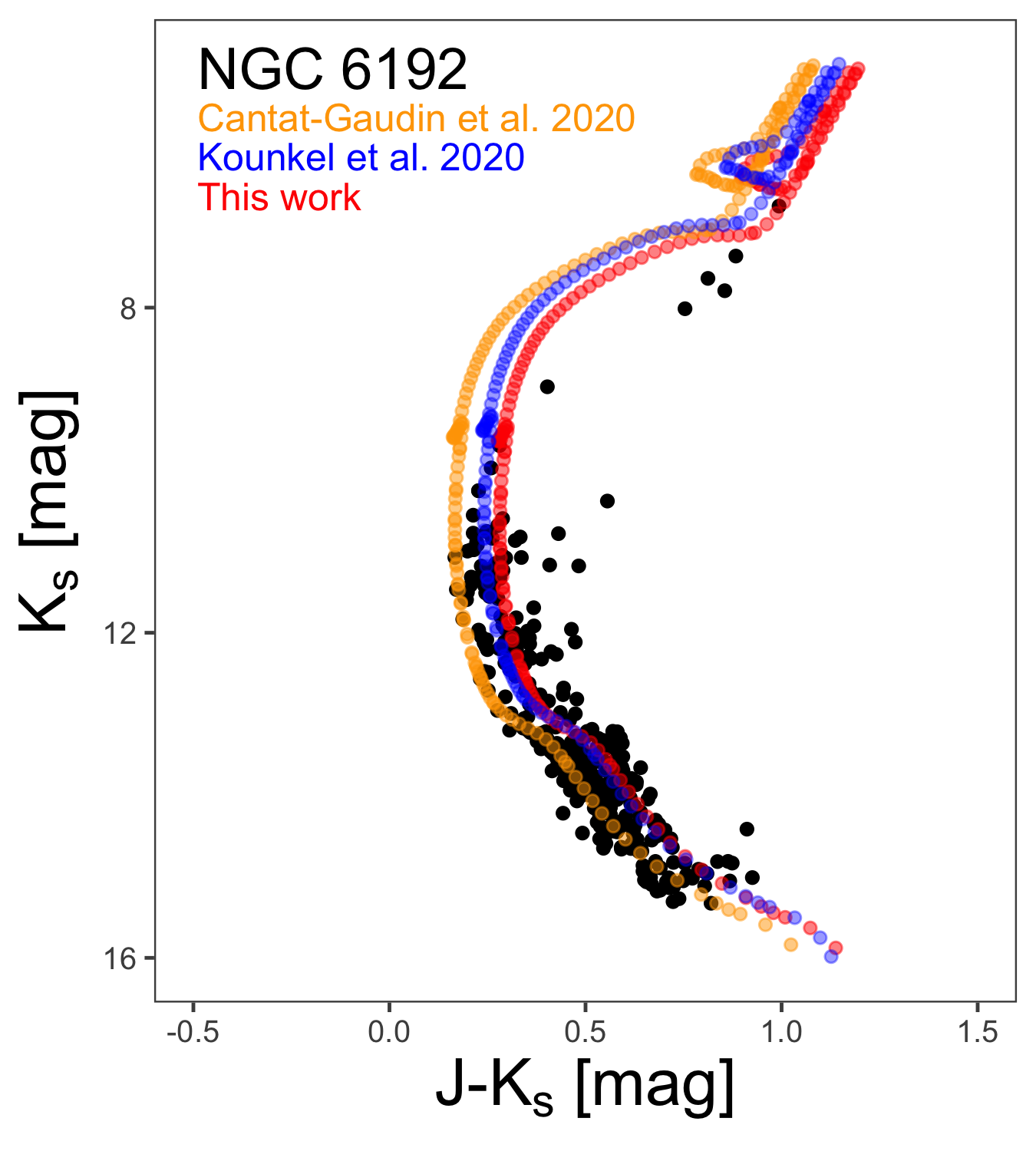}
\includegraphics[scale=0.075]{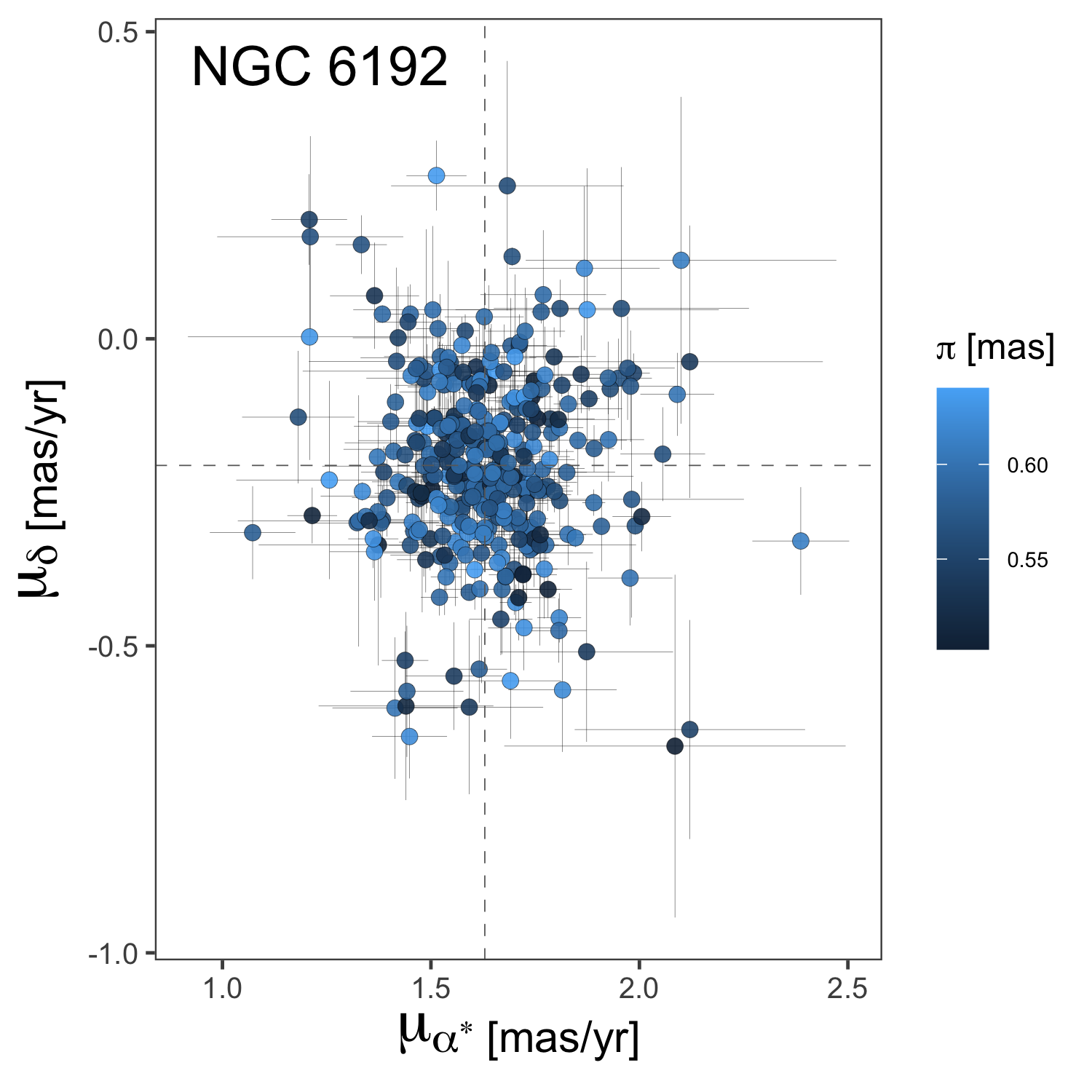}
\includegraphics[scale=0.075]{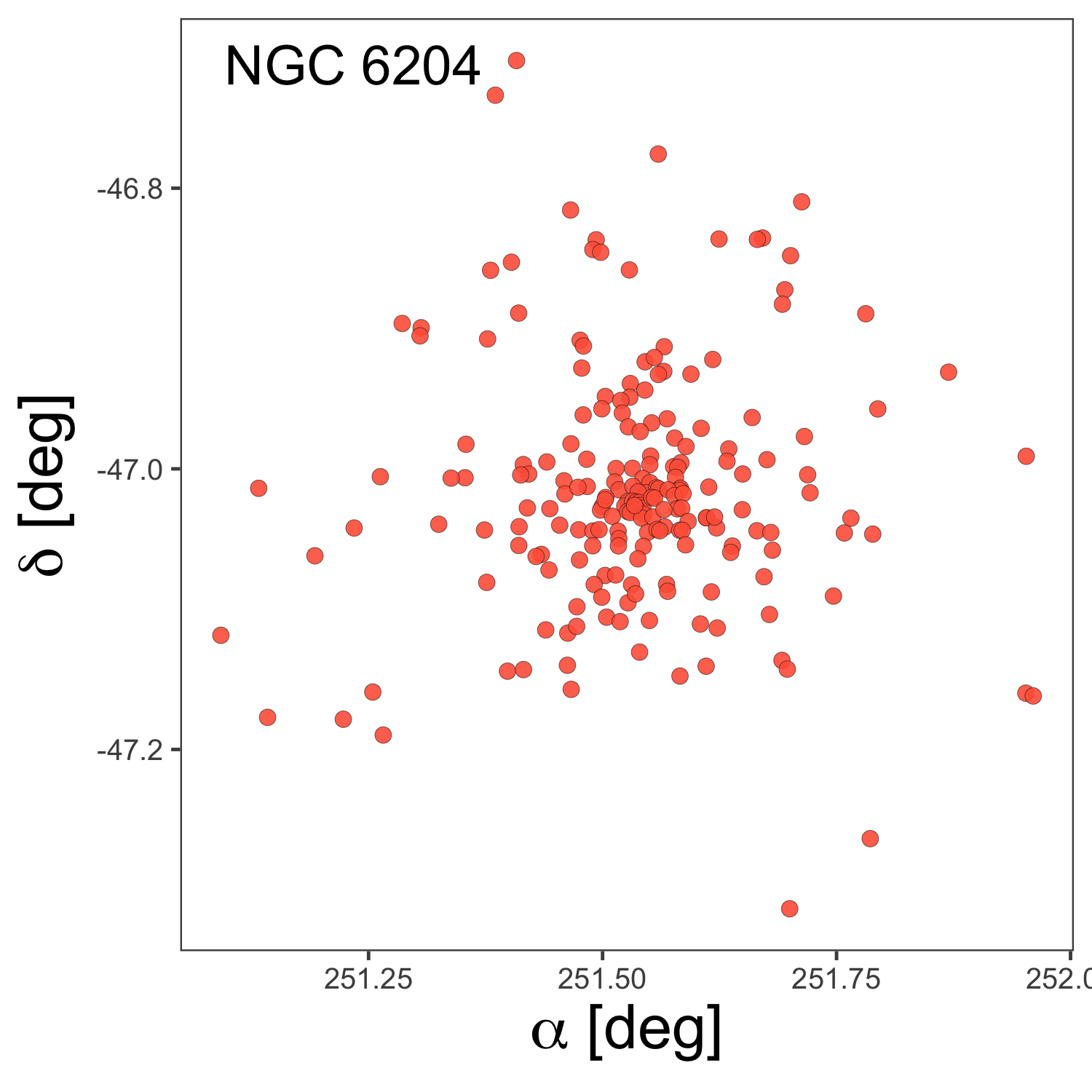}
\includegraphics[scale=0.075]{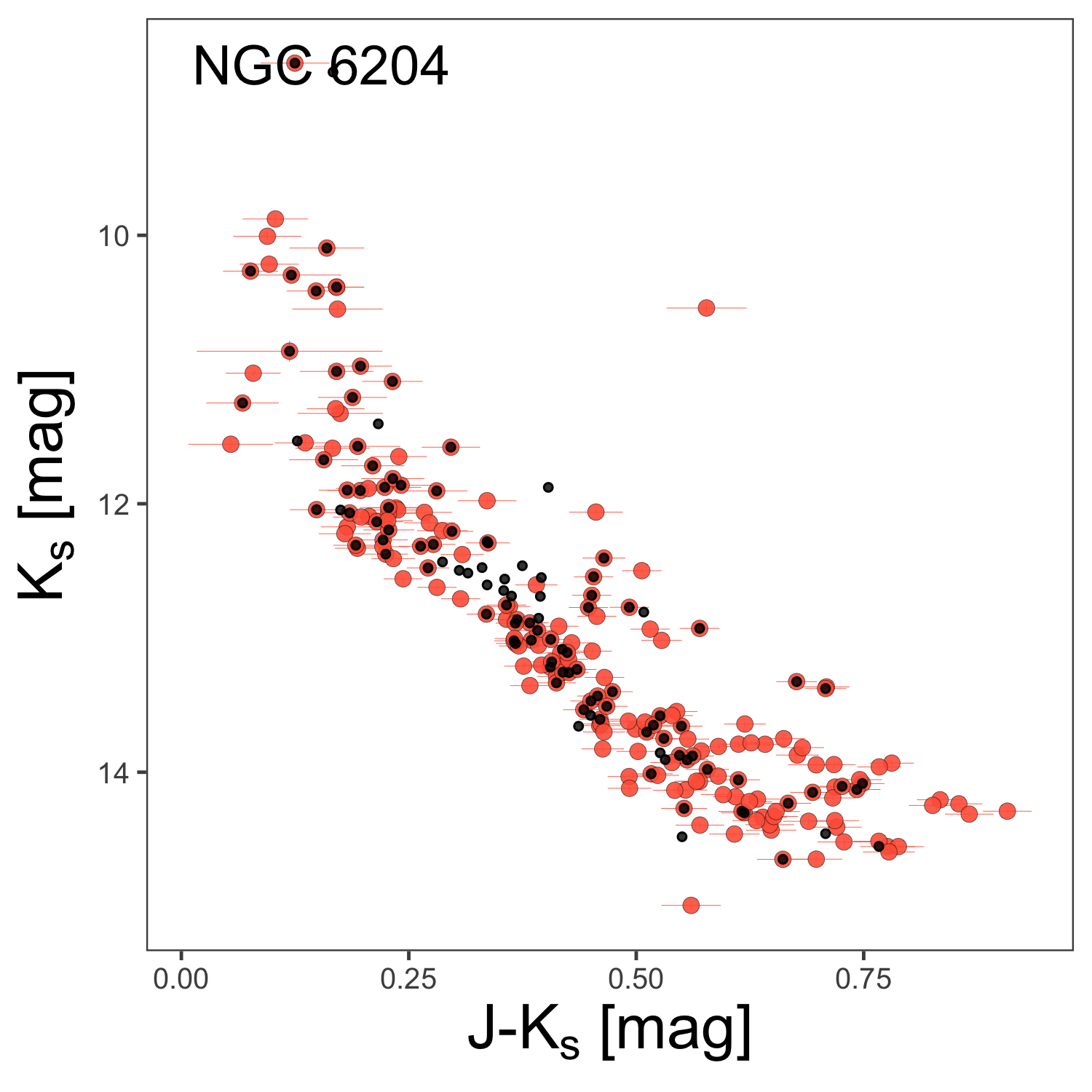}
\includegraphics[scale=0.075]{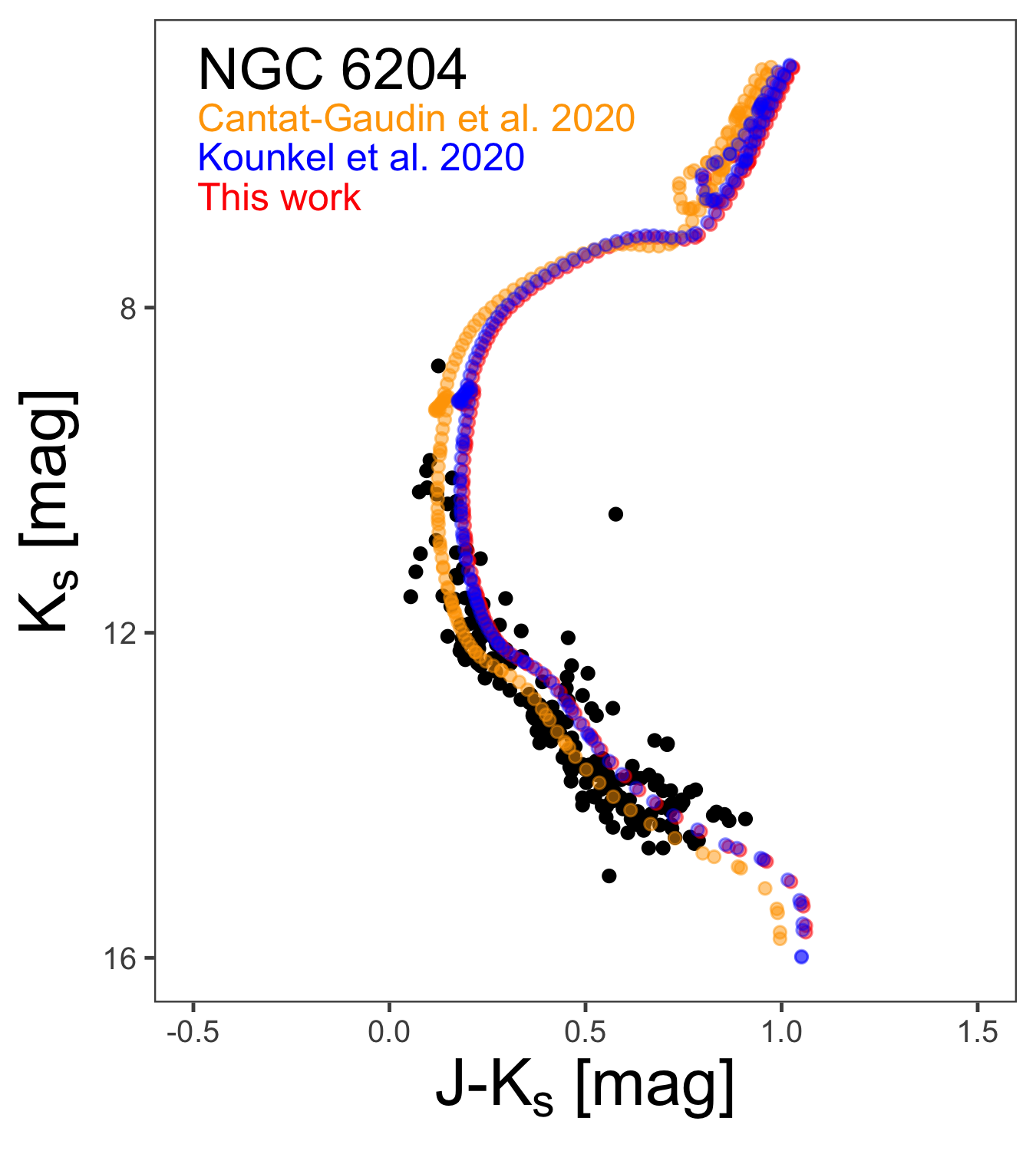}
\includegraphics[scale=0.075]{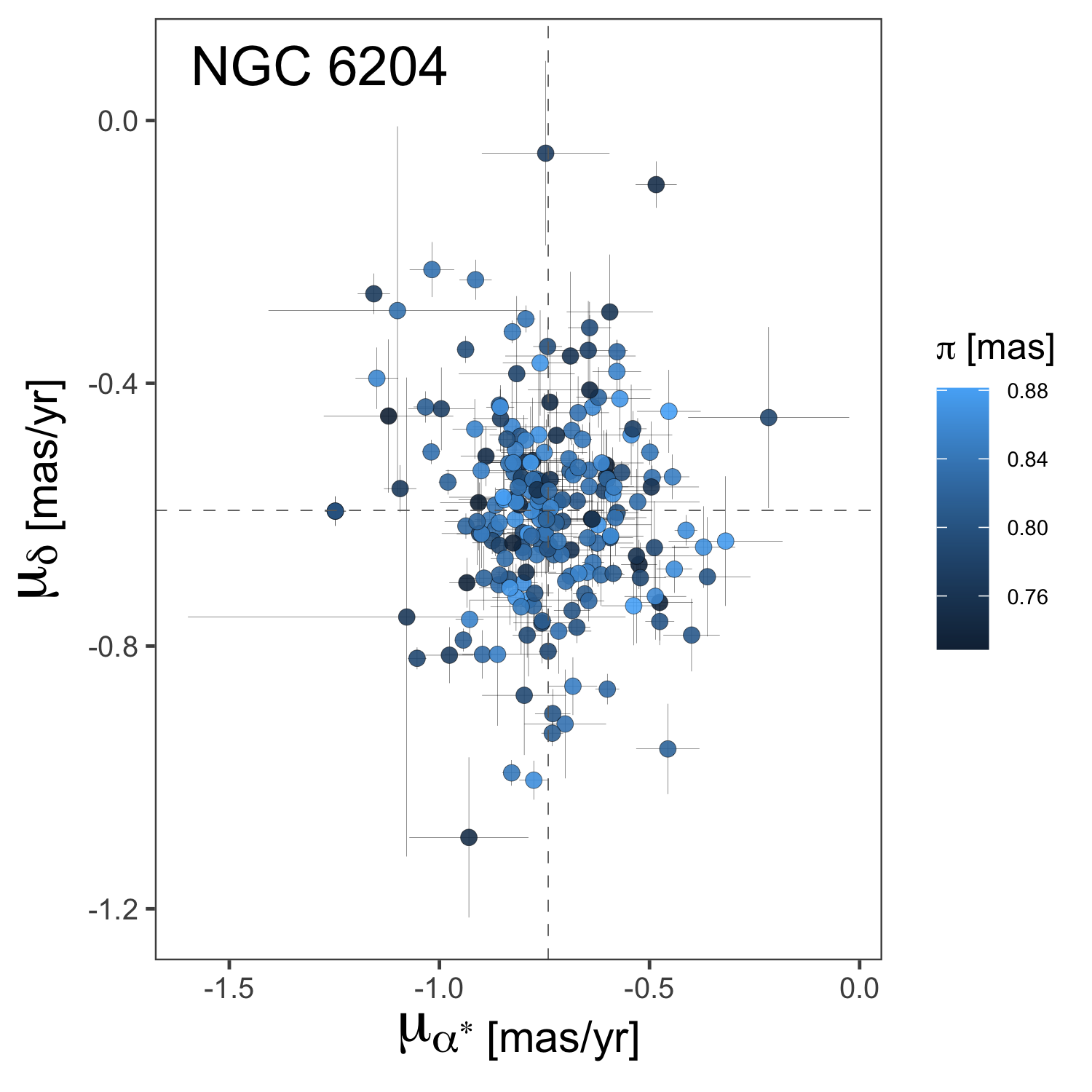}
\includegraphics[scale=0.075]{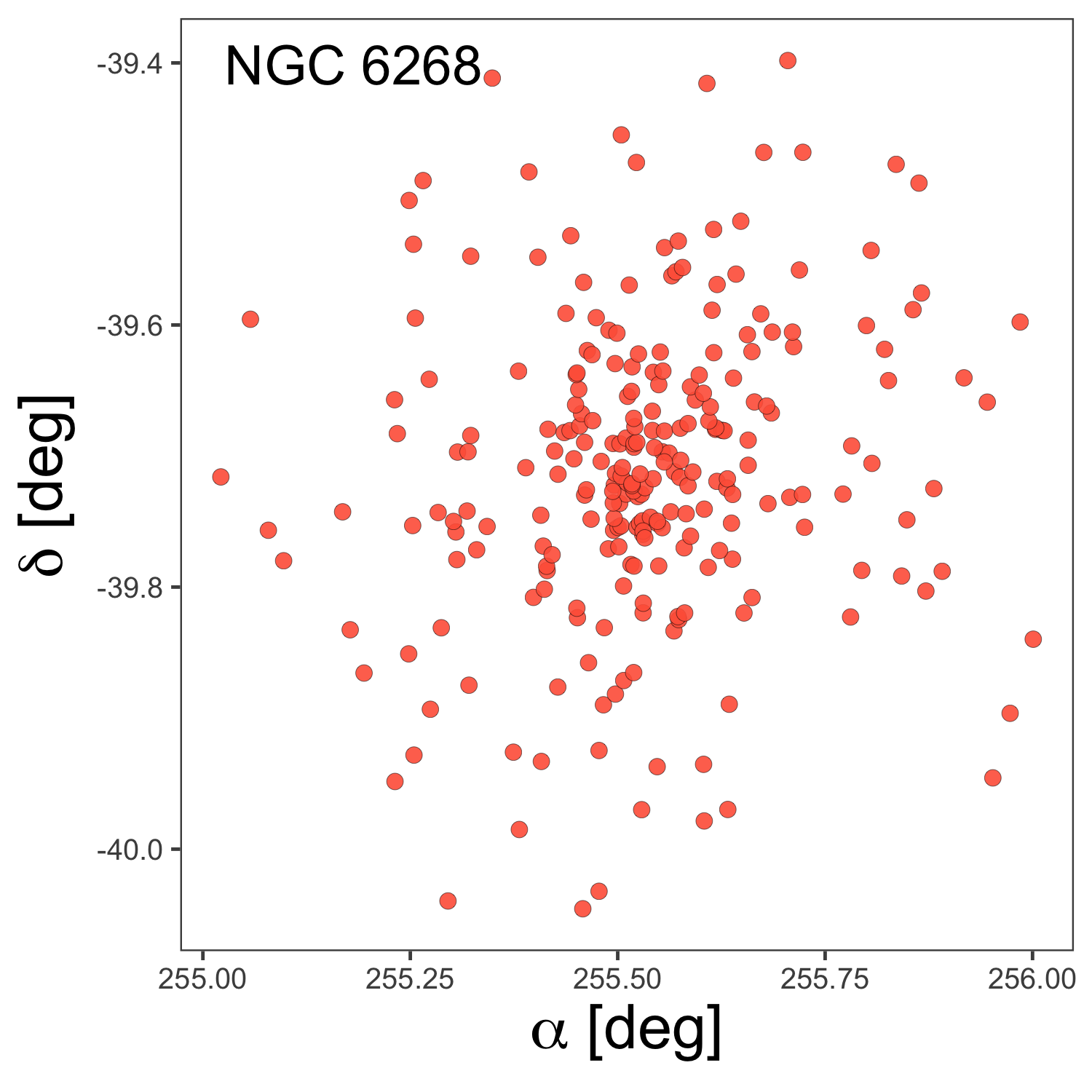}
\includegraphics[scale=0.075]{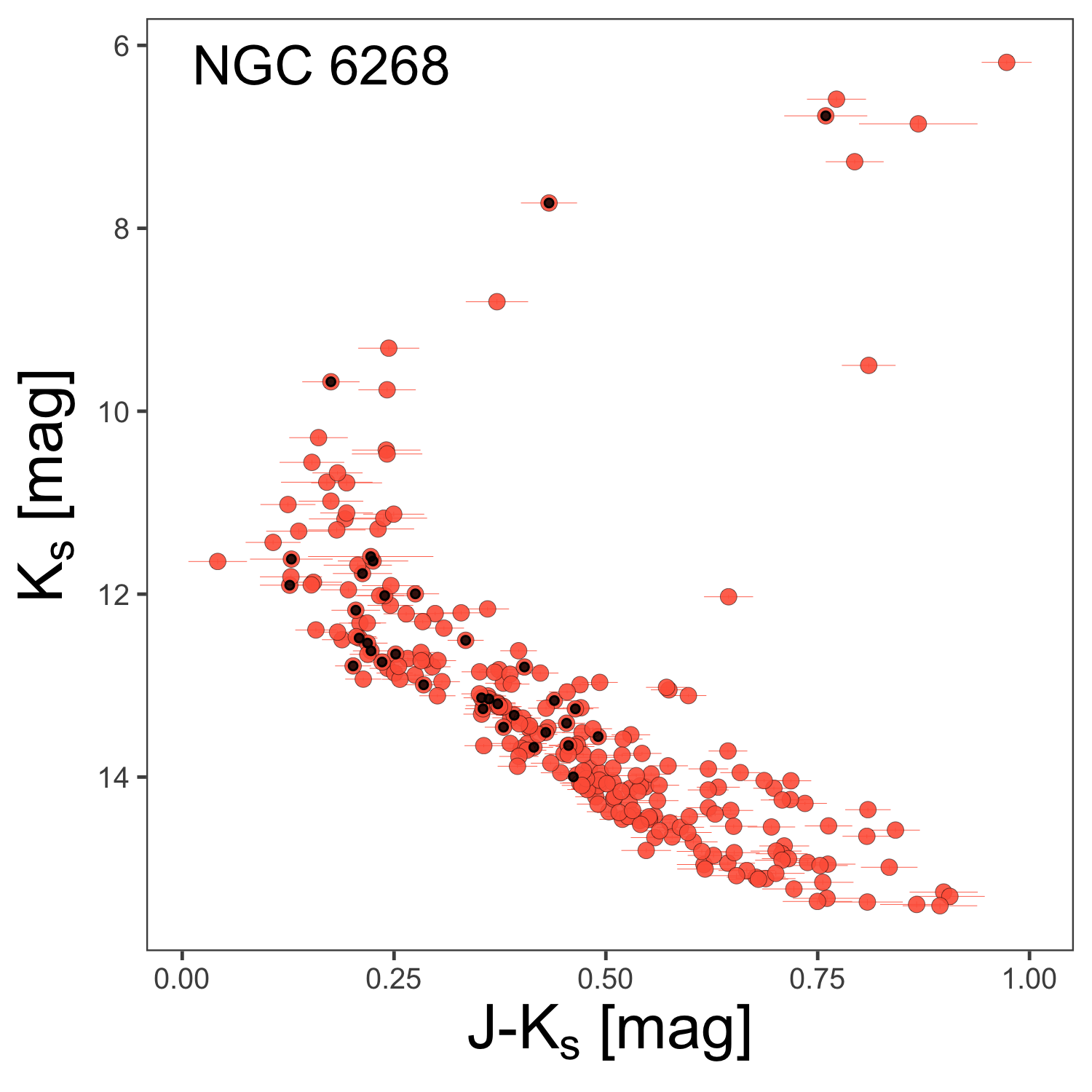}
\includegraphics[scale=0.075]{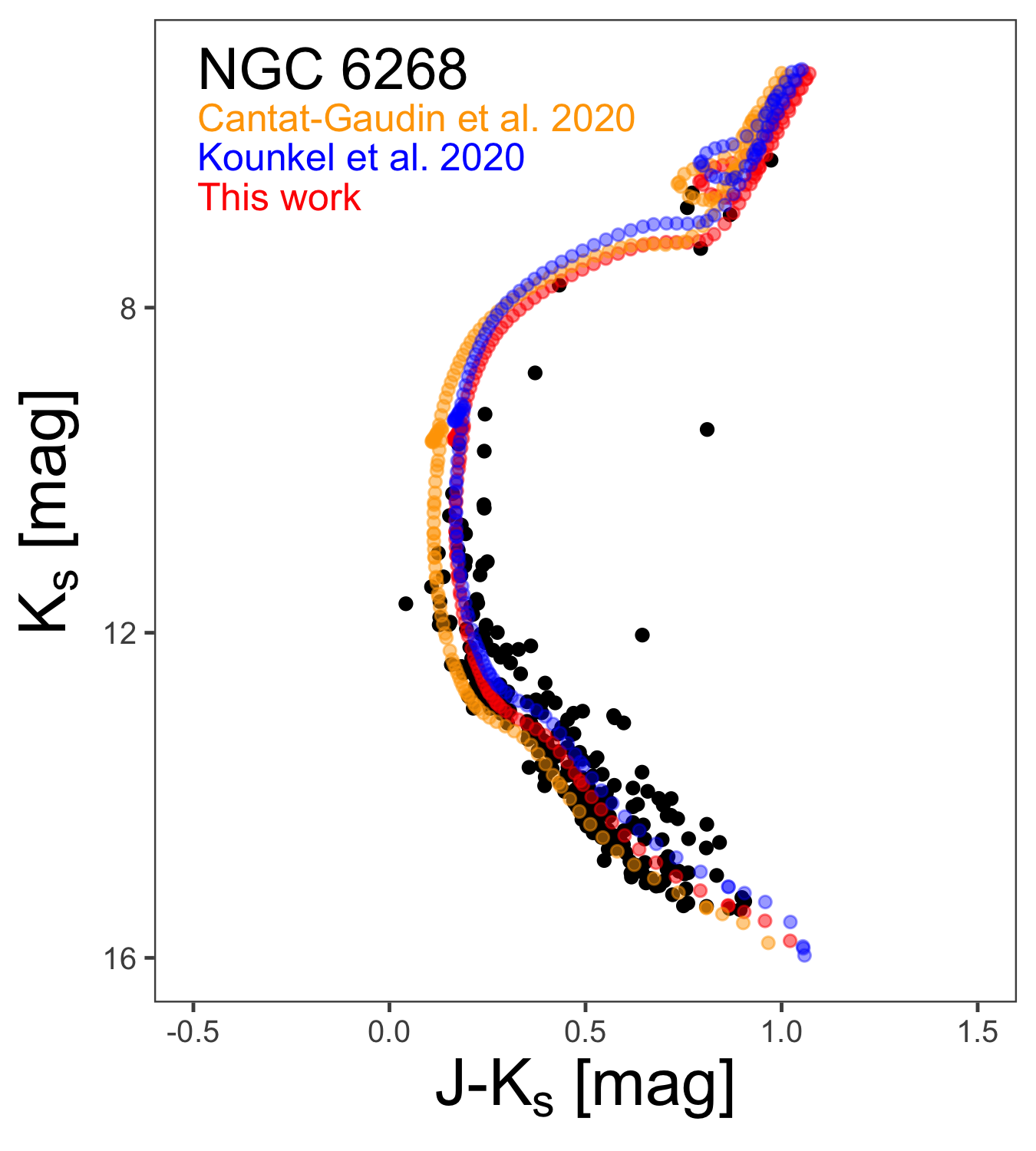}
\includegraphics[scale=0.075]{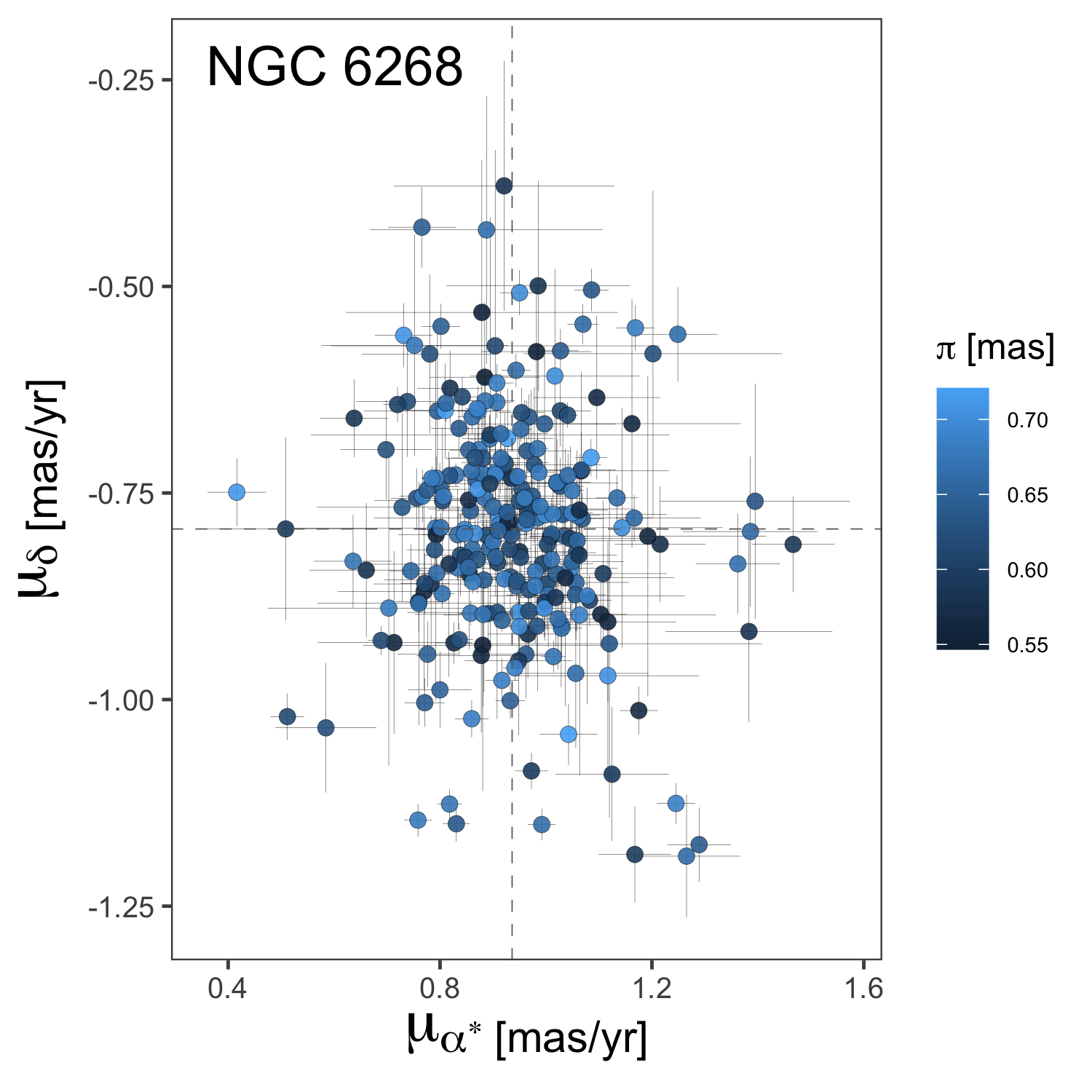}
\includegraphics[scale=0.075]{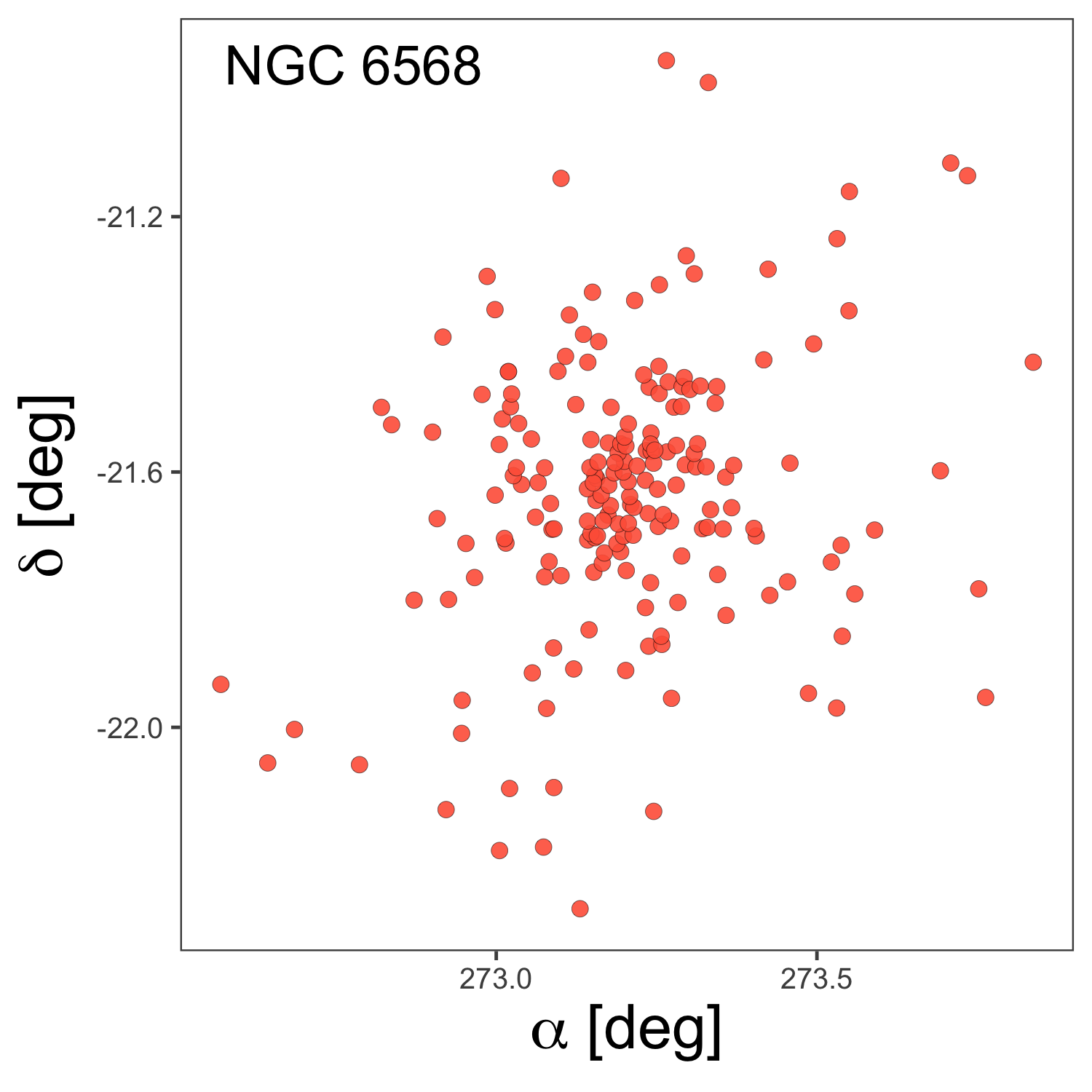}
\includegraphics[scale=0.075]{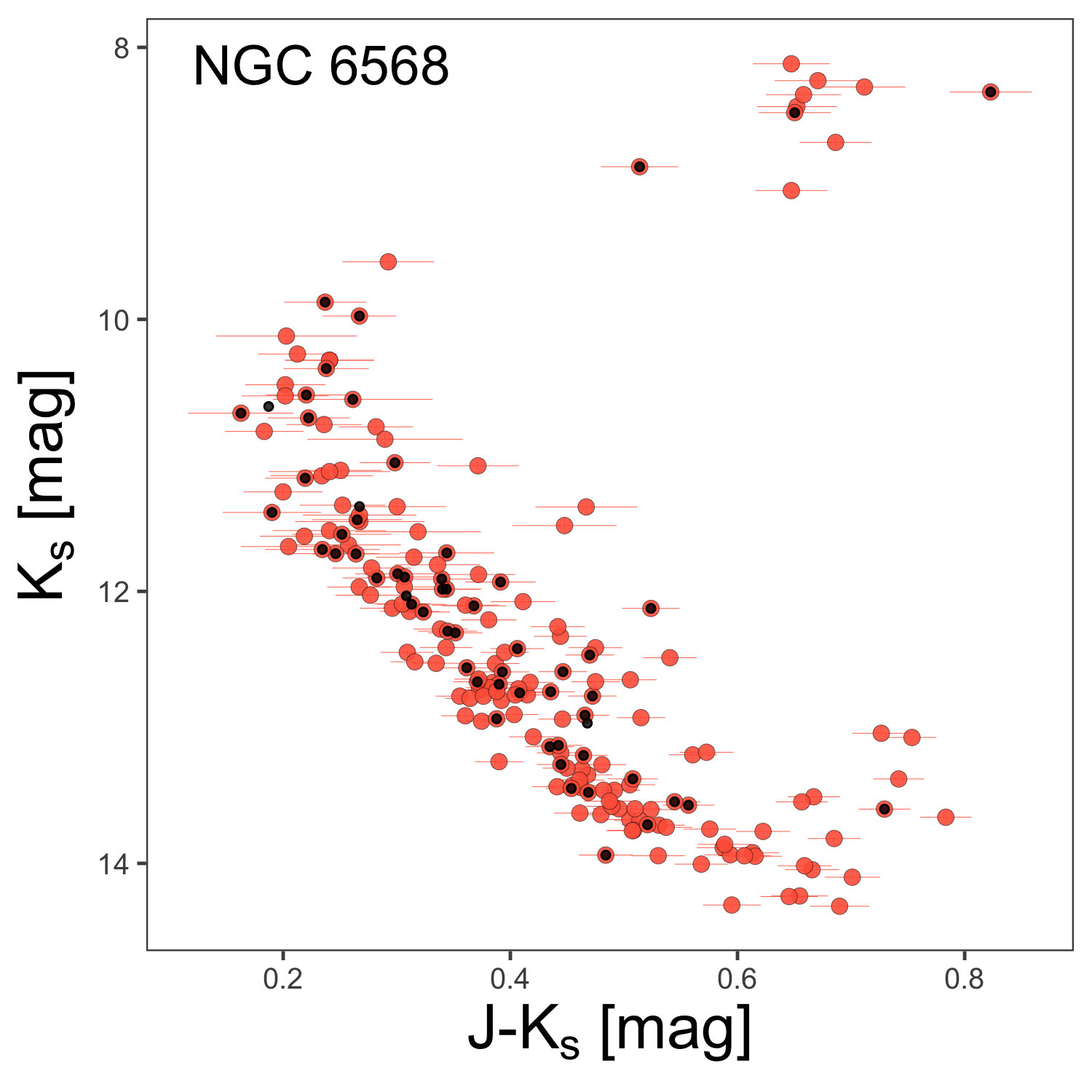}
\includegraphics[scale=0.075]{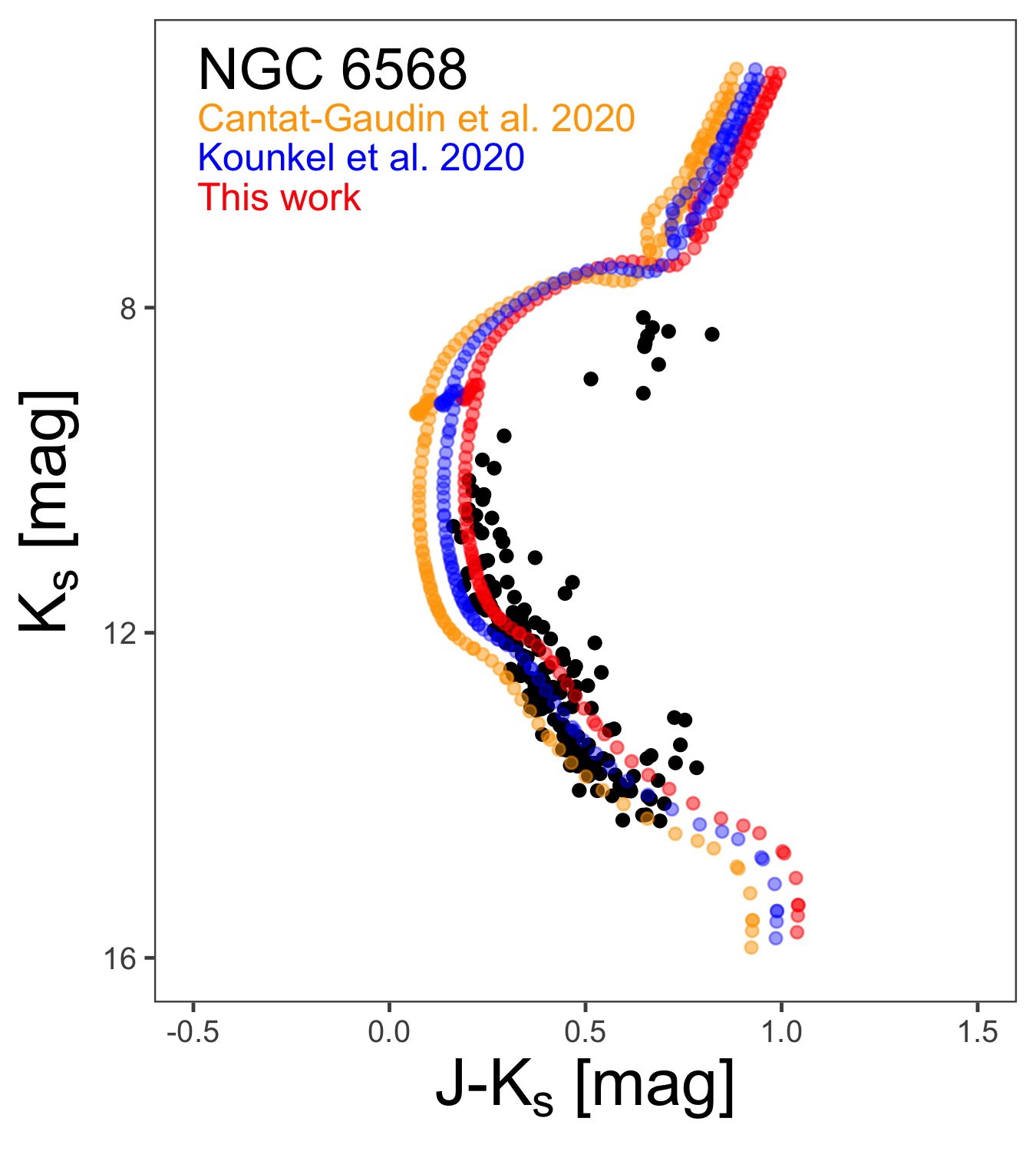}
\includegraphics[scale=0.075]{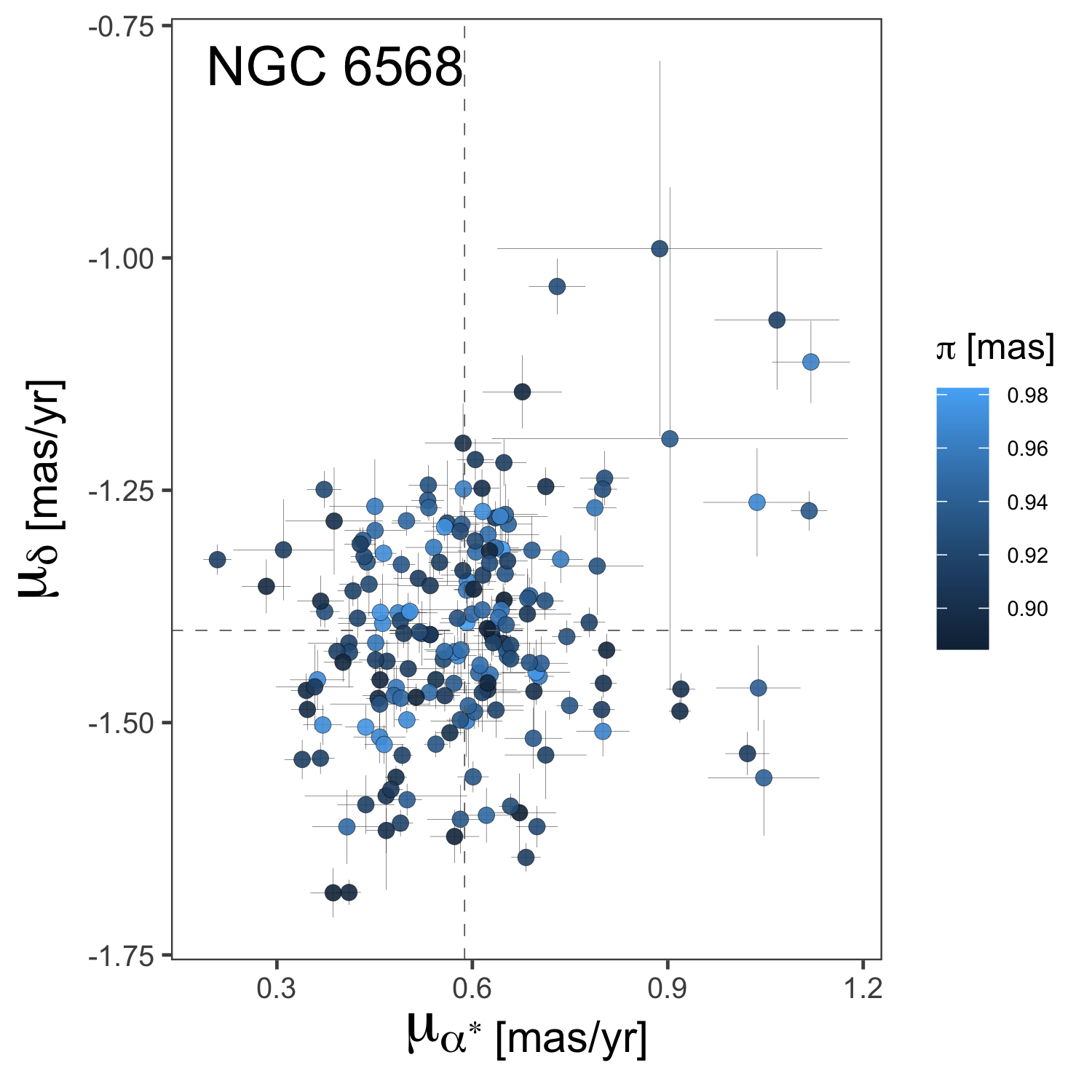}
\caption{Same as Figure\,\ref{fig:clusters}}
\end{figure*}

\begin{figure*}
\includegraphics[scale=0.075]{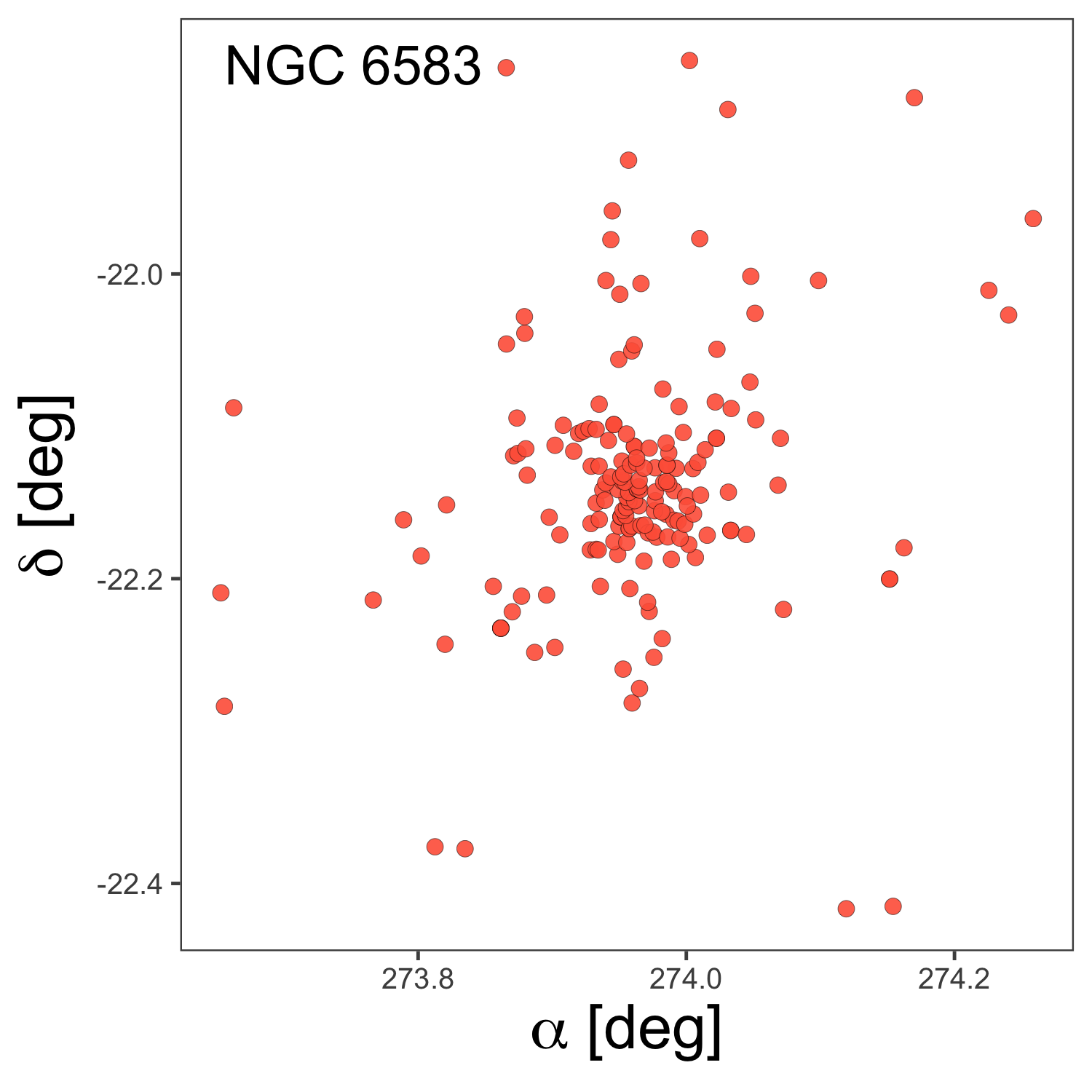}
\includegraphics[scale=0.075]{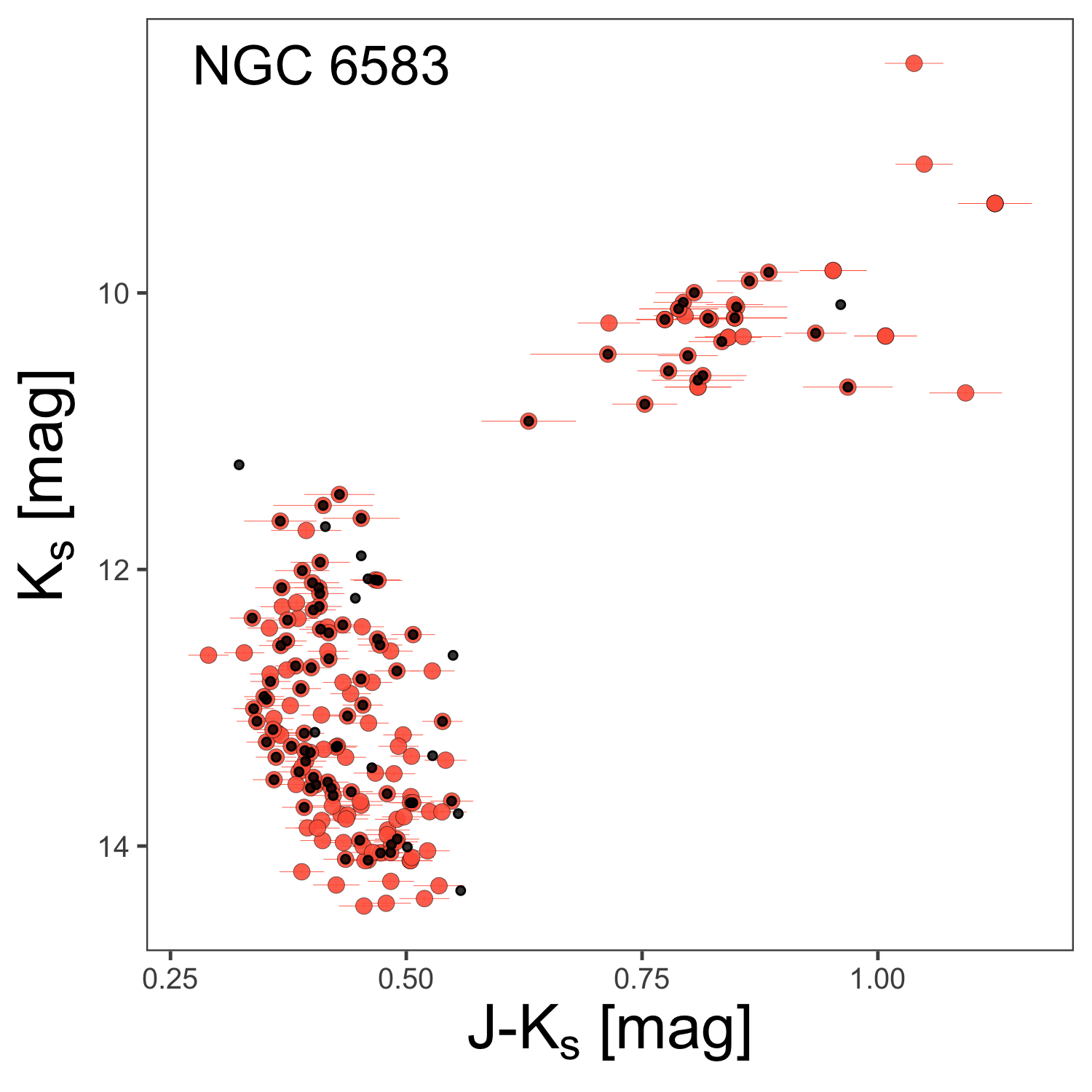}
\includegraphics[scale=0.075]{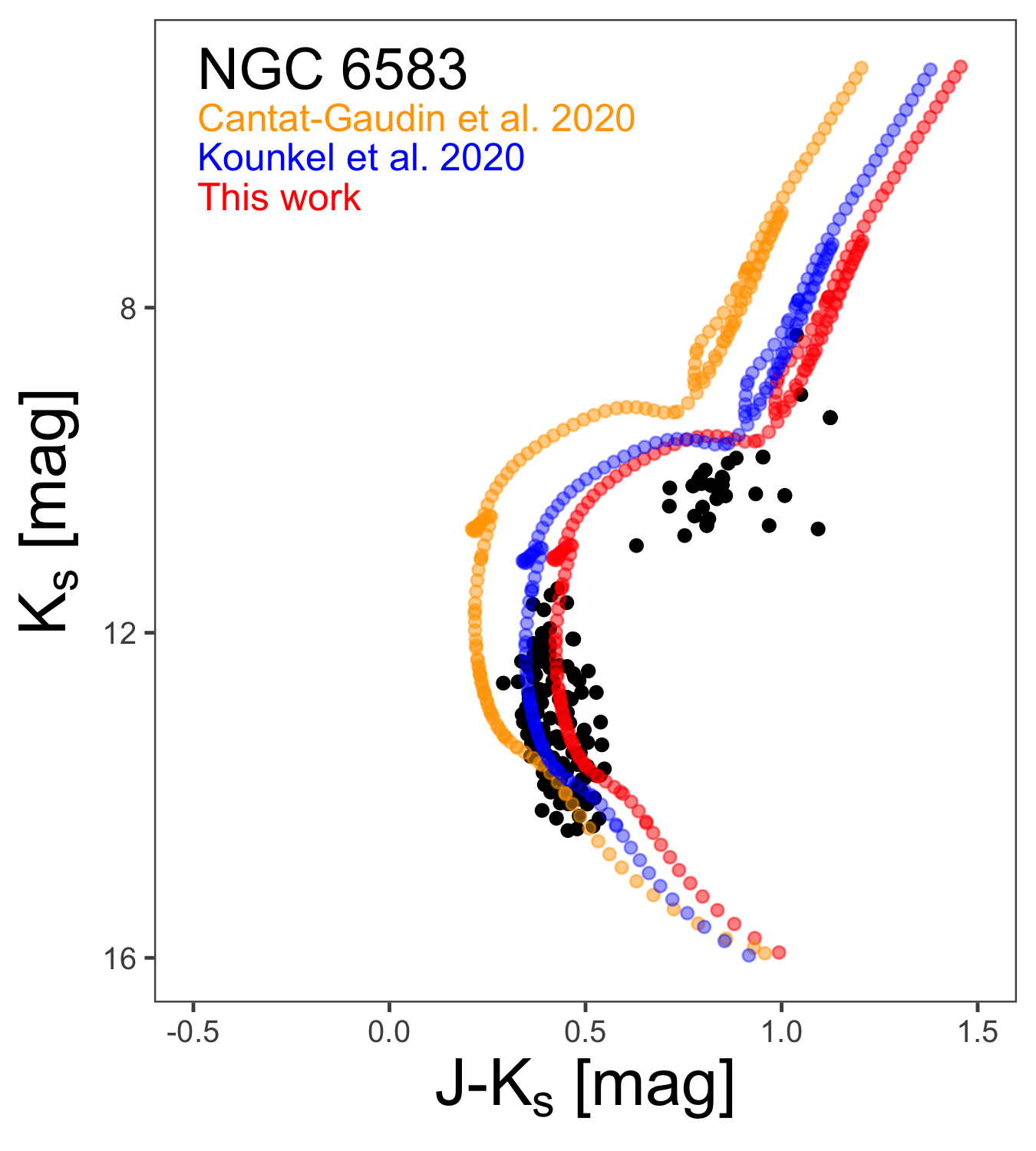}
\includegraphics[scale=0.075]{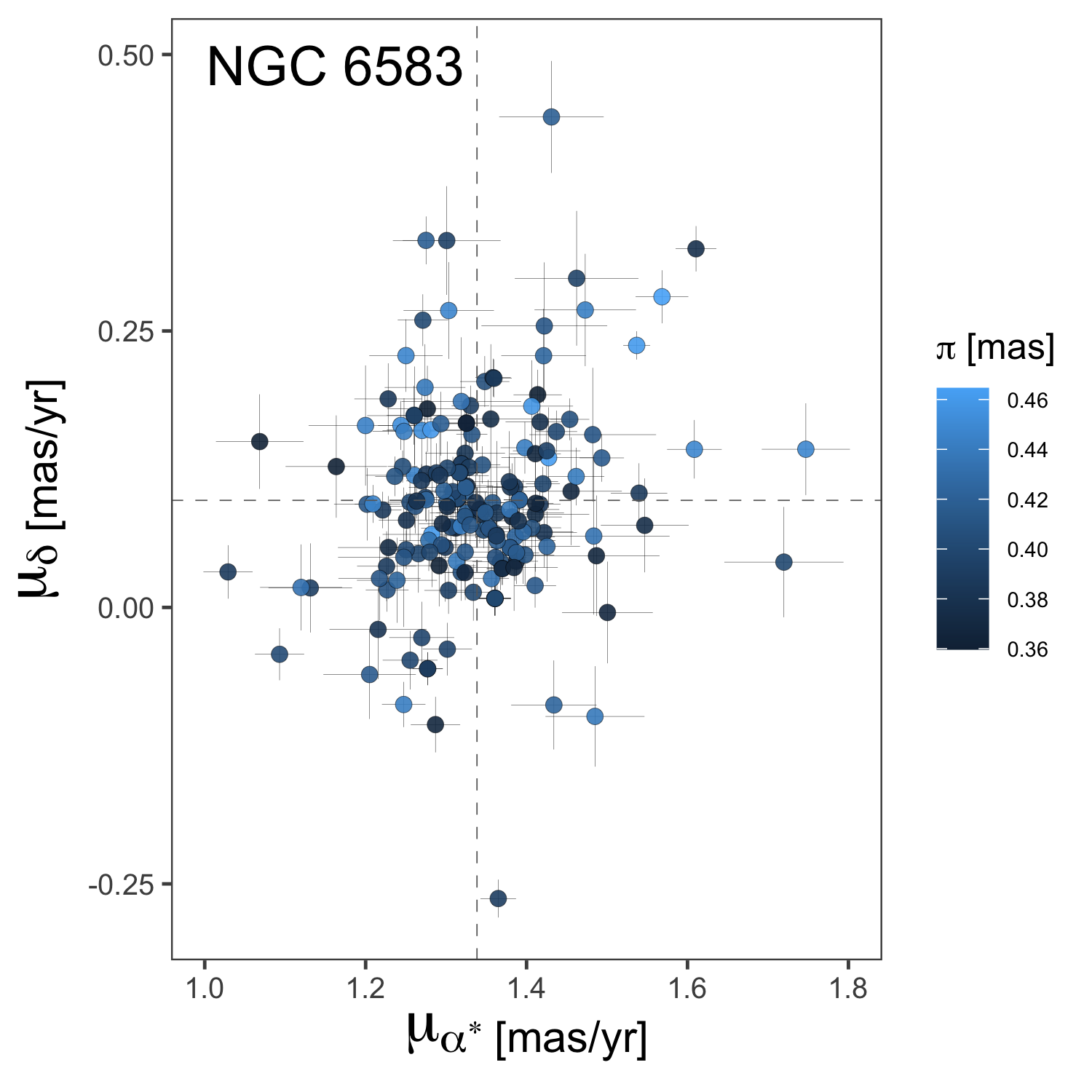}
\includegraphics[scale=0.075]{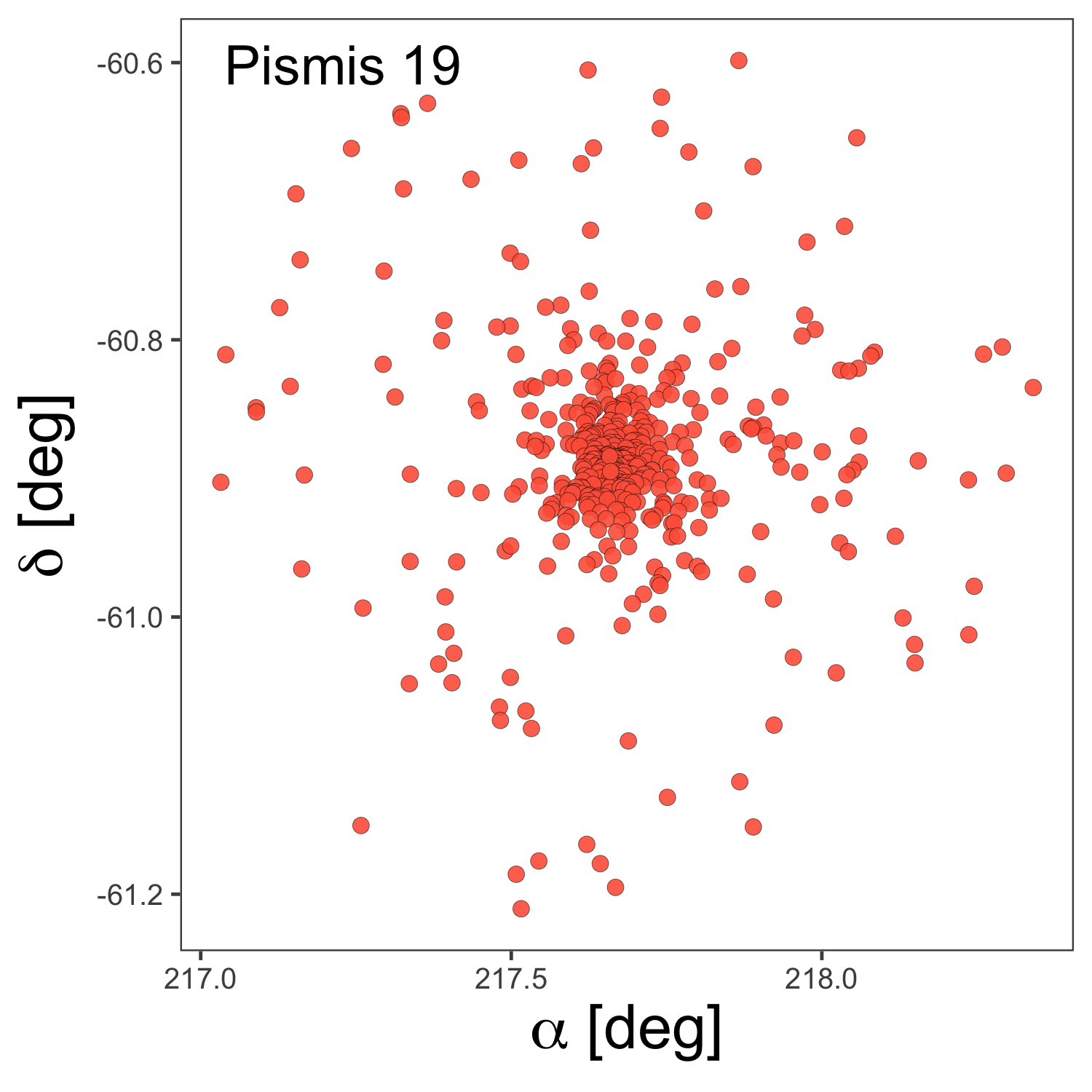}
\includegraphics[scale=0.075]{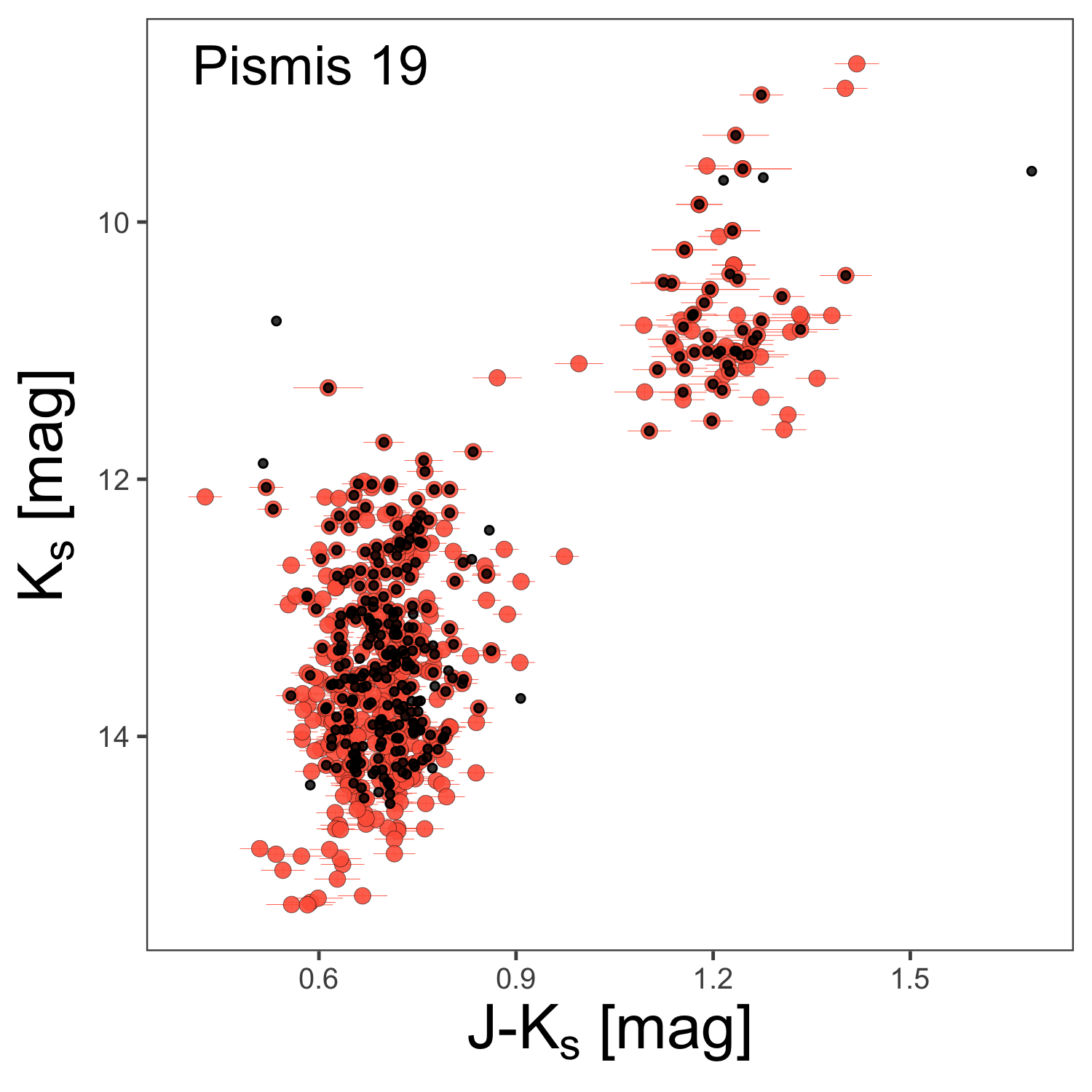}
\includegraphics[scale=0.075]{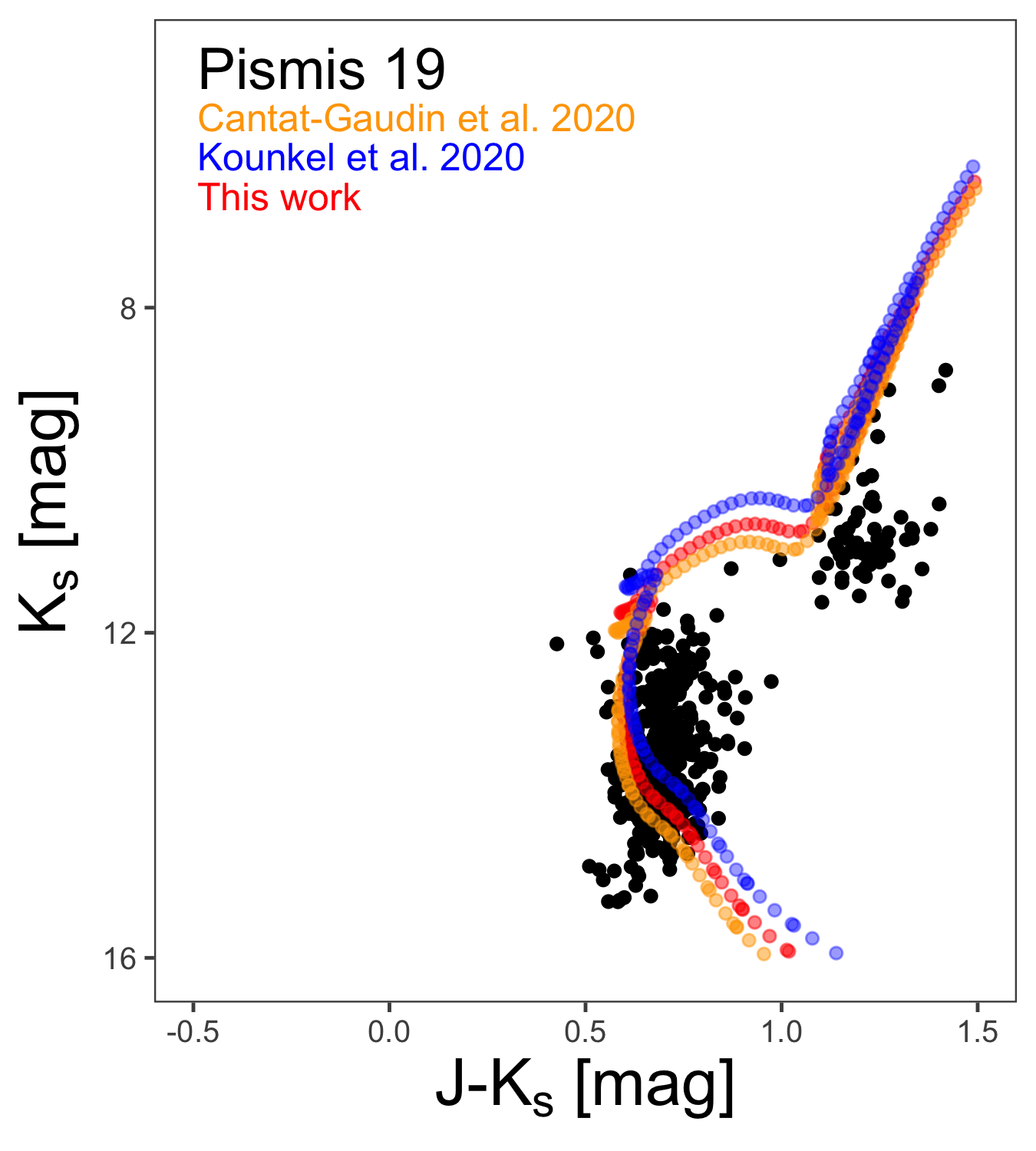}
\includegraphics[scale=0.075]{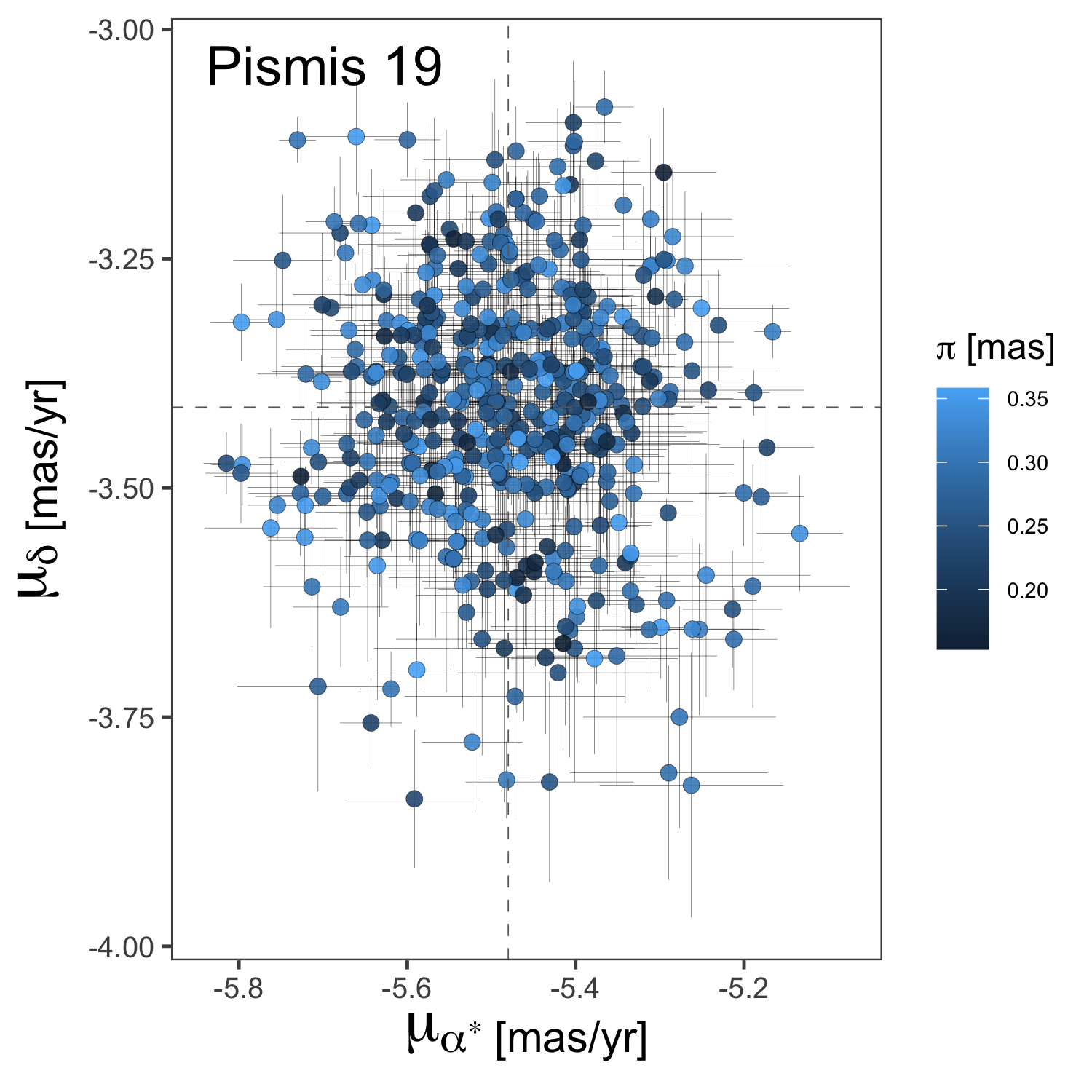}
\includegraphics[scale=0.075]{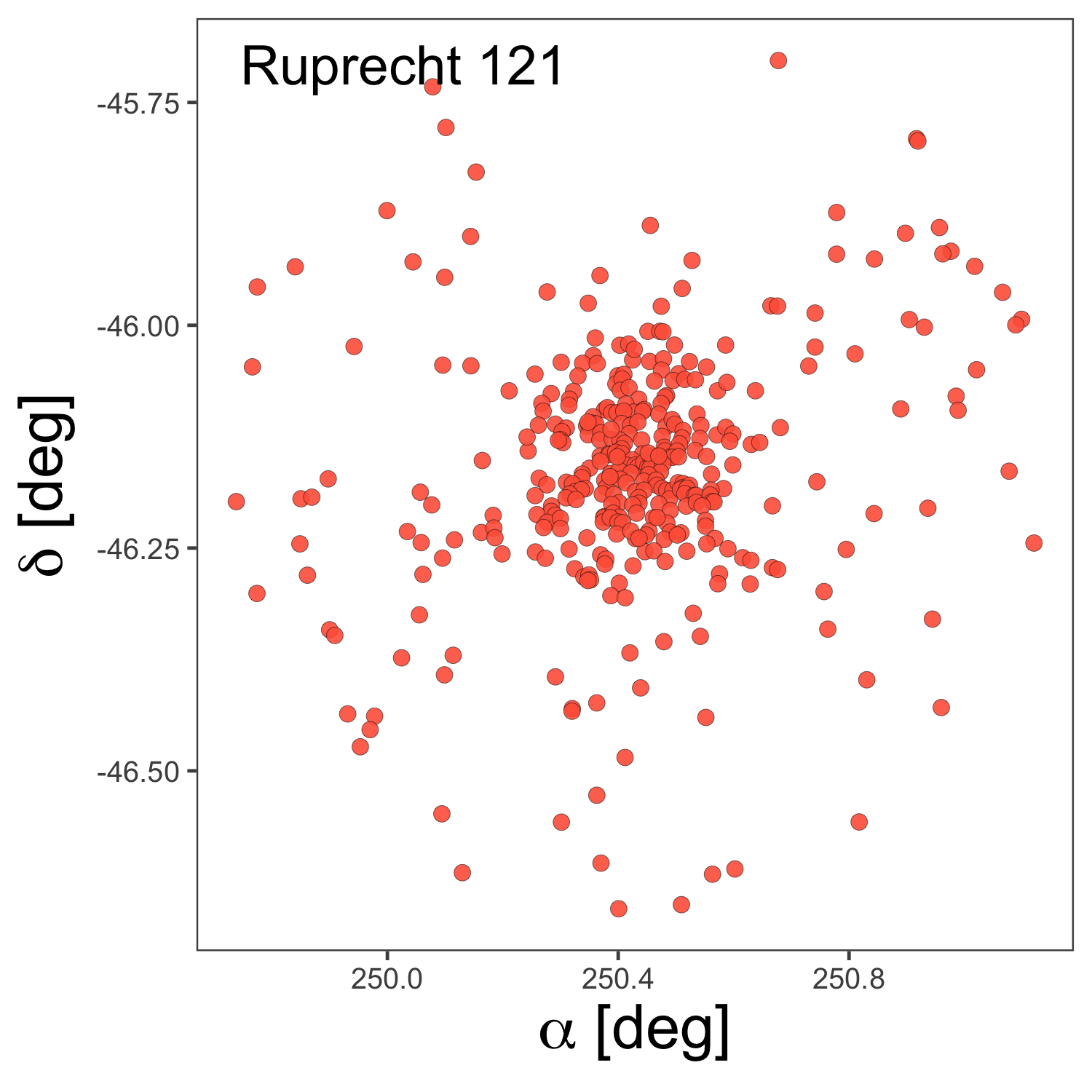}
\includegraphics[scale=0.075]{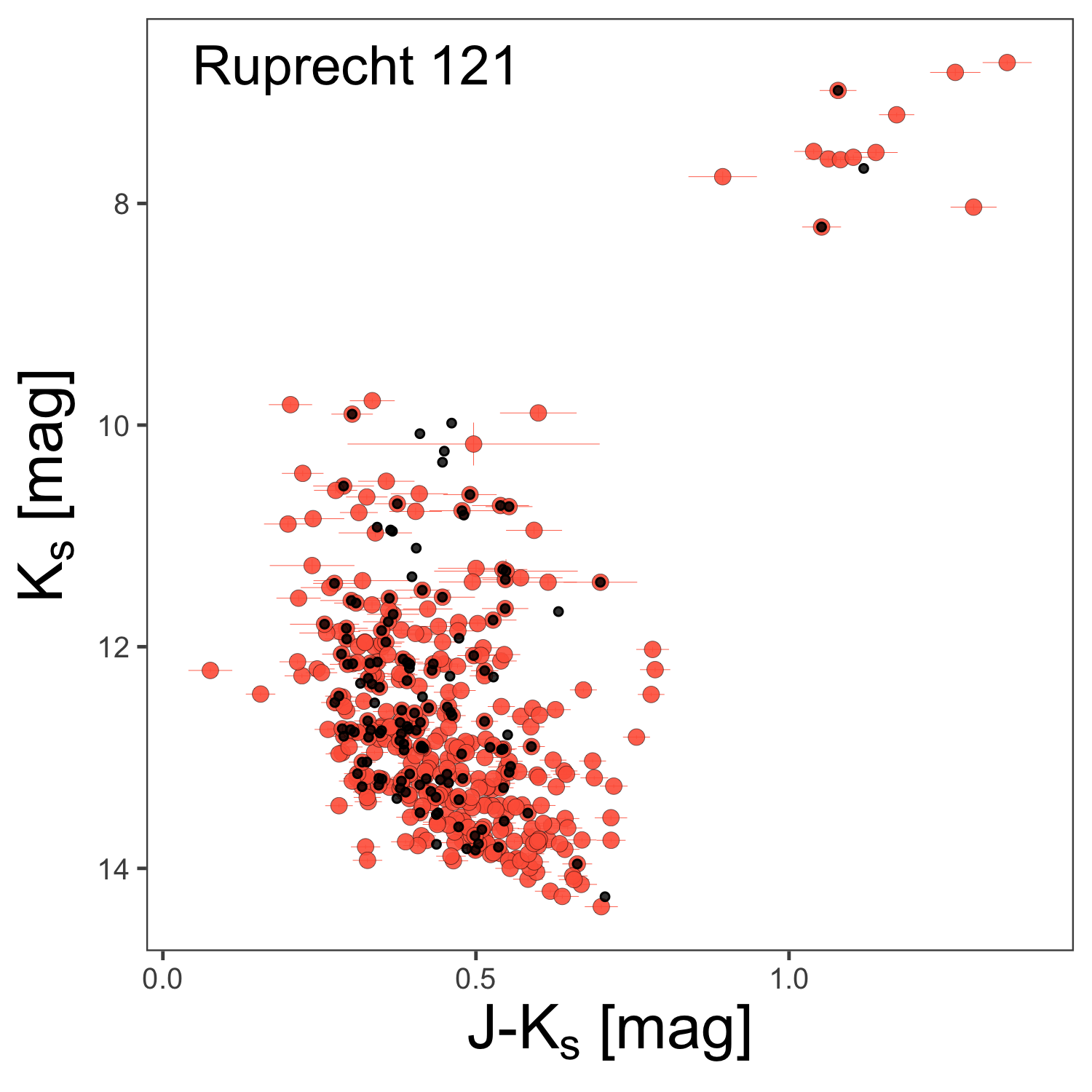}
\includegraphics[scale=0.075]{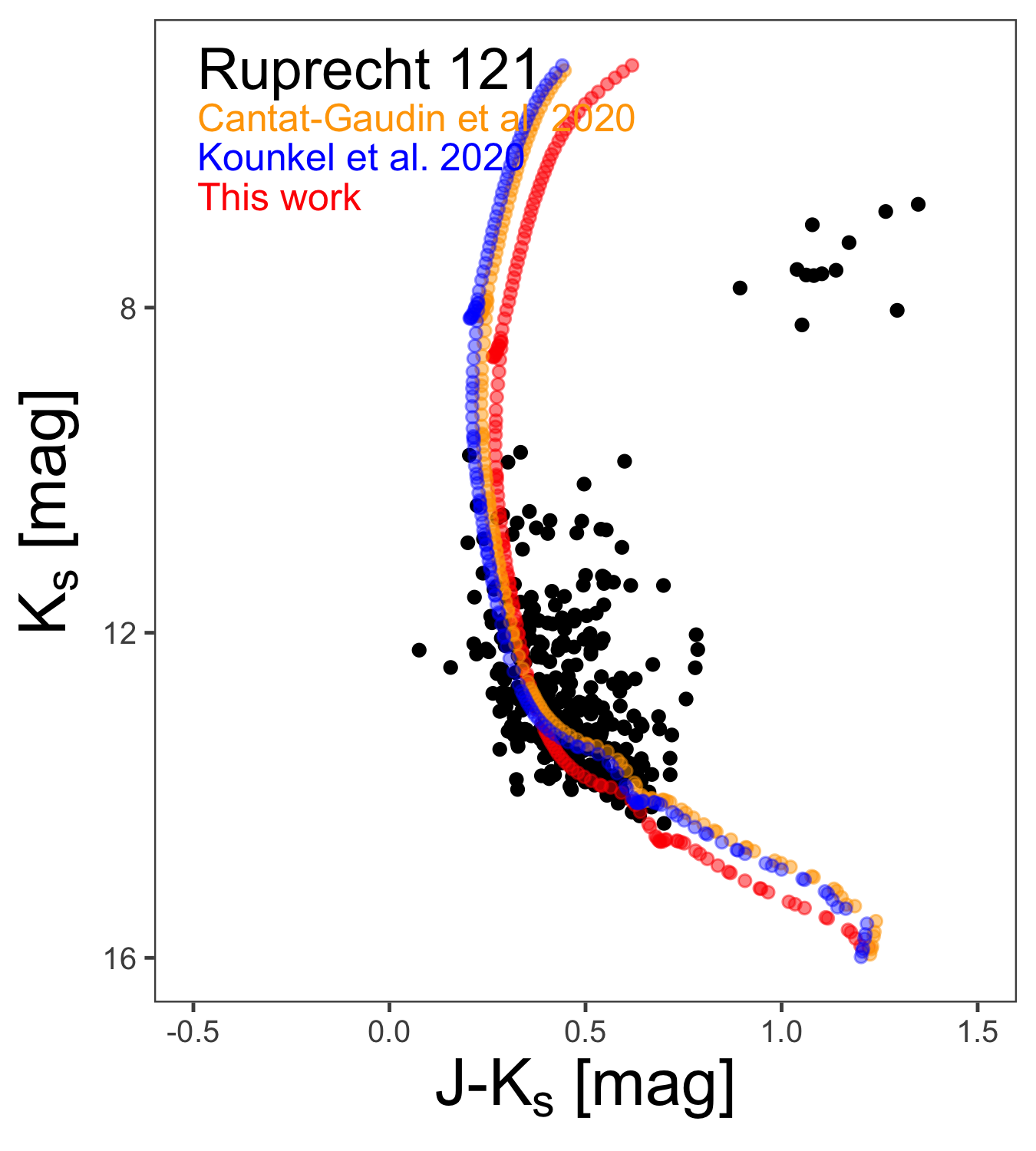}
\includegraphics[scale=0.075]{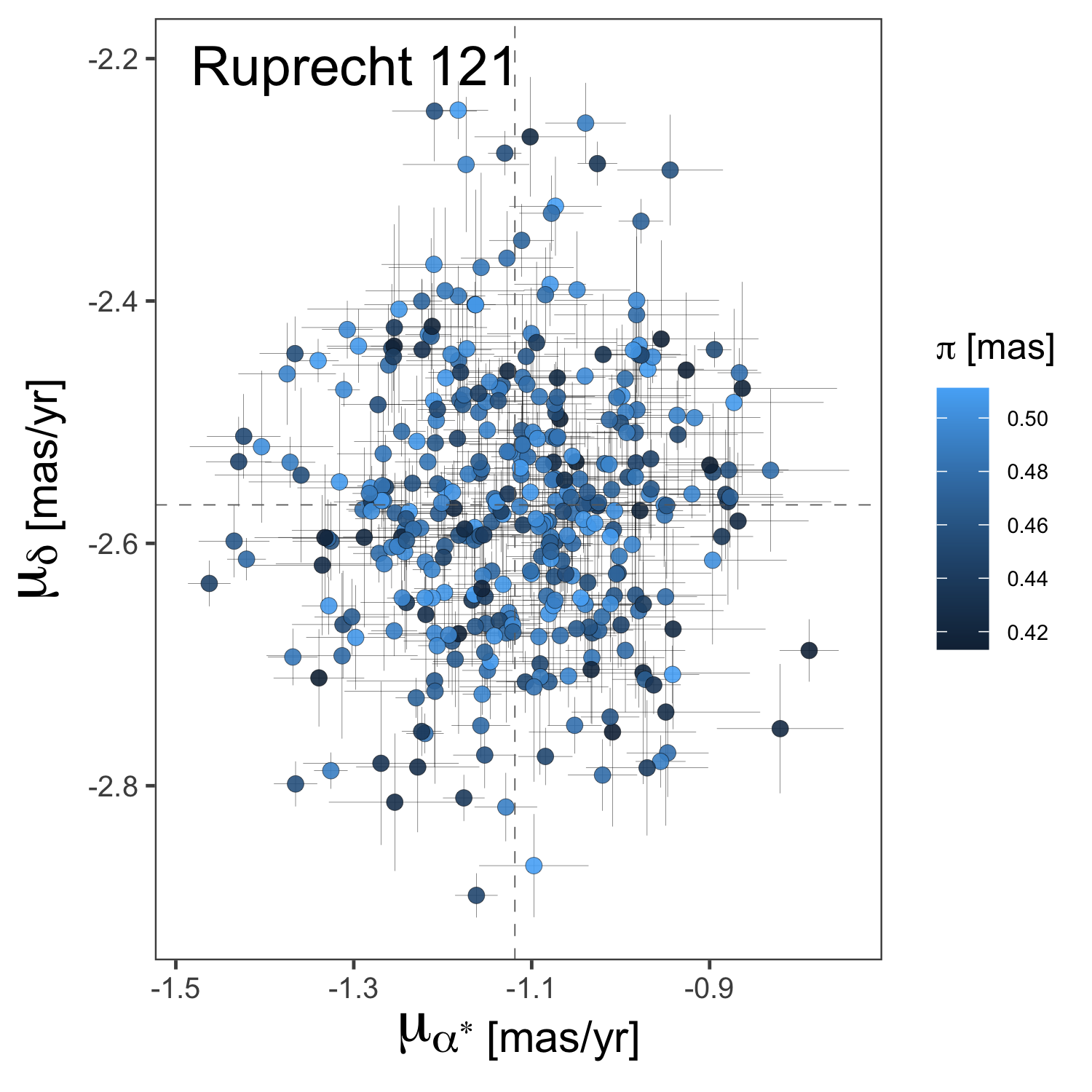}
\includegraphics[scale=0.075]{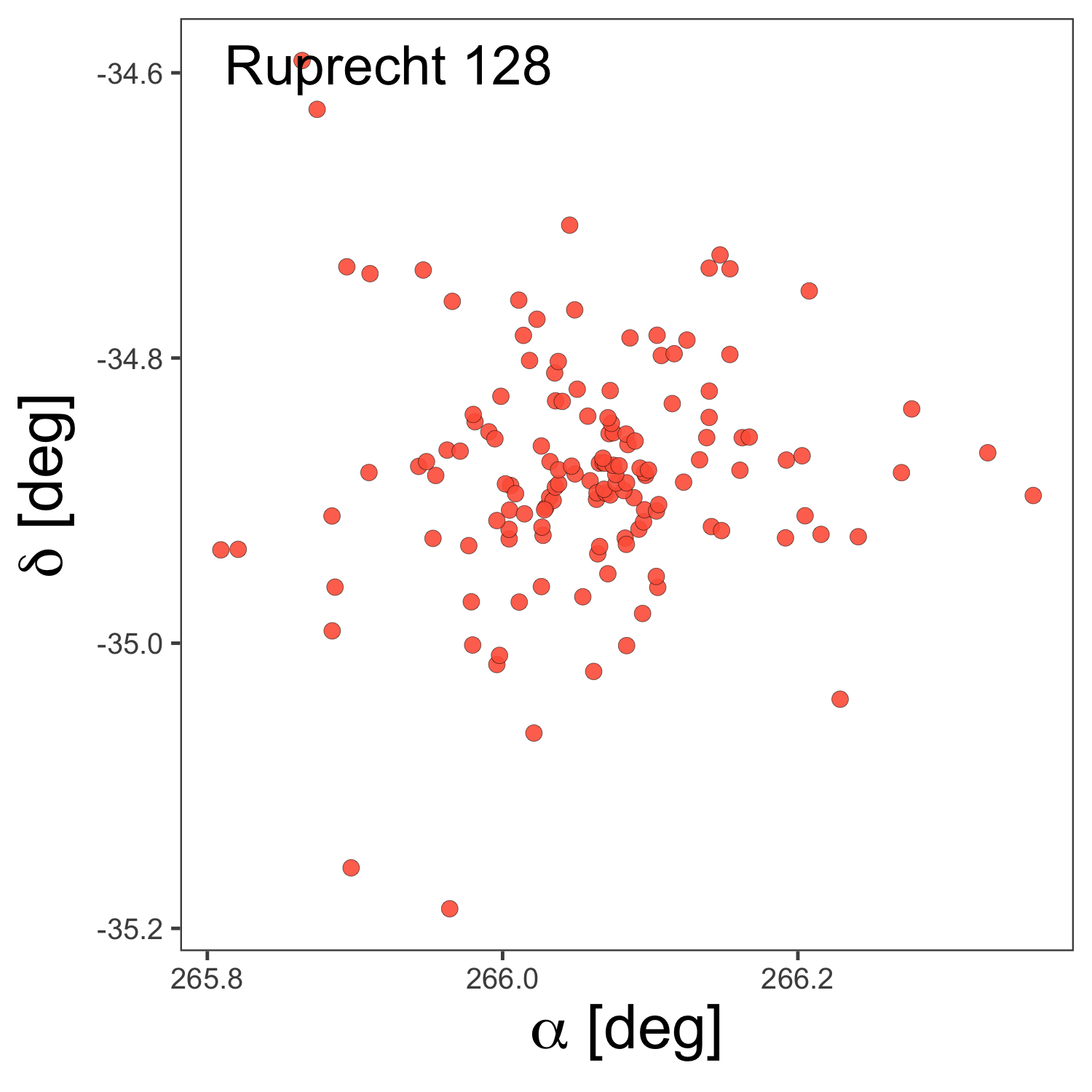}
\includegraphics[scale=0.075]{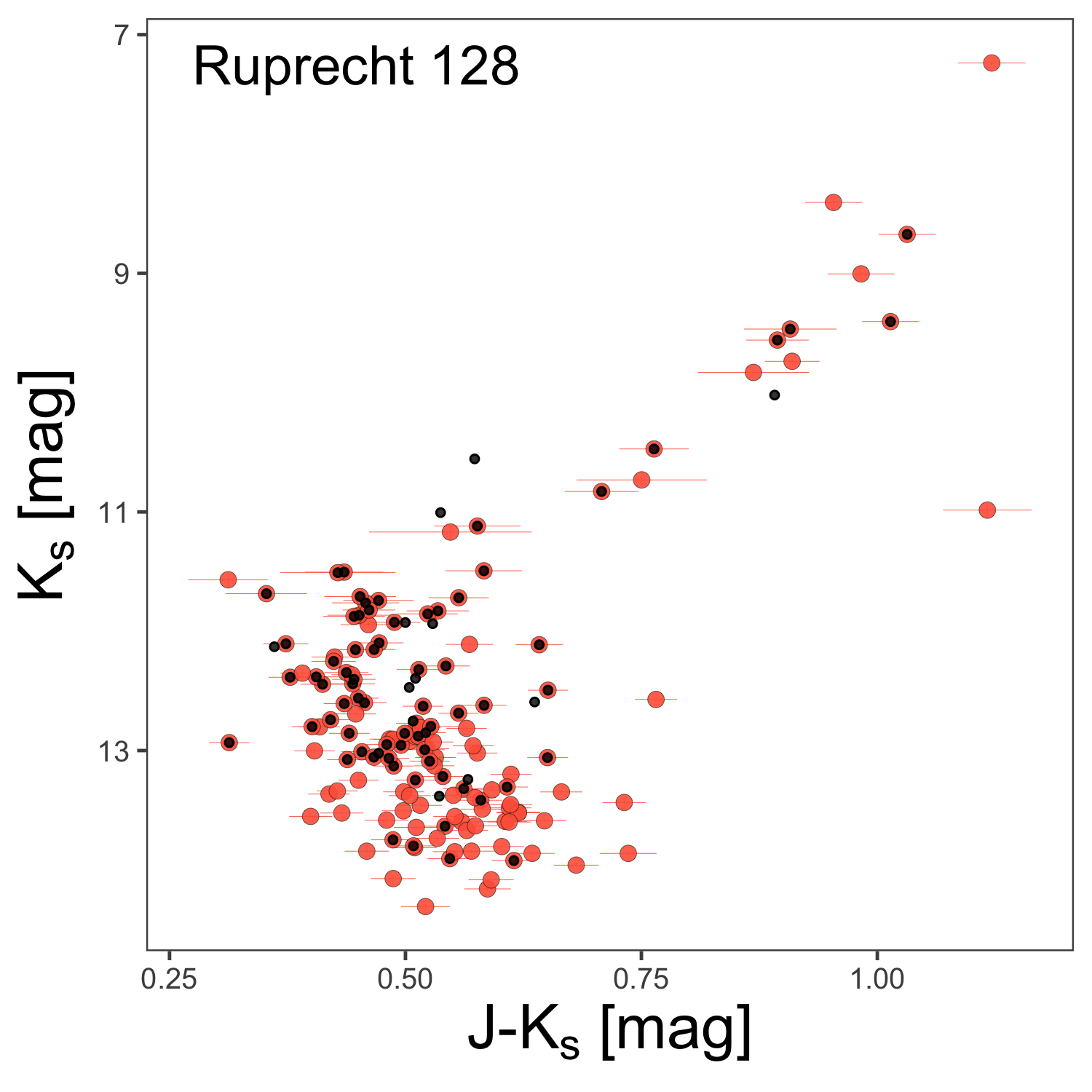}
\includegraphics[scale=0.075]{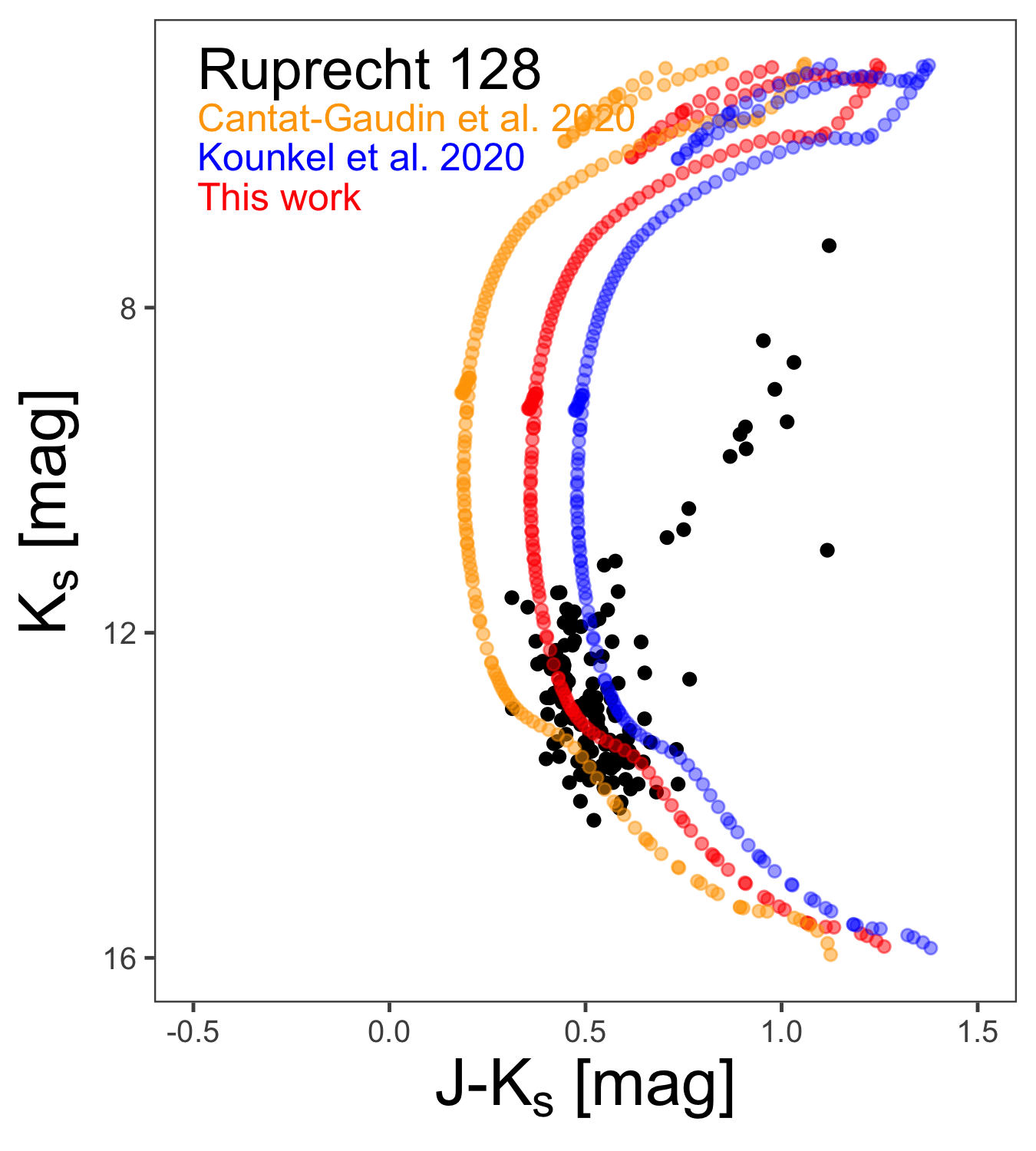}
\includegraphics[scale=0.075]{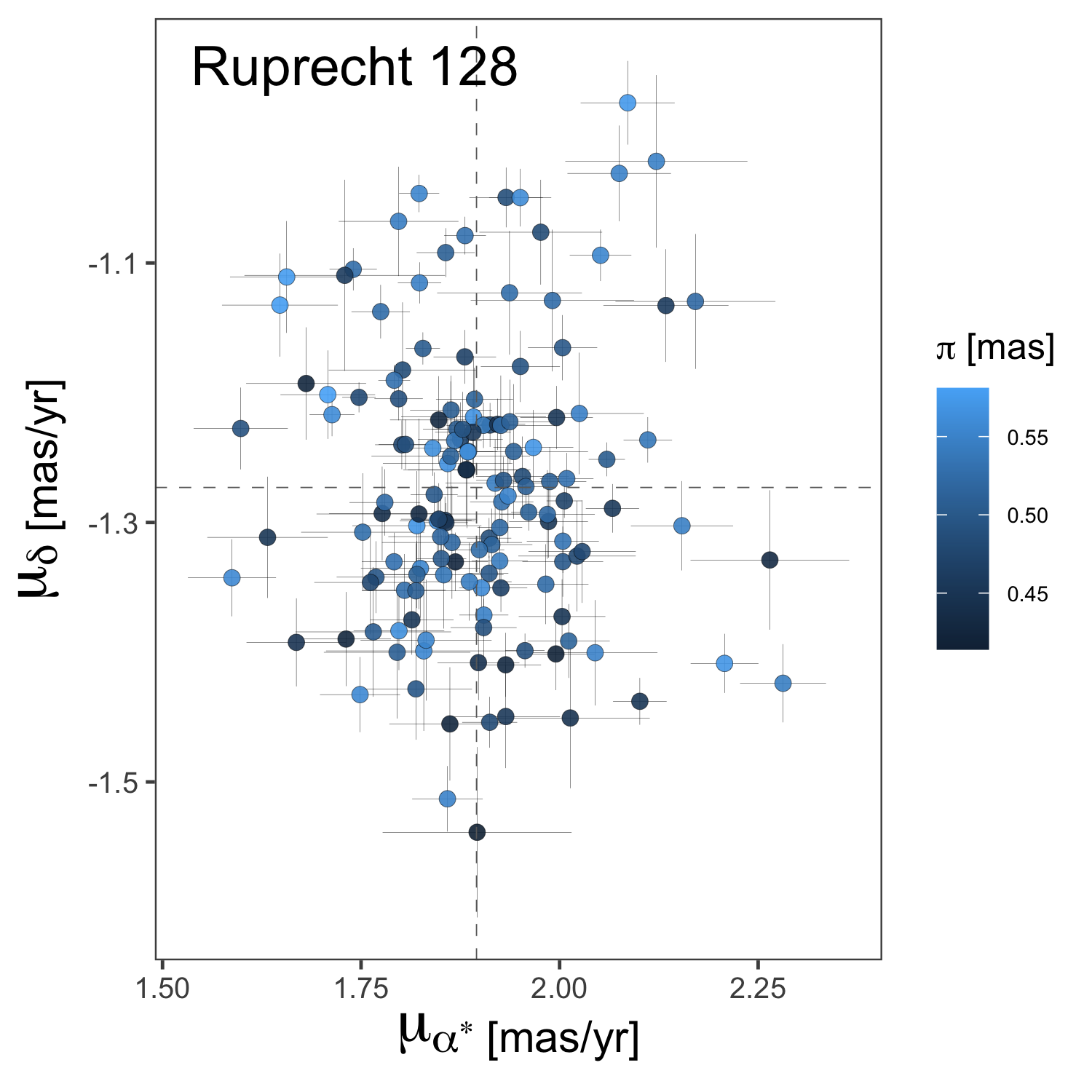}
\includegraphics[scale=0.075]{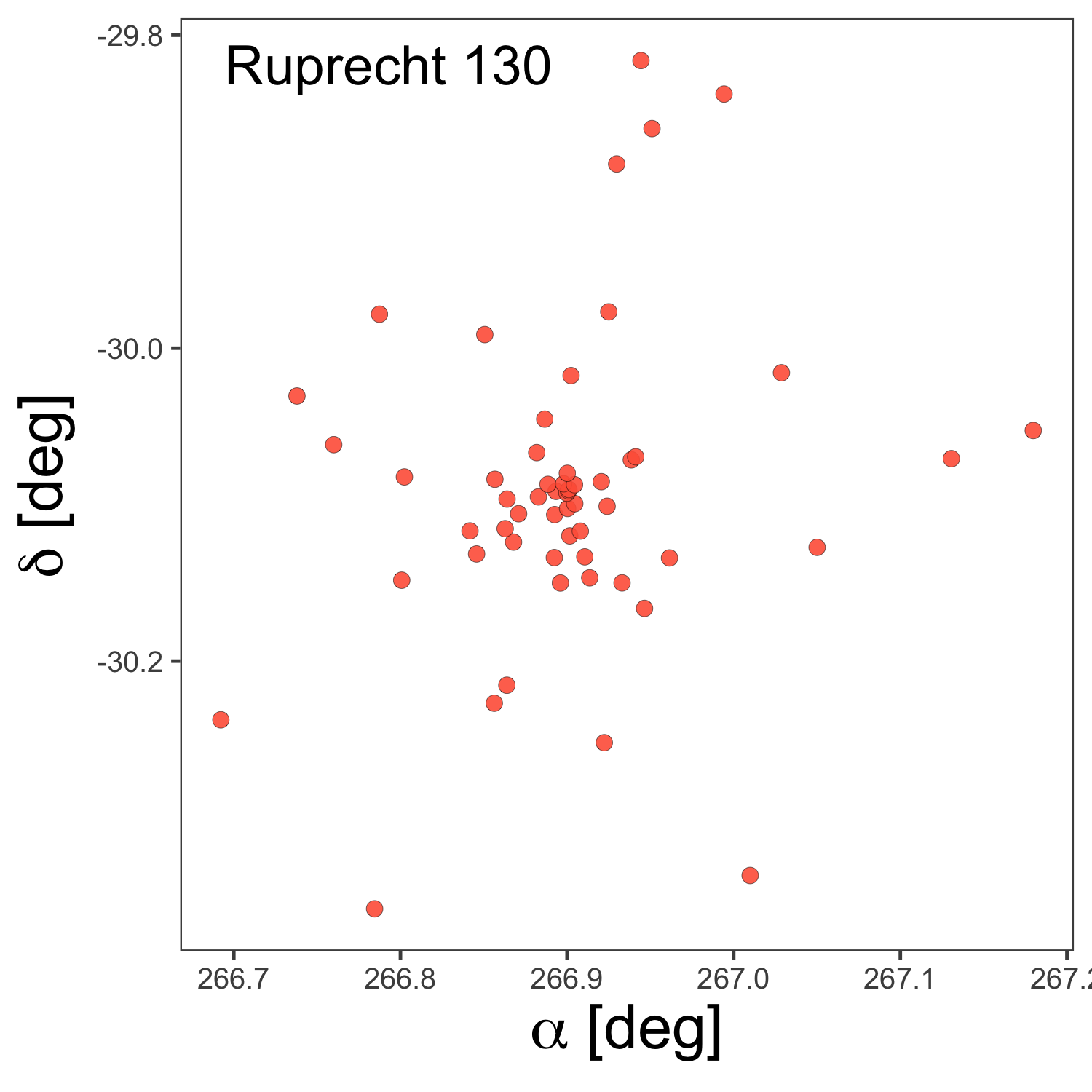}
\includegraphics[scale=0.075]{Ruprecht_130pub_cmd.png}
\includegraphics[scale=0.075]{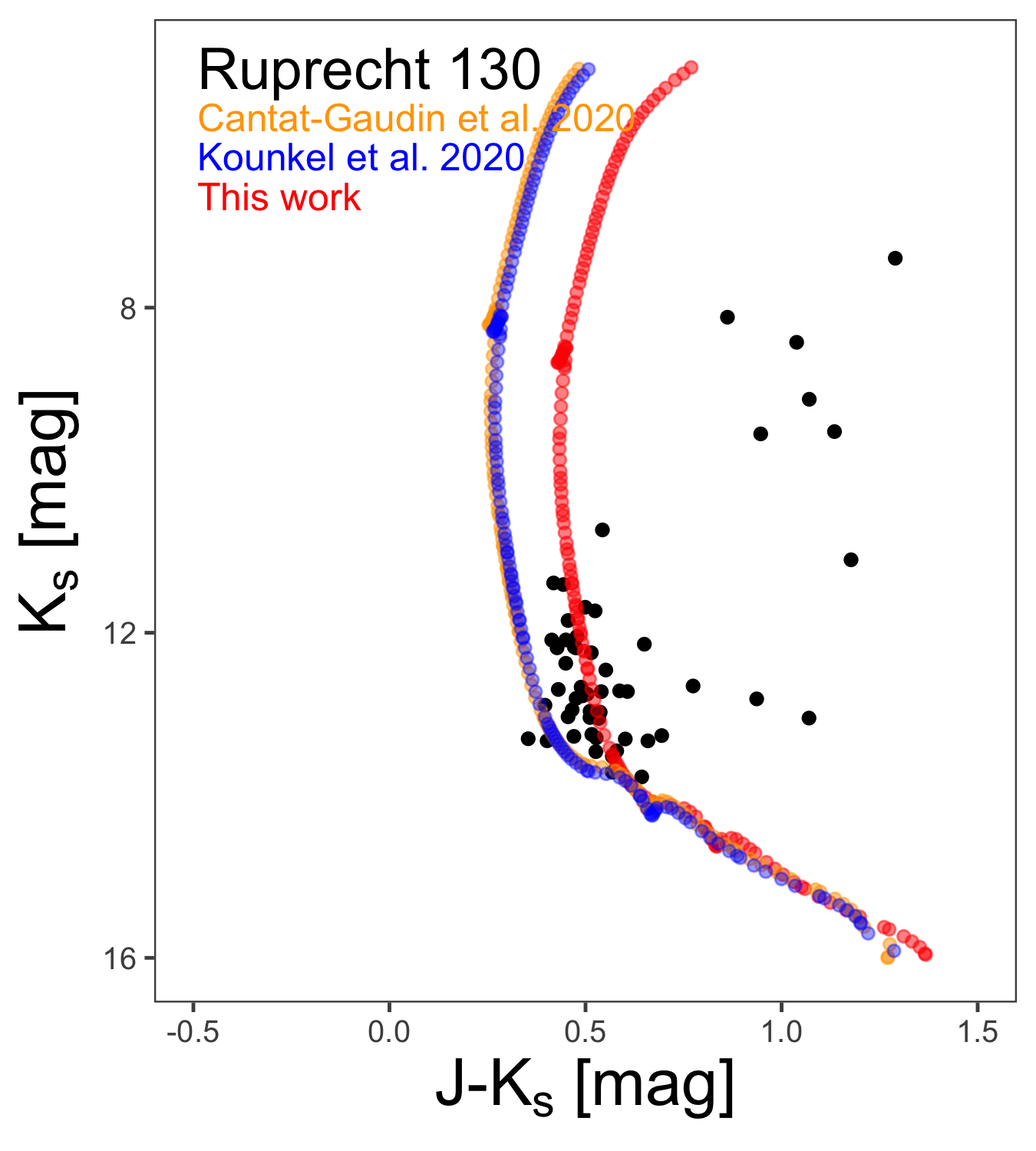}
\includegraphics[scale=0.075]{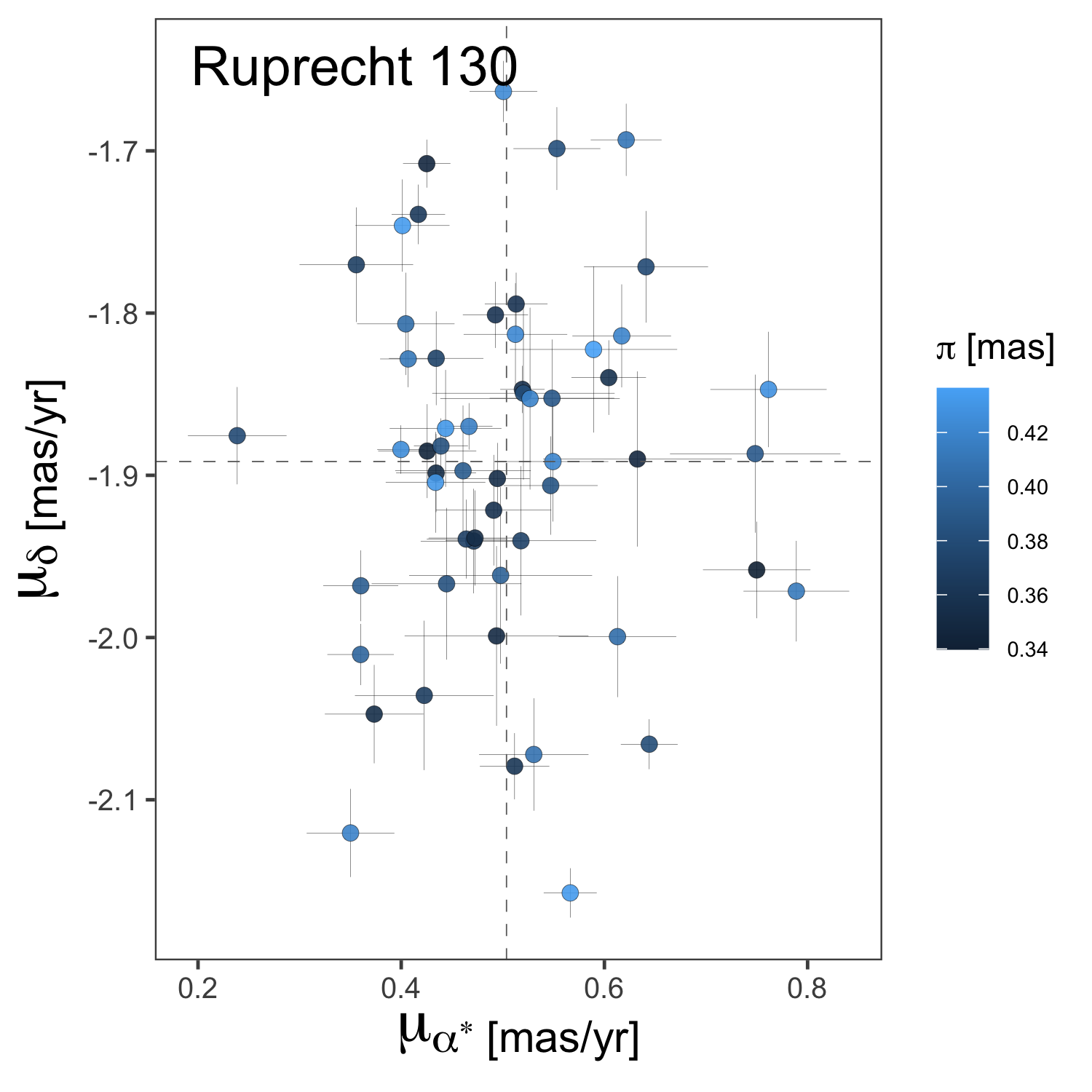}
\includegraphics[scale=0.075]{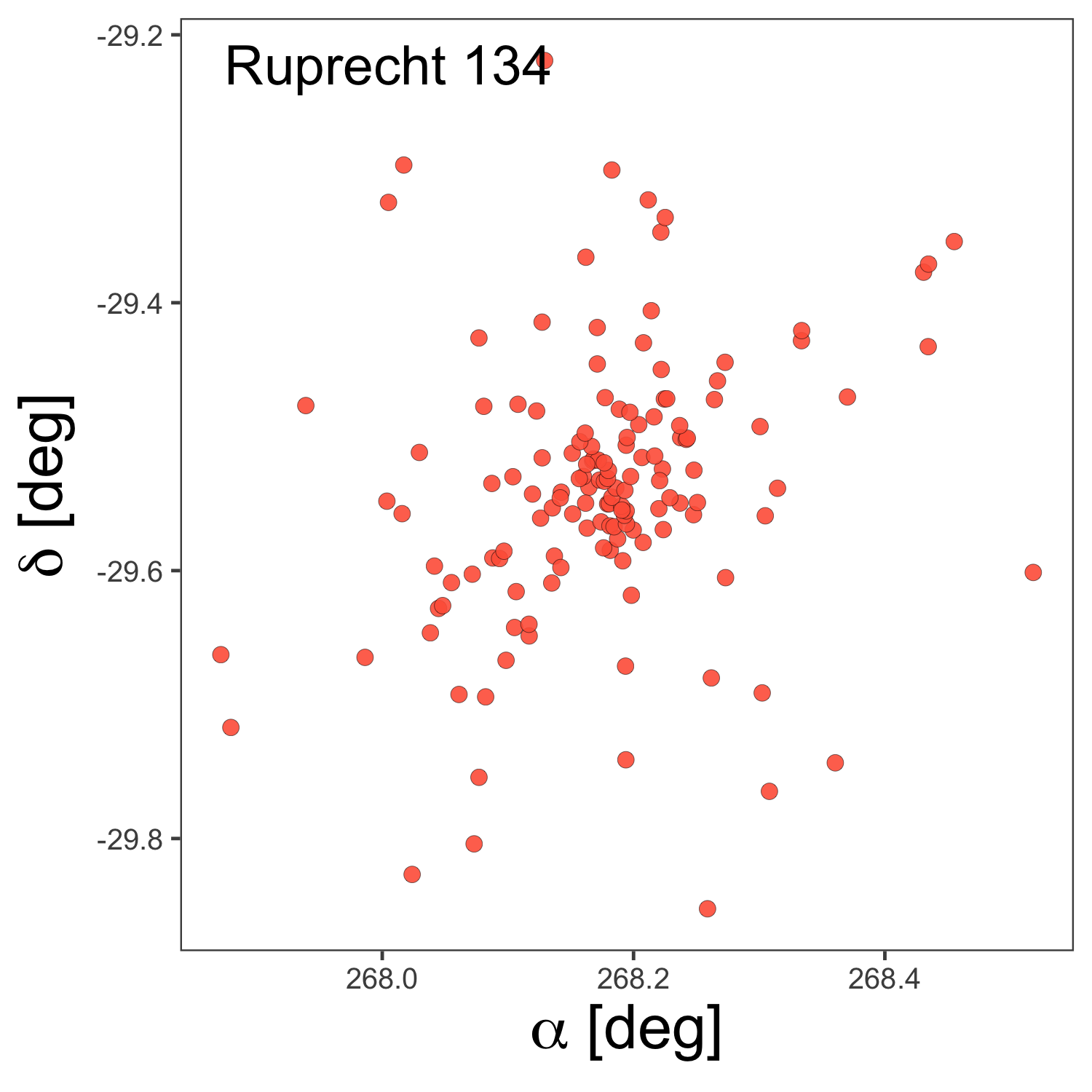}
\includegraphics[scale=0.075]{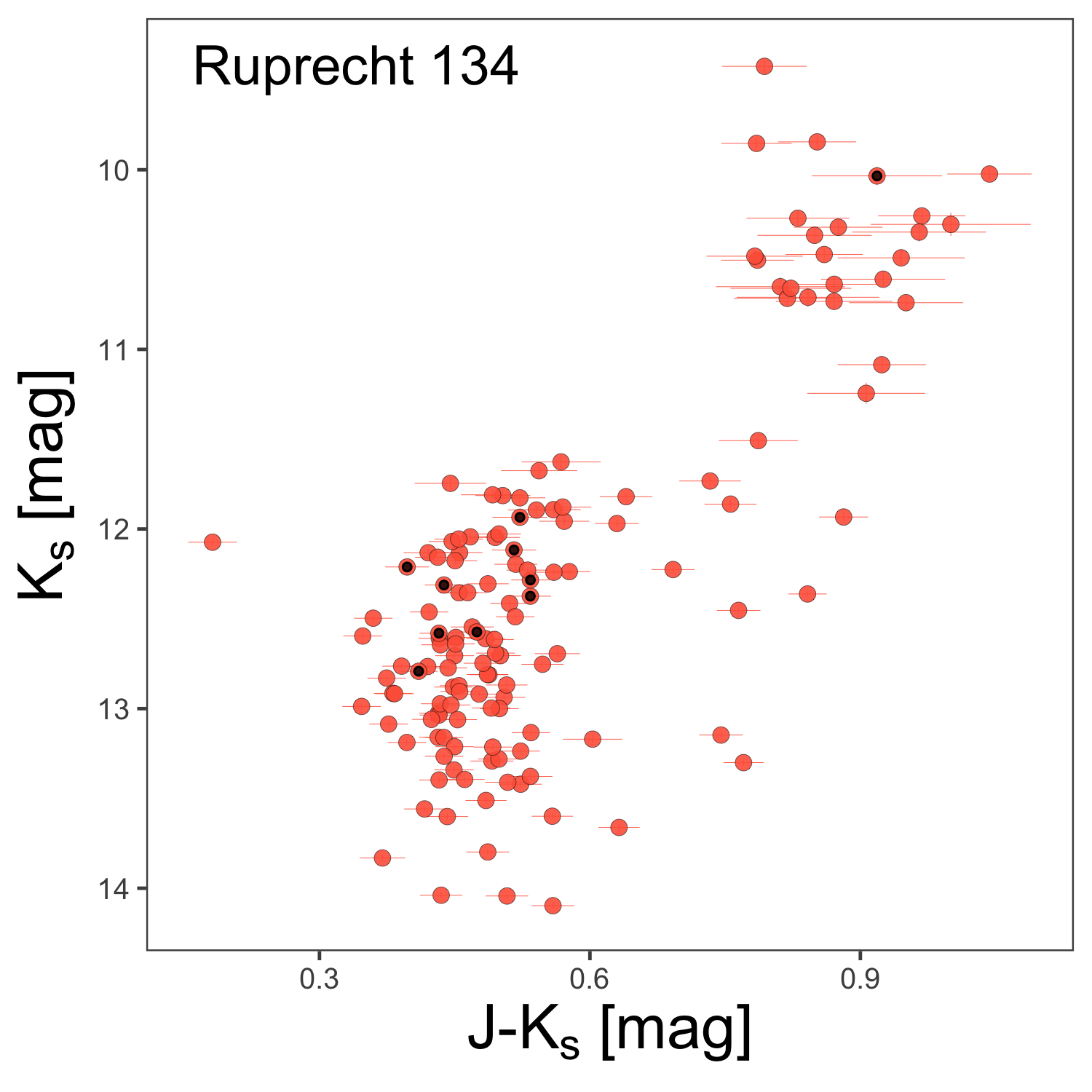}
\includegraphics[scale=0.075]{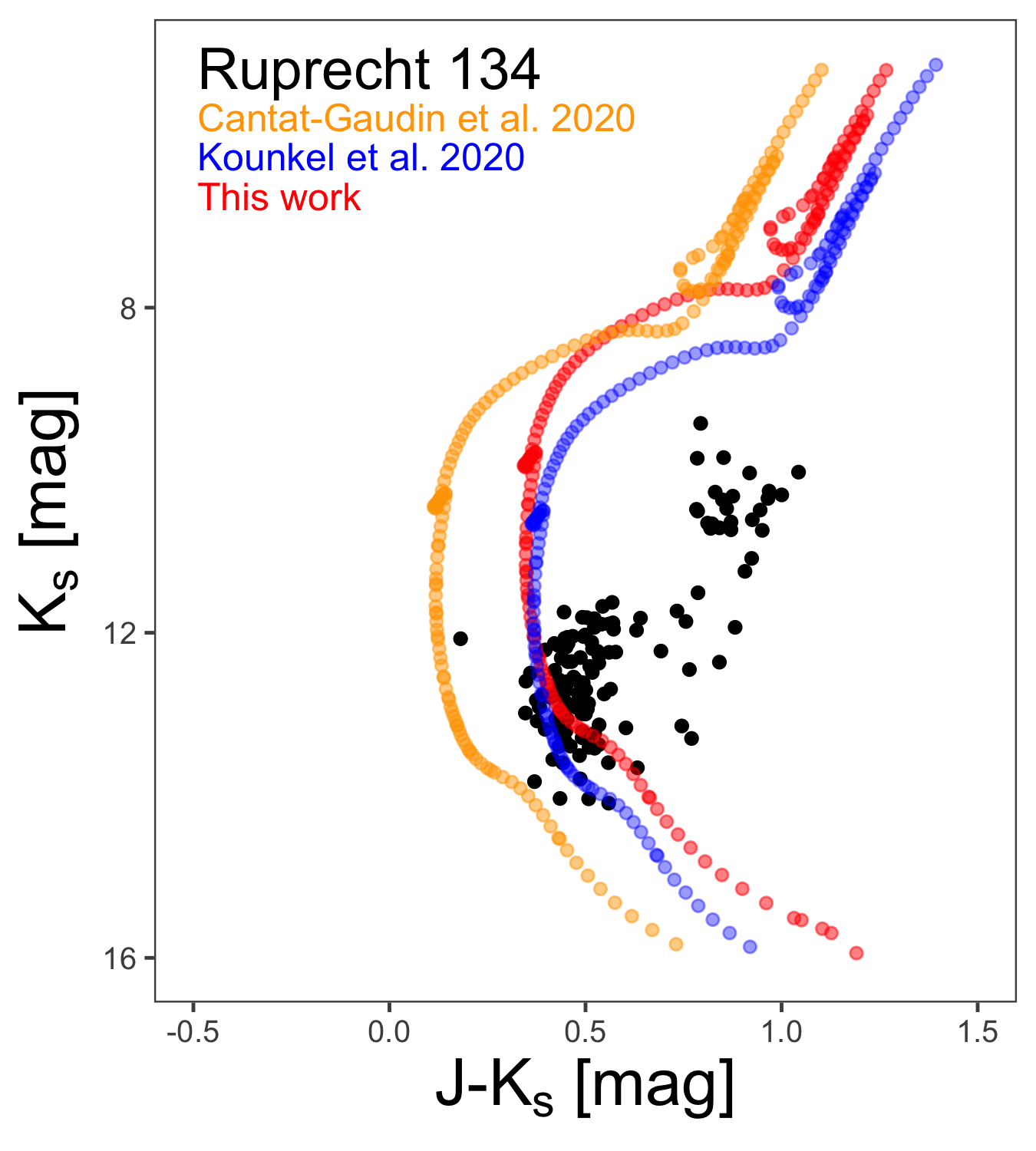}
\includegraphics[scale=0.075]{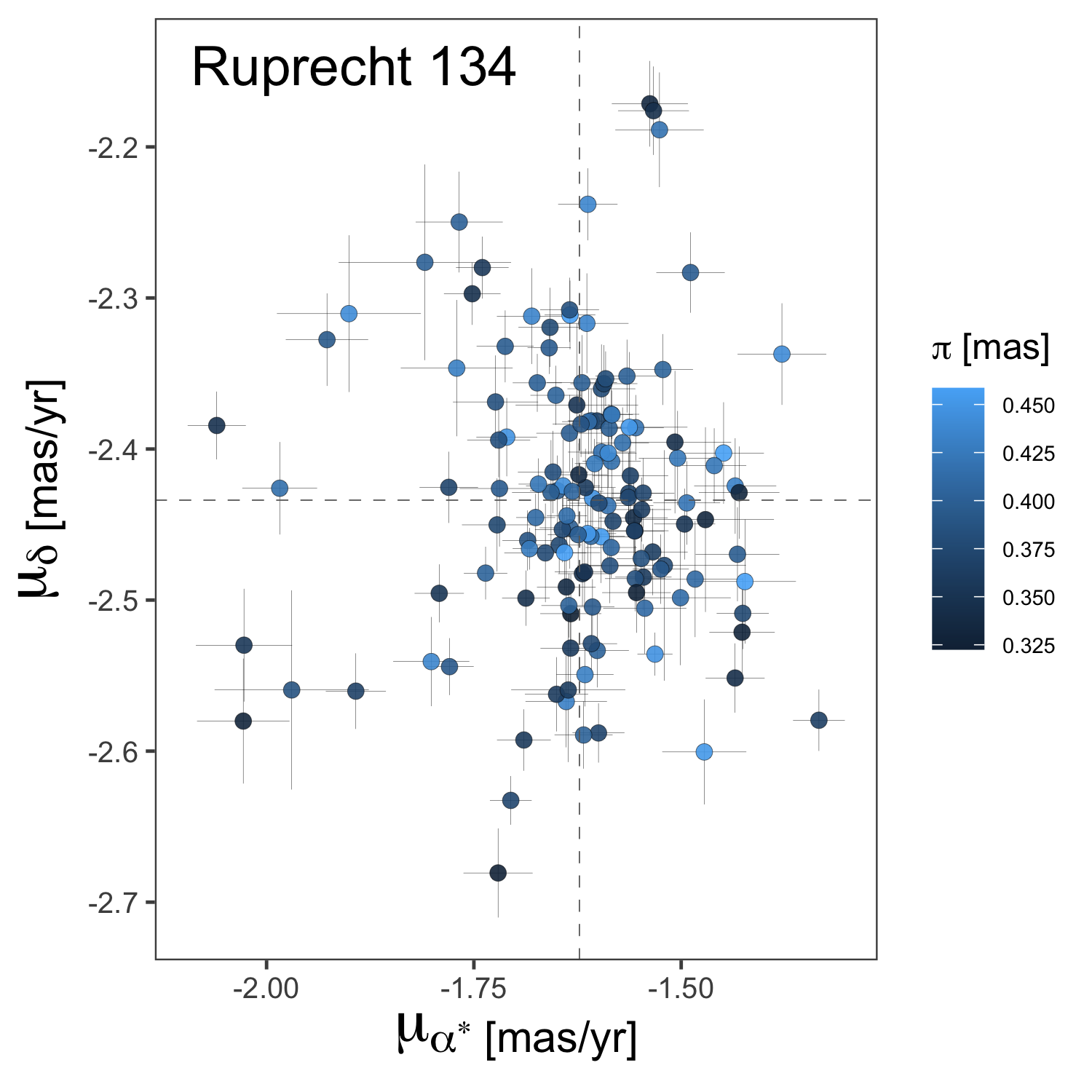}
\caption{Same as Figure\,\ref{fig:clusters}}
\end{figure*}

\begin{figure*}
\includegraphics[scale=0.075]{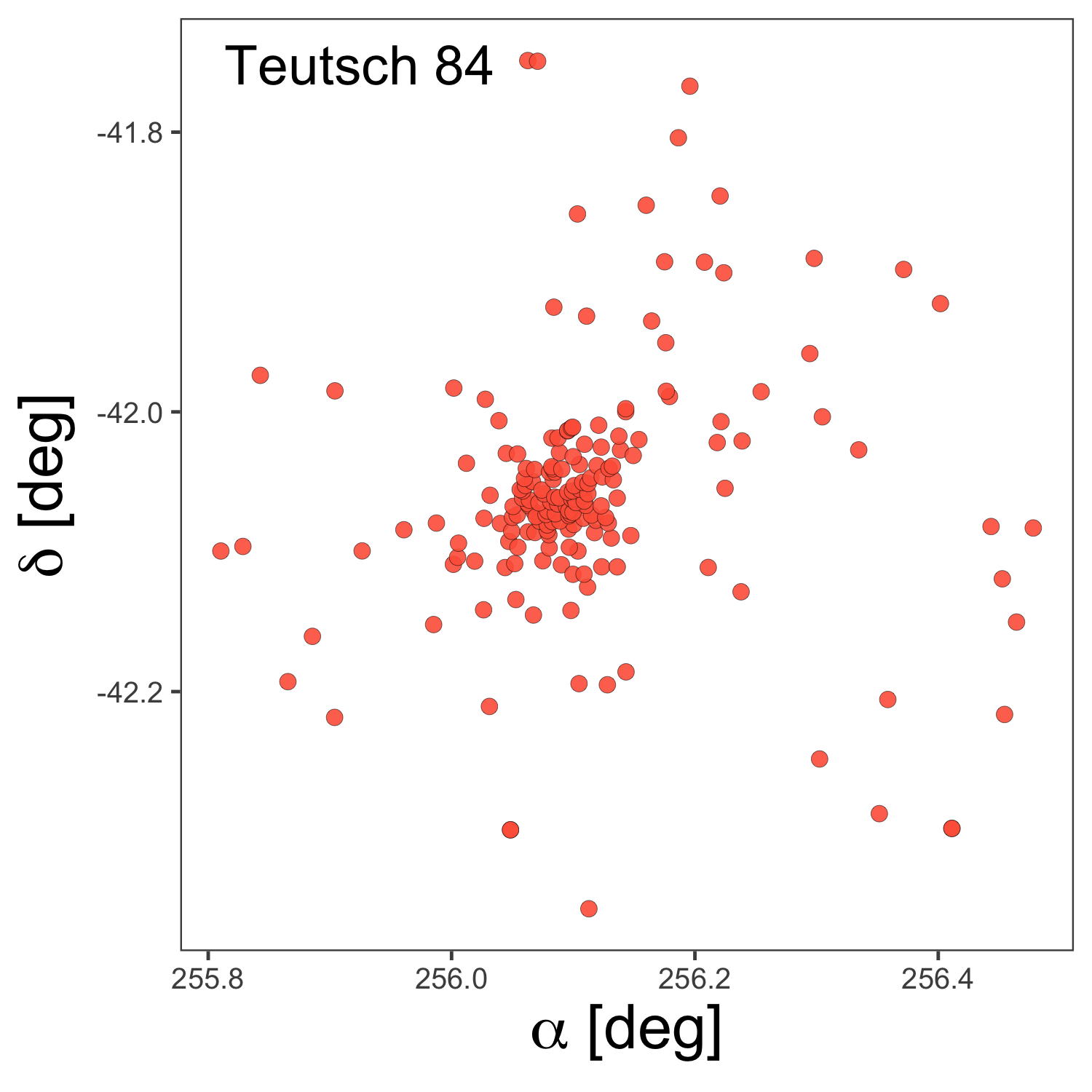}
\includegraphics[scale=0.075]{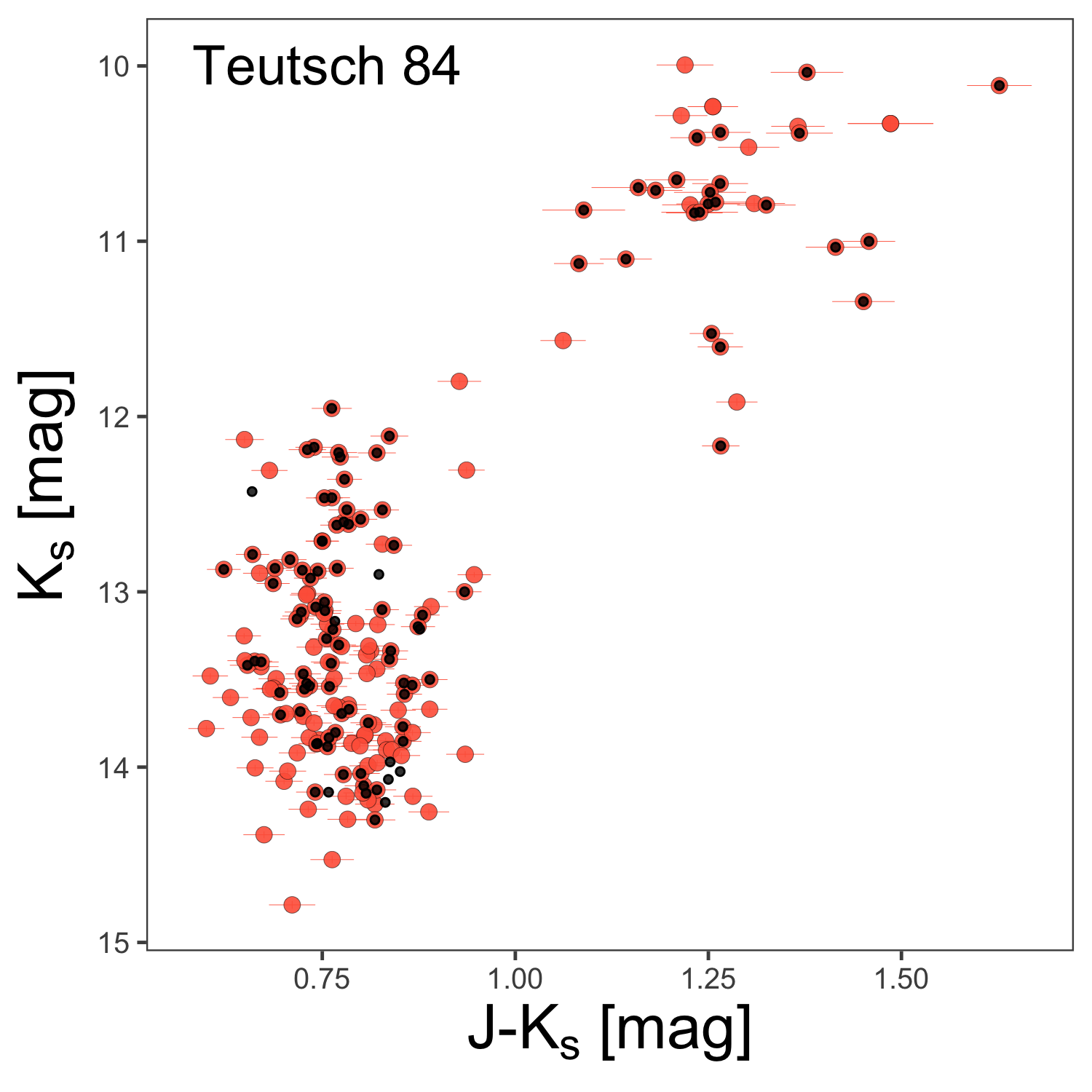}
\includegraphics[scale=0.075]{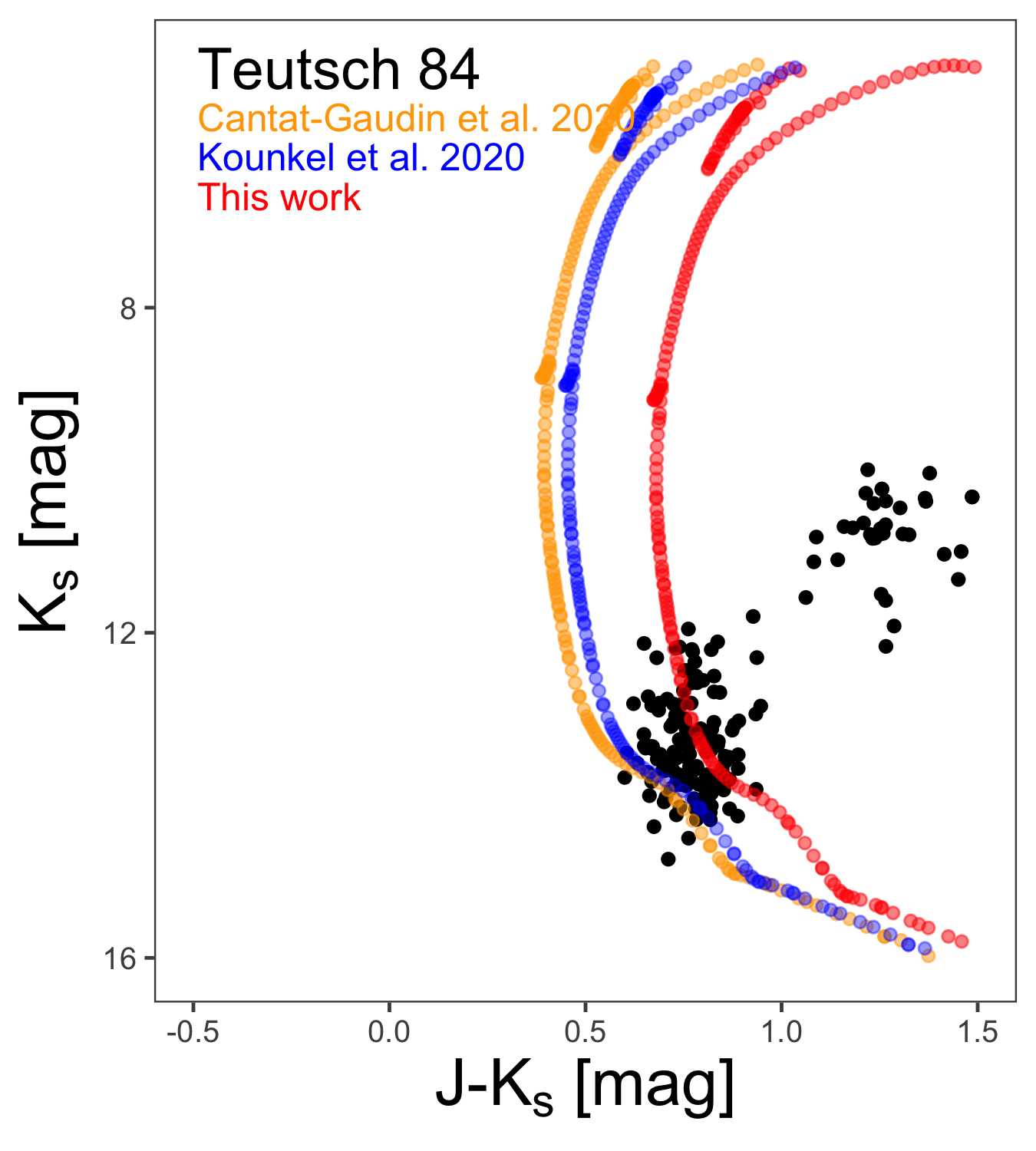}
\includegraphics[scale=0.075]{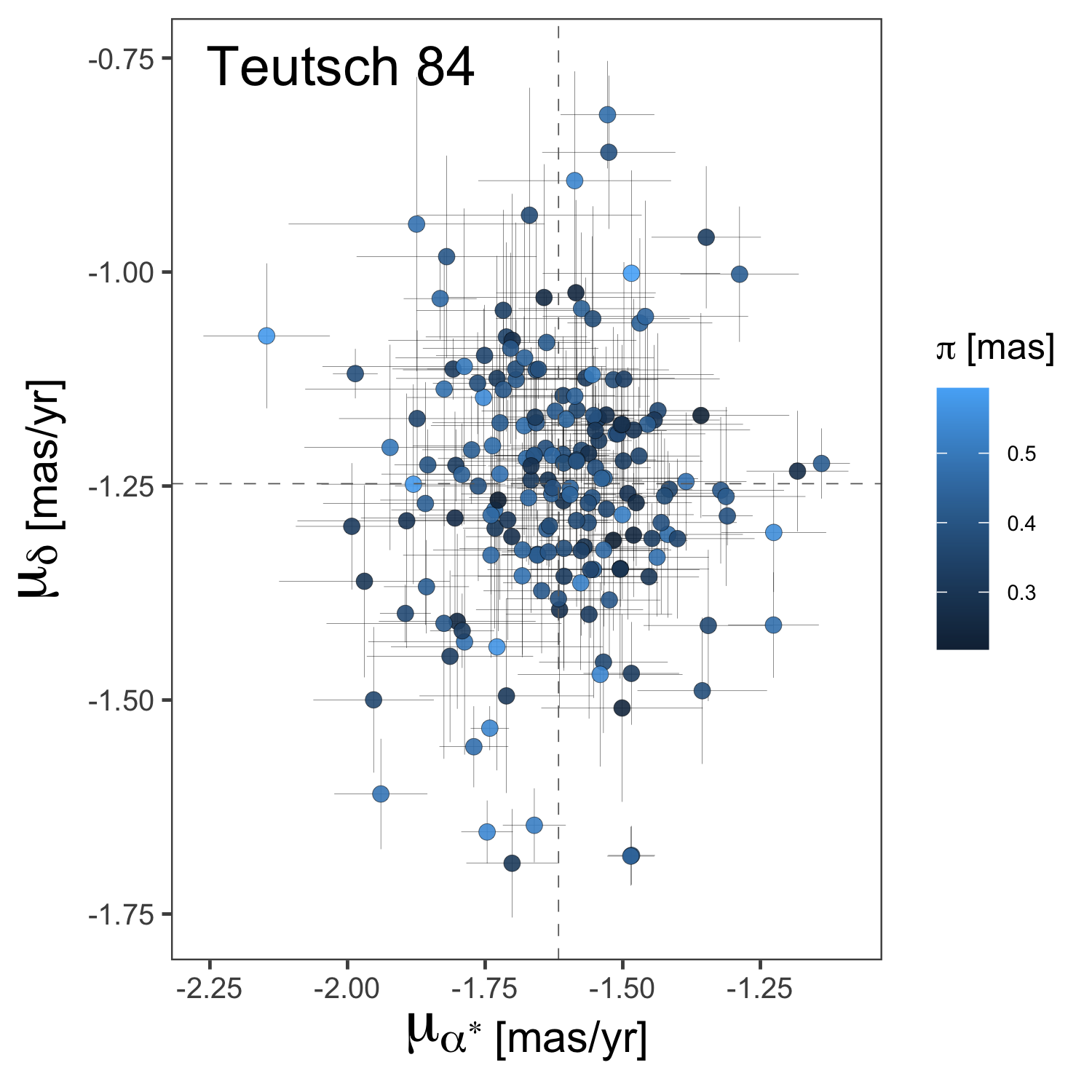}
\includegraphics[scale=0.075]{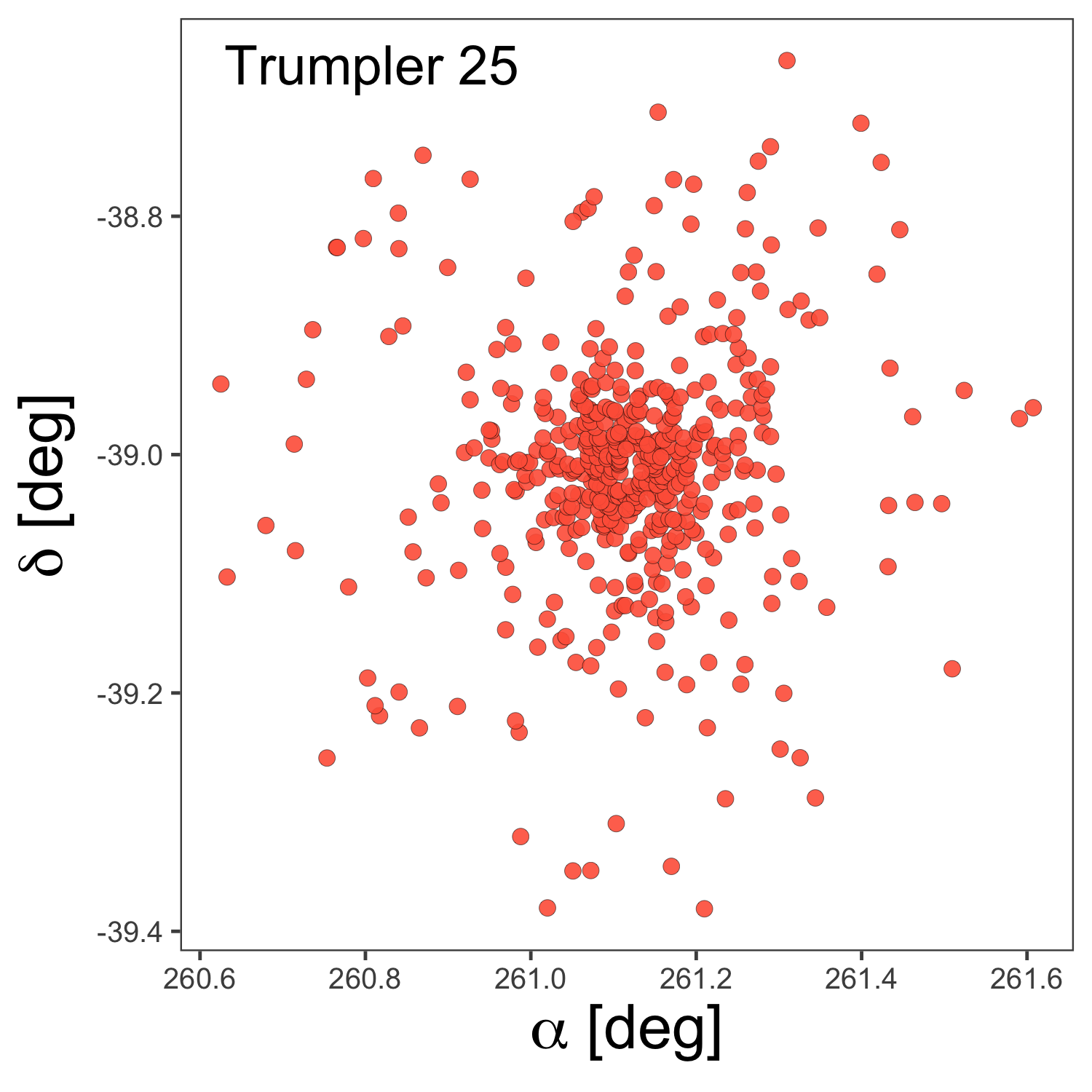}
\includegraphics[scale=0.075]{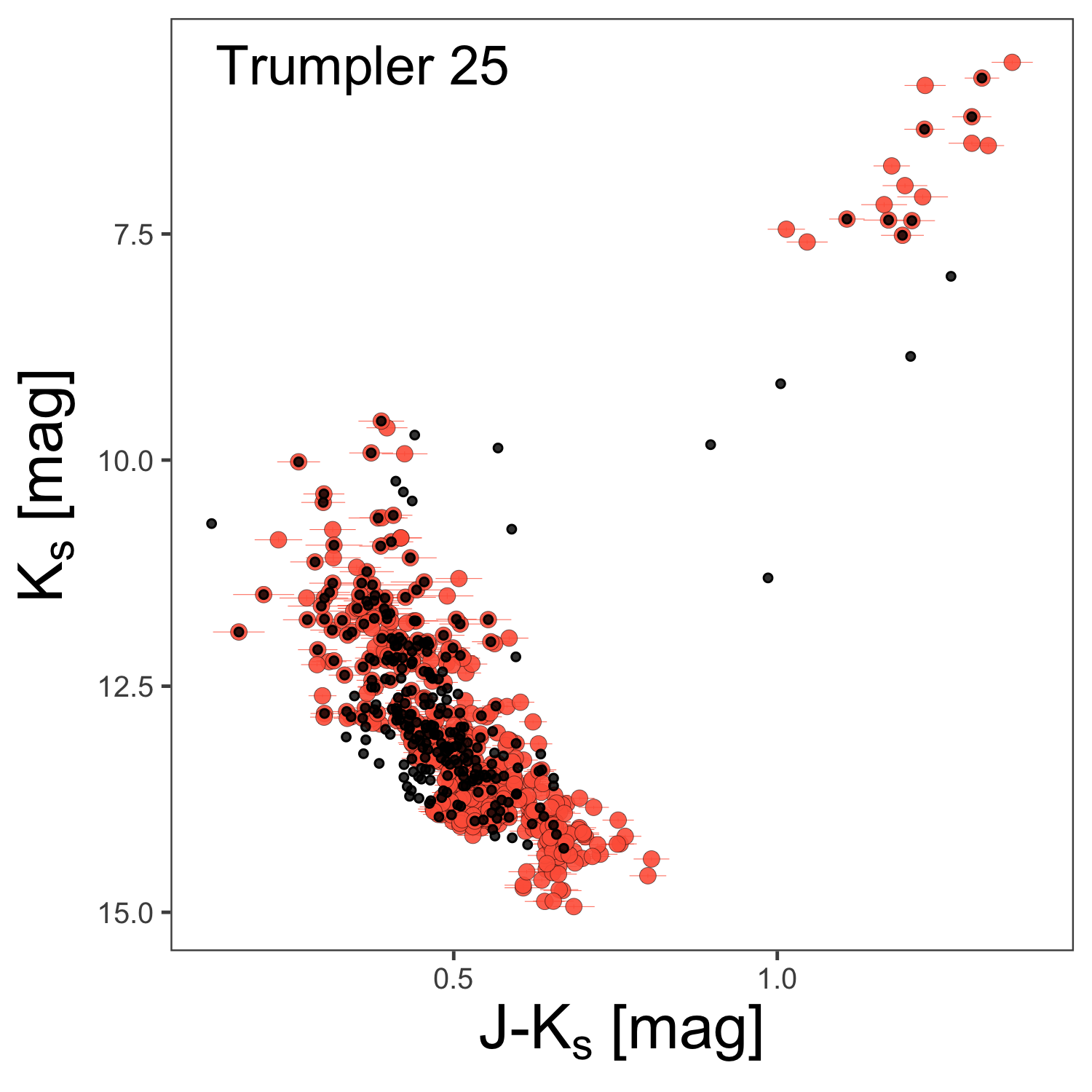}
\includegraphics[scale=0.075]{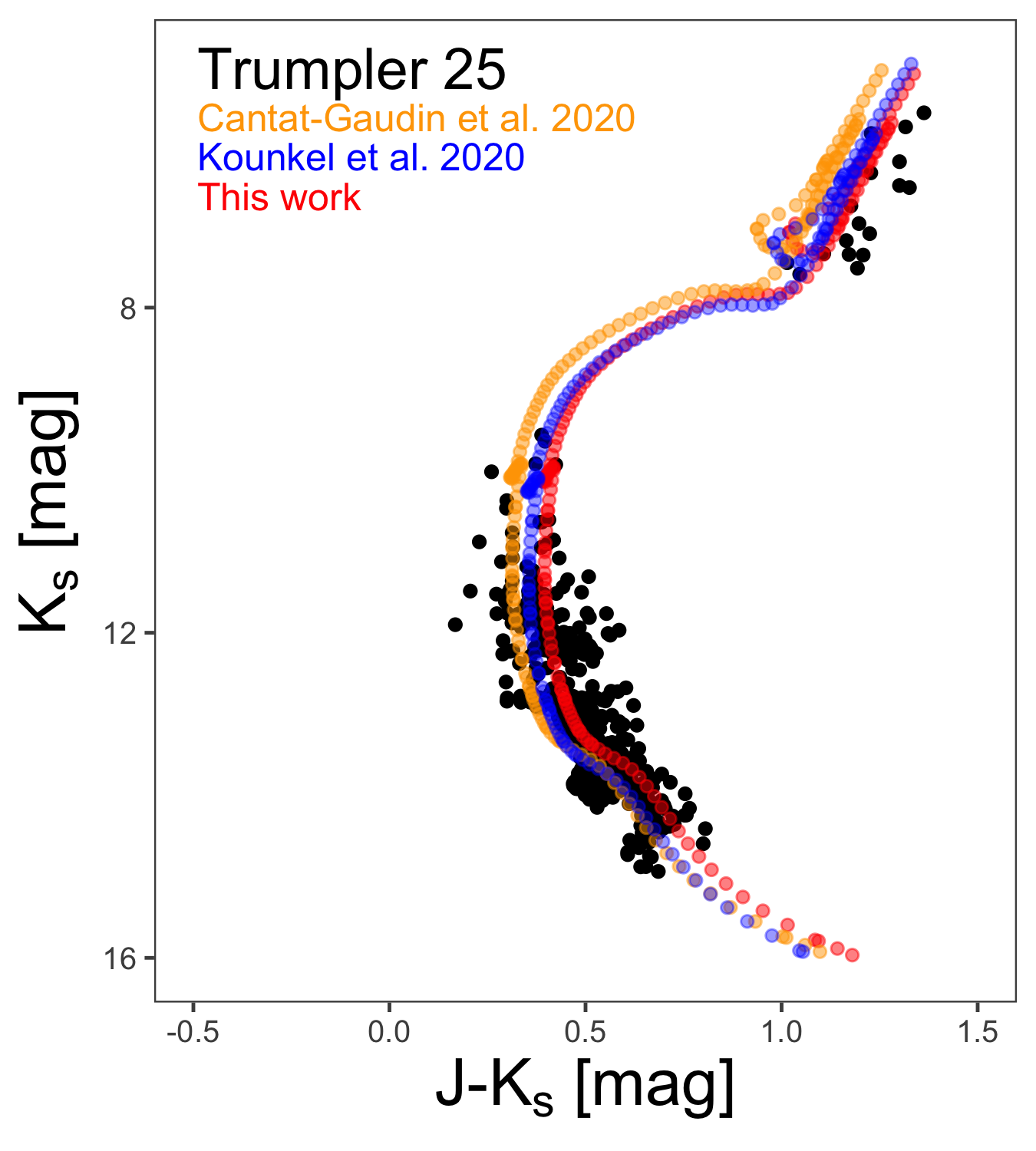}
\includegraphics[scale=0.075]{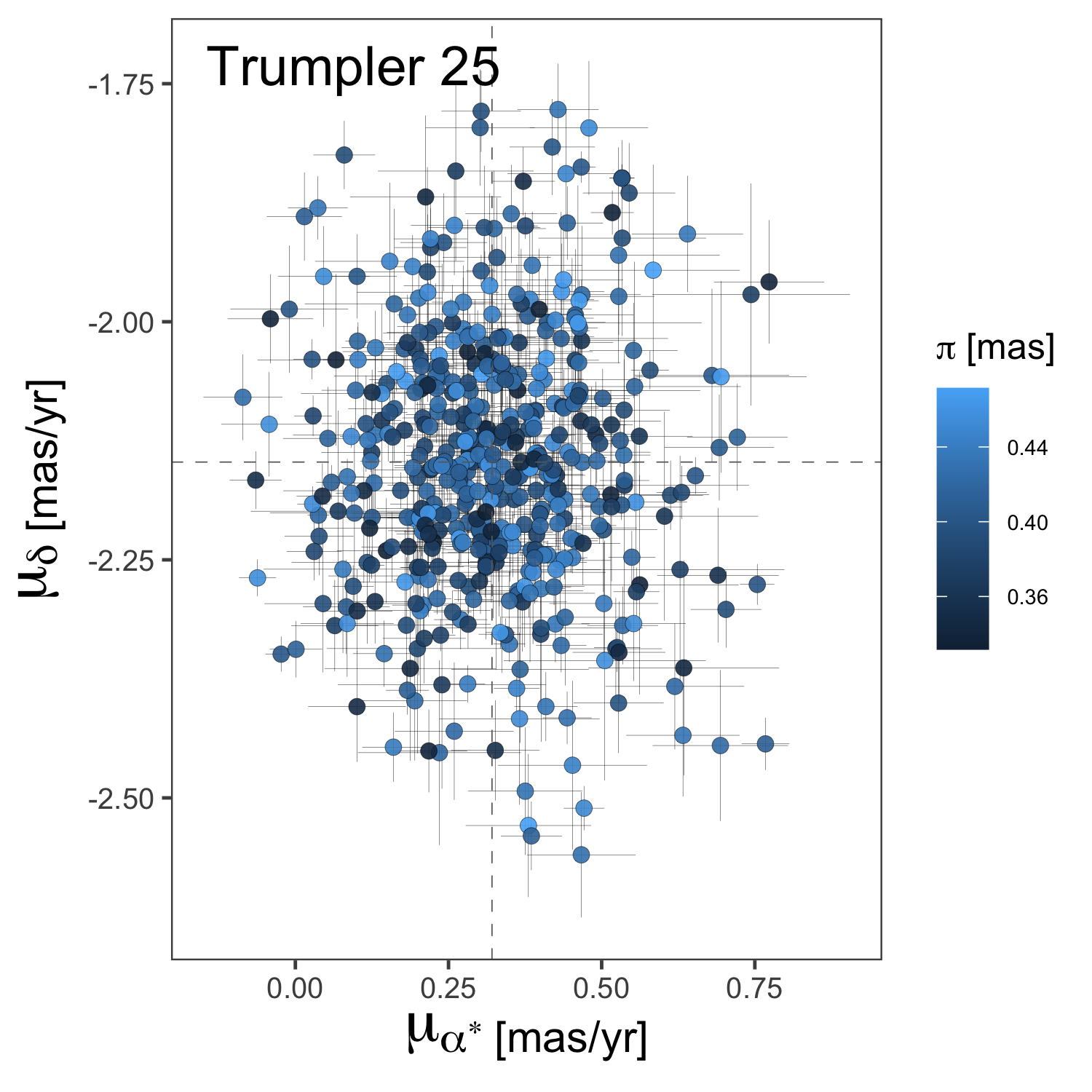}
\includegraphics[scale=0.075]{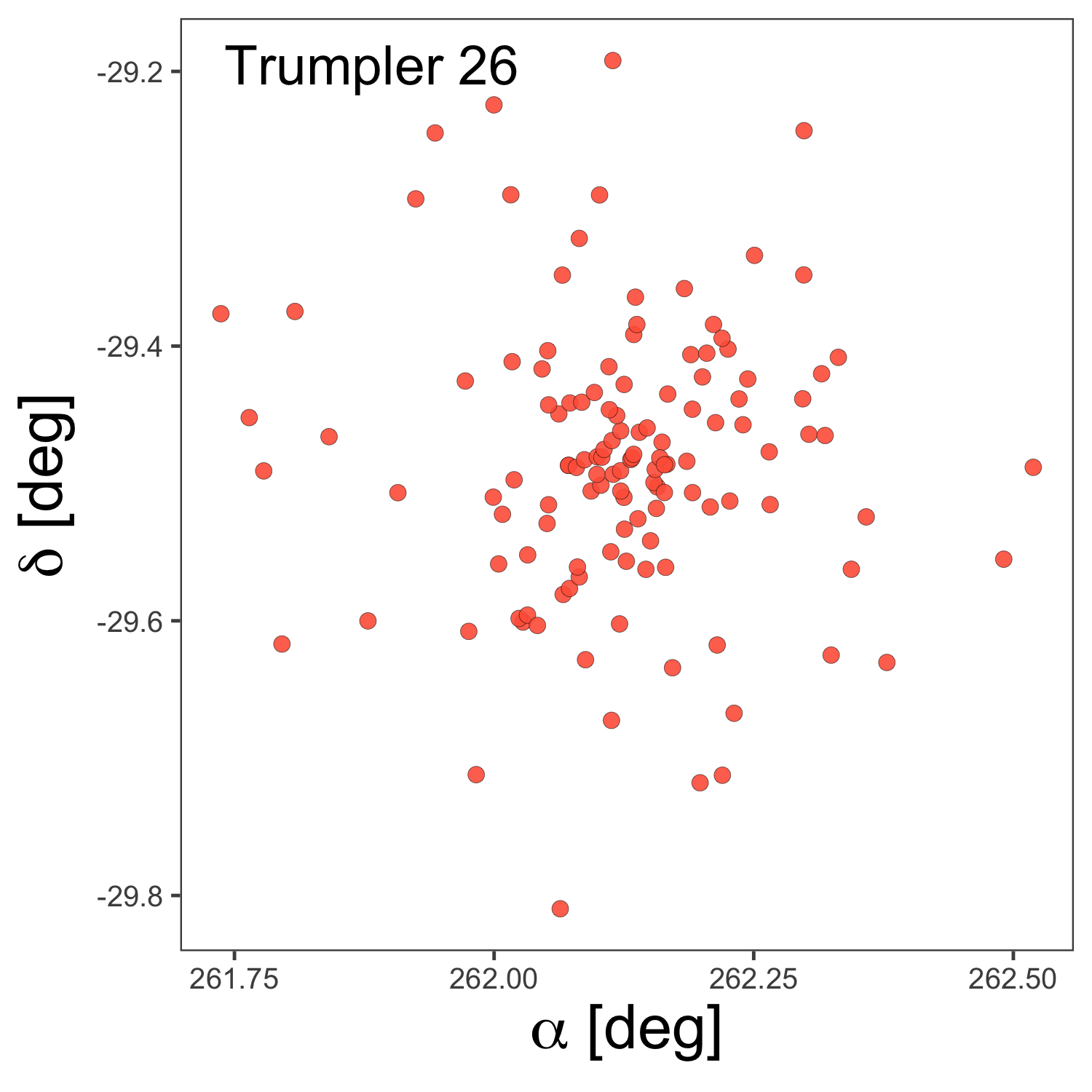}
\includegraphics[scale=0.075]{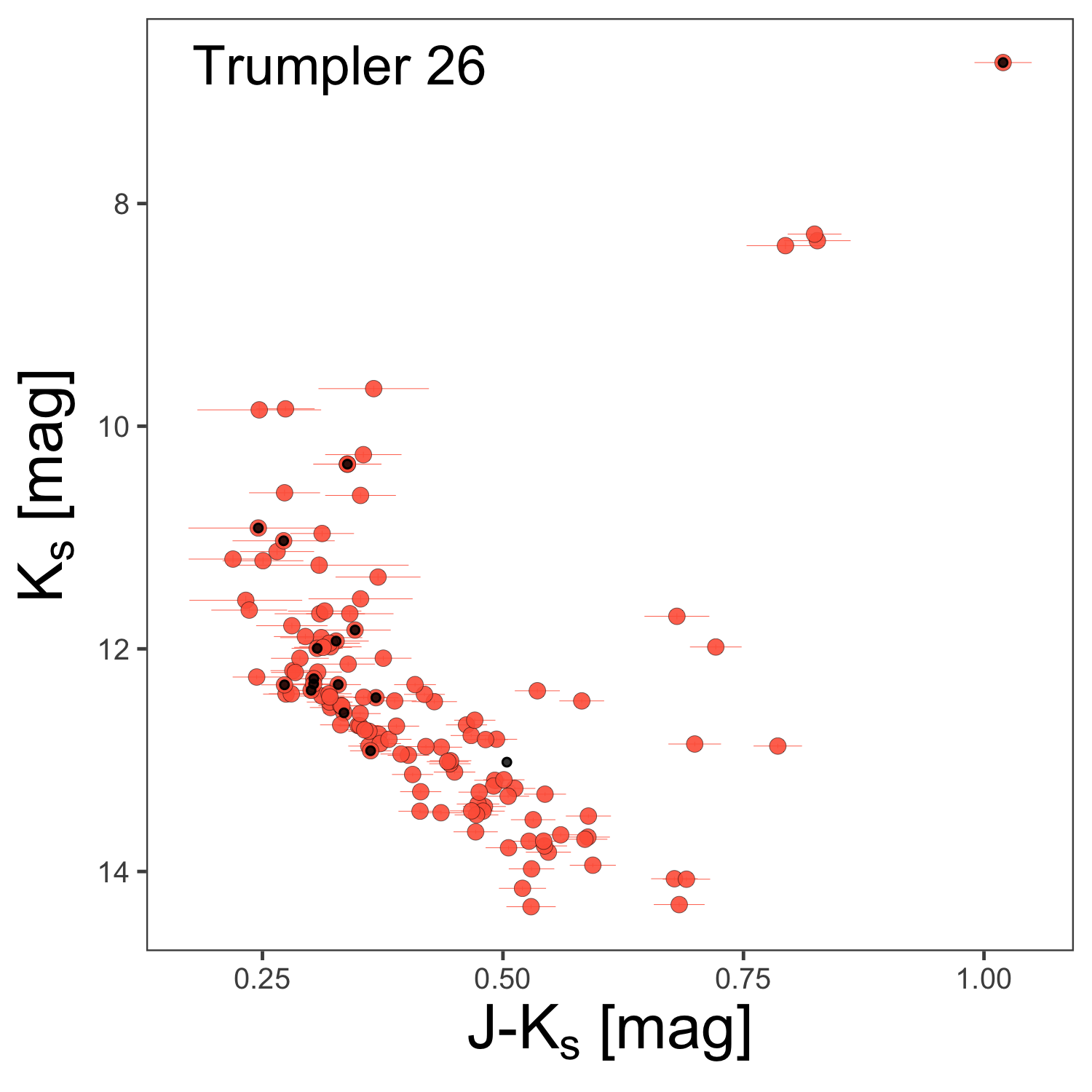}
\includegraphics[scale=0.075]{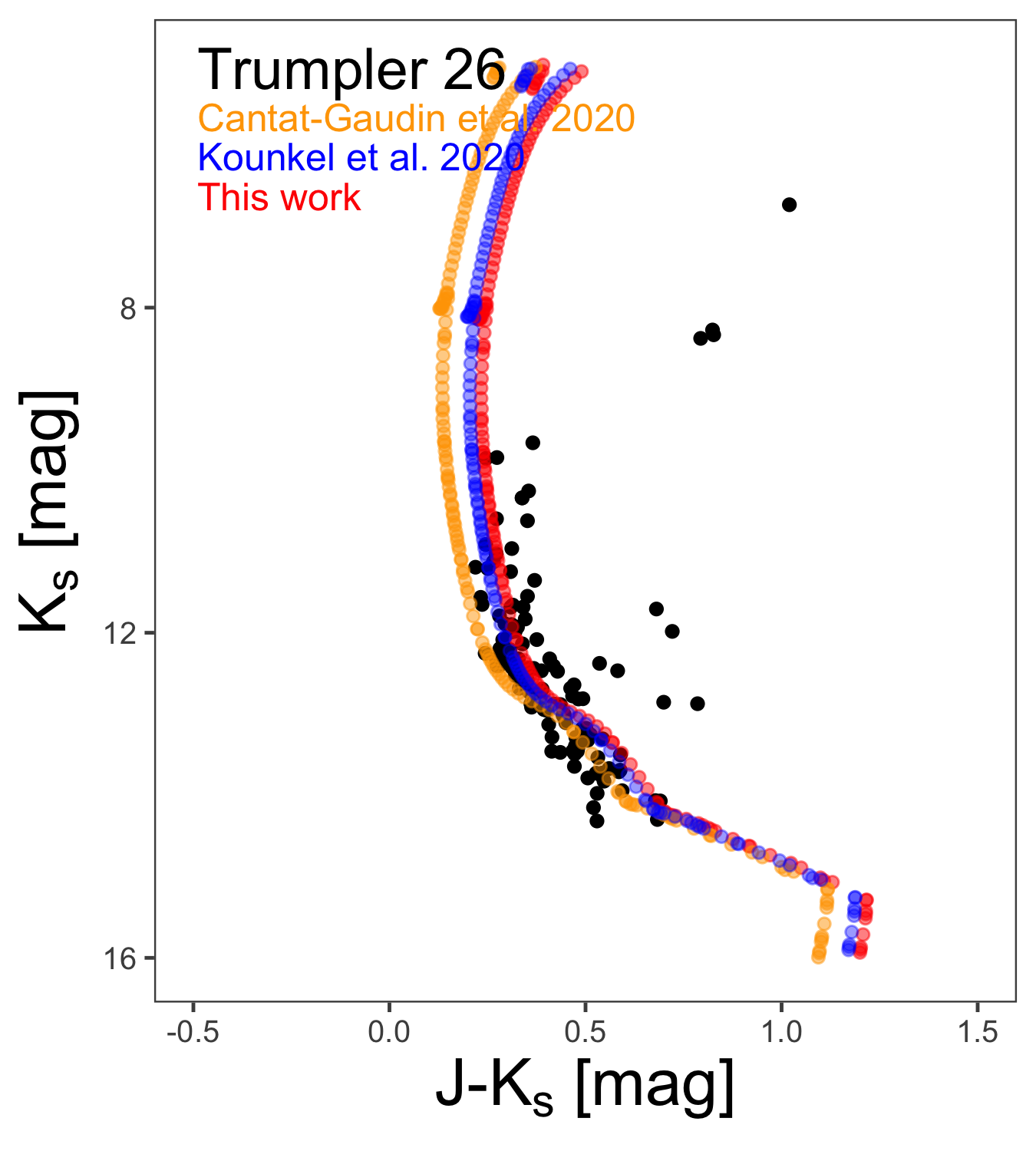}
\includegraphics[scale=0.075]{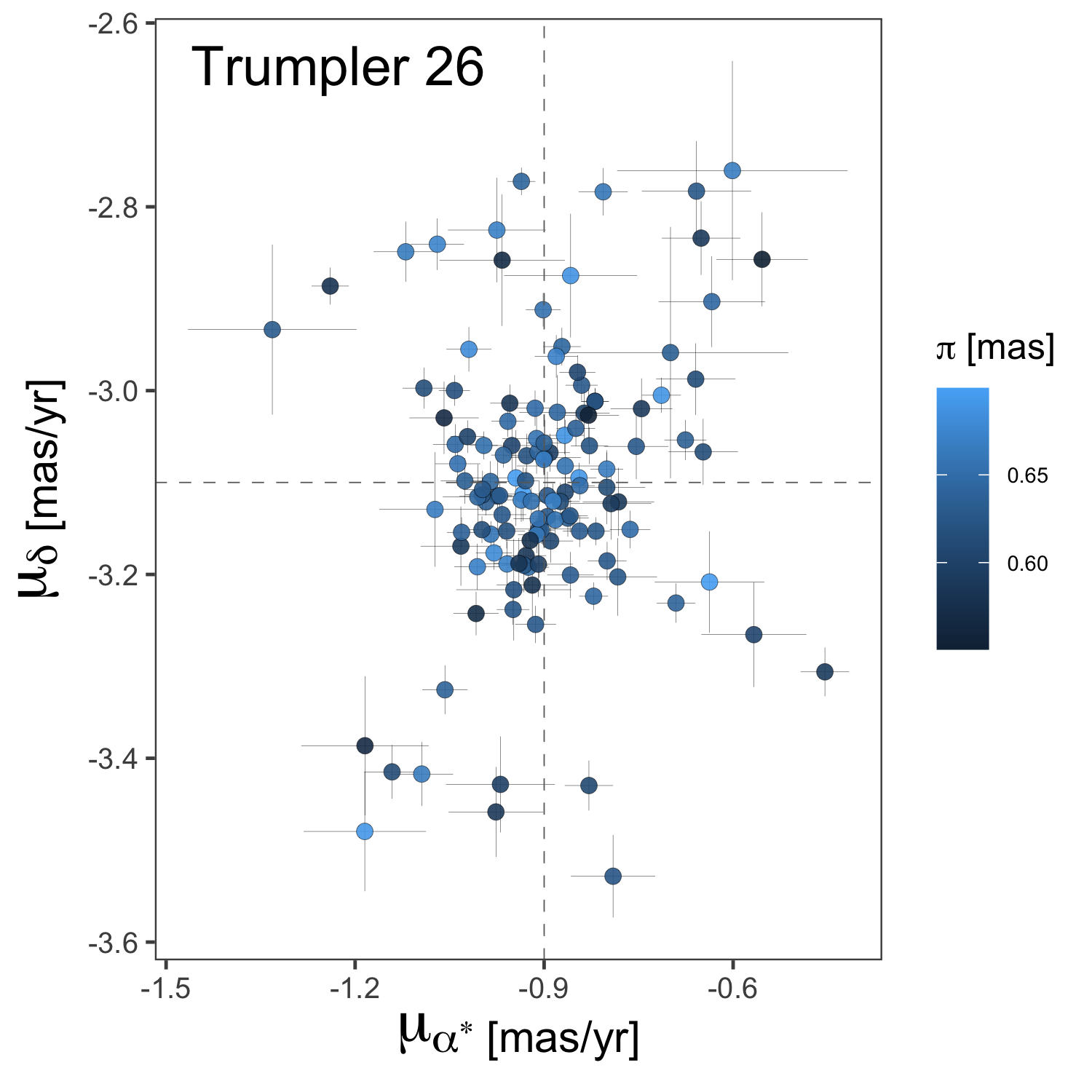}
\includegraphics[scale=0.075]{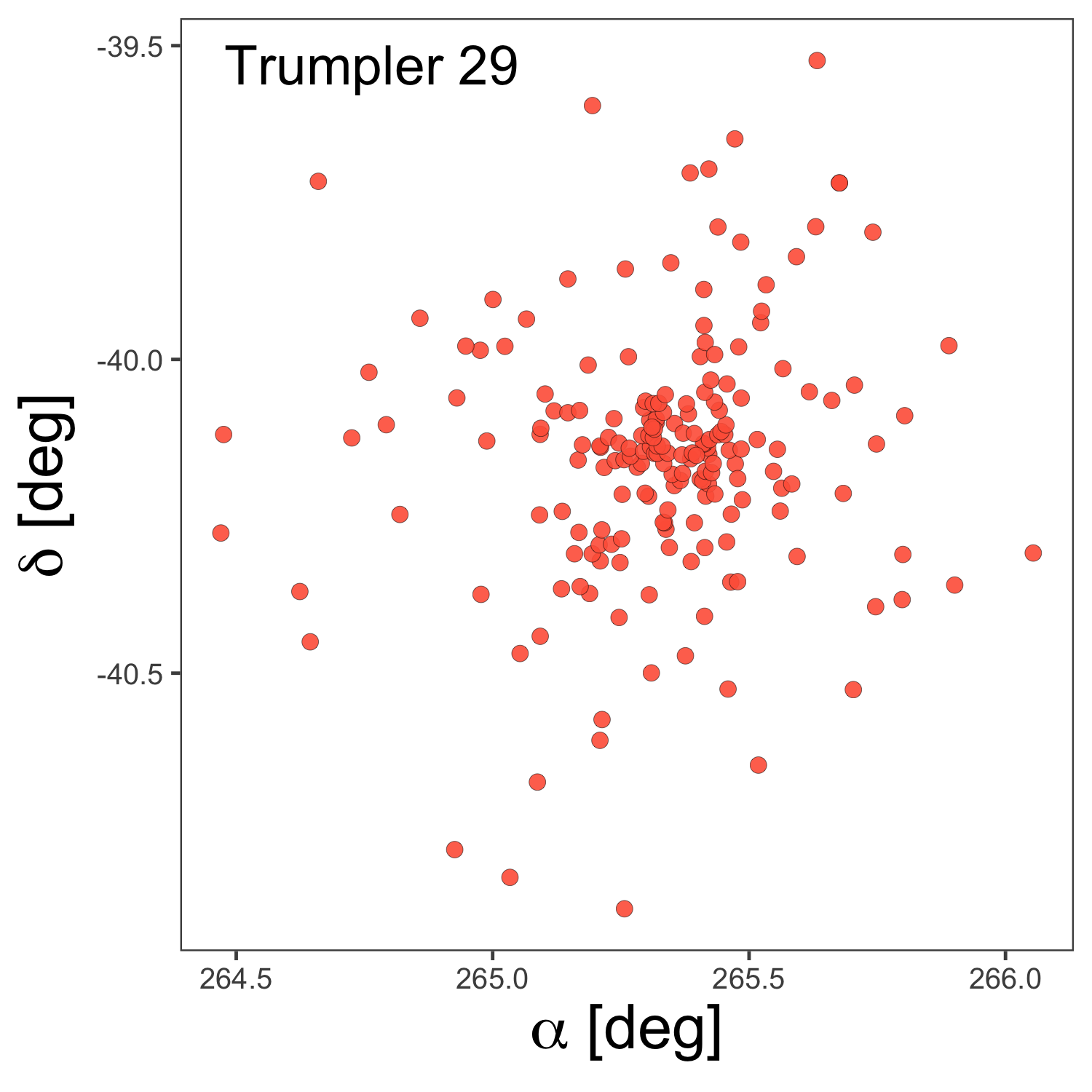}
\includegraphics[scale=0.075]{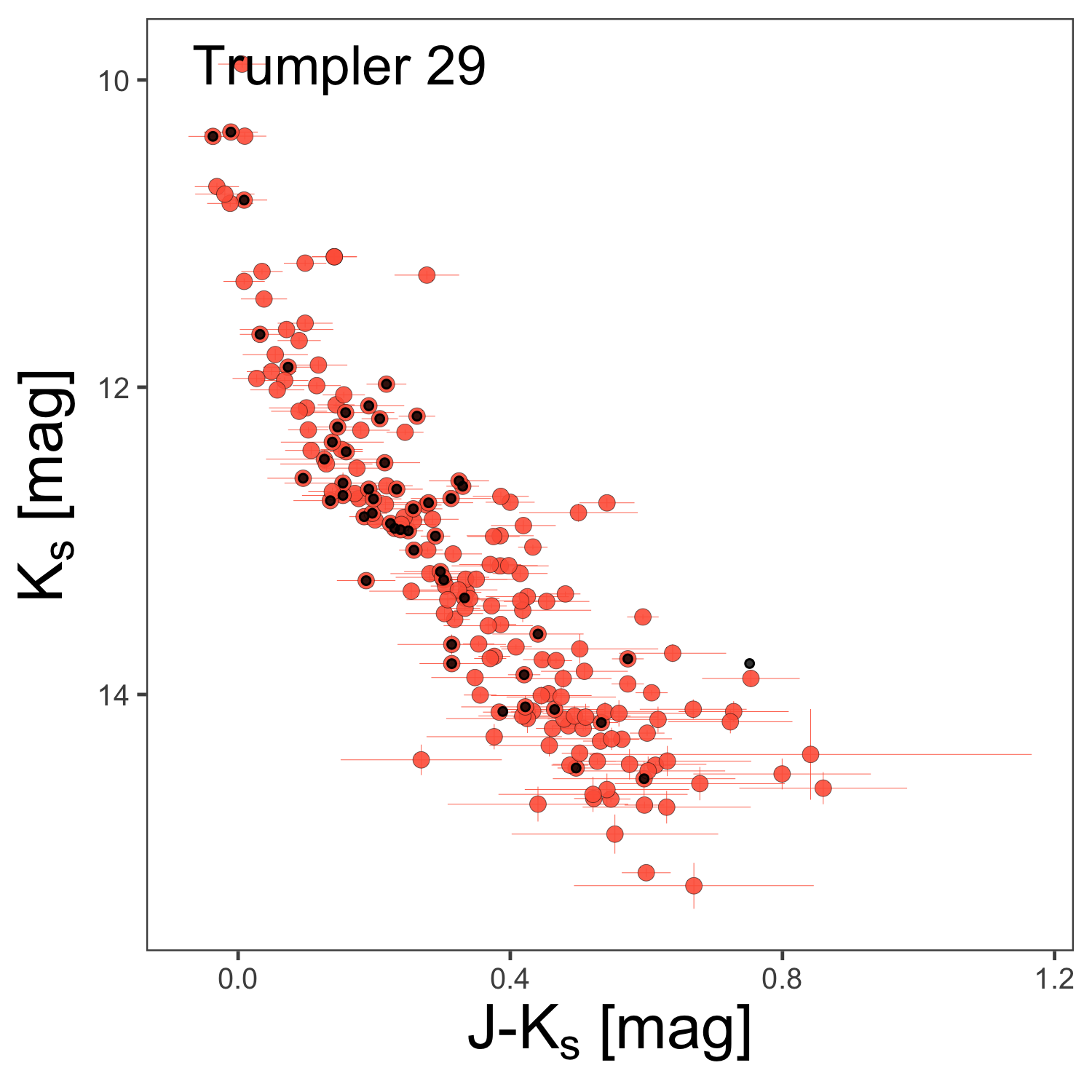}
\includegraphics[scale=0.075]{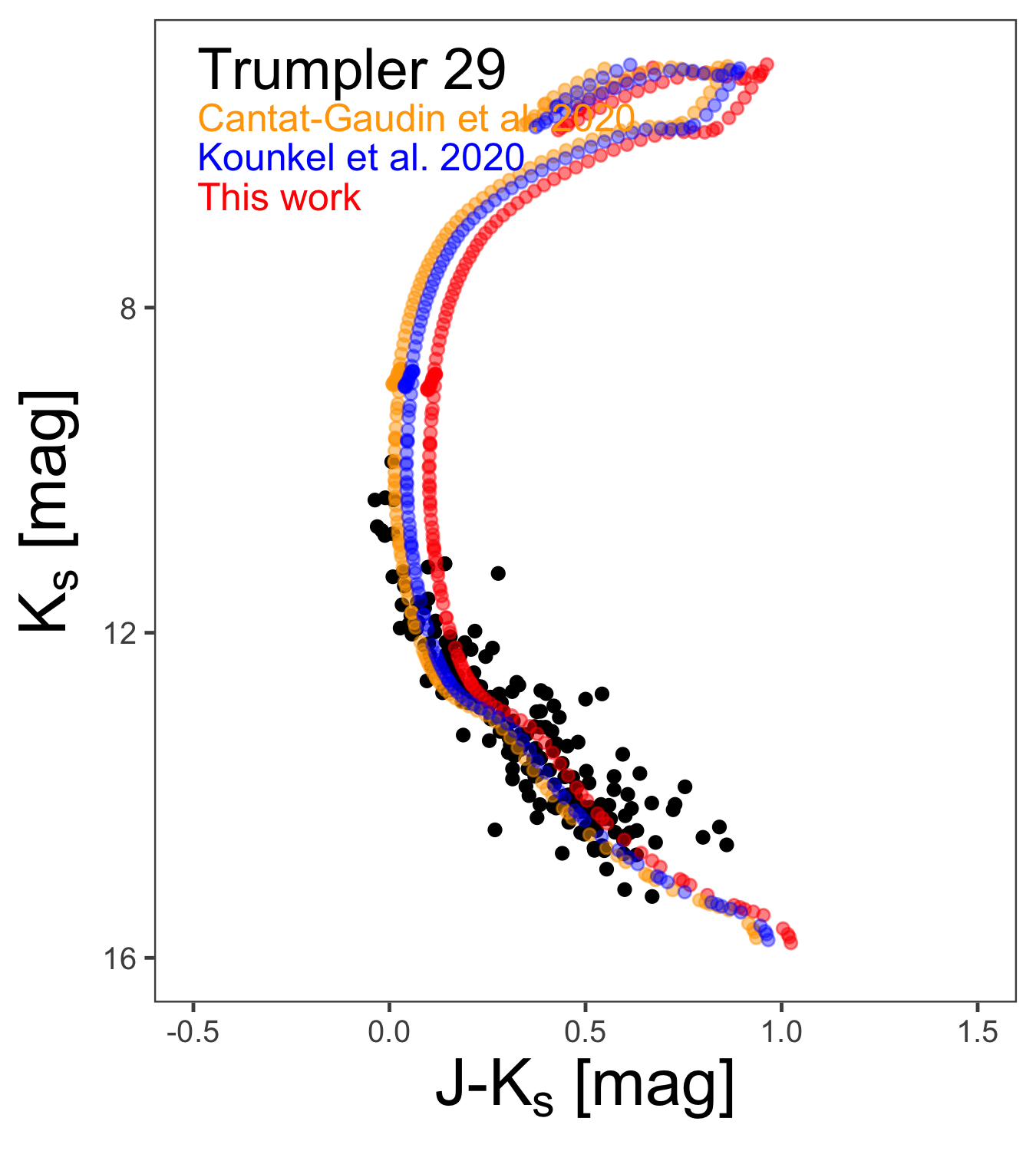}
\includegraphics[scale=0.075]{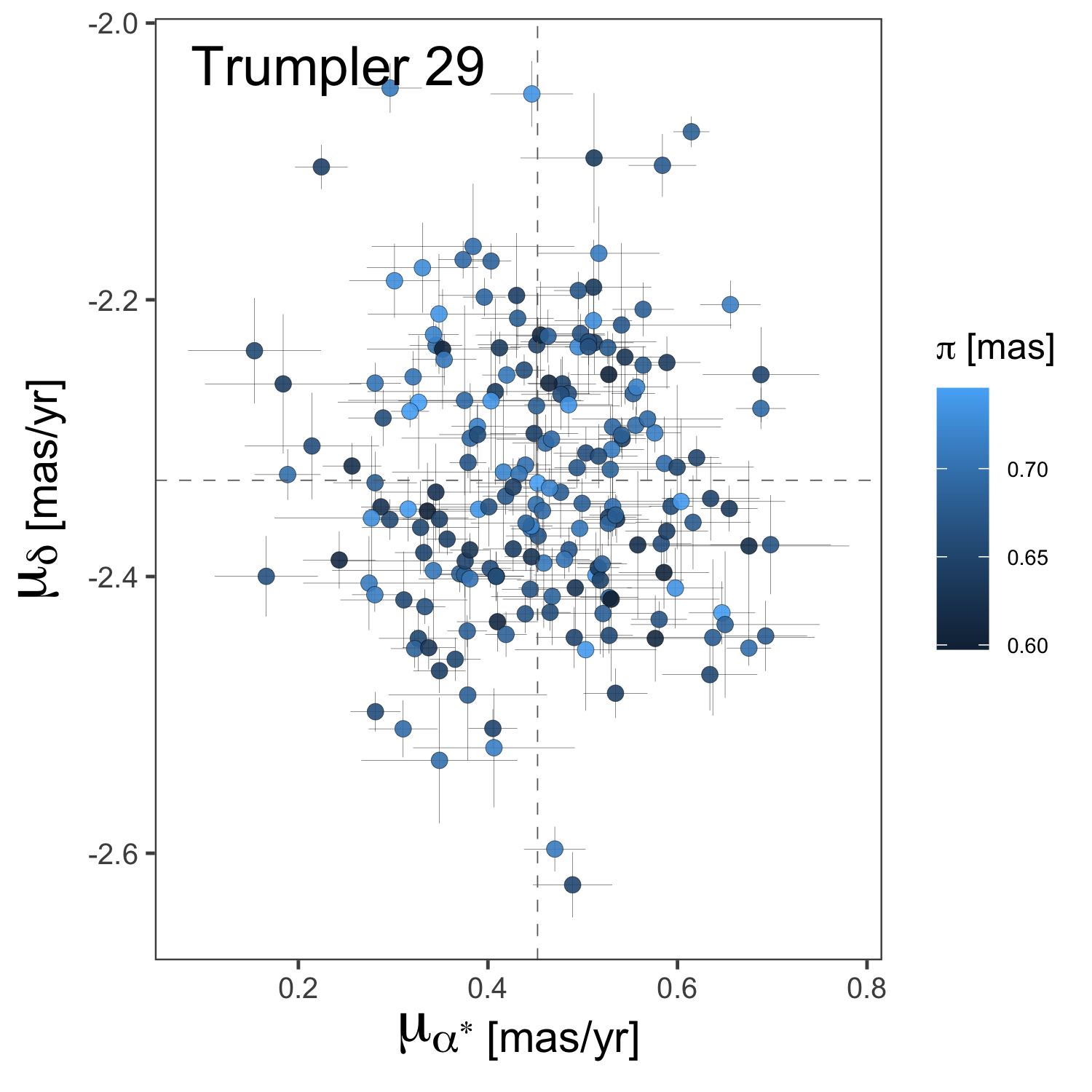}
\includegraphics[scale=0.075]{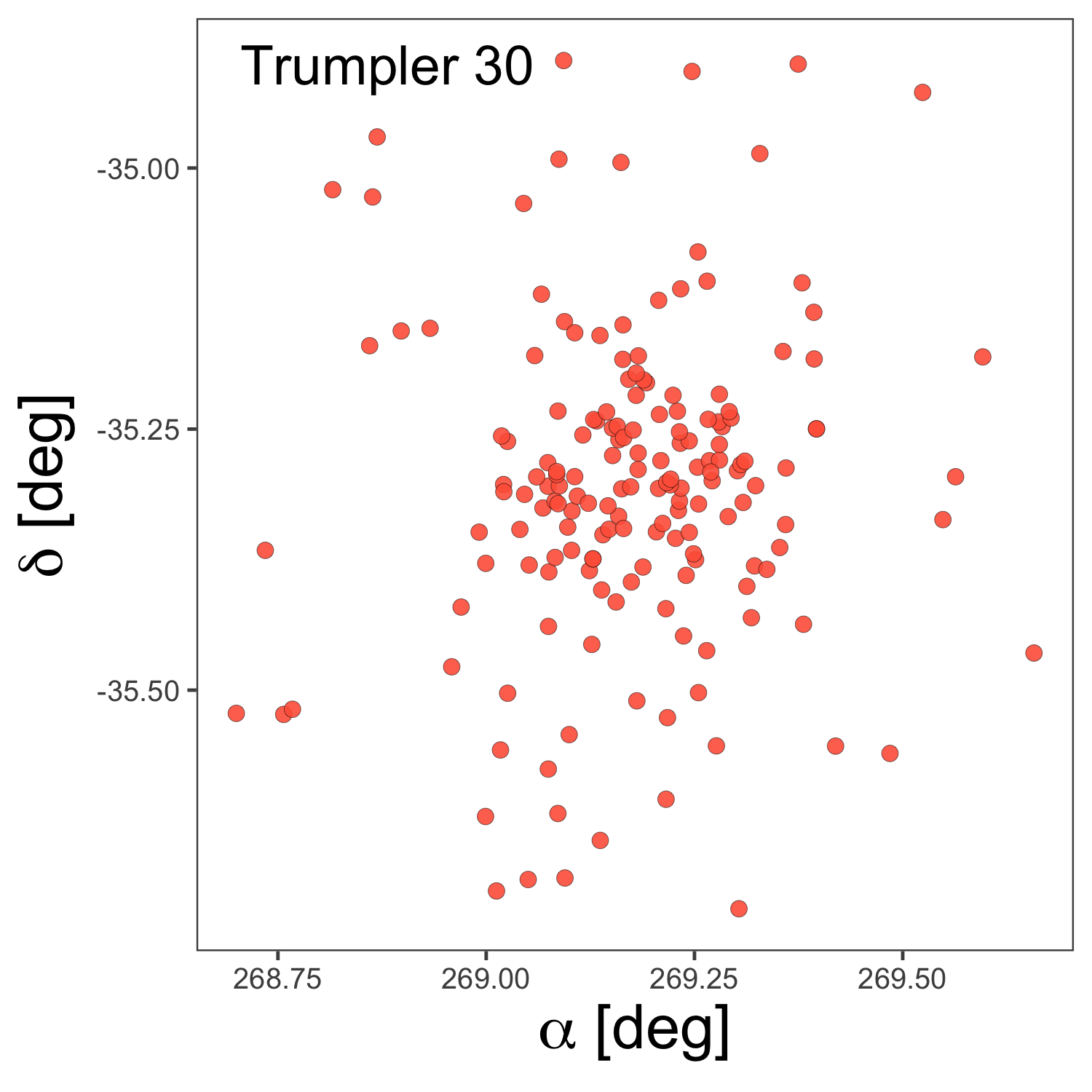}
\includegraphics[scale=0.075]{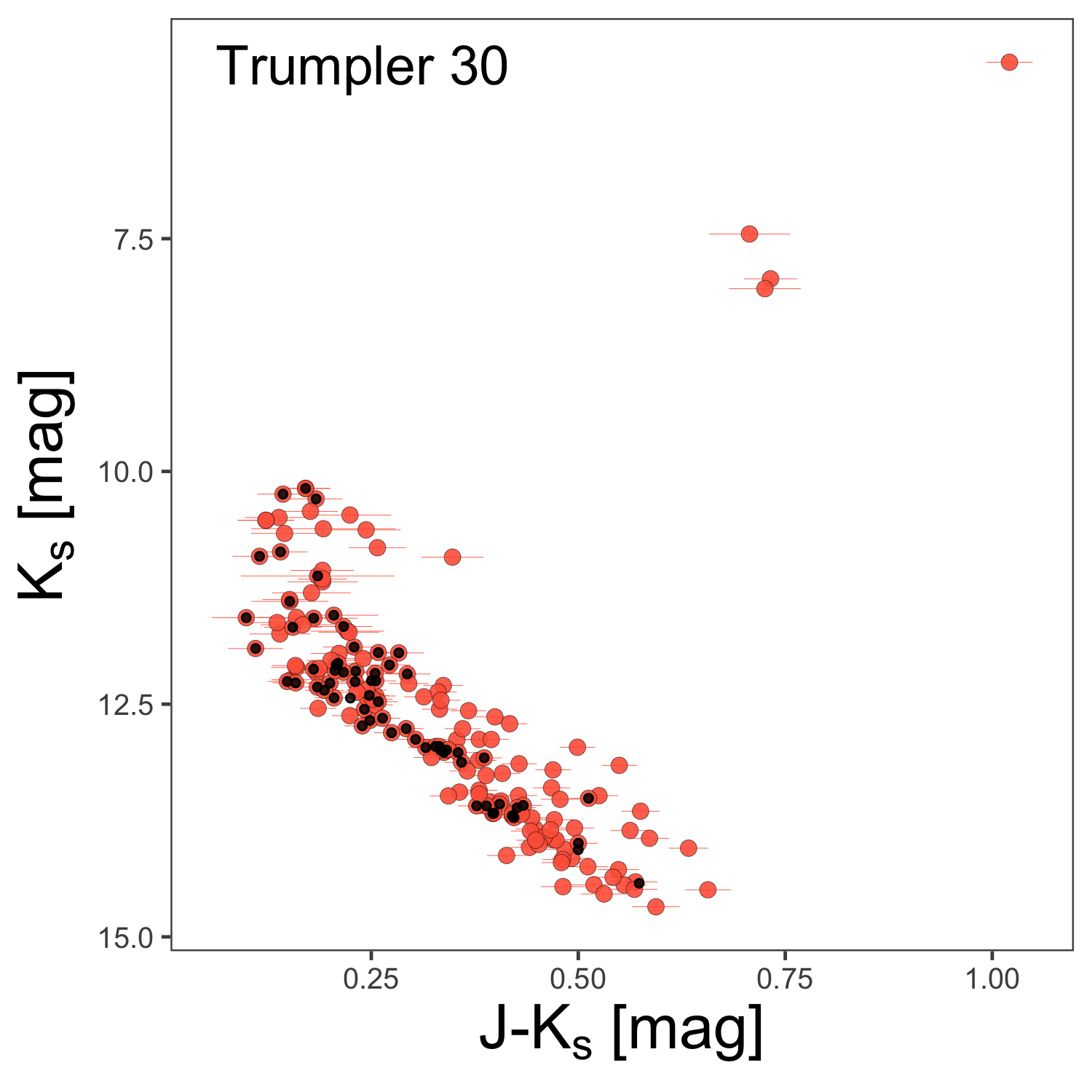}
\includegraphics[scale=0.075]{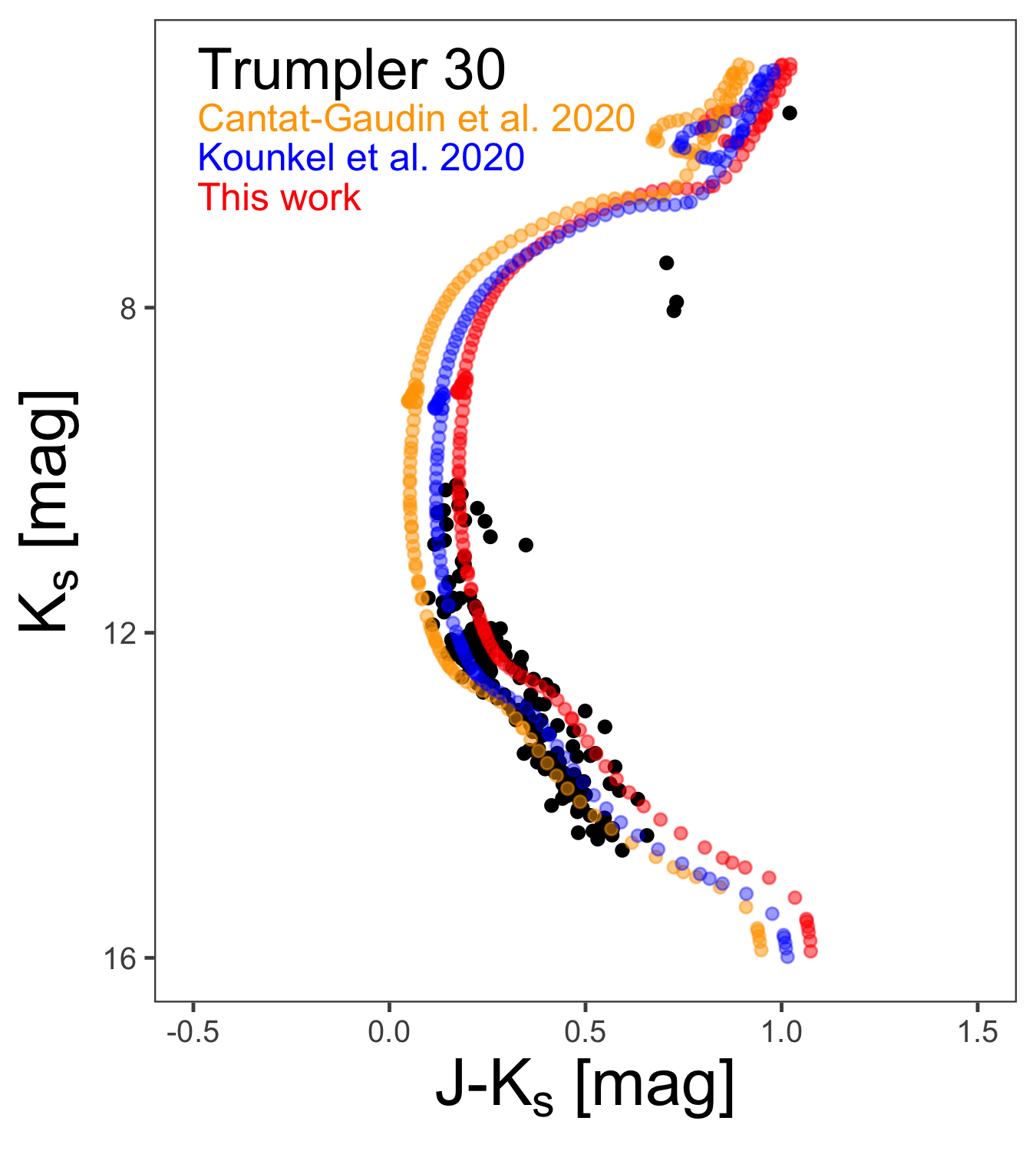}
\includegraphics[scale=0.075]{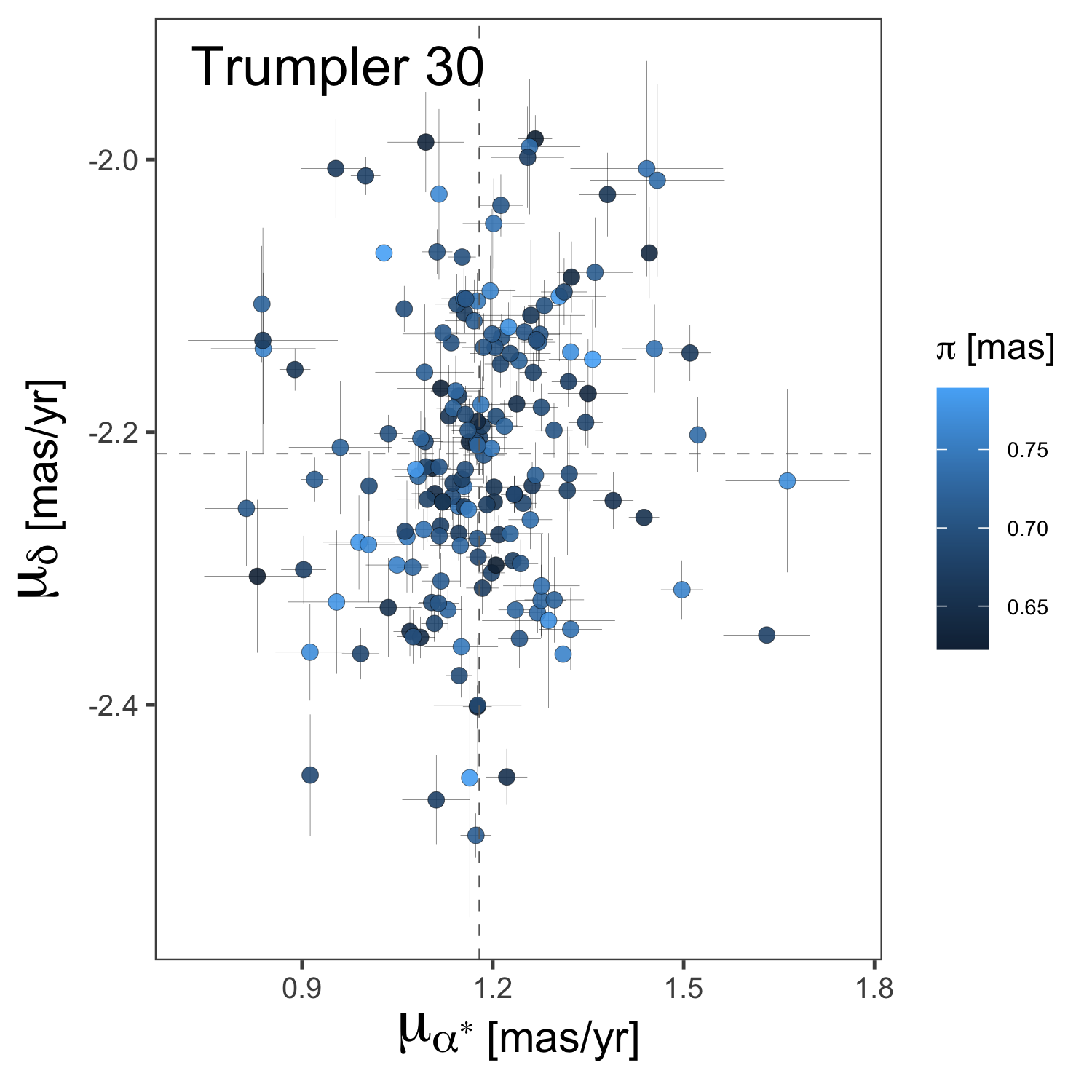}
\includegraphics[scale=0.075]{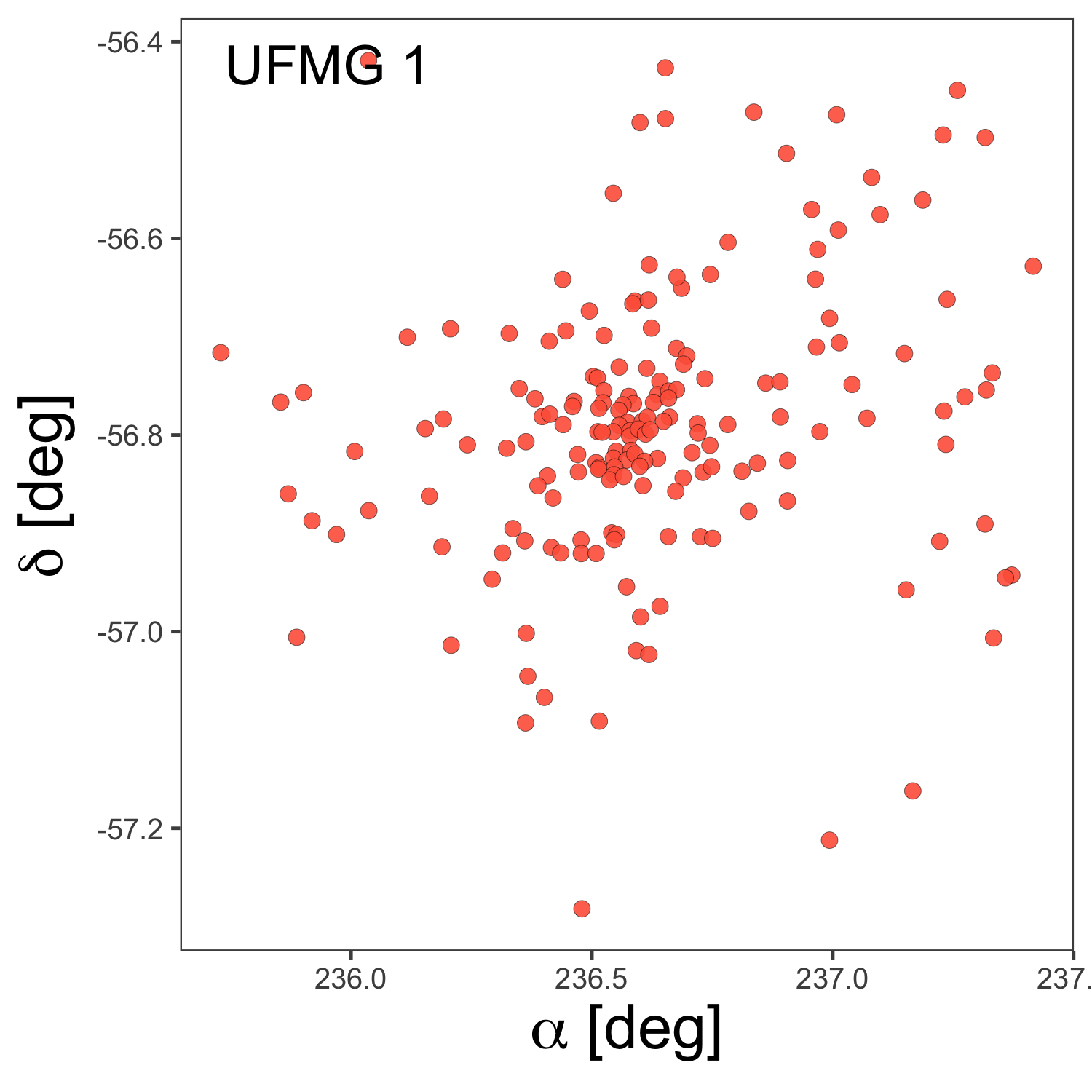}
\includegraphics[scale=0.075]{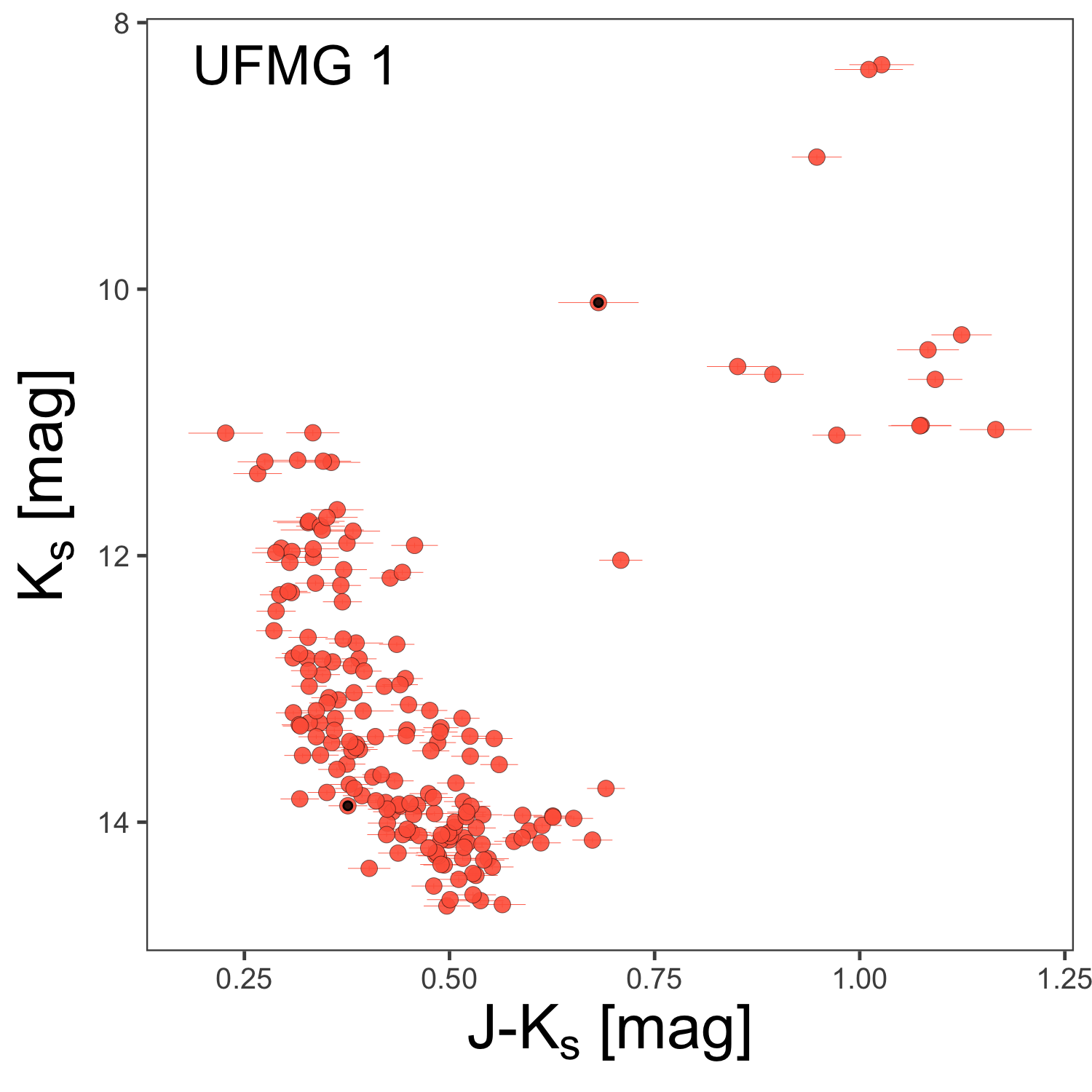}
\includegraphics[scale=0.075]{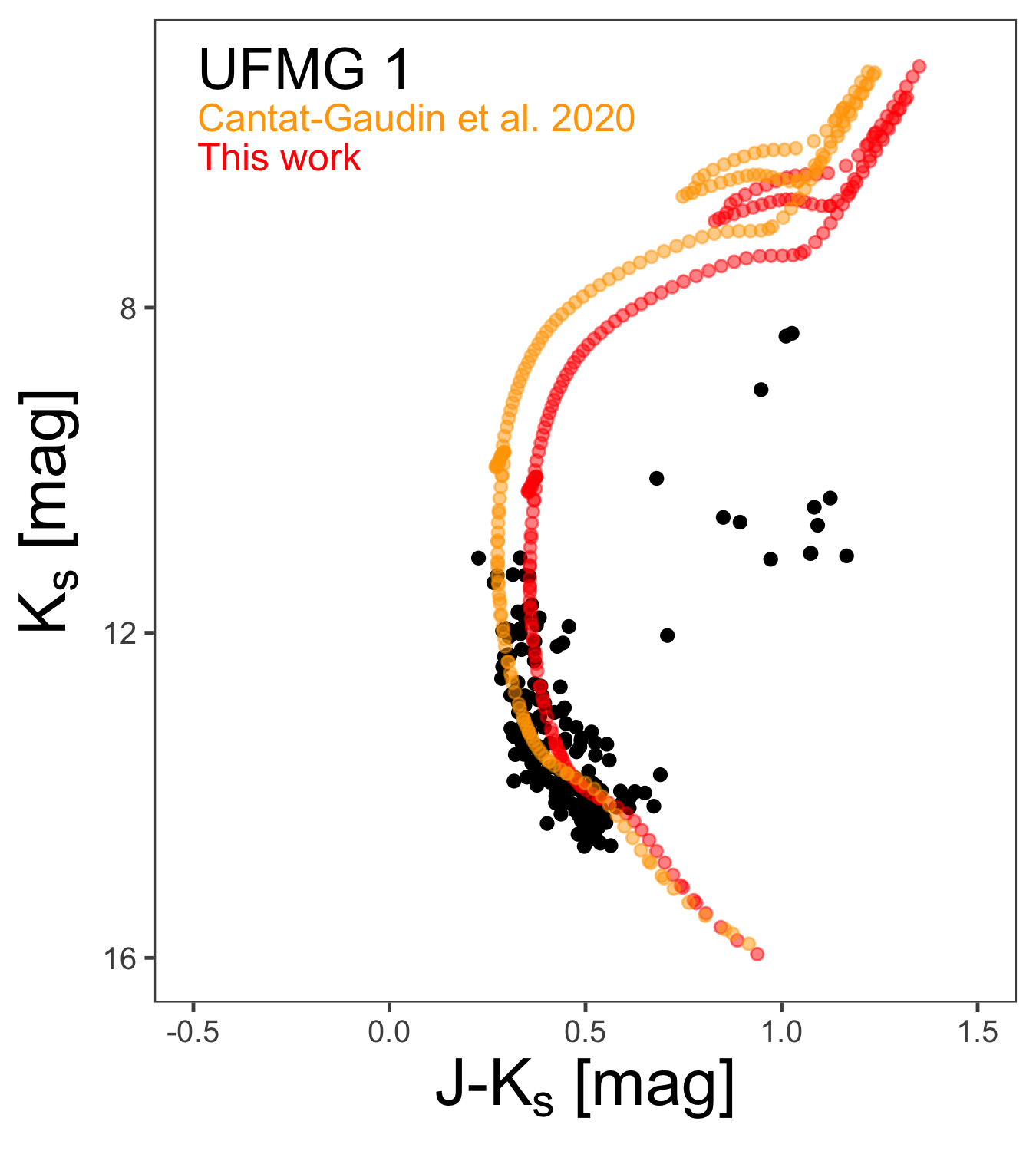}
\includegraphics[scale=0.075]{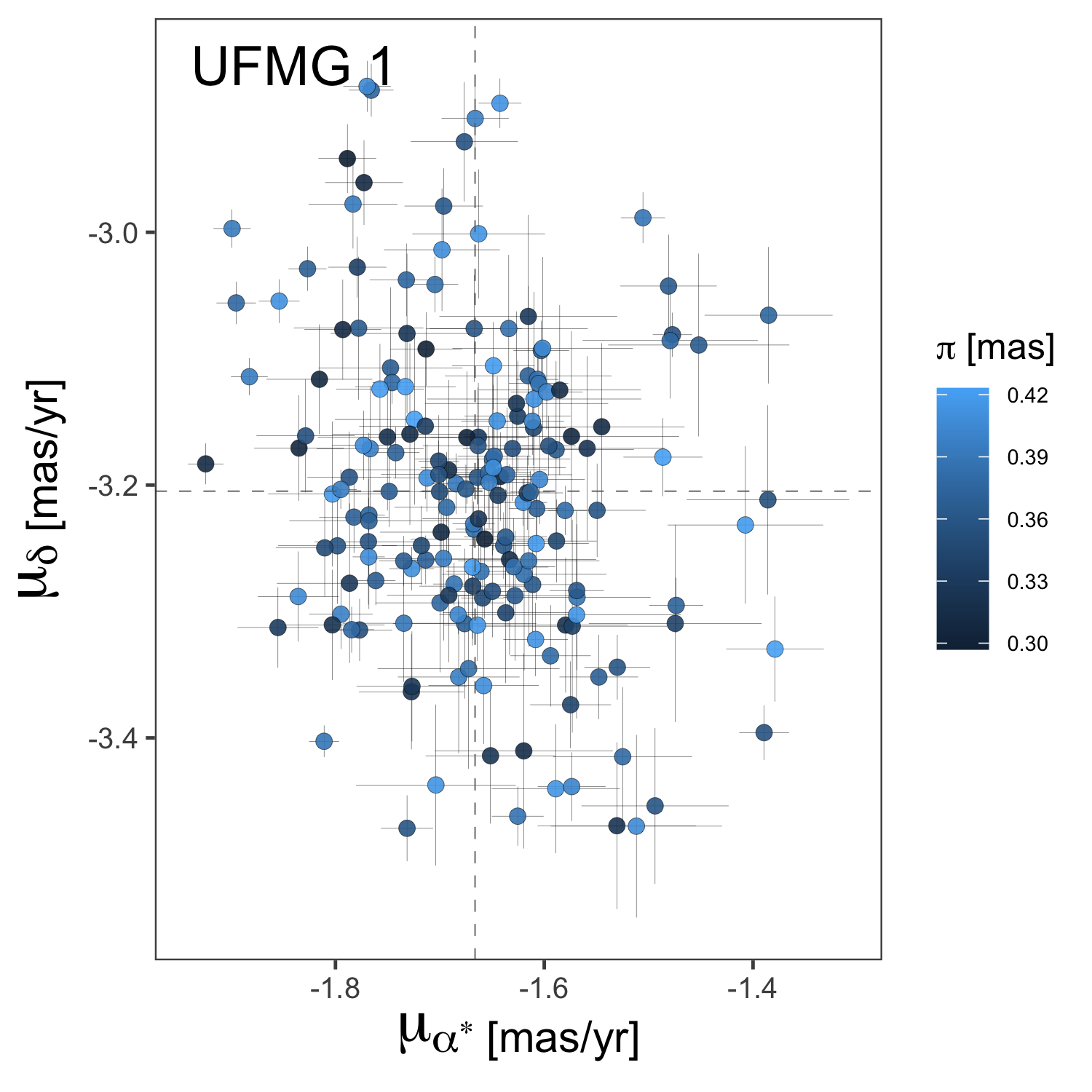}
\caption{Same as Figure\,\ref{fig:clusters}}
\end{figure*}

\begin{figure*}
\includegraphics[scale=0.075]{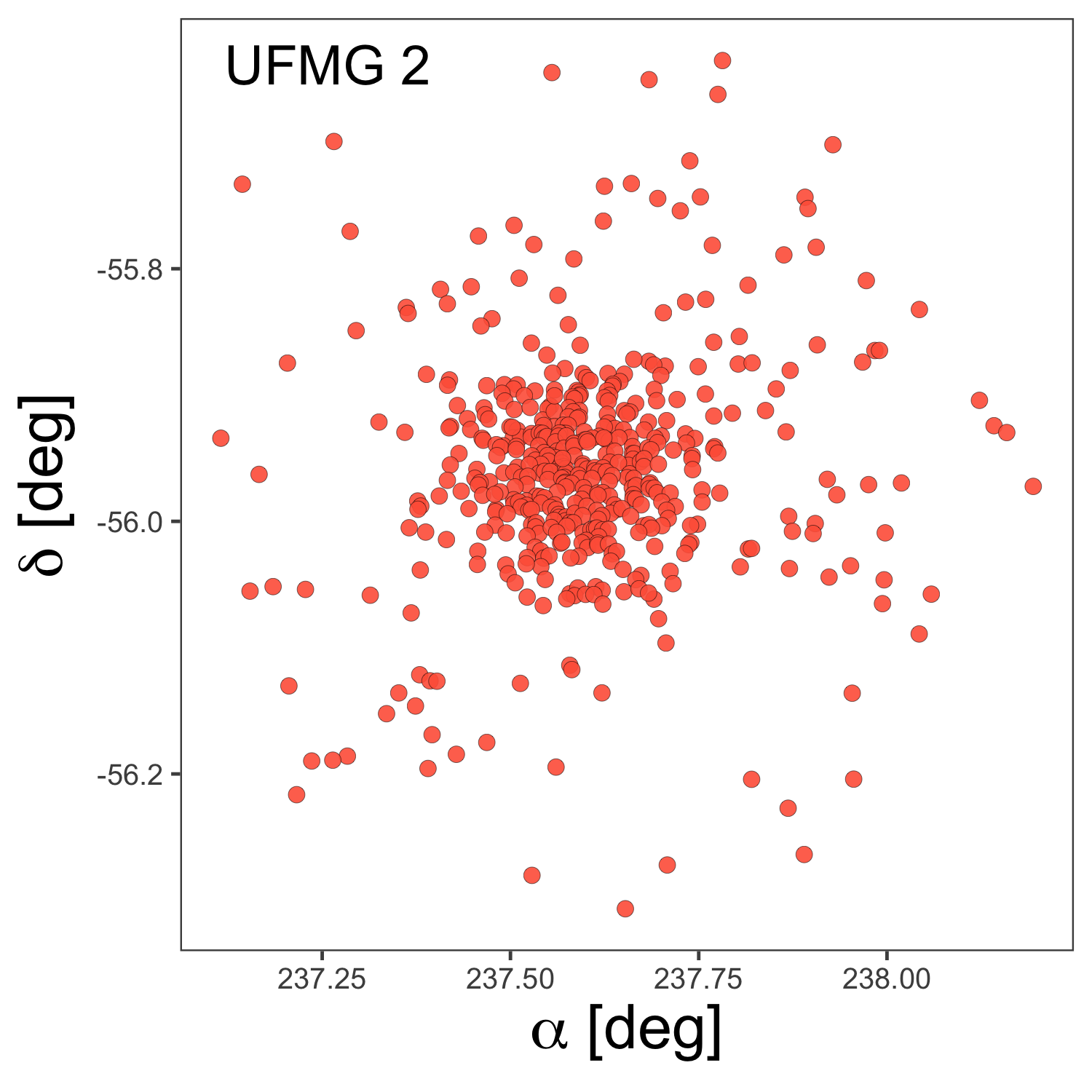}
\includegraphics[scale=0.075]{UFMG_2pub_cmd.png}
\includegraphics[scale=0.075]{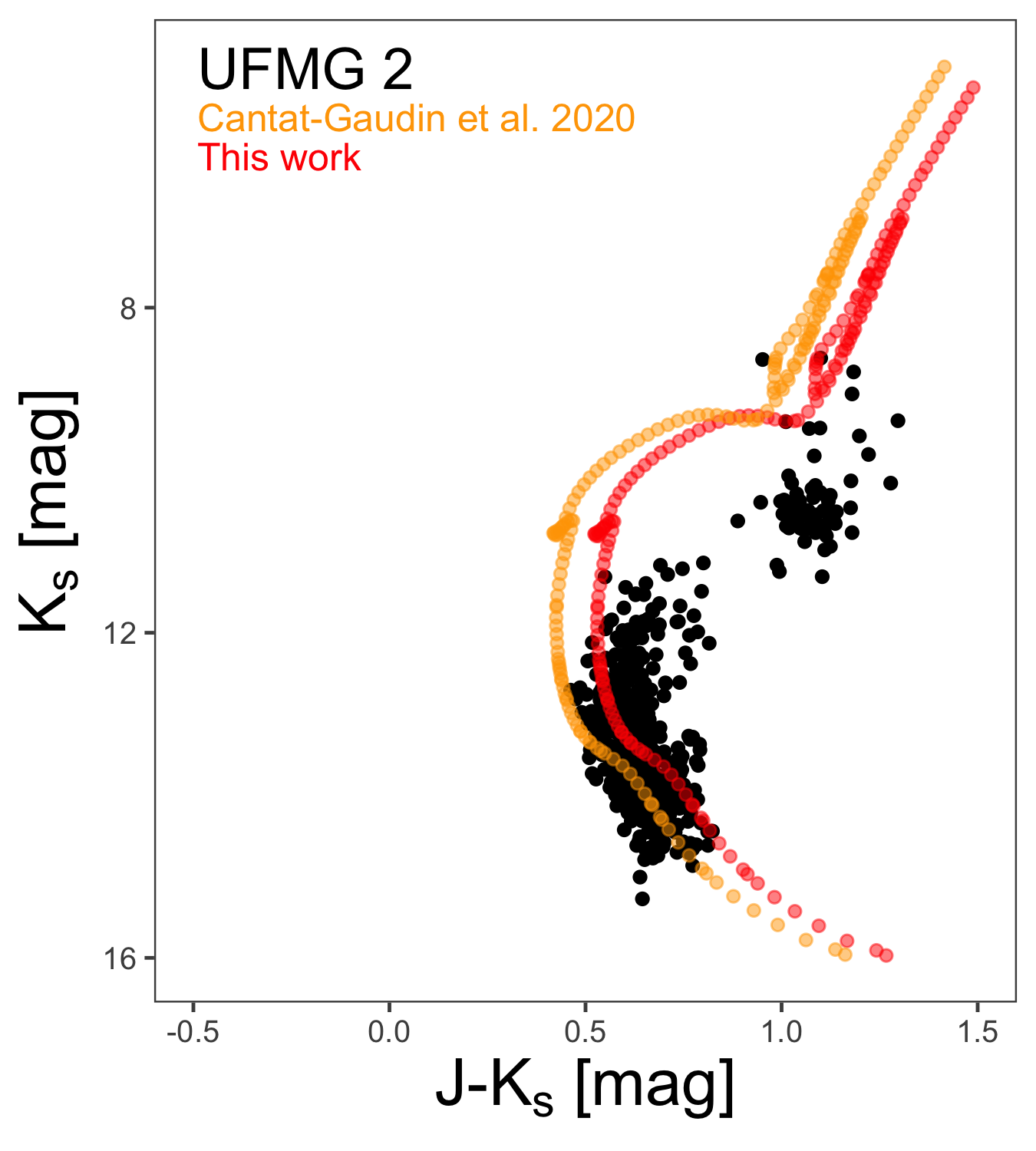}
\includegraphics[scale=0.075]{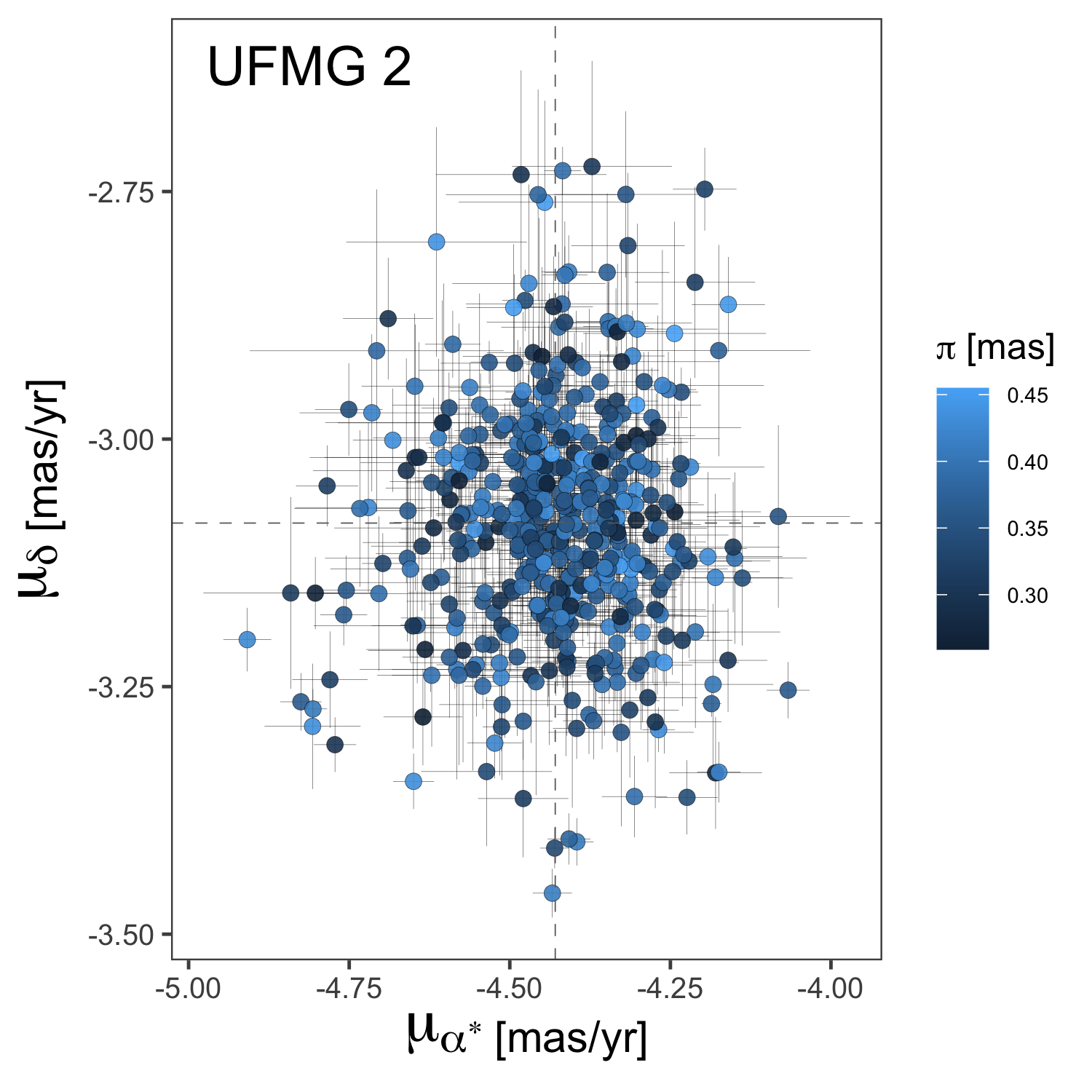}
\includegraphics[scale=0.075]{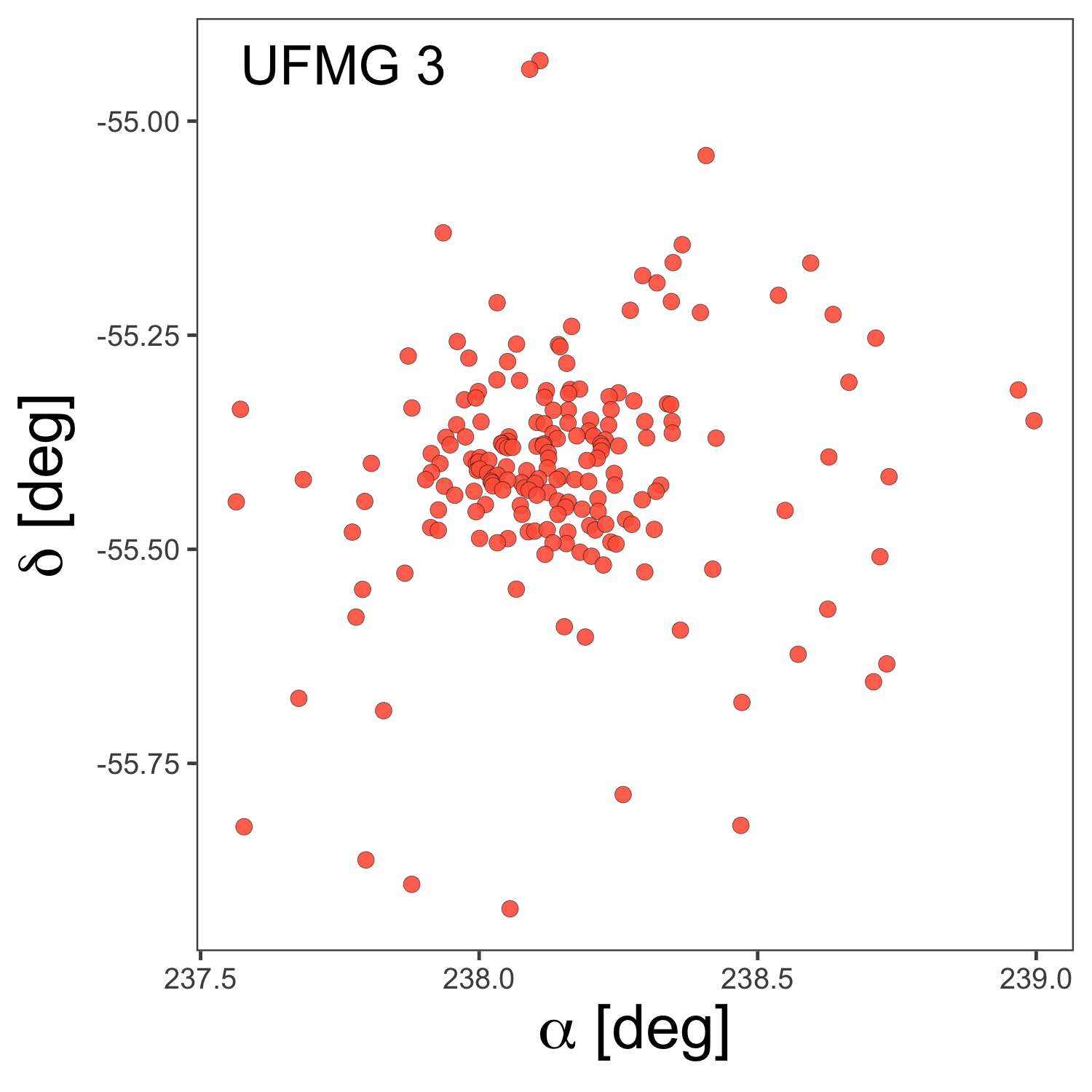}
\includegraphics[scale=0.075]{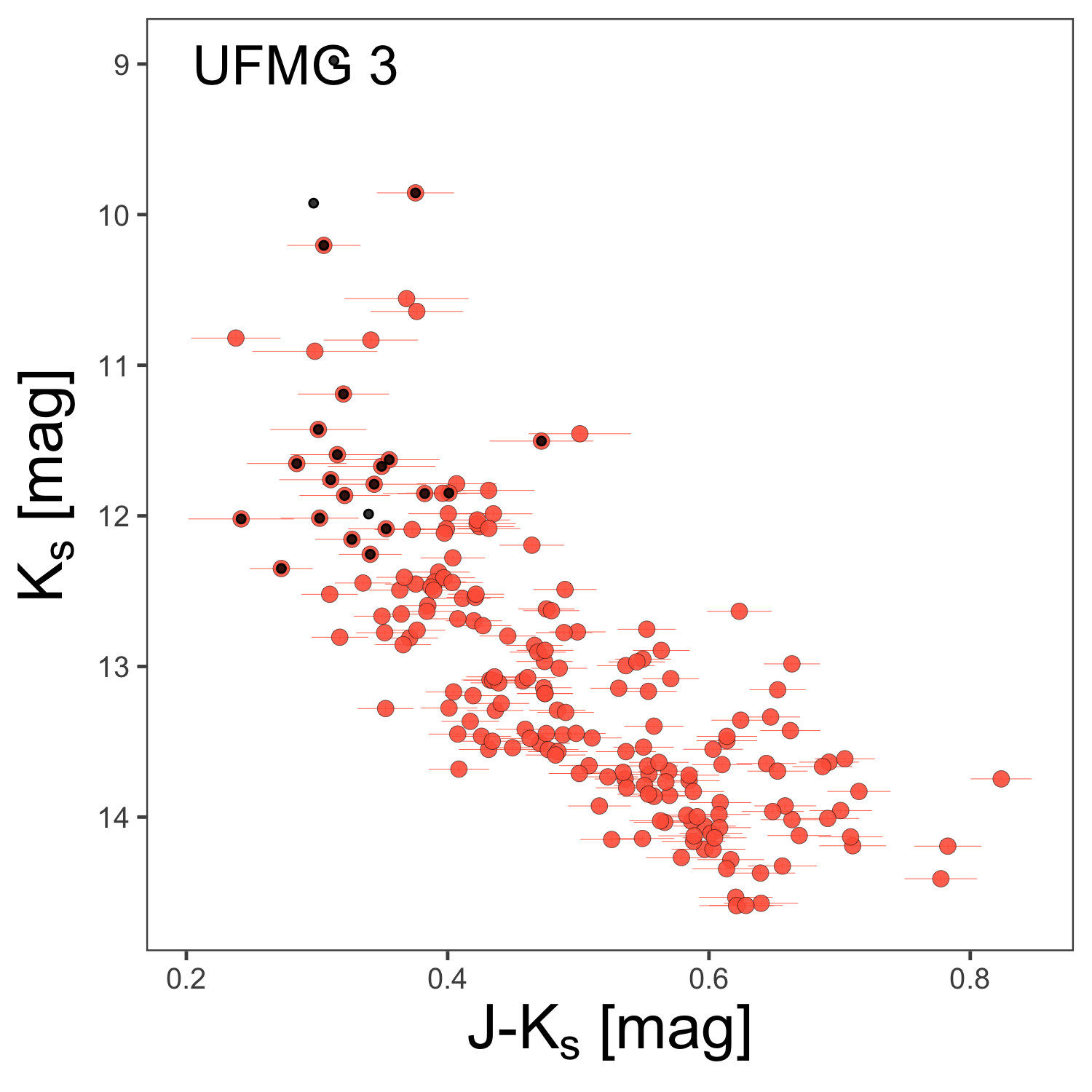}
\includegraphics[scale=0.075]{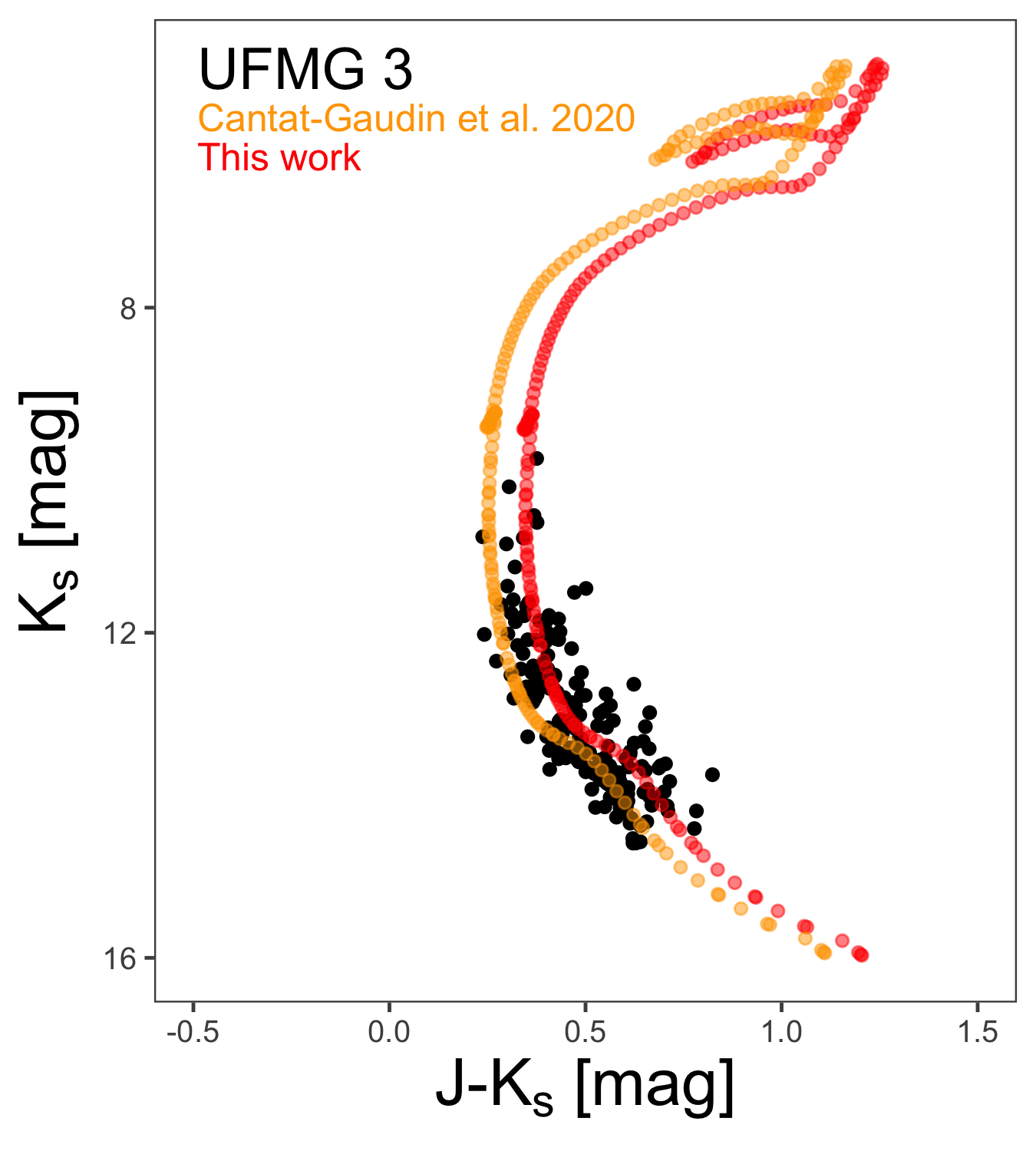}
\includegraphics[scale=0.075]{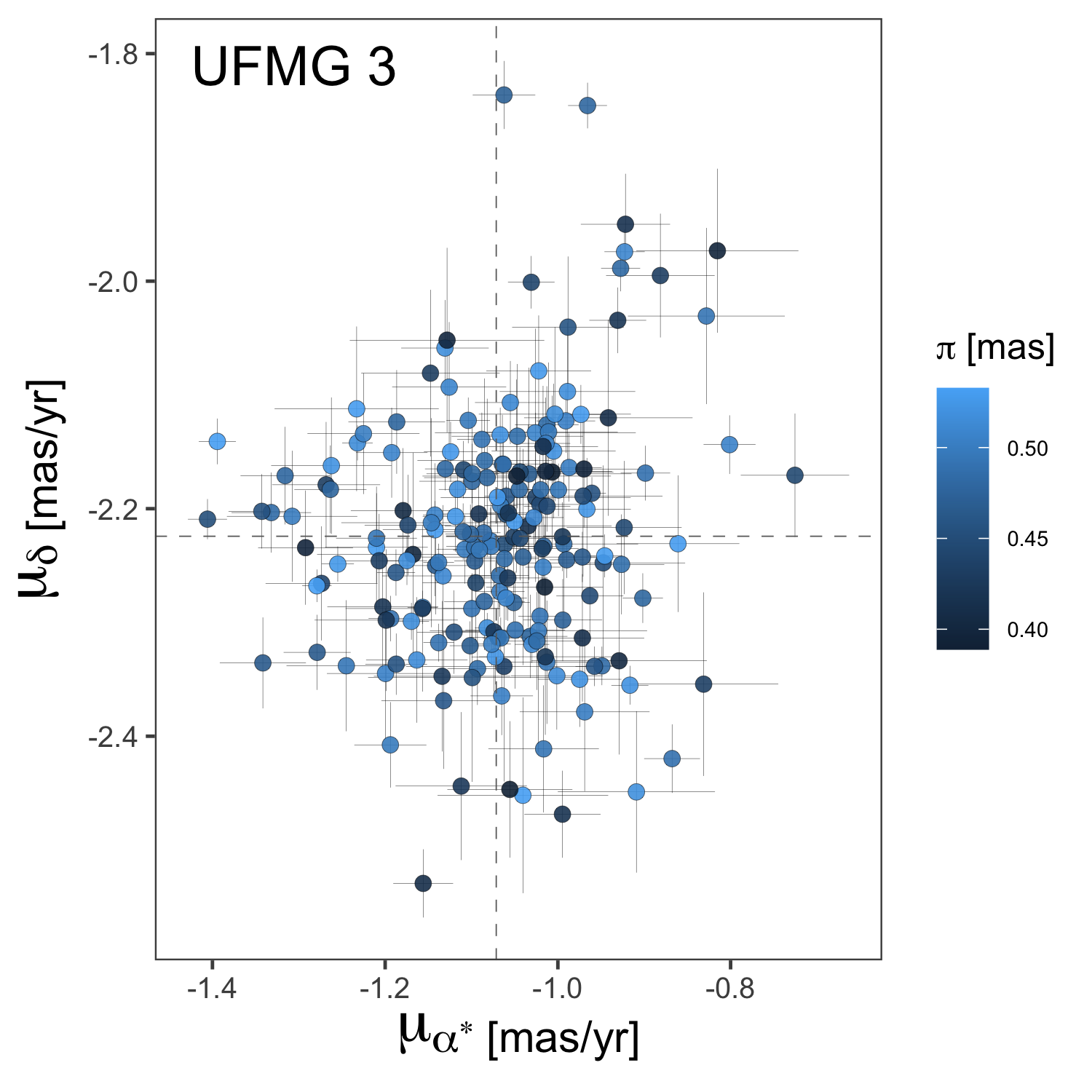}
\caption{Same as Figure\,\ref{fig:clusters}}
\includegraphics[scale=0.07]{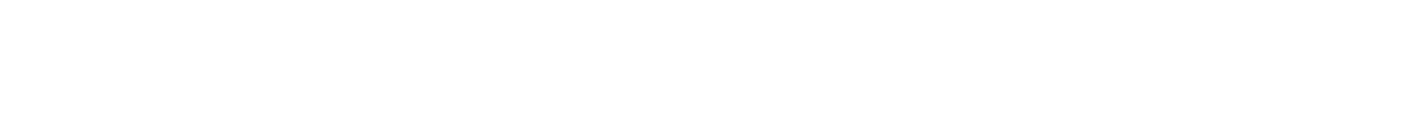}
\end{figure*}

\label{lastpage}
\end{document}